\documentclass[12pt,letterpaper,titlepage,final]{report}
\pdfoutput=1
\usepackage{graphicx}
\usepackage[utf8]{inputenc}
\usepackage{amsmath}
\usepackage{amsfonts}
\usepackage{amssymb}
\usepackage{setspace}
\usepackage{breqn}
\usepackage{cite}
\usepackage[numbers]{natbib}
\usepackage{hyperref}
\usepackage[notcite,notref]{showkeys}
\usepackage[nottoc,notlof,notlot]{tocbibind}
\usepackage{placeins}
\usepackage{notoccite}
\usepackage[toc,page]{appendix}

\usepackage[margin=1.0in]{geometry}

\newcommand{\vc}{\boldsymbol} 
\newcommand{\la}{\left\langle} 
\newcommand{\ra}{\right\rangle} 
\makeatletter
\newsavebox\myboxA
\newsavebox\myboxB
\newlength\mylenA

\newcommand{\partkey}[1]{%
  \renewcommand{\part@key}{#1}%
}
\newcommand{\part@key}{???} 
\renewcommand{\thepart}{\part@key}

\newenvironment{dedication}
     {\vspace{6ex}\begin{quotation}\begin{center}\begin{em}}
     {\par\end{em}\end{center}\end{quotation}}
  
\newcommand*\overbar[2][0.75]{%
    \sbox{\myboxA}{$\m@th#2$}%
    \setbox\myboxB\null
    \ht\myboxB=\ht\myboxA%
    \dp\myboxB=\dp\myboxA%
    \wd\myboxB=#1\wd\myboxA
    \sbox\myboxB{$\m@th\overline{\copy\myboxB}$}
    \setlength\mylenA{\the\wd\myboxA}
    \addtolength\mylenA{-\the\wd\myboxB}%
    \ifdim\wd\myboxB<\wd\myboxA%
       \rlap{\hskip 0.5\mylenA\usebox\myboxB}{\usebox\myboxA}%
    \else
        \hskip -0.5\mylenA\rlap{\usebox\myboxA}{\hskip 0.5\mylenA\usebox\myboxB}%
    \fi}
\makeatother

\title{\textbf{ {\fontsize{16pt}{1em}\selectfont Turbulence, Gravity, and Multimessenger Asteroseismology
} }}
\author{\textbf{by} \bigskip \\ \textbf{John Ryan Westernacher-Schneider} \bigskip\bigskip\bigskip \\ \textbf{A Thesis} \\ \textbf{presented to} \\ \textbf{The University of Guelph} \bigskip\bigskip\bigskip \bigskip\bigskip\bigskip \\ \textbf{In partial fulfilment of requirements} \\ \textbf{for the degree of} \\ \textbf{Doctor of Philosophy} \\ \textbf{in} \\ \textbf{Physics} \bigskip\bigskip\bigskip }
\date{\textbf{Guelph, Ontario, Canada} \bigskip \\ \textbf{\copyright} \, \textbf{John Ryan Westernacher-Schneider, October, 2018}}

\begin{document}

\pagenumbering{roman}

\maketitle


\begin{center}

\Large{\textsc{Abstract}}

\thispagestyle{empty}

\bigskip\bigskip\bigskip

\Large{\textsc{Turbulence, Gravity, and Multimessenger Asteroseismology}}

\bigskip\bigskip\bigskip\bigskip

\begin{minipage}[b]{0.45\linewidth}
\normalsize{John Ryan Westernacher-Schneider, MSc \\
University of Guelph, 2018}
\end{minipage}
\hspace{0.5cm}
\begin{minipage}[b]{0.45\linewidth}
\flushright
\normalsize{Advisor: \hspace{2.1cm} \strut \\
Professor Luis Lehner}
\end{minipage}

\normalsize

\bigskip\bigskip

\end{center}

\doublespacing{{\bf Part IA:} We present numerical measurements of relativistic scaling relations in $(2+1)$-dimensional conformal fluid turbulence, which perform favourably over their non-relativistic versions. As seen with incompressible turbulence in past studies, we find that the energy spectrum exhibits $k^{-2}$ scaling rather than the Kolmogorov/Kraichnan expectation of $k^{-5/3}$.

{\bf Part IB:} We compute the fractal dimension $D$ of a turbulent anti-deSitter black brane reconstructed from boundary fluid data using the fluid-gravity duality. Our value of $D=2.584(1)$ is consistent with the upper bound $D\leq 3$, resolving a recent claim that $D=3+1/3$. We describe how to covariantly define the fractal dimension of spatial sections of the horizon, and we speculate on assigning a `bootstrapped' value to the entire horizon.

{\bf Part II:} We report progress implementing a fluid code with post-Newtonian (PN) gravity in spherical symmetry. The PN formalism couples a fluid, its self-gravity, and a black hole via elliptic equations. This eliminates radiative modes, allowing larger time steps, which is helpful for studying systems with very long time scales, eg. tidal disruption events.

{\bf Part III:} Asteroseismology of rotating core-collapse supernovae is possible with a multimessenger strategy. We show an $l=2$, $m=0$, $n\gtrsim 2$, $f\lesssim 280$ Hz mode of the core is responsible for emission in gravitational waves and neutrinos. The angular harmonics of the neutrino emission is consistent with the mode energy around the neutrinospheres, where $r\sim 70$ km. Thus, neutrinos carry information about the mode in the outer region of the core, whereas gravitational waves probe the deep inner core $r \lesssim 30$ km.
}
\setcounter{page}{2}
\addcontentsline{toc}{section}{Abstract}

\newpage
\clearpage
\begin{dedication}
    \thispagestyle{plain}
    \vspace*{\fill}
    To my mother, whose patience, strength, and love knows no bounds;

and my father, who is sorely missed.
    \vspace*{\fill}
\addcontentsline{toc}{section}{Dedication}
\setcounter{page}{3}
\end{dedication}
\clearpage




\newpage
\singlespacing
\begin{center}

\Large{\textsc{Acknowledgments}}
\addcontentsline{toc}{section}{Acknowledgements}
\setcounter{page}{4}

\end{center}

\normalsize

\doublespacing{It is a pleasure to thank my advisor, Luis Lehner, for his guidance and support in the completion of this thesis. I would also like to thank the remainder of my committee members Eric Poisson, Erik Schnetter, and Alexandros Gezerlis for their support.

Also, thank you to the following: Michael Waite for detailed discussions on dealiasing and other matters, Guido Boffetta for sharing unpublished statistical values from the study~\cite{Boffetta:2000}, Gregory Falkovich for an interesting discussion on turbulence, Neil Cornish, Zachary Vernon, Robie Hennigar, and Eric Poisson for helpful discussions pertaining to my fractal dimension work. 

I am especially indebted to my collaborator Evan O'Connor who performed all of the core-collapse supernova simulations in Part~\eqref{part:IV} of this dissertation. That work was initiated in the summer of 2017 at the Niels Bohr Institute, during the Kavli Summer Program in Astrophysics; I would like to thank the hosts for their hospitality, and the organizers for running a great program.


}

\singlespacing

\tableofcontents

\listoftables

\listoffigures

\newpage

\pagenumbering{arabic}

\singlespacing

\begin{table} \footnotesize
\centering
\begin{tabular}{|cllc|}
\hline
& & & \\
& \multicolumn{2}{c}{\textsc{Notation Key}} & \\
& & & \\
& \textbf{Symbol} & \textbf{Description} & \\
& & & \\
& $d$ & Number of spatial dimensions & \\
& $a,b,c,...$ (indices) & Spacetime indices $0\ldots d$ & \\
& $i,j,k,...$ (indices) & Spatial indices $1\ldots d$ & \\
& $\vc{u}$, $u^i$, $v^i$ & Fluid 3-velocity & \\
& $D$ & Fractal dimension, conserved relativistic rest mass density & \\
& $u^a$ & Fluid 4-velocity & \\
& $\nu$, $\alpha$, $\beta$ & Kinematic viscosity, friction coefficient, forcing strength, respectively (Part~\eqref{part:I}) & \\
& $\psi$ & Stream function (Part~\eqref{part:I}), post-Newtonian potential (Part~\eqref{part:PN}) & \\
& $\rho_0$, $\epsilon$, $p$ & Rest mass density ($\rho$ in Part~\eqref{part:IV}), specific internal energy, pressure, respectively & \\
& $\rho$ & Total energy density (Parts~\eqref{part:I} \&~\eqref{sec:fracdim}), rest mass density (Part~\eqref{part:IV}) & \\
& $\omega$, $\vc{\omega}$ & Vorticity in 2D, 3D, respectively& \\
& $\vc{f}$, $f$ & External force 3-vector & \\
& $f^a$, $F^a$ & External force 4-vector, friction 4-vector, respectively & \\
& $E$, $P$ & Average specific energy, palinstrophy, respectively & \\
& $E(k)$ & Isotropic specific energy spectrum & \\
& $\Omega$ & Average specific enstrophy& \\
& $\left\langle.\right\rangle$ & Spatial average, ensemble average & \\
& $\epsilon_{\nu}$, $\eta_{\nu}$ & Average specific energy/enstrophy dissipated by viscosity, respectively & \\
& $L_f$, $L_{\alpha}$, $L_{\nu}$ & Forcing, friction, and viscous length scales, respectively & \\
& $k$ & Magnitude of wavevector & \\
& $\epsilon$, $\eta$ & Isotropic specific spectral energy/enstrophy flux, respectively (Parts~\eqref{part:I} \&~\eqref{sec:fracdim}) & \\
& $\delta v_{\parallel}$, $\delta v_{\perp}$ & Velocity difference in directions parallel and perpendicular to $\vc{r}$ & \\
& $S_n$ & Velocity structure function of order $n$ & \\
& $\eta_{ab}$, $g_{ab}$ & Minkowski metric, general spacetime metric, respectively & \\
& $\gamma_{ij}$ & Intrinsic metric of spacelike hypersurfaces & \\
& $g$ & Determinant of general spacetime metric & \\
& $T_{ab}$ & Energy-momentum tensor& \\
& $\delta^a_b$ & Kronecker delta & \\
& $c, c_s$ & Speed of light and speed of sound, respectively & \\
& $\gamma$ & Lorentz gamma factor & \\
& $\vc{r}$, $r_i$ & Separation 3-vector & \\
& $S_i$, $\mathcal{E}$, $\tau$, & Conservative fluid variables (in addition to $D$) & \\
& $t_{LC}$, $L_{\text{box}}$ & Light-crossing time and linear box size, respectively & \\
& $T,L$ (indices) & transverse, longitudinal directions, respectively & \\
& $L_\nu$, $R_\nu$ & Neutrino luminosity, neutrinosphere radius & \\
& $\Omega$ & Pre-collapse central rotation rate & \\
& $\nu_e$, $\bar{\nu}_e$, $\nu_x$ & Electron, anti-electron, and heavy lepton neutrinos, respectively & \\
& $\phi$, $\omega^i$, $\psi$, $\chi_{ij}$ & Post-Newtonian metric potentials (Part~\eqref{part:PN}) & \\
& $F_\phi$, $f_\omega$ & Auxiliary post-Newtonian variables (Part~\eqref{part:PN}) & \\
& $Y_{lm}$ & Spherical harmonic & \\
& $\eta_r$, $\eta_\theta$, $\eta_\phi$ & Components of infinitesimal Eulerian displacement & \\
& $\Delta$ & Measure of eigenfunction match & \\
& $\sigma$ & Mode frequency & \\
& $F$, $H_1$, $H_2$... & Fundamental mode and its overtones & \\
& ${}^l f$, ${}^l p_1$, ${}^l p_2$... & Fundamental $l$ mode and its overtones & \\
& & & \\ \hline
\end{tabular}
\caption[Notation key]{A summary of the some of the notation used in this dissertation.}
\label{notationkey}
\end{table}

%
%
\partkey{IA}
\part{Relativistic turbulence in $2+1$ dimensions} \label{part:I}

\section*{Executive summary}

Scaling relations have been a central pillar of the statistical understanding of turbulence for many decades. They have found uses in astrophysics for example via size-linewidth relations, where the Doppler broadening of spectral lines from a turbulent cloud of matter is related to its size via a turbulent scaling relation. Extending these relations to the relativistic case is a logical next step, with recent progress being made in this direction. In this part, we present measurements of relativistic scaling relations in $(2+1)$-dimensional conformal fluid turbulence from direct numerical simulations, in the weakly compressible regime. We analytically derived these scaling relations previously in~\cite{WS2015} for a relativistic fluid; this work is a continuation of that study. We first explicitly demonstrate that the non-relativistic limit of these scaling relations reduce to known results from the statistical theory of incompressible Navier-Stokes turbulence. In simulations of the inverse-cascade range, we find the relevant relativistic scaling relation is satisfied to a high degree of accuracy. We observe that the non-relativistic versions of this scaling relation underperform the relativistic one in both an absolute and relative sense, with a progressive degradation as the rms Mach number increases from $0.14$ to $0.19$. In the direct-cascade range, the two relevant relativistic scaling relations are satisfied with a lower degree of accuracy in a simulation with rms Mach number $0.11$. We investigate the poorer agreement with further simulations of an incompressible Navier-Stokes fluid. Finally, as has been observed in the incompressible Navier-Stokes case, we show that the energy spectrum in the inverse-cascade of the conformal fluid exhibits $k^{-2}$ scaling rather than the Kolmogorov/Kraichnan expectation of $k^{-5/3}$, and that it is not necessarily associated with compressive effects.

\begin{figure}[h!]
\centering
\hbox{\hspace{1cm}\includegraphics[width=0.85\textwidth]{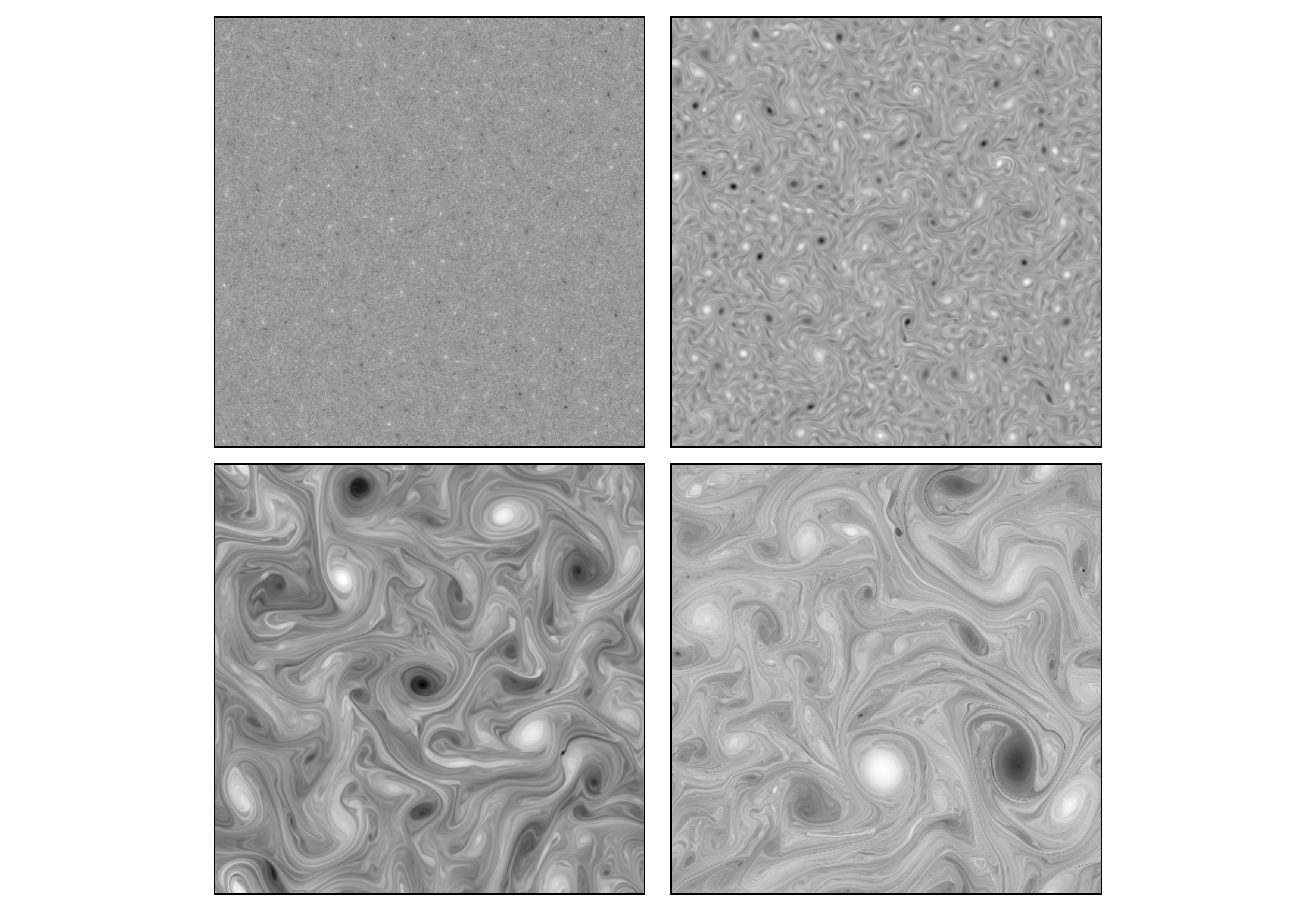}}
\raggedright
Figure IA: Snapshots of vorticity from our simulations. See the text at the end of Part~\eqref{part:I} for details.
\label{fig:allvorts}
\end{figure}

\chapter{Introduction}\label{sec:intro}
Relativistic hydrodynamics has become a subject of increased interest
in recent years. Beyond its relevance in astrophysical scenarios 
(e.g.~\cite{rezzolla2013relativistic,Gammie:2003rj,Shibata:2005gp,Baumgarte:2010:NRS:2019374,Lehner:2014asa}), 
it has become relevant to the description of quark-gluon plasmas (e.g.~\cite{Luzum:2008cw,vanderSchee:2013pia}) and, 
through the fluid-gravity correspondence, it has found its way into the realm of fundamental gravity 
research~\cite{Baier:2007ix,Bhattacharyya:2008jc}. Intriguingly, this correspondence
has revealed that gravity can exhibit turbulent behavior, and studies of its possible consequences are gaining
interesting momentum~\cite{Eling:2010vr,Carrasco:2012nf,Chesler:2013lia, Galtier:2017mve,FIXINGTHEORIES}.
The understanding of turbulence in any regime is a difficult task given its intrinsic complexity, and despite
a long history of efforts in the subject, our knowledge of this rich phenomenon is still incomplete. Important headways into
this subject have been made thanks to statistical analysis complemented with 
numerical simulations (e.g.~\cite{Benzi,Frisch:1996,Dobler:2003,Boffetta:2012,cardy2008non}).
Of particular interest is to understand the possible onset of turbulence, and especially to derive scaling relations in fully-developed scenarios, since those cases are amenable to a statistical description from which
generic statements can be drawn and then tested numerically and observationally.

To date however, only limited attention has been placed on the relativistic turbulent regime, and most of what is
known restricts to the behavior of turbulent, incompressible flows  in the non-relativistic regime.
There has been some work on the analytical front~\cite{Fouxon:2009rd,WS2015,Drivas2017} and several numerical 
investigations~\cite{Carrasco:2012nf,Radice:2012pq,Green:2013zba,WS2015, Chandra:2015iza,MacFadyen2012,MacFadyen2013,East2016,Zhdankin2017, Uzdensky2017}. Because correlation functions can indeed be measured in relevant 
scenarios -- perhaps even in  
quark-gluon plasma\cite{Romatschke:2007mq,Luzum:2008cw,Heinz:2011kt,vanderSchee:2013pia,Fukushima:2016xgg} --  and interesting implications for the gravitational field follow from holography, it is of interest to further investigate 
relativistic turbulence. 

\emph{In the current work we measure scaling relations in $(2+1)$-dimensional
relativistic conformal fluids in the weakly-compressible turbulent regime and compare them to the predictions in~\cite{WS2015} and various limits thereof. The $(2+1)$-dimensional
case is especially relevant to draw intuition for related phenomena in $(3+1)$-dimensional gravity with the help of the fluid-gravity correspondence (e.g.~\cite{Eling:2010vr,Green:2013zba,FIXINGTHEORIES})}. This work is largely a continuation and completion of~\cite{WS2015}, which contained a numerical study which was inconclusive at the time.

This part is organized as follows. Chapter~\eqref{sec:P1background} provides some background material, discussing both
the inverse- and direct-cascade ranges that could ensue in fully-developed turbulence and, in particular,
relevant scaling relations which we have measured. In Chapter~\eqref{sec:P1implementation} we provide details of our numerical implementations. This includes the use of a random external force to generate the turbulent flow, as well as considerations specific to simulating either a conformal fluid or an incompressible Navier-Stokes fluid. The equivalence between previously known results and the incompressible limit of the scaling relations derived in~\cite{WS2015} is explicitly demonstrated. We give our results in Chapter~\eqref{sec:P1results}, where our numerical measurements of the scaling relations derived in~\cite{WS2015} are presented. In Chapter~\eqref{sec:P1discussion}, we provide
additional discussion and ancillary numerical results, including the demonstration of a $k^{-2}$ energy spectrum in the inverse-cascade of a turbulent conformal fluid which is not due to compressive effects.

In this part, angle brackets $\la . \ra$ will refer to ensemble averages. Letters at the beginning of the alphabet $(a,b,c,\ldots)$ will represent spacetime indices $(0,1,2)$, while letters in the middle of the alphabet $(i,j,k,\ldots)$ will represent spatial indices $(1,2)$. We follow Einstein summation convention. In the context of correlation functions, which often depend on two points $\vc{r}_2$ and $\vc{r}_1$, we define $\vc{r} = \vc{r}_2-\vc{r}_1$. To avoid cumbersome notation, we denote quantities evaluated at $\vc{r}_2$ with a prime (eg. $T_{ij}^\prime$) and quantities evaluated at $\vc{r}_1$ without one (eg. $T_{ij}$). The metric signature is $(-,+,+)$ for our $(2+1)$-dimensional setup. 

%
%
\chapter{Background}\label{sec:P1background}
We will make extensive connections with the work presented in~\cite{WS2015}, where
specific scaling relations were derived analytically for $(2+1)$-dimensional relativistic hydrodynamic turbulence. 
We will compare these relations with suitable limits in order to make contact with previously known results, as well as to gauge the importance of relativistic vs compressible contributions in our simulations of the specific case of a conformal fluid.

\section{Incompressible non-relativistic limit of the scaling relations}
In this section we explicitly demonstrate that the incompressible Navier-Stokes limit of the relativistic scaling relations 
presented in~\cite{WS2015} can be written in terms of known results. We will use the particular form of a barotropic perfect fluid stress-energy tensor with equation of state $P=\rho/w$, where $P$ is the pressure, $\rho$ is the energy density, and $w$ is the equation of state parameter. In doing so, we obtain incompressible counterparts to the relativistic scaling relations we measure in simulations, which act as a point of reference against which to gauge the relative performance of the relations derived in~\cite{WS2015}.

\subsection{Inverse-cascade range} \label{sec:inverseincompressiblelimit}

The first scaling relation, which is valid in the inverse-cascade range, reads~\cite{WS2015}
\begin{eqnarray}
\la T^\prime_{0i} T^{i}_{j} \ra = \frac{1}{2} \epsilon r_j,\label{eq:inversecascade_WS}
\end{eqnarray}
where $\epsilon = \partial_0 \la T_{0i}T_{0}^{i}/2 \ra$. For a $P=\rho/w$ perfect fluid, where $T^{ab} = ((1+w)/w)\rho u^a u^b + (\rho/w) \eta^{ab}$, the scaling relation expands to
\begin{eqnarray}
\la \frac{1+w}{w} \rho^\prime \gamma^{\prime 2} v^\prime_i \left( \frac{1+w}{w}\rho \gamma^2 v^i v_j + \frac{\rho}{w} \delta^i_j \right) \ra = \frac{1}{2} \partial_0 \la \left(\frac{1+w}{w}\right)^2 \rho^2 \gamma^4 \frac{v_i v^i}{2} \ra r_j, \label{eq:inversecascade_WS_expanded}
\end{eqnarray}
where $\gamma$ is the Lorentz factor and $v^i$ is the spatial velocity ($u^a = (-1,v^i)$). In the extreme incompressible non-relativistic limit, $\rho \rightarrow \text{constant}$ and $\gamma \rightarrow 1$. Thus $\rho^\prime = \rho$, and Eq.~\eqref{eq:inversecascade_WS_expanded} becomes
\begin{eqnarray}
\left( \frac{1+w}{w} \right)^2 \rho^2 \la v^\prime_i v^i v_j \ra + \frac{\rho^2}{w} \la v^\prime_i \delta^i_j \ra = \left( \frac{1+w}{w} \right)^2 \rho^2 \frac{1}{2} \epsilon_{\text{NS}} r_j,
\end{eqnarray}
where we have defined $\epsilon_{\text{NS}} \equiv \partial_0 \la v_i v^i \ra/2$ as the incompressible Navier-Stokes version of $\epsilon$. Note that the second term on the left-hand side vanishes due to statistical isotropy, yielding the final result
\begin{eqnarray}
\la v^\prime_i v^i v_j \ra = \frac{1}{2}\epsilon_{NS} r_j .\label{eq:inversecasadelimit}
\end{eqnarray}
Since Eq.~\eqref{eq:inversecasadelimit} is the incompressible Navier-Stokes limit of the relativistic scaling relation in Eq.~\eqref{eq:inversecascade_WS}, they can be compared in the inverse-cascade range of relativistic or compressible turbulence in order to gauge their relative performance.

Notice one can also
arrive at Eq.~\eqref{eq:inversecasadelimit} using known results in the theory of $(2+1)$-dimensional incompressible Navier-Stokes turbulence. In the derivations presented in~\cite{Bernard:1999}, an intermediate result is displayed as
\begin{eqnarray}
\la \delta v_j \delta v_i \delta v^i \ra = 2 \epsilon_{NS} r_j, \label{eq:inversecascade_bernard}
\end{eqnarray}
valid in the inverse-cascade range. Here, $\delta$ denotes a difference, i.e. $\delta v_j \equiv v_j^\prime - v_j$. Expanding out the left-hand side of Eq.~\eqref{eq:inversecascade_bernard} and using statistical homogeneity yields 
\begin{eqnarray}
\la \delta v_j \delta v_i \delta v^i \ra = 4 \la v^\prime_i v^i v_j \ra + 2 \la v_j^\prime v^i v_i \ra.
\end{eqnarray}
By incompressibility, the second term on the right-hand side is divergence-free. Thus, assuming isotropy and regularity at $r=0$, it must vanish~\cite{Landau:1987} (this argument will be used repeatedly in Sec.~\eqref{sec:direct-cascade-limit}). Therefore, Eq.~\eqref{eq:inversecascade_bernard} becomes
\begin{eqnarray}
 \la v^\prime_i v^i v_j \ra = \frac{1}{2} \epsilon_{NS} r_j,
\end{eqnarray}
which is the incompressible non-relativistic limit we obtained in Eq.~\eqref{eq:inversecasadelimit}.

%
%

\subsection{Direct-cascade range}\label{sec:direct-cascade-limit}

The second relativistic scaling relation that we consider, which is valid instead in the direct-cascade range, reads~\cite{WS2015}
\begin{eqnarray}
\la \omega^\prime \bar{\omega}_j \ra = -\frac{1}{2}\varepsilon r_j, \label{eq:directcascade_WS}
\end{eqnarray}
where $\omega \equiv \epsilon^{ik} \partial_i T_{0k}$, $\bar{\omega}_j = \epsilon^{ik} \partial_i T_{jk}$, and $\varepsilon \equiv \la \mathcal{F} \omega \ra \equiv \la (\epsilon^{ik} \partial_i f_k) \omega \ra$. Note that $f_k$ is a random external force. In the incompressible non-relativistic limit, $\varepsilon$ becomes proportional to the \emph{enstrophy dissipation rate} $\epsilon_{\omega}$~\cite{Bernard:1999}, namely $\varepsilon \rightarrow ((1+w)/w)^2 \rho^2\epsilon_{\omega}$. Expanding the left-hand side of Eq.~\eqref{eq:directcascade_WS} and setting $\rho=\text{constant}$ and $\gamma=1$ as before, we obtain
\begin{eqnarray}
\la \omega^\prime \bar{\omega}_j \ra &=& \la \epsilon^{ik} \partial_i T_{kj} \epsilon^{mn} \partial^\prime_m T^\prime_{0n} \ra \nonumber\\
&=& \la \epsilon^{ik}\partial_i \left( \frac{1+w}{w}\rho \gamma^{ 2} v_k v_j + \frac{1}{w}\rho \delta_{kj} \right) \epsilon^{mn} \partial^\prime_m \left( \frac{1+w}{w}\rho^\prime \gamma^{\prime 2} v^\prime_n \right) \ra \nonumber\\
&=& \left(\frac{1+w}{w}\rho\right)^2 \la \epsilon^{ik}\partial_i \left( v_k v_j \right) \epsilon^{mn} \partial^\prime_m v^\prime_n \ra + \left(1+w\right) \left( \frac{\rho}{w} \right)^2 \la \delta_{kj} \epsilon^{mn} \partial_m^\prime v_n^\prime \ra.
\end{eqnarray}
The second term on the right-hand side is proportional to the average vorticity, which vanishes by parity invariance. Thus Eq.~\eqref{eq:directcascade_WS} becomes
\begin{eqnarray}
\la \epsilon^{ik}\partial_i \left( v_k v_j \right) \epsilon^{mn} \partial^\prime_m v^\prime_n \ra = -\frac{1}{2}\epsilon_\omega r_j,
\end{eqnarray}
where the non-relativistic vorticity is $\epsilon^{mn} \partial_m v_n \equiv \omega_{\text{NR}}$. The left-hand side needs to be manipulated further in order to compare with standard results (e.g.~\cite{Bernard:1999}). First, notice that the ensemble average on the left-hand side expands under the product rule to
\begin{eqnarray}
\la \epsilon^{ik}\partial_i \left( v_k v_j \right) \epsilon^{mn} \partial^\prime_m v^\prime_n \ra &=& \la \omega_{\text{NR}}^\prime \omega_{\text{NR}} v_j \ra + \la \epsilon^{ik} v_k \partial_i v_j \epsilon^{mn} \partial^\prime_m v^\prime_n \ra.
\end{eqnarray}
We can show that the second term on the right-hand side is zero as follows:
\begin{eqnarray}
\la \epsilon^{ik} v_k \partial_i v_j \epsilon^{mn} \partial^\prime_m v^\prime_n \ra &=& \left( \delta^{im}\delta^{kn}-\delta^{in}\delta^{km} \right) \la v_k \partial_i v_j \partial^\prime_m v^\prime_n \ra \nonumber\\
&=& \la v_k \partial_i v_j \left( \partial^{\prime i} v^{\prime k} - \partial^{\prime k} v^{\prime i} \right) \ra \nonumber\\
&=& \la \left( v_y \partial_x - v_x \partial_y \right) v_j \left(\partial^{\prime x} v^{\prime y} - \partial^{\prime y} v^{\prime x} \right) \ra \nonumber\\
&=& -\la \omega_{\text{NR}}^\prime \left(\vc{v}\times \vc{\nabla}\right) v_j \ra \nonumber\\
&=& -\la \omega_{\text{NR}}^\prime \epsilon^{ik} v_i \partial_k v_j \ra ,
\end{eqnarray}
where we used the identity $\epsilon^{ik} \epsilon^{mn} = \delta^{im}\delta^{kn} - \delta^{in}\delta^{km}$ in the first line. Again, isotropy and regularity at the origin will imply this vanishes, provided it is divergence-free~\cite{Landau:1987}. Thus, we can compute its divergence and show that it vanishes:
\begin{eqnarray}
-\partial^{\prime j} \la \omega_{\text{NR}}^\prime \epsilon^{ik} v_i \partial_k v_j \ra &=& \partial^{j} \la \omega_{\text{NR}}^\prime \epsilon^{ik} v_i \partial_k v_j \ra \nonumber\\
&=& \la \omega_{\text{NR}}^\prime \epsilon^{ik} \partial^j v_i \partial_k v_j \ra \nonumber\\
&=& \la \omega_{\text{NR}}^\prime \left( \partial^x v_x \partial_y v_x - \partial^x v_y \partial_x v_x + \partial^y v_x \partial_y v_y - \partial^y v_y \partial_x v_y \right) \ra \nonumber\\
&=& \la \omega_{\text{NR}}^\prime \left( \partial_y v_x - \partial_x v_y \right) \partial_i v^i \ra \nonumber\\
&=& 0,
\end{eqnarray}
where we used incompressibility in the second and last lines. The relativistic scaling relation Eq.~\eqref{eq:directcascade_WS} thus reduces in the incompressible Navier-Stokes limit to 
\begin{eqnarray}
\la \omega_{\text{NR}}^\prime \omega_{\text{NR}} v_j \ra = -\frac{1}{2}\epsilon_\omega r_j. \label{eq:directcascadeWS_limit}
\end{eqnarray}
As before, this relation is equivalent to an intermediate standard result from~\cite{Bernard:1999}, namely
\begin{eqnarray}
\la \delta v_j \left( \delta \omega \right)^2 \ra = -2\epsilon_\omega r_j. \label{eq:directcascade_bernard}
\end{eqnarray}
To see this, expand the left-hand side and use statistical symmetries to obtain
\begin{eqnarray}
\la \delta v_j \left( \delta \omega \right)^2 \ra &=& 4 \la \omega_{\text{NR}}^\prime \omega_{\text{NR}} v_j \ra + 2\la v_j^\prime \omega_{\text{NR}}^2 \ra,
\end{eqnarray}
and then note that the second term on the right-hand side vanishes by incompressibility, isotropy, and regularity at $r=0$~\cite{Landau:1987}. Thus Eq.~\eqref{eq:directcascade_bernard} is the same as Eq.~\eqref{eq:directcascadeWS_limit}. Since Eq.~\eqref{eq:directcascadeWS_limit} is the incompressible Navier-Stokes limit of Eq.~\eqref{eq:directcascade_WS}, it can be compared in the direct-cascade range of relativistic or compressible turbulence in order to gauge their relative performance.

Finally, we demonstrate that the relativistic correlation derived in~\cite{WS2015}, which reads
\begin{eqnarray}
\la T^\prime_{0T} T_{LT} \ra = \frac{\varepsilon}{24} r^3, \label{eq:directcascadeWS_weird}
\end{eqnarray}
also reduces to a known result in the incompressible non-relativistic limit. Note that the subscripts $(L,T)$ refer to the longitudinal ($\parallel \vc{r}$) and transverse ($\perp \vc{r}$) directions, respectively. Once again, setting $\rho=\text{constant}$ and $\gamma=1$ yields
\begin{eqnarray}
\la v^\prime_T v_L v_T \ra = \frac{\epsilon_\omega}{24} r^3. \label{eq:directcascadeWS_weird_limit}
\end{eqnarray}
Since Eq.~\eqref{eq:directcascadeWS_weird_limit} is the incompressible Navier-Stokes limit of Eq.~\eqref{eq:directcascadeWS_weird}, they can also be compared in the direct-cascade of relativistic or compressible turbulence in order to gauge their relative performance.

Again, Eq.~\eqref{eq:directcascadeWS_weird_limit} can be obtained from standard results in~\cite{Bernard:1999}. The first intermediate result for the direct-cascade range that we use reads
\begin{eqnarray}
\la \delta v_j \delta v_i \delta v^i \ra = \frac{1}{4}\epsilon_\omega x_j r^2. \label{eq:directcascade2_Bernard}
\end{eqnarray}
Using statistical symmetries, the left-hand side expands to $4 \la v^\prime_i v^i v_j \ra + 2 \la v^\prime_j v^i v_i \ra$, and the second term vanishes due to incompressibility, isotropy, and regularity at the origin~\cite{Landau:1987}. Thus, setting $j=L$ in Eq.~\eqref{eq:directcascade2_Bernard}, we obtain
\begin{eqnarray}
\la v^\prime_L v_L v_L \ra + \la v^\prime_T v_L v_T \ra = \frac{1}{16}\epsilon_\omega r^3. \label{eq:firstpiece}
\end{eqnarray}
We can eliminate the first term on the left-hand side using the well-known $1/8$-law, also derived in~\cite{Bernard:1999} and valid in the direct-cascade range, $\la (\delta v_L)^3 \ra = 6 \la v^\prime_L v_L v_L \ra = (1/8) \epsilon_\omega r^3$. This substitution finally yields Eq.~\eqref{eq:directcascadeWS_weird_limit}.
%
%
\chapter{Implementation}\label{sec:P1implementation}
As stated, our goal is to explore scaling relations in conformal fluid turbulence. To ensure a clean inertial
regime is established for the computation of the appropriate quantities, we include a driving source. Additionally, we 
ensure the numerical methods employed are consistent with the statistical properties of the flow we want to
study. In this section we describe key aspects of our numerical implementation, beginning with general considerations in Sec.~\eqref{sec:implementation-general}. Following this, we present specific considerations for the incompressible and relativistic cases in Secs.~\eqref{sec:implementation-NS} and~\eqref{sec:implementation-rel}, respectively.
%
%
\section{General considerations}\label{sec:implementation-general}
\subsection{Stochastic Runge-Kutta integrator}\label{sec:SRKII}
In order to implement a random white noise force in a simulation, a special integration algorithm must be used. Based on the work of Honeycutt~\cite{Honeycutt:1992}, we use a second-order Stochastic Runge-Kutta algorithm (SRKII). The Gaussian random force we use, defined later in Eq.~\eqref{eq:Rspacecorr}, is homogeneous, which means the average and variance of the force at every point in space is the same. Thus the prescription described in~\cite{Honeycutt:1992} is applied to each real space point, producing control over the injection rates in an aggregate sense.
%
%
\subsection{Pseudorandom number generation}
The random force we employ requires pseudorandom number generation at every time step. For this purpose, we implement the Intel MKL Vector Statistical Library. In particular, we use the Mersenne Twister~\footnote{With BRNG parameter VSL\_BRNG\_MT19937}~\cite{INTEL} and block-splitting for parallel applications~\cite{survivalguideRNG}. We have checked that the energy spectrum $E(k)$ in steady-state is unaffected by the choice of random number generator by comparing the Mersenne Twister (VSL\_BRNG\_MT19937) and the 59-bit multiplicative congruential generator (VSL\_BRNG\_MCG59). We also checked that the output of our code is system-independent~\cite{survivalguideRNG} by running it on two independent clusters.
%
%
\subsection{Defining an injection length scale} \label{sec:injectionlength}
In studies of turbulence, the energy/enstrophy injection and scale play a crucial role in establishing and identifying
particularly relevant dynamical ranges. 
One can define an injection length scale associated with the external force in terms of the injection rates of energy and enstrophy as follows. Given Kraichnan-Batchelor~\cite{Kraichnan:1967} scaling of the energy spectrum in the inverse and direct cascades, $E(k) \sim \epsilon_0^{2/3} k^{-5/3}$, $\eta_0^{2/3} k^{-3}$, respectively, one can take the injection scale to be the wavenumber at which $E(k)$ transitions between these two scalings. Thus, set $\epsilon_0^{2/3} k_f^{-5/3} = \eta_0^{2/3} k_f^{-3}$ and solve to find $k_f = \sqrt{\eta_0/\epsilon_0}$. This definition will accurately represent the injection scale up to a numerical factor of order $\sim 1$, so long as the energy spectrum transitions between these two behaviours over a short range of wavenumbers.
%
%
\section{Incompressible case}\label{sec:implementation-NS}
\subsection{Formulation}
In the {\em incompressible} Navier-Stokes case in $2$D, the entire dynamics is determined by a single pseudo-scalar quantity, the vorticity $\omega = \vc{\nabla} \times \vc{v}$. Thus, it is computationally more efficient to evolve the vorticity equation directly, rather than the components of the velocity. We write the vorticity equation in ``flux-conservative form",
\begin{eqnarray}
\partial_t \omega + \partial_i (v^i \omega) = f_\omega - \nu_4 \partial^4 \omega,
\end{eqnarray}
where $f_\omega$ is the random force defined in the next section, and the dissipative term $- \nu_4 \partial^4 \omega \equiv - \nu_4 \nabla^4 \omega$ on the right-hand side is often referred in the turbulence literature as ``hyperviscosity of order $4$''. Hyperviscosity is frequently used in simulations of an incompressible Navier-Stokes fluid~\cite{Boffetta:2012}, since it limits the range of scales over which dissipation is active (yielding wider inertial ranges for a given grid resolution).
%
%
\subsection{Random force and injection rates}\label{sec:force}

The external force appears as $f_{\omega}\equiv \vc{\nabla}\times\vc{f}$, and we wish to construct $f_\omega$ directly
with the appropriate statistical properties. Given a Gaussian random force with a two-point correlation in real space given by
\begin{eqnarray}
\la f_\omega(t,0) f_\omega(t^\prime,r) \ra = g(r) \delta(t-t^\prime), \label{eq:Rspacecorr}
\end{eqnarray}
for some function $g(r)$, the injection rate of enstrophy will be given by $g(0)/2\equiv \eta_0$~\cite{Novikov:1965}, owing to the delta function (i.e. white noise) and to the choice of Gaussian randomness. Ignoring the temporal part of the correlation, we have in Fourier space
\begin{eqnarray}
\la \hat{f}_\omega (\vc{k}) \hat{f}_\omega^* (\vc{k}) \ra = \hat{g} (k), \label{eq:Fspacecorr}
\end{eqnarray}
where reality of the force in real space requires $f_\omega (-\vc{k}) = f^*_\omega(\vc{k})$. 

In order to specify the enstrophy injection rate $\eta_0$, we use a rescaling strategy as follows. First, define two random scalar fields $A(\vc{k})$, $B(\vc{k})$, with zero average $\la A \ra = \la B \ra = 0$ and unit variance $\la A^2 \ra = \la B^2 \ra = 1$ at all wavenumbers, and set $\hat{f}_\omega (\vc{k}) = A(\vc{k}) + i B(\vc{k})$. We first seek an isotropic rescaling $\hat{f}_\omega \rightarrow \tilde{g}(k) \hat{f}_\omega$ that gives the profile of Eq.~\eqref{eq:Fspacecorr} up to a constant factor. Under this rescaling, $A,B \rightarrow \tilde{g}A,\tilde{g}B$, so the zero average is unchanged but the variance transforms to $\la A^2 \ra, \la B^2 \ra \rightarrow \tilde{g}^2\la A^2\ra, \tilde{g}^2\la B^2\ra = \tilde{g}^2$. Thus,
\begin{eqnarray}
\la \hat{f}_\omega (\vc{k}) \hat{f}_\omega^*(\vc{k}) \ra &=& (A+iB)(A-iB) \, , \nonumber\\
&=& A^2 + B^2 \, , \nonumber\\
&\rightarrow & \tilde{g}^2 (A^2+B^2) \, , \nonumber\\
&=& 2\tilde{g}^2 (k) \, . \label{eq:1strescale}
\end{eqnarray}
Thus choosing $\tilde{g} \propto \sqrt{\hat{g}/2}$ gives the desired spatial profile up to a constant factor. To fix the enstrophy injection rate (as $\eta_0 = g(0)/2$), we seek a second rescaling $\hat{f}_\omega \rightarrow R\hat{f}_\omega$ with $R=$ constant determined as follows. As it stands, Eq.~\eqref{eq:1strescale} will produce an enstrophy injection rate given by half of its inverse Fourier transform evaluated at $r=0$, 
\begin{eqnarray}
\tilde{\eta}_0 \equiv \frac{1}{2} FT^{-1}(2\tilde{g}^2(k))\vert_{r=0}.
\end{eqnarray}
Under the second rescaling, Eq.~\eqref{eq:1strescale} becomes $2 R^2 \tilde{g}^2(k)$. Thus the appropriate rescaling is 
\begin{eqnarray}
R=\sqrt{\eta_0 / \tilde{\eta}_0}.
\end{eqnarray}
If one wishes instead to specify the energy injection rate, simply note that for a solenoidal force $\vc{\nabla}\cdot\vc{f}=0$, we have the spatial part of Eq.~\eqref{eq:Rspacecorr} given by
\begin{eqnarray}
\la f_\omega (0) f_\omega (r) \ra &\equiv & \la f_\omega f^\prime_\omega \ra \nonumber\\
&=& \la \epsilon^{ij} \partial_i f_j \epsilon^{mn} \partial^\prime_m f^\prime_n \ra \nonumber\\
&=& \epsilon^{ij}\epsilon^{mn} \partial_i \partial^\prime_m \la f_j f^\prime_n \ra \nonumber\\
&=& (\delta^{im}\delta^{jn} - \delta^{in}\delta^{jm}) \partial_i \partial^\prime_m \la f_j f^\prime_n \ra \nonumber\\
&=& \partial^i \partial^\prime_i \la f^j f^\prime_j \ra \nonumber\\
&=& -\partial^i \partial_i \la f^j f^\prime_j \ra \nonumber\\
&=& -\nabla^2 \la \vc{f} \cdot \vc{f}^\prime \ra .
\end{eqnarray}
So by solving the Poisson equation $\nabla^2 \la \vc{f}(0) \cdot \vc{f}(r) \ra = -g(r)$ one finds the energy injection rate $\epsilon_0$ from the relation $\la \vc{f}(0) \cdot \vc{f}(r) \ra\vert_{r=0} = 2\epsilon_0$. The rescaling factor $R$ can be chosen appropriately in this case. Extracting these a priori injection rates of energy and enstrophy allows one to define an injection length scale as per Sec.~\eqref{sec:injectionlength}.

For our incompressible simulations of the direct-cascade we use a `rectangular' profile, namely $\hat{g}(k) = 1$ in a narrow range of wavenumbers around $k_f$, zero otherwise.
%
%
\subsection{Dealiasing}\label{sec:dealiasing}
The Navier-Stokes equation has a quadratic nonlinearity. Thus, two wavenumbers $k_1$, $k_2$ can interact to populate a third wavenumber $k_3=k_1+k_2$. Since we have a finite range of scales resolved in any simulation, $k_3$ could exceed the largest resolved wavenumber, and thus would become represented on the grid as a lower wavenumber $\mathcal{N}-k_3$ (where $\mathcal{N}$ is the grid resolution). In this case, we say $k_3$ has been \emph{aliased}. Prescriptions exist to avoid such aliasing errors. For a quadratically nonlinear term $F\times G$, if we filter out all wavenumber modes with $k>\mathcal{N}/3$ in $F$ and $G$ prior to multiplication, then filter $F\times G$ in the same manner, we will eliminate all aliasing errors. Such a prescription is known as the $2/3$-dealiasing rule, since one retains $2/3$ of the domain in Fourier space. Analogous dealiasing rules exist for higher-order nonlinearities, with less and less of the domain being retained as the order increases. Thus, full dealiasing becomes computationally prohibitive for higher-order nonlinearities, such as for a relativistic fluid flow.
%
%

\section{Relativistic conformal fluid case}\label{sec:implementation-rel}
\subsection{Formulation}
The system of equations is given by $\partial_a T^{ab} = f^b$ and the conformal perfect fluid stress-energy tensor $T^{ab} = (3/2)\rho u^a u^b + (1/2)\rho\eta^{ab}$, which uses the conformal equation of state $P=\rho/2$ in $(2+1)$ dimensions. Defining the conservative variables as $(D,S^i) = (T^{00},T^{0i})$, they appear in terms of the primitive variables as
\begin{eqnarray}
(D,S^i) = \left( \frac{3}{2}\rho\gamma^2-\frac{1}{2}\rho, \frac{3}{2}\rho\gamma^2 v^i \right),
\end{eqnarray}
where $v^i$ is the spatial velocity and $\gamma$ is the Lorentz factor. In terms of these variables, the equations of motion appear in flux-conservative form as
\begin{eqnarray}
\partial_t D + \partial_i S^i &=& 0 \label{eq:Deom} \\
\partial_t S^i + \partial_j (S^j v^i + \frac{1}{2}\rho\delta^{ij}) &=& f^i. \label{eq:Seom}
\end{eqnarray}
We use finite differences to discretize the derivatives, with RK4 in space and SRKII (see Sec.~\eqref{sec:SRKII}) in time. The system is damped at short wavelengths using a 4th-order dissipation scheme discussed in Sec.~\eqref{sec:treatlargek}.

\subsection{Random force and injection rates}
We choose the Gaussian white-noise force $f^i$ to be divergence-free by deriving it from a stream function $\psi$, $(f_x, f_y) = (\partial_y \psi, -\partial_x \psi)$. Thus, numerically we build $\psi$ directly in the manner described in Sec.~\eqref{sec:force}. For simulations of the inverse-cascade, we choose
\begin{eqnarray}
\la \psi^\prime \psi \ra = \epsilon l_f^2 \exp{(-r^2/2l_f^2)} \delta(t-t^\prime), \label{eq:relforcestatistics}
\end{eqnarray}
where $l_f$ is the characteristic length scale of the correlation, and  $\epsilon = \la T^{0i} f_i \ra$~\cite{WS2015} is a constant. One can verify the equality $\epsilon = \la T^{0i} f_i \ra$ by applying the 2-dimensional Laplacian to Eq.~\eqref{eq:relforcestatistics}, then noting that the spatial part of $\la f_i (\vc{r}) f^i(0) \ra$, written as $F^i_i \equiv \text{tr}F$, is given by $\text{tr}F = -\nabla^2 \la \psi(\vc{r}) \psi(0) \ra$ and $\text{tr}F = 2\la T^{0i}f_i \ra$~\cite{WS2015}. In the weakly compressible regime, $\epsilon$ is approximately the injection rate of $(1/2)\la T^{0i}T^0_i \ra$, whereas in the incompressible regime it fixes the Newtonian kinetic energy injection rate. 

For simulations of the direct-cascade, we instead choose
\begin{eqnarray}
\la \hat{\psi} \hat{\psi}^{*} \ra \propto 
{\begin{cases}
 1 \:\:\:\: k \sim k_f \, , \\
 0 \:\:\:\: \text{otherwise} \, . \end{cases}}\label{eq:spikeprofile}
\end{eqnarray}
%
%

\subsection{Dealiasing} \label{sec:treatlargek}
As alluded to in Sec.~\eqref{sec:dealiasing}, in the relativistic case a full dealiasing is computationally prohibitive. Since the computation of the velocity from the conservative hydrodynamic variables, followed by the computation of the flux, amounts to forming a product of up to $5$ fields, there is a quintic nonlinearity. In the weakly-compressible regime, however, a $2/3$-dealiasing rule would likely eliminate a satisfactory amount of aliasing, since the density and Lorentz factor have a small amount of power at all wavenumbers $k\neq 0$. However, in a future study we wish to explore the strongly compressible and ultrarelativistic regimes where a $2/3$-rule would be inadequate. Thus we opt instead to use a 4th-order numerical dissipation scheme to suppress large wavenumber modes (since we want to explore the suitability of alternative dealiasing strategies for that future study) and employ a sufficiently high resolution (so that possibly
spurious effects stay mainly confined at very high frequencies). For a variable $U$, this scheme amounts to including a term $-\nu_{\text{num}} (\partial_x^4 + \partial_y^4) U$ on the right-hand side of its evolution equation, where $\nu_{\text{num}}>0$ is the strength of the dissipation. It is numerically convenient to write this term as $-\kappa (dx^3 \partial_x^4 + dy^3 \partial_y^4)U$ and control the dissipation strength $\kappa$, as its magnitude will be closer to $1$ and the dissipation length scale will move with the resolution~\cite{gustafsson1995time}.
%
%
\chapter{Results}\label{sec:P1results}
In all simulations we use periodic boundary conditions with a box size of $L=2\pi$ and resolution of $\mathcal{N}^2=2048^2$, with a variable step size determined by a CFL condition~\cite{courant1928partiellen}. This resolution has proven quite adequate for studying correlation functions in both the inverse-cascade (eg.~\cite{Boffetta:2000}) and direct-cascade (eg.~\cite{Pasquero2002,Chen2003}) in incompressible fluid turbulence. We find it is also adequate for the weakly compressible regime studied here, since we have obtained the same inverse-cascade range spectral slope with $\mathcal{N} = 4096$.

The time scale over which a turbulent flow is presumed to erase knowledge of its initial conditions is the large-eddy turnover time, which has various interpretations in the literature. Borue~\cite{Borue1994} estimates it as $T=2\pi/\omega_{\text{rms}}$, where $\omega_{\text{rms}}$ is the root-mean-squared vorticity. More generally, we have $T=L/U$ where $L$ is the scale of the largest eddies and $U$ is a characteristic speed at that scale. $L$ is estimated as $2\pi/k_i$, where $k_i$ is the infrared ``cutoff" ($\sim$ largest energy-containing scale), and we estimate $U$ as the root-mean-square of the velocity. In our simulations, these time scales will be quoted for reference. 

Averages will be computed over time, or over independent simulations, or both. The adequacy of the sample sizes is gauged via comparison of the average with the statistical error $\sigma/\sqrt{N}$, where $\sigma$ is the sample standard deviation and $N$ is the sample size. For example, a correlation function $f(r)$ will have an ensemble of values for each $r$, and $\sigma(r)$ is computed as the standard deviation of that collection of values.
%
%
\section{Inverse-cascade simulations} \label{sec:inversecascaderesults}
We simulate the inverse-cascade of a $(2+1)$-dimensional conformal fluid with an external force described by Eq.~\eqref{eq:relforcestatistics}, and an injection scale $k_f \equiv 2\pi/L_f \sim 203$ defined by $k_f=\sqrt{\eta_0/\epsilon_0}$, as in Sec.~\eqref{sec:injectionlength}. We consider three cases with the numerical dissipation strength given by $\kappa = (0.05,0.03,0.02)$ (so as to compare results among them) and when quoting properties of each case we will present them in this order. Since the force is somewhat broadband, it has power in the dissipation range of scales. Thus, decreasing the dissipation strength is enough to increase the energy growth rate, and thus the rms Mach number of the flow, $v_{rms}/c_s$, where $c_s$ is the sound speed ($1/\sqrt{2}$ of the speed of light, in our case). Statistical quantities are averaged over ensembles of independent simulations, as well as averaged over an interval of time after the energy passes $k=10$ and before it reaches the box size. Table~\eqref{table:icparams} contains various parameters of the flows, as well as the sample sizes for the joint average over an ensemble and over time.

\begin{figure}[h!]
\centering
\hbox{\hspace{0.8cm}\includegraphics[width=0.9\textwidth]{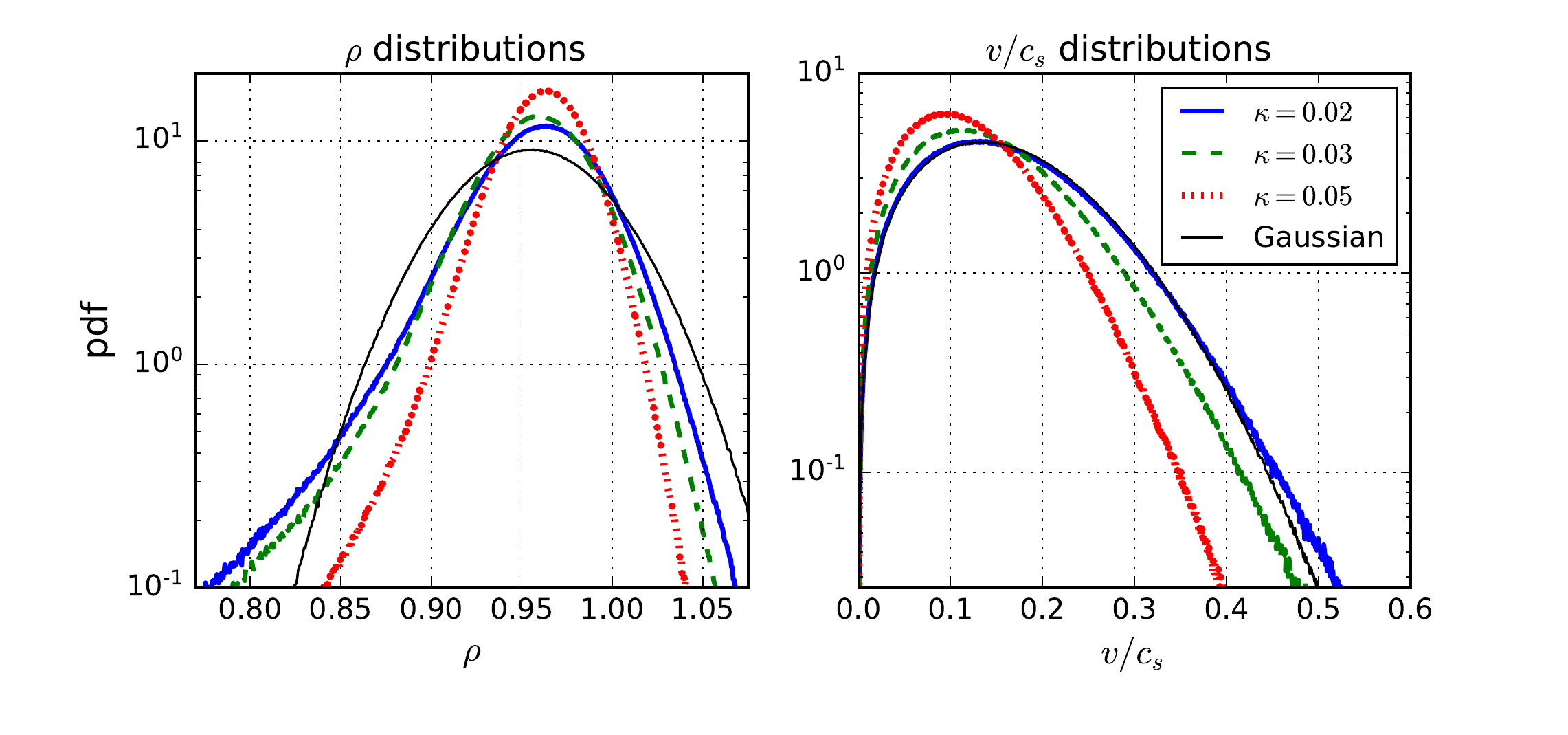}}
\caption{Probability distribution functions for the inverse-cascade simulations. The pdfs of the energy density $\rho$ (Left) and the Mach number $v/c_s$ (Right) are displayed, where $c_s = c/\sqrt{2}$, plotted on a semi-log scale. All dissipation cases are overlaid for ease of comparison. The density $\rho$ and velocity field $(v_x,v_y)$ are high-pass filtered ($k>10$) for a more sensible comparison in this quasi-steady regime (i.e. no large-scale dissipation). The cutoff $k=10$ is chosen based on the maxima of the spectra in Fig.~\eqref{fig:spectra_inverse_cascade} occurring at $k<10$. For comparison, purely Gaussian distributions are plotted (black, solid, thinner lines) with its average and standard deviation matched to data from the dissipation case $\kappa=0.02$. In the order of increasing energy growth rate, the standard deviations of pdf($\rho$) and pdf($v/c_s$) in each case are $(0.0286,0.0395,0.0438)$ and $(0.122,0.147,0.167)$, respectively. In the same order, the rms Mach numbers are $(0.1386,0.1674,0.1893)$. These properties indicate a weakly compressible flow.} \label{fig:pdfs_inverse_cascade}
\end{figure}

In Fig.~\eqref{fig:pdfs_inverse_cascade} we characterize the flows by presenting the probability distributions
functions (pdfs) of the energy density and Mach number. The pdfs are observed to widen as the energy growth rate increases, as one would expect. For comparison, in both cases we also plot Gaussian distributions (black, dashed) with average and standard deviation matched to the data from the $\kappa = 0.02$ case. The Gaussian provides a good fit to the Mach number pdf (although with a slight hint of non-Gaussianity in the tail), whereas the energy density pdf exhibits a stronger, exponential tail towards smaller values.

\begin{figure}[h!]
\centering
\hbox{\hspace{0.8cm}\includegraphics[width=0.9\textwidth]{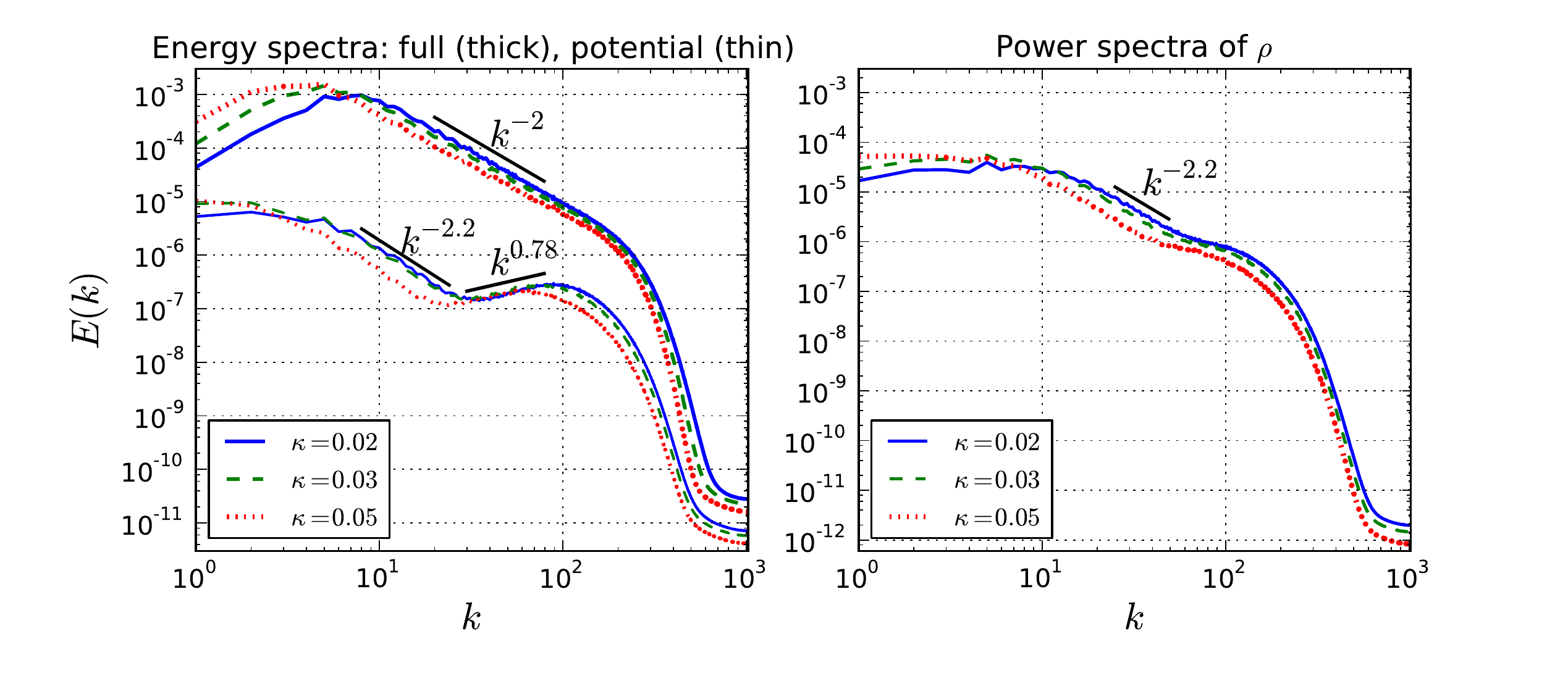}}
\caption{(Left): Newtonian energy spectra of the inverse-cascade simulations plotted on a log-log scale. The spectra corresponding to the full velocity field (thicker lines) and the curl-free potential part (thinner lines) are displayed. All energy growth rate cases are displayed, with the same colour coding and line styles as in Fig.~\eqref{fig:pdfs_inverse_cascade}. The best-fit power laws for the full spectra over the range $k=20-80$ in order of decreasing dissipation strength are $(-1.80,-1.89,-1.96)$. (Right): Power spectra of the energy density $\rho$ for all energy growth rates, plotted on a log-log scale.} \label{fig:spectra_inverse_cascade}
\end{figure}

In Fig.~\eqref{fig:spectra_inverse_cascade} (Left) we display the angle-averaged Newtonian kinetic energy spectra $E(k) \equiv \pi \la \hat{v}^2(\vc{k}) \ra$ (both the full spectrum and the potential part, obtained by projecting the velocity onto $\hat{\vc{k}}$ in Fourier space). We observe a steepening of the inertial range scaling towards $E(k) \sim k^{-2}$, which we note is steeper than the Kolmogorov/Kraichnan power law of $k^{-5/3}$. The spectra are not changed significantly ($<1\%$) by instead using density-weighted velocities $\rho^{1/3} \vc{v}$ or $\rho\vc{v}$, the former having been suggested in the $(3+1)$-dimensional context in~\cite{Kritsuk:2007} to restore Kolmogorov/Kraichnan scaling from the observed spectral exponent of $k^{-2}$. The spectrum of the potential component of the velocity exhibits a bump beginning at $k\sim 30$, with scaling of $k^{-2.2}$ and $k^{0.78}$ on either side. Such a bump towards large $k$ is commonly observed in spectra in simulated compressible flows in $(3+1)$ dimensions, eg.~\cite{Radice:2012pq,Kritsuk:2007,Dobler:2003,Federrath:2013}, and is attributed in those cases to an artefact of high-order numerical dissipation known as the bottleneck effect~\cite{Falkovich:1994}. This effect has also been observed in simulated compressible 2D flows which exhibit transfer of energy to small scales~\cite{Biskamp:1998}. The bump we observe in Fig.~\eqref{fig:spectra_inverse_cascade} is likely due to the same effect, although we cannot make a conclusive statement since we have not performed the specific resolution studies necessary to do so, nor have we used dissipation of a different order. The late-time spectra obtained from our simulations of the direct-cascade (not shown) also exhibit such a bump, and in that case we note that reducing the time step by half does not change the bump perceptibly. 

With regard to the full inverse-cascade spectra in Fig.~\eqref{fig:spectra_inverse_cascade} (Left), it is worth noting that there is no large-scale friction. In~\cite{Scott:2007}, it was shown that the presence of large-scale friction can affect the inertial range spectrum in the incompressible Navier-Stokes case. In the same study it was also shown that measurements of the inertial range spectrum are not reliable without a sufficiently resolved enstrophy cascade ($k_{\text{max}}/k_f \sim 16$, where $k_{\text{max}}$ is defined as $\mathcal{N}/3$). We do not have the direct-cascade range resolved to this degree in Fig.~\eqref{fig:spectra_inverse_cascade} ($k_{\text{max}}/k_f \sim 3.4$). The approach of the full spectrum towards $k^{-2}$ is generally expected for compressible turbulence in both $(3+1)$ dimensions (see eg.~\cite{Federrath:2013}) and $(2+1)$ dimensions (see eg.~\cite{Passot1995}), although usually for much larger Mach numbers than our current simulations. With that said, $(2+1)$-dimensional conformal fluids are special (eg. having a very large sound speed and no mass density), and its turbulent regime is seldom studied (see eg.~\cite{Carrasco:2012nf,Green:2013zba}), so one may not expect the same energy spectra a priori. We elaborate more on this in Chapter~\eqref{sec:P1discussion}, where we demonstrate that the $k^{-2}$ spectrum is not necessarily associated with compressive effects.

\begin{table}[]
\centering
\caption{Parameters of inverse-cascade simulations: $\kappa$ is the dissipation parameter described in Sec.~\eqref{sec:treatlargek}; $\epsilon$ is the growth rate of $(1/2)\la T_{0i}T^{i}_0\ra$; $v_{rms}/c_s$ is the rms Mach number; $N_t$ is the number of snapshots averaged over time; $N_{ens}$ is the number of independent runs (ensemble size); $2\pi/\omega_{rms}$ is the eddy turnover time defined by the rms vorticity; $L/v_{rms}$ is the eddy turnover time defined by $v_{rms}$ and $L=2\pi/10$; $\delta T$ is the time interval between snapshots of the flow that are averaged over; $T_1$ and $T_2$ are respectively the first the last times over which the temporal average is computed. For comparison, note that the light-crossing time is $2\pi$.}
\label{table:icparams}
\begin{tabular}{llllllllll}
\hline
$\kappa$ & $\epsilon\times 10^{4}$ & $v_{rms}/c_s$ & $N_t$ & $N_{ens}$ & $2\pi/\omega_{rms}$ & $L/v_{rms}$ & $\delta T$ & $T_1$ & $T_2$ \\ \hline
0.05 & 3.3 & 0.1386        & 7     & 20        & 1.15                & 6.411       & 5          & 80    & 110   \\ \hline
0.03 & 5.3 & 0.1674        & 5     & 60        & 0.99                & 5.308       & 5          & 60    & 80    \\ \hline
0.02 & 7.0 & 0.1893        & 4     & 20        & 0.90                & 4.694       & 5          & 40    & 55    \\ \hline
\end{tabular}
\end{table}

\begin{figure}[h!]
\centering
\hbox{\hspace{-1cm}\includegraphics[width=1.1\textwidth]{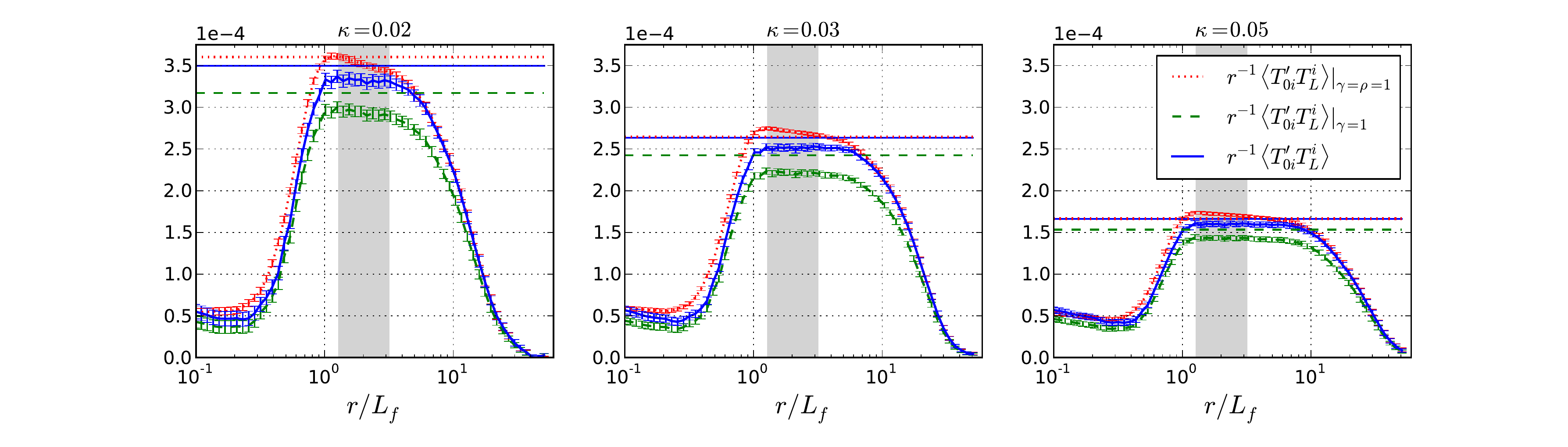}}
\caption{The relativistic correlation function $\la T_{0i}^\prime T^i_L \ra $ (solid blue) and its non-relativistic compressible and incompressible counterparts, $\la T_{0i}^\prime T^i_L \ra\vert_{\gamma=1} $ (dashed green) and $\la T_{0i}^\prime T^i_L \ra\vert_{\gamma=\rho=1} $ (dotted red), respectively, compensated by $r^{-1}$. From left to right: cases with dissipation strength $\kappa=0.02,\: 0.03, \: 0.05$, respectively. Each prediction for the inverse-cascade range $r/L_f \sim (10^0,10^1)$ is plotted as a horizontal line with matching line style. The predictions follow from Eq.~\eqref{eq:inversecascade_WS} and evaluations at $\gamma=1$ or $\gamma=\rho=1$ thereof. Note that the centre and right plots have a nearly indistinguishable prediction for the relativistic and incompressible correlation functions (solid blue and red dotted lines, respectively). Error bars correspond to the statistical uncertainty $\sigma/\sqrt{N}$ for each value of $r/L_f$, where $N$ is the sample size and $\sigma$ is the sample standard deviation. The shaded grey area indicates the range of $r/L_f$ over which we fit a power-law, and we use the same range across all cases to ensure a fair comparison. Note the linear vertical scale, which accentuates deviations from the expected power law.} \label{fig:corrs_inverse_cascade_compensated}
\end{figure}

In Fig.~\eqref{fig:corrs_inverse_cascade_compensated} we plot the relativistic correlation function appearing in Eq.~\eqref{eq:inversecascade_WS}, $\la T_{0i}^\prime T^i_L \ra$, compensated for the expected scaling $r^{-1}$, with a linear vertical scale to help distinguish different power laws. We also plot the incompressible limit of that correlation function, obtained by setting $\rho=\gamma=1$ (herein ``the incompressible correlation"), as well as a non-relativistic but compressible version obtained by setting $\gamma=1$ only (herein ``the compressible correlation"). The former is equivalent to known results from incompressible Navier-Stokes turbulence (see Sec.~\eqref{sec:inverseincompressiblelimit}), while the latter can be obtained from the left-hand side of Eq.~\eqref{eq:inversecascade_WS} using the non-relativistic perfect fluid energy-momentum tensor, which is just the relativistic one with $\gamma=1$. We use these comparisons to separately gauge the degree to which compressive and relativistic effects are important. In addition,  we also include the predictions for each case, in matching colour, obtained from Eq.~\eqref{eq:inversecascade_WS} and evaluations at $\gamma=1$ and $\gamma=\rho=1$ thereof. Error bars correspond to the statistical uncertainty $\sigma/\sqrt{N}$.

\begin{figure}[h!]
\centering
\hbox{\hspace{4.5cm}\includegraphics[width=0.5\textwidth]{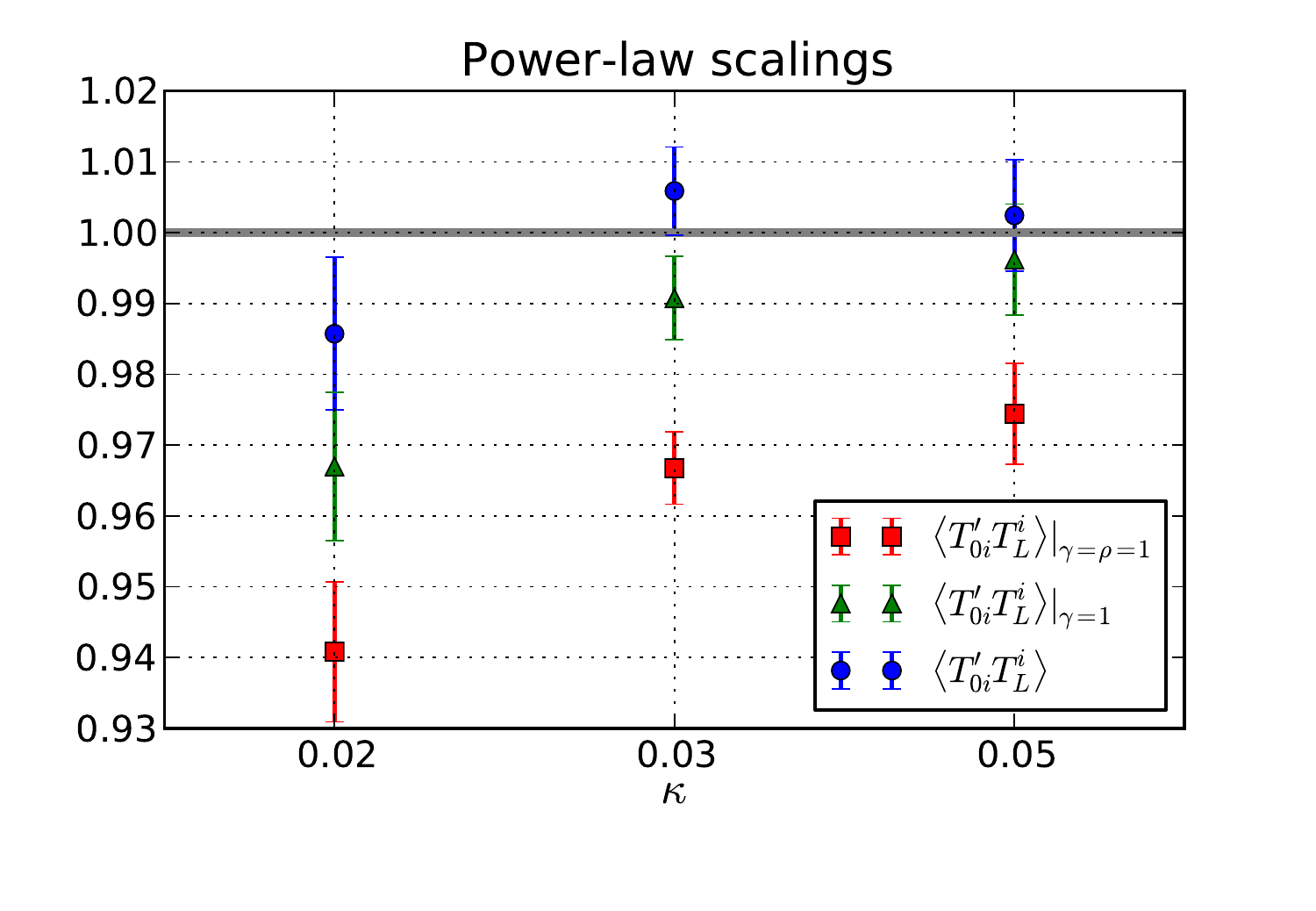}}
\caption{Least-squares power law fits for the relativistic correlation function $\la T_{0i}^\prime T^i_L \ra $ (circles) and its non-relativistic compressible and incompressible counterparts, $\la T_{0i}^\prime T^i_L \ra\vert_{\gamma=1} $ (triangles) and $\la T_{0i}^\prime T^i_L \ra\vert_{\gamma=\rho=1} $ (squares), respectively. All three dissipation cases are displayed. Error bars correspond to the standard deviation of the fitted power law scaling, obtained via random resampling with replacement ($10^3$ trials). The relativistic correlation function outperforms its compressible and incompressible counterparts across all cases, with a monotonic degradation observed for the latter two as the dissipation strength is decreased (and correspondingly, as the rms Mach number is increased). } \label{fig:powerlaw_fits_inverse_cascade}
\end{figure}

For ease of comparison across cases, each plot has the same vertical axis range. As it is clear from the figure, we observe a progressive degradation of the scaling of the incompressible and compressible correlation functions as dissipation is weakened (and thus Mach number grows), while the relativistic one predicted in~\cite{WS2015} outperforms. This is shown quantitatively in Fig.~\eqref{fig:powerlaw_fits_inverse_cascade}, where we display power law fits performed over the shaded interval of Fig.~\eqref{fig:corrs_inverse_cascade_compensated}. The shaded interval is the same across all cases in order to make a fair comparison, and is chosen to capture the power law observed in the $\kappa=0.02$ case (which has the narrowest scaling range). As dissipation is weakened, a monotonic shallowing of the best-fit power law is observed for both the incompressible and compressible correlation functions. This trend is more significant for the incompressible correlation function. The absolute performance of the relativistic correlation function is superior to the compressible and incompressible correlation functions across all cases, and its relative performance improves as dissipation is decreased (i.e. differences in best-fit power law become larger).

The relativistic correlation function, although exhibiting power-law scaling $\sim r$ in all cases, nonetheless exhibits an increasing disagreement with the \emph{magnitude} of the prediction in Eq.~\eqref{eq:inversecascade_WS} (see Fig.~\eqref{fig:corrs_inverse_cascade_compensated}). In the most extreme case ($\kappa =0.02$, $v_{rms}/c_s = 0.19$), the overall magnitude is less than the prediction by $\sim 4\%$. Our numerical ensemble of flows may be biased towards lower magnitudes, since runs with sufficiently large fluctuations from the random force can become numerically unstable and fail. As dissipation is weakened, this occurs more often. Thus, it is possible that the increasing disagreement of the magnitude of $\la T_{0i}^\prime T^i_L \ra$ and $(1/2)\epsilon r$ is an artefact of this bias. A high-resolution shock-capturing implementation could determine whether this increasing disagreement is a real effect.

%
%
\section{Direct-cascade simulation} \label{sec:directcascaderesults}
\begin{figure}[h!]
\centering
\hbox{\hspace{0cm}\includegraphics[width=\textwidth]{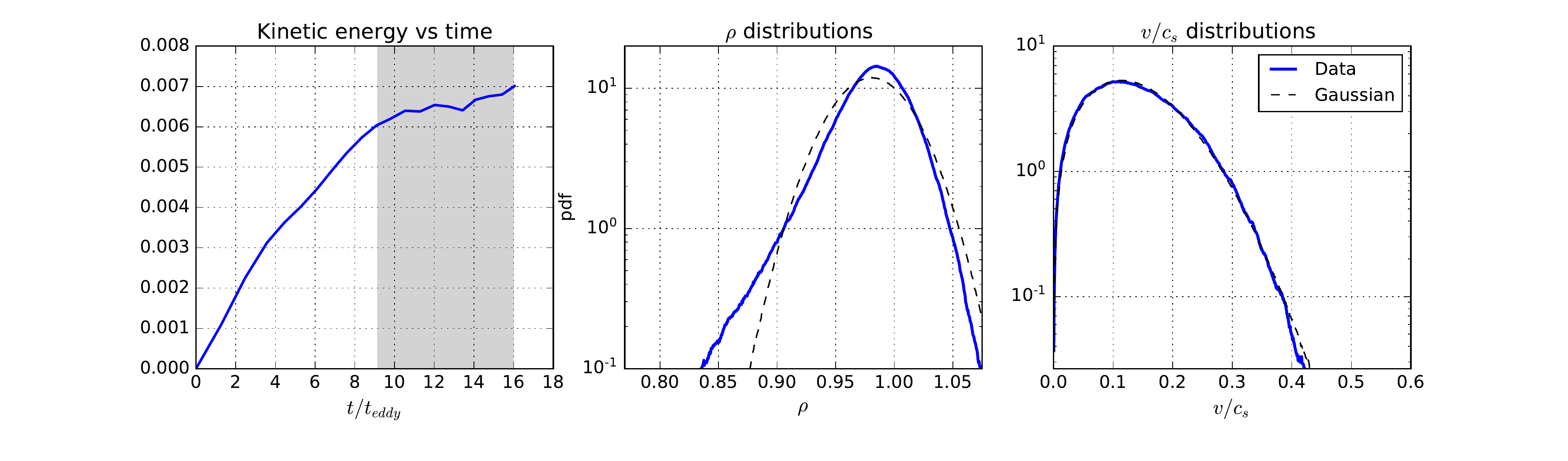}}
\caption{(Left): Average Newtonian specific kinetic energy plotted as a function of dimensionless time for the direct-cascade simulation. The eddy turnover time is defined as $t_{eddy} = 2\pi/\omega_{rms}$. The shaded region indicates the interval of time over which all other quantities are averaged. (Centre and Right): Probability distribution functions for the energy density $\rho$ (Centre, blue, solid) and the Mach number $v/c_s$ (Right, blue, solid), where $c_s = c/\sqrt{2}$, plotted on a semi-log scale. For comparison, purely Gaussian distributions are plotted (black, dashed) with the average and standard deviation matched to the data. The standard deviation of the $\rho$ and $v/c_s$ distributions are $(0.0336,0.0758)$, respectively. The rms Mach number is $0.11$. These properties indicate a weakly compressible flow.} \label{fig:pdfs_direct_cascade}
\end{figure}

To simulate the direct-cascade of a $(2+1)$-dimensional conformal fluid, we instead use an external force with support only around $k_f=7$, as described by Eq.~\eqref{eq:spikeprofile}. As in our inverse-cascade simulations, we use 4th-order numerical dissipation as in Sec.~\eqref{sec:dealiasing}, with the choice $\kappa = 0.01$. However, in contrast to our inverse-cascade simulations, here we use a large-scale dissipation mechanism known as 4th-order \emph{hypofriction}, which takes the form of a term $-\mu \nabla^{-4}S^{i}$ on the right-hand side of Eq.~\eqref{eq:Seom}. We compute the inverse Laplacians spectrally, setting constant modes to zero. Such a term has power restricted to large scales, and terminates the brief inverse cascade from $k=7$ towards $k=0$. We find the value $\mu=0.15$ to be adequate for preventing a build-up of energy (and eventual condensation) at large scales. An energy condensate would be characterized by continued energy growth and the emergence of two dominant vorticies of opposing parity superposed on a noisy flow (see eg.~\cite{Chertkov:2007}). Statistical quantities are averaged over the shaded interval of time indicated in Fig.~\eqref{fig:pdfs_direct_cascade} (Left), which consists of $N=10$ snapshots separated by $\delta T \sim 0.68$. The interval is chosen to maximize the number of snapshots available (to minimize statistical fluctuations), while remaining in a regime that roughly resembles a steady-state. Note that the measured correlations are not significantly affected if the temporal average begins slightly earlier or later. The average time step over this interval is $10^{-3} t_{eddy}$, where $t_{eddy}=2\pi/\omega_{rms} \sim 0.5$ is also averaged over the shaded interval. The injection rate of $(1/2)\la T_{0i}T_0^i\ra$ is measured initially to be $\epsilon = 2\times 10^{-4}$. For reference, the characteristic time as per $v_{rms}$ is $L/v_{rms} \sim 26$, where we take $L=2\pi/3$ since the maximum of the energy spectrum occurs at $k=3$. The rms Mach number over the shaded interval is $\sim 0.11$, once again indicating the weakly compressible regime.

\begin{figure}[h!]
\centering
\hbox{\hspace{0cm}\includegraphics[width=\textwidth]{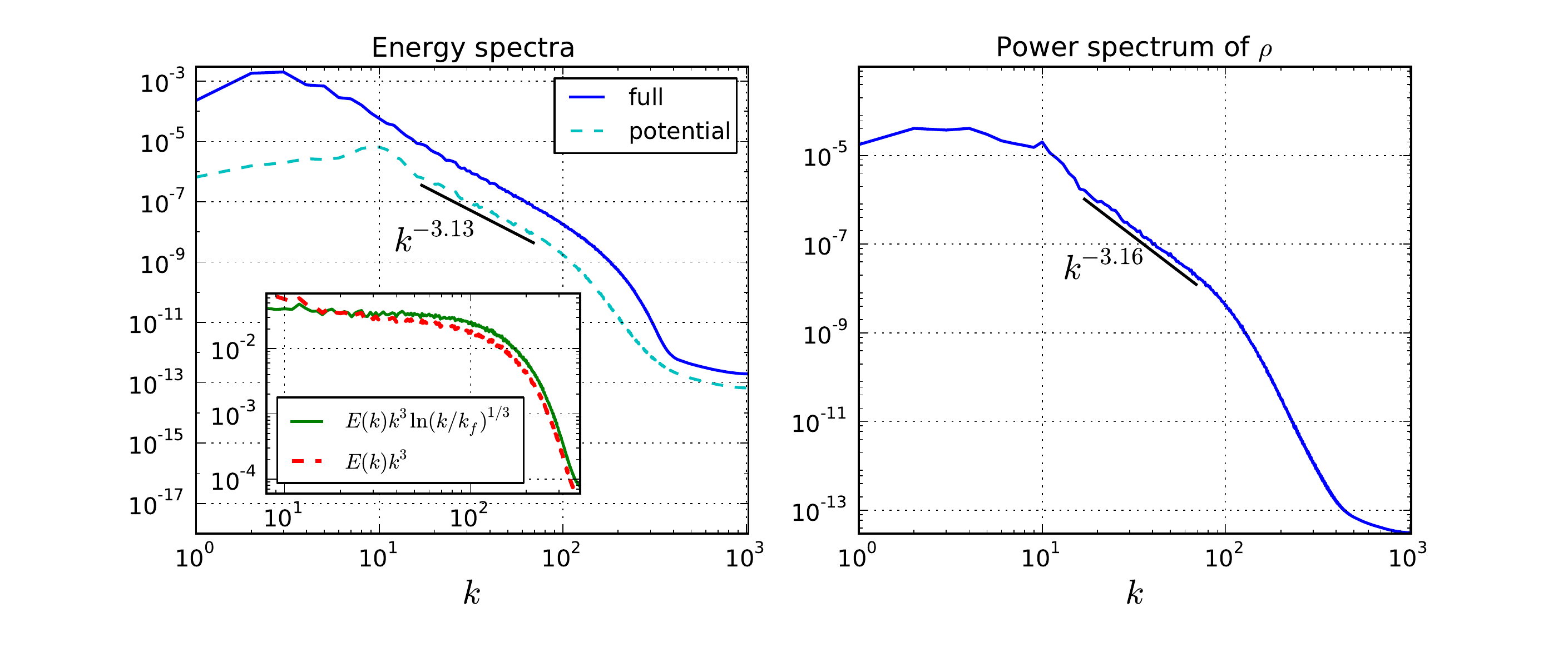}}
\caption{(Left): Newtonian energy spectra of the inverse-cascade simulations plotted on a log-log scale. The spectra corresponding to the full velocity field (solid, blue) and the curl-free potential part (dashed, cyan) are displayed. A least-squares best fit power law of the potential part $\sim k^{-3.13}$ over the range $k\in [17,70]$ is displayed. The inset displays the full spectrum compensated by $k^3 \ln{(k/k_f)}^{1/3}$ (solid, green) or $k^3$ only (dashed, red), showing that the logarithmic correction provides a better fit than the pure power law. (Right): The power spectrum of the energy density $\rho$, with the best fit power law of $k^{-3.16}$ over the range $k\in [17,70]$ displayed.} \label{fig:direct_cascade_all_spectra}
\end{figure}

In Fig.~\eqref{fig:pdfs_direct_cascade} (Left), we display the average Newtonian specific kinetic energy of the fluid as a function of time. As mentioned, we average various quantities over the shaded interval of time. The energy is beginning to plateau over this interval, however it continues to grow slowly. If evolved longer, the compressive component of the velocity begins to dominate over the curl-free part. To study such a regime more accurately, a Riemann solver would be desirable in order to more faithfully capture the dominant shockwave phenomena. Since we are instead using artificial high-order numerical dissipation, we choose to restrict our analysis to earlier times, when the compressive component of the velocity field is still subdominant ($\sim 10\%$ of the total energy at a given scale $k$ - see Fig.~\eqref{fig:direct_cascade_all_spectra} (Left)). Our high-order dissipation also results in large bottleneck effects at later times, which contaminate a rather large portion of the inertial range.

In Fig.~\eqref{fig:pdfs_direct_cascade} (Centre and Right), we display the probability distributions of the energy density (Centre, blue, solid) and Mach number (Right, blue, solid). For comparison, in both cases we also plot Gaussian distributions (black, dashed) with average and standard deviation matched to the data. Similar to the inverse-cascade simulations, the Gaussian provides a good fit to the Mach number pdf (although with weaker hints of a non-Gaussian tail in this case), whereas the energy density pdf exhibits a stronger, exponential tail towards smaller values.

In Fig.~\eqref{fig:direct_cascade_all_spectra} we display the power spectra of the velocity (Left) and energy density (Right). The full energy spectrum of the flow (blue, solid), together with the energy spectrum of the compressive, curl-free, potential part of the velocity (cyan, dashed). The latter is seen to be subdominant by a factor of $\sim 10$ over the range $k\in [10,100]$, which qualitatively corresponds to the direct-cascade interial range. The potential spectrum is fit by a power law $k^{-3.13}$ over this range, while for the full spectrum we observe $k^{-3}$ scaling with the multiplicative logarithmic correction $\ln(k/k_f)^{-1/3}$. The inset shows the full spectrum compensated by $k^{3}$ with and without the logarithmic correction, with the presence of the logarithmic correction being favoured (flatter curve). In the literature, the presence of this correction seems to depend on several factors, including the length of time over which the average is taken, and the presence of large-scale dissipation~\cite{Pasquero2002,Alvelius2000,Chen2003,Vallgren2011}. 

\begin{figure}[h!]
\centering
\hbox{\hspace{1.7cm}\includegraphics[width=0.8\textwidth]{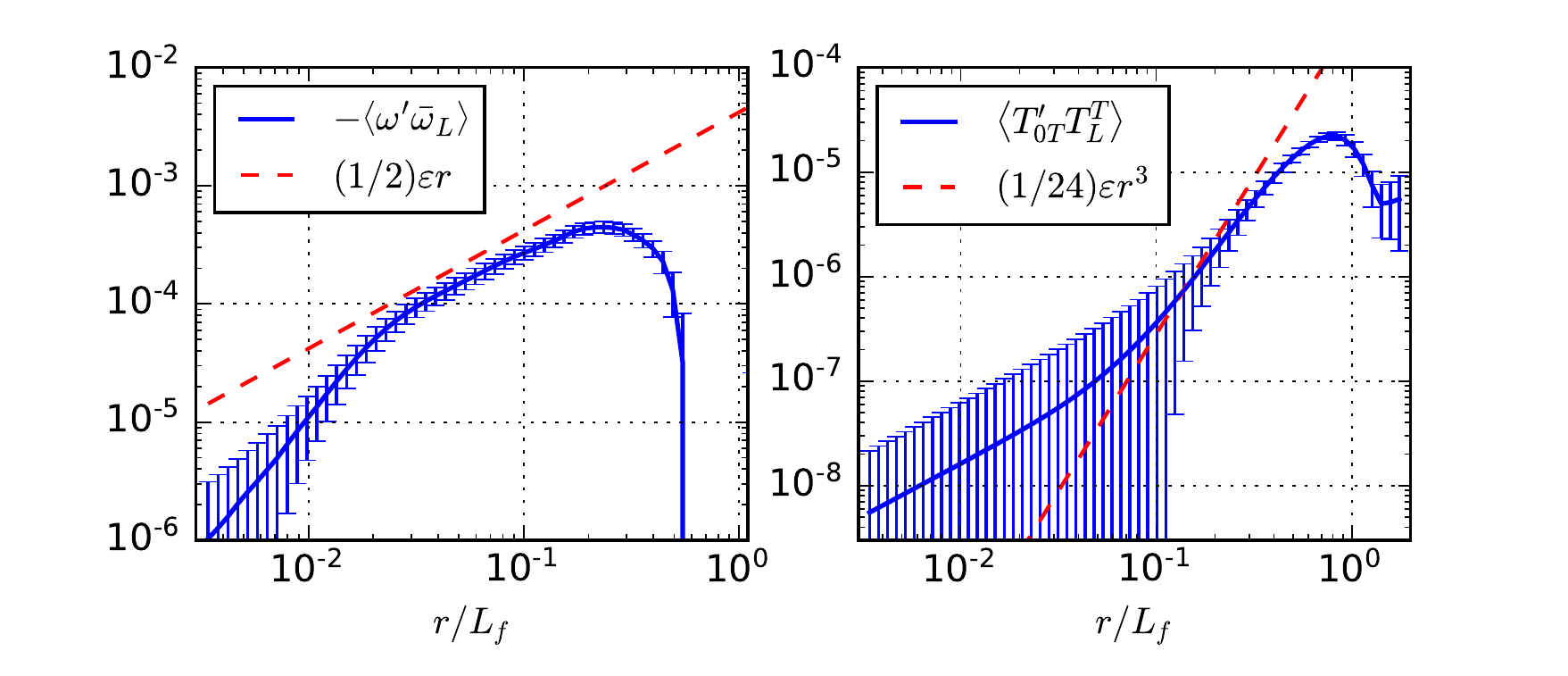}}
\caption{(Left): The relativistic correlation function $-\la \omega^\prime \bar{\omega}_L \ra$ (blue, solid) plotted with its prediction in the direct-cascade range, $(1/2)\varepsilon r$ (red, dashed), as per Eq.~\eqref{eq:directcascade_WS}. The correlation function $-\la \omega^\prime \bar{\omega}_L \ra$ is fit by $\sim r^{2.24}$ at shorter distances and $\sim r^{0.840}$ at larger distances, with the transition occurring near $r/L_f \sim 2\times 10^{-2}$. (Right): The relativistic correlation function $\la T^\prime_{0T} T^T_L \ra$ (blue, solid) plotted with its prediction in the direct-cascade range, $(1/24)\varepsilon r^3$ (red, dashed), as per Eq.~\eqref{eq:directcascadeWS_weird}. The correlation function $\la T^\prime_{0T} T^T_L \ra$ is fit by $\sim r^{1.05}$ at shorter distances and $\sim r^{2.40}$ at longer distances, with the transition occurring near $r/L_f \sim 10^{-1}$. Error bars correspond to the statistical uncertainty $\sigma/\sqrt{N}$ for each value of $r/L_f$, where $N$ is the sample size and $\sigma$ is the sample standard deviation. Note that the statistical uncertainty at short distances is sufficiently large that the sign of the correlation functions is uncertain there. The corresponding compressible and incompressible limits of these correlation functions (obtained by setting $\gamma=1$ or $\gamma=\rho=1$, as in Sec.~\eqref{sec:inversecascaderesults}) do not behave significantly differently. This allows us to approximate $\varepsilon$ by its incompressible limit $((1+w)/w)^2 \rho^2 \epsilon_\omega$ as in Sec.~\eqref{sec:direct-cascade-limit}, and we take $\epsilon_\omega$ to be the initial enstrophy growth rate (before dissipation mechanisms become important). However, we substitute $\la \rho^2 \ra \sim 0.96$ in place of $\rho^2$, which slightly improves agreement with the predictions.} \label{fig:corrs_direct_cascade}
\end{figure}

In Fig.~\eqref{fig:corrs_direct_cascade}, we display the two measured correlation functions for which we have predictions in the direct cascade (Eqs.~\eqref{eq:directcascade_WS} and~\eqref{eq:directcascadeWS_weird}). Errors again correspond to the statistical uncertainty $\sigma/\sqrt{N}$. We find reasonable agreement in the case of Eq.~\eqref{eq:directcascade_WS} (Left), and less so in the case of Eq.~\eqref{eq:directcascadeWS_weird} (Right). The measured power laws are $r^{0.84}$ (Left) and $r^{2.4}$, as compared to the predictions of $r$ and $r^3$, respectively. However, we note that the non-trivial factors of $1/2$ (Left) and $1/24$ (Right) yield marked agreement in magnitude. We suspect that by decreasing contamination from our modified large- and small-scale dissipation mechanisms, agreement with our predictions would improve, since in our inverse-cascade simulations we found that removing large-scale dissipation altogether improved agreement with our predictions significantly. We also note that the statistical uncertainty at short length scales is large enough that the sign of the correlation functions is uncertain there. We estimate the quantity $\varepsilon$ by its incompressible limit $((1+w)/w)^2 \rho^2 \epsilon_\omega$, with a further substitution of $\la \rho^2 \ra = 0.96$ in place of $\rho^2$ (which improves agreement slightly). This estimate is justified by the fact that the correlation functions we measure do not differ significantly from their incompressible counterparts (i.e. setting $\rho=\gamma=1$) in this regime.

%
%
\chapter{Discussion \& Conclusions}\label{sec:P1discussion}
\begin{figure}[h!]
\centering
\hbox{\hspace{3.5cm}\includegraphics[width=0.6\textwidth]{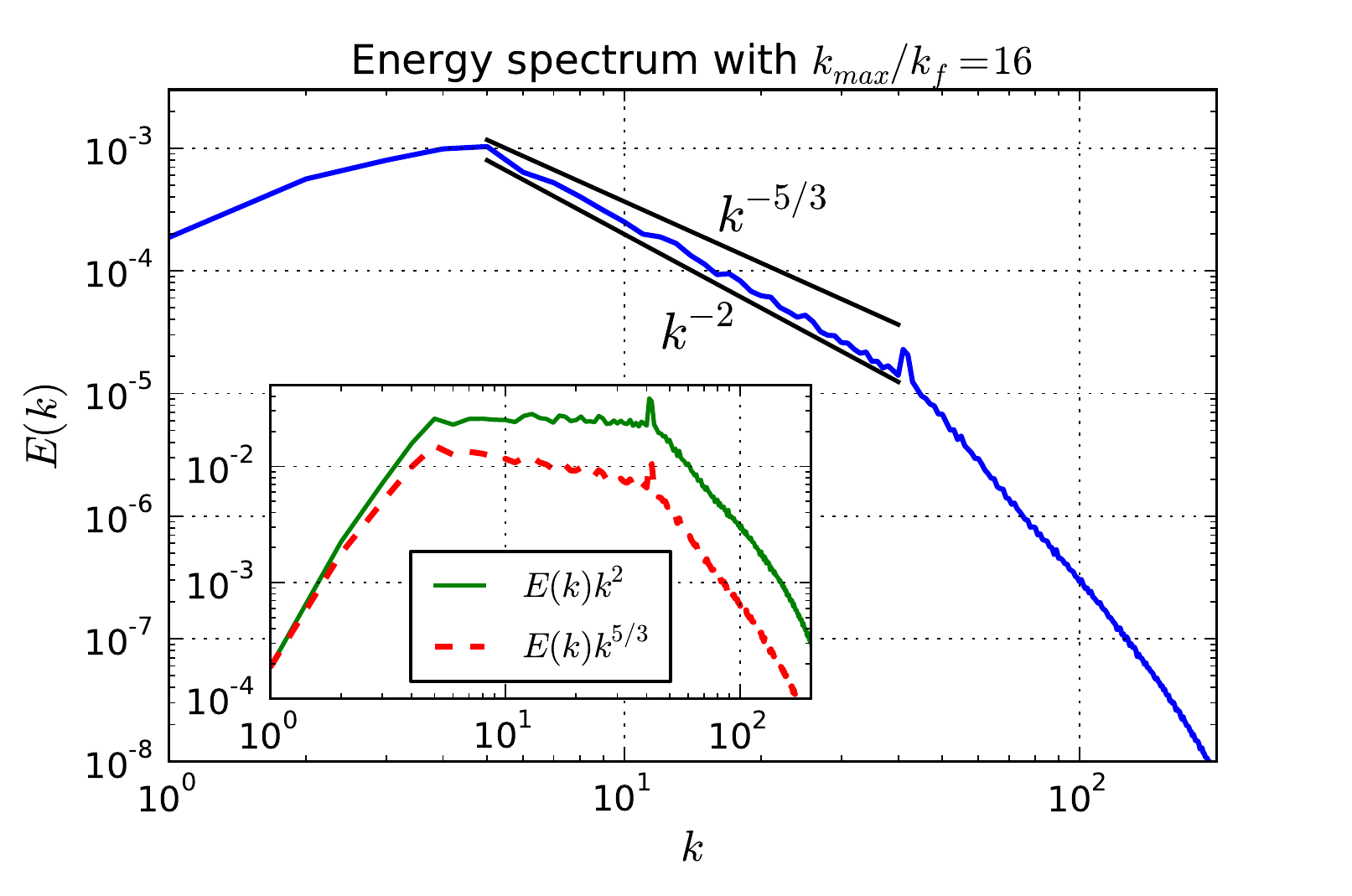}}
\caption{Energy spectrum $E(k)$ (blue, solid) with forcing active at $k_f=42$, such that $k_{max}/k_f = 16$, where $k_{max} = N/3$ and $N=2048$. Power laws $k^{-2}$ and $k^{-5/3}$ (black, solid) are shown for comparison. The inset displays compensated spectra $E(k)k^2$ (green, solid) and $E(k)k^{5/3}$ (red, dashed). The spectrum is well-represented by $k^{-2}$, rather than $k^{-5/3}$, consistent with~\cite{Scott:2007} (albeit for a conformal fluid in our case). The best fit power-law slope over the range $k\in [5,40]$ is $-2.047$.} \label{fig:k42_spectrum}
\end{figure}

As mentioned in Sec.~\eqref{sec:inversecascaderesults}, we observed that our energy spectra approach a $k^{-2}$ scaling in the inverse-cascade range. We point out that it was observed in~\cite{Scott:2007} that the incompressible case exhibits the same scaling provided large-scale friction is absent and the direct cascade is sufficiently resolved ($k_{\text{max}}/k_f \geq 16$, where we define $k_{\text{max}}=\mathcal{N}/3$). It thus becomes a prescient question whether a sufficiently resolved direct cascade in the conformal fluid case will yield the same result. To answer this, we perform an ensemble of $20$ simulations with $\mathcal{N}=2048$ and $k_{\text{max}}/k_f = 16$, with a forcing profile given by Eq.~\eqref{eq:spikeprofile}. The resulting energy spectrum is displayed in Fig.~\eqref{fig:k42_spectrum}, with the inset displaying the same spectrum compensated by either $k^{2}$ or $k^{5/3}$. After filtering out the modes $k \in [0,5]$ (i.e. all modes less than the maximum of the spectrum), the rms Mach number for this flow is $\sim 0.11$. We perform this filtering so as to have a more fair comparison of rms Mach number with our inverse-cascade simulations in Sec.~\eqref{sec:inversecascaderesults}, which we remind were $\sim 0.14$, $0.17$, and $0.19$. As evident from Fig.~\eqref{fig:k42_spectrum}, the spectrum clearly favours a $k^{-2}$ description in the inverse-cascade range, and not the Kolmogorov/Kraichnan $k^{-5/3}$ description. The best-fit power law over the range $k \in [5,40]$ yields $k^{-2.047}$. 

Interestingly, we note in passing that this $k^{-2}$ scaling of the energy spectrum would change the result of the purported calculation of the fractal dimension of a turbulent $(3+1)$-dimensional AdS-black brane presented in~\cite{Adams:2013vsa} to $D=3$ (rather than $D=3+1/3$). An analysis of this will be the central topic in Part~\eqref{sec:fracdim}.

We also point out that, despite the narrow inverse-cascade range, we nonetheless observe a similarly narrow $\sim r^{0.95}$ power law scaling in the same correlation functions analyzed in Sec.~\eqref{sec:inversecascaderesults} (not shown). This suggests that the scaling relation Eq.~\eqref{eq:inversecascade_WS} continues to hold with a more resolved direct-cascade range.

\begin{figure}[h!]
\centering
\hbox{\hspace{0cm}\includegraphics[width=\textwidth]{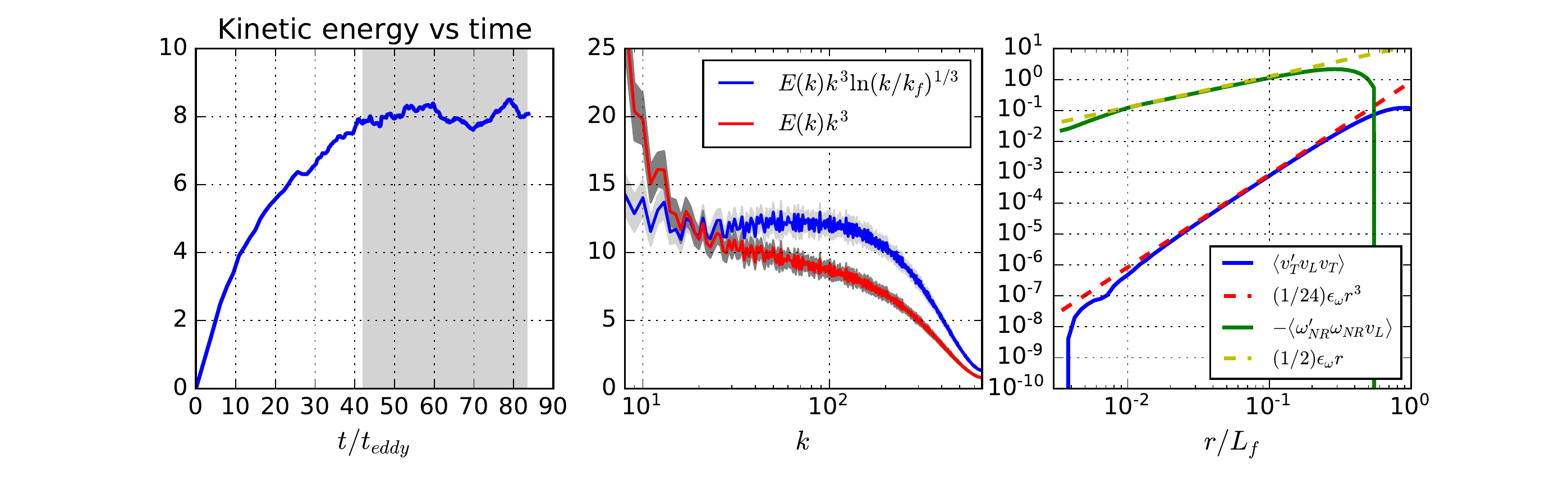}}
\caption{Data from our simulation of an incompressible Navier-Stokes fluid. (Left): The average specific Newtonian kinetic energy plotted as a function of time. The shaded interval corresponds to the interval over which we compute statistical averages. (Centre): The energy spectrum $E(k)$ compensated by either $k^3 \ln{k/k_f}^{1/3}$ or $k^3$. The shaded envelopes correspond to $5\times$ the statistical error $\sigma/\sqrt{N}$. The logarithmic correction is evidently favoured. (Right): The correlation functions (solid) and their predictions (dashed) corresponding to the incompressible counterparts of Eqs.~\eqref{eq:directcascade_WS} and~\eqref{eq:directcascadeWS_weird}.} \label{fig:NSstuff}
\end{figure}

In Sec.~\eqref{sec:directcascaderesults}, we observed hints of the predicted scaling of the correlation functions displayed in Fig.~\eqref{fig:corrs_direct_cascade}. By instead simulating an incompressible fluid (as per Sec.~\eqref{sec:implementation-NS}), for which shockwave phenomena are not present, we can measure the incompressible limits of Eqs.~\eqref{eq:directcascadeWS_limit} and~\eqref{eq:directcascadeWS_weird_limit} with greater statistical significance. Conformal fluids have been shown to possess a scaling limit to an incompressible Navier-Stokes fluid~\cite{Bhattacharyya2009,Fouxon2008conformal}. We present the results of our simulation in Fig.~\eqref{fig:NSstuff} (Right). The enstrophy injection rate is set a priori to $\epsilon_\omega = 28$. We use 4th-order hypofriction $-\lambda \nabla^{-4} \omega$ and hyperviscosity $-\nu_4 \nabla^4 \omega$, with $\lambda=0.15$ and $\nu_4 = 10^{-10}$. The forcing profile is given by 
\begin{eqnarray}
\la f_\omega f^\prime_\omega \ra \propto {\begin{cases}
 1 \:\:\:\: k \sim k_f \\
 0 \:\:\:\: \text{otherwise} \end{cases}},
\end{eqnarray}
and the injection scale is set to $k_f = 7$. The average specific kinetic energy is plotted as a function of time in Fig.~\eqref{fig:NSstuff} (Left), with the averaging interval shaded gray. The energy spectrum compensated by $k^3$ and $k^3\ln{(k/k_f)}^{1/3}$ is displayed in Fig.~\eqref{fig:NSstuff} (Centre), with the shaded envelopes indicating $5\times$ the statistical uncertainty $\sigma/\sqrt{N}$. The logarithmic correction is clearly favoured. In Fig.~\eqref{fig:NSstuff} (Right), we plot the correlation functions $\la v_T^\prime v_L v_T \ra$ and $-\la \omega^\prime_{\text{NR}} \omega_{\text{NR}} v_L \ra$, together with their respective predictions in the direct cascade $(1/24)\epsilon_\omega r^3$ and $(1/2)\epsilon_\omega r$. The agreement is very much improved over Fig.~\eqref{fig:corrs_direct_cascade}, which suggests that despite the low Mach number in our direct cascade simulation of the conformal fluid, that situation is nonetheless quite different from the incompressible case (at least insofar as the numerical challenges are greater in the former case, eg. small scales being contaminated by bottleneck effects).

Finally, at the beginning of this part in Fig.~\eqref{fig:allvorts} we display snapshots of the vorticity from several of our simulations. By doing so, we intend to provide intuition as to how the inverse and direct cascades appear in real space. In particular, the stretching and mixing of vorticity isolines characteristic of the direct-cascade range are readily identified as `turbulence' qualitatively, where coherent features are seen over a variety of scales (see Fig.~\eqref{fig:allvorts} (Bottom Left) and (Bottom Right)). By contrast, the inverse-cascade range has a much noisier appearance, as in Fig.~\eqref{fig:allvorts} (Top Left). We have not observed an explicit acknowledgement of this qualitative fact in the literature, since numerical studies of the inverse-cascade range seldom include plots of the vorticity (eg.~\cite{Boffetta:2000}). Unless the direct-cascade range is resolved, the turbulent flow qualitatively appears as random noise -- but even if it is resolved, a clear hierarchy of scales is not apparent in the inverse-cascade range. In Fig.~\eqref{fig:allvorts} (Top Right) we show a mixed case with the forcing acting at an intermediate scale $k_f=42$. This case is a simulation targeting the inverse-cascade range, but the direct-cascade range is just beginning to be resolved as well. Consequently, vorticity isoline mixing is beginning to be apparent, superposed on top of a more noisy structure.



\partkey{IB}
\part{Fractal dimension of turbulent black holes} \label{sec:fracdim}

\section*{Executive summary}

The possibility of turbulent ringing of black holes has spurred recent interest and investigations. One question is how can we characterize their states, and~\cite{Adams:2013vsa} proposed using the concept of \emph{fractal dimension}. We present numerical measurements of the fractal dimension of a turbulent asymptotically anti-deSitter black brane reconstructed from simulated boundary fluid data at the perfect fluid order using the fluid-gravity duality. We argue that the boundary fluid energy spectrum scaling as $E(k)\sim k^{-2}$ is a more natural setting for the fluid-gravity duality than the Kraichnan-Kolmogorov scaling of $E(k) \sim k^{-5/3}$, but we obtain fractal dimensions $D$ for spatial sections of the horizon $H\cap\Sigma$ in both cases: $D=2.584(1)$ and $D=2.645(4)$, respectively. These results are consistent with the upper bound of $D\leq 3$, thereby resolving the tension with the recent claim in~\cite{Adams:2013vsa} that $D=3+1/3$. We offer a critical examination of the calculation which led to their result, and show that their proposed definition of fractal dimension performs poorly as a fractal dimension estimator on $1$-dimensional curves with known fractal dimension. Finally, we describe how to define and in principle calculate the fractal dimension of spatial sections of the horizon $H\cap \Sigma$ in a covariant manner, and we speculate on assigning a `bootstrapped' value of fractal dimension to the entire horizon $H$ when it is in a statistically quasi-steady turbulent state.

\begin{figure}[h!]
\centering
\hbox{
\hspace{2cm}\includegraphics[width=0.7\textwidth]{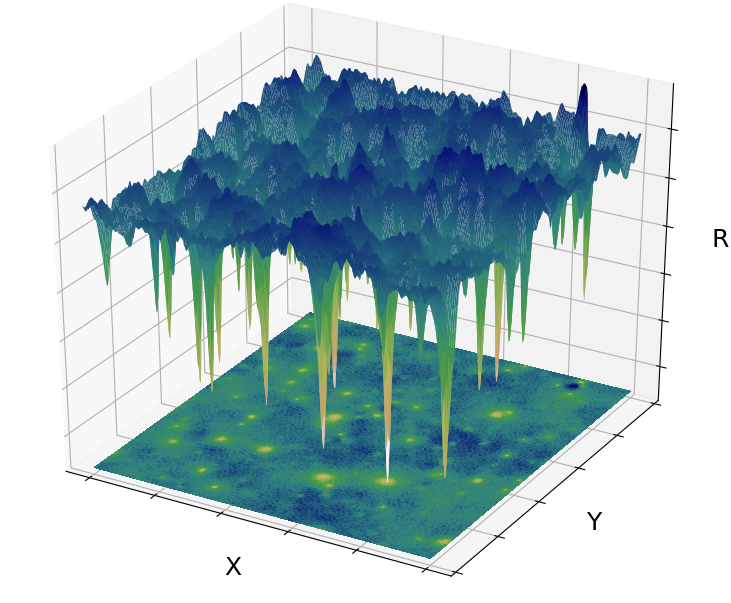}}
\raggedright

Figure IB: Snapshot of a reconstructed black brane horizon from boundary fluid data.
\label{fig:BHreconstructed}
\end{figure}

\chapter{Introduction}\label{sec:intro}
In a certain regime, the existence of turbulence in the gravitational field was recently demonstrated in numerical simulations of a perturbed black brane in asymptotically anti-deSitter (AAdS) spacetime in~\cite{Adams:2013vsa}. Such behavior was expected on the basis of the work of~\cite{Carrasco:2012nf} and the fluid-gravity duality, which gives an approximate dual description of the bulk geometry in terms of a conformal fluid living on the conformal boundary of the spacetime (see eg.~\cite{Bhattacharyya:2008jc,Bhattacharyya:2008mz, VanRaamsdonk:2008fp}, or~\cite{Hubeny:2011hd} for a review and further references). This duality has opened the door to cross-pollination between the fields of gravity and fluid dynamics (eg.~\cite{Oz:2010wz, Eling:2010vr,Carrasco:2012nf,Green:2013zba,Eling:2013sna,Adams:2012pj, Adams:2013vsa,adams2014dynamical,Eling:2015mxa,WS2015,JRWS:2017}), and even resulting in insights with potential relevance to gravitational wave astrophysics~\cite{Yang:2014tla}.

Interestingly, in~\cite{Adams:2013vsa} it was argued that a $(3+1)$-dimensional AAdS-black brane spacetime in a turbulent quasi-steady state has an event horizon with fractal dimension $D = 3 + 1/3$. Although the intersection of the horizon $H$ with a spacelike slice $\Sigma$ has dimension $2$, in a turbulent state one expects a bumpy horizon exhibiting approximate self-similarity over some range of scales, and therefore a fractal dimension $D$ in the range $2 \leq D \leq 3$. Since the result $D=3+1/3$ of~\cite{Adams:2013vsa} lies above this range, it is in tension with this basic expectation. Indeed, since $\Sigma$ is Riemannian and connected, it can be regarded as a metric space where its distance function is defined as the infimum of lengths of paths connecting any two points. Therefore, since $H\cap \Sigma$ is embedded in it, its fractal dimension cannot exceed the dimension of $\Sigma$~\cite{falconer1986geometry}.

In this work we begin in Chapter~\eqref{sec:background} with a review of the relevant calculation in~\cite{Adams:2013vsa}, then in Sec.~\eqref{sec:criticism} we provide a critical examination. We argue that their calculation does not use their proposed definition of fractal dimension, and so the fact that their result exceeds the upper bound $D=3$ does not necessarily invalidate their definition. Nonetheless, in Sec.~\eqref{sec:testcases} we argue using well-understood test cases of statistically self-similar 1-dimensional curves embedded in the Euclidean plane that their proposed definition of fractal dimension is not reliable as a fractal dimension estimator. Next, in Chapter~\eqref{sec:results} we present an alternative numerical calculation of the fractal dimension of a turbulent black brane using simulated data of the dual turbulent fluid and the fluid-gravity duality at lowest (perfect fluid) order. We do so over the inverse-cascade range of a weakly-compressible conformal fluid with two sets of data corresponding Kraichnan-Kolmogorov scaling of the energy spectrum $E(k) \sim k^{-5/3}$ as well as the scaling $E(k)\sim k^{-2}$ which emerges as the direct-cascade becomes well-resolved (and in the absence of large-scale friction)~\cite{Scott:2007,JRWS:2017}. \emph{We obtain fractal dimensions of $D\approx 2.58$ and $D\approx 2.65$ for each case, respectively, thus removing the aforementioned tension concerning the claim that $D=3+1/3$}. Lastly, in Sec.~\eqref{sec:discussion} we describe what would be required to define, and in principle compute, the fractal dimension covariantly.
%
%
\chapter{Background}\label{sec:background}
In this section we briefly review the argument presented in~\cite{Adams:2013vsa}, specializing to the case of a $(3+1)$-dimensional bulk spacetime. Further details can be found in that work.

In~\cite{Adams:2013vsa} it is proposed that the fractal dimension of the horizon be defined via the scaling of the horizon area course-grained on a scale $\delta x$. Writing the course-grained area as a Riemann sum of the intrinsic area elements $A \approx \Sigma_i \sqrt{\gamma(x_i)} \Delta^2 x_i$, with $\Delta^2 x_i \approx (\delta x)^2$ and $\gamma(x_i)$ the intrinsic metric determinant evaluated at the point $x_i$, one extracts the purported fractal dimension $D$ from the scaling $A \sim (\delta x)^{2-D}$. One can immediately see that this definition has some of the expected behaviour: i) if the intrinsic metric determinant is constant over the surface, then the course-grained area does not depend on $\delta x$ so we must have $D=2$ (a smooth surface), ii) as the surface becomes rough ($D>2$), the area grows faster as $\delta x$ decreases (and indeed becomes infinite as $\delta x \rightarrow \infty$).

The calculation of the fractal dimension begins by considering the congruence of null geodesics $n$ generating the horizon. This is described by the Raychaudhuri equation in vacuum, which governs the evolution of the horizon area element $\sqrt{\gamma}$,
\begin{eqnarray}
\kappa \mathcal{L}_n \sqrt{\gamma} + \frac{1}{2}\frac{1}{\sqrt{\gamma}} (\mathcal{L}_n \sqrt{\gamma})^2 - \mathcal{L}^2_n \sqrt{\gamma} = \sqrt{\gamma} \Sigma^i_j \Sigma^j_i,
\end{eqnarray}
where $n$ is the null normal to the horizon, $\mathcal{L}_n$ is a Lie derivative along $n$, $\kappa$ is the non-affinity of $n$ measured by the right-hand side of the geodesic equation $n_a \nabla^a n^b = \kappa n^b$, and $\Sigma^i_j$ is the shear. The regime of validity of the fluid-gravity duality is that of slowly-varying fields, so one expects the higher-order derivative terms $ - \mathcal{L}^2_n \sqrt{\gamma}$ and $\frac{1}{2}\frac{1}{\sqrt{\gamma}} (\mathcal{L}_n \sqrt{\gamma})^2$ to be subleading with respect to both the right-hand side $\sqrt{\gamma}\Sigma^i_j \Sigma^j_i$ and the lower-order derivative term $\kappa \mathcal{L}_n \sqrt{\gamma}$. Dropping those higher-order terms and integrating over the spatial section of the horizon $H\cap\Sigma$ yields an expression for the time rate of change of the horizon area $A$,
\begin{eqnarray}
\frac{dA}{dt} = \int d^2x \frac{\sqrt{\gamma}}{\kappa} \Sigma^i_j \Sigma^j_i = \int_0^\infty dk \mathcal{A}(t,k), \label{eq:integratearea}
\end{eqnarray}
where $\mathcal{A}$ is the isotropic power spectrum of the `rescaled' shear $\theta^i_j \equiv \sqrt[4]{\gamma/\kappa^2} \Sigma^i_{\phantom{i}j}$, and the last equality follows from the Plancherel theorem. In~\cite{Adams:2013vsa} it was observed in full numerical simulations of a (3+1)-dimensional turbulent AAdS-black brane spacetime that $\mathcal{A} \sim k^2 E(k)$, at least in the regime of the simulations, where $E(k)$ is the isotropic velocity power spectrum of the boundary fluid. By assuming Kraichnan-Kolmogorov scaling $E(k) \sim k^{-5/3}$ over the inverse-cascade range, one has $\mathcal{A} \sim k^{1/3}$ there. By inserting a large wavenumber cutoff $k_{\textrm{max}} \sim 1/\delta x$ in the wavenumber integral in Eq.~\eqref{eq:integratearea} one finds $dA/dt \sim k_{\textrm{max}}^{4/3} \sim (\delta x)^{-4/3}$ for $k_{\text{max}}$ in a sufficiently wide inverse-cascade range. Matching this to the scaling $(\delta x)^{2-D}$ finally yields $D=3+1/3$. 

One may object that the scaling $(\delta x)^{2-D}$ is to be applied to $A$, not $dA/dt$. However, in the quasi-steady state of a turbulent fluid forced at scale $k_f$ with inertial range scaling extending to a large scale $k_{\text{IR}}$, the spectrum $E(k)$ is well-approximated by piecewise power laws with the inertial range portion over $k \in (k_{IR},k_f)$ unchanging except that $k_{\text{IR}}$ decreases with time (see eg. \cite{fontane2013}). I.e. the inertial range becomes larger with time, but the spectrum over that range does not change. Plugging such a piece-wise power-law into the right-hand side of Eq.~\eqref{eq:integratearea} with UV cutoff $k_{\text{max}} \in (k_{IR},k_f)$ allows one to perform both the wavenumber and time integration explicitly. Thus the piecewise power-law model of $E(k)$ relevant to turbulent flows in quasi-steady state implies that $A$ and $dA/dt$ scale in the same way with the UV cutoff $k_{\text{max}}$, for $k_{\text{max}}$ sufficiently large. In the following section, we identify other possible sources of problems with the calculation.
%
%
\section{Critical examination of~\cite{Adams:2013vsa}}\label{sec:criticism}
We begin by noting what it means in position space to insert the small-scale cutoff $k_{\text{max}}$ in the wavenumber integral in Eq.~\eqref{eq:integratearea}. In order to do this we must write $\mathcal{A}(k,t)$ explicitly~\cite{Adams:2013vsa}:
\begin{eqnarray}
\mathcal{A}(t,k) \equiv \frac{\partial}{\partial k} \int_{|\boldsymbol{k}^\prime| \leq k} \frac{d^2k^\prime}{(2\pi)^2} \bar{\theta}^{*i}_{\phantom{*i}j} (t,\boldsymbol{k}^\prime) \bar{\theta}^j_{\phantom{j}i} (t,\boldsymbol{k}^\prime), \label{eq:curvaturespectrum}
\end{eqnarray}
where $\bar{\theta}^i_{\phantom{i}j} (t,\boldsymbol{k})$ is the Fourier transform of the rescaled shear, $\bar{\theta}^i_{\phantom{i}j} (t,\boldsymbol{k}) = \int d^2x e^{-i\boldsymbol{k} \cdot \boldsymbol{x}} \theta^i_{\phantom{i}j} (t,\boldsymbol{x})$. Eq.~\eqref{eq:curvaturespectrum} can be rewritten as
\begin{eqnarray}
\mathcal{A}(t,k) &=& \frac{\partial}{\partial k} \int_0^k dk^\prime \int_0^{2\pi} d\phi \frac{k^\prime}{(2\pi)^2} \bar{\theta}^{*i}_{\phantom{*i}j} (t,\boldsymbol{k}^\prime) \bar{\theta}^j_{\phantom{j}i} (t,\boldsymbol{k}^\prime) \nonumber\\
&=& \left[ \int_0^{2\pi} d\phi \frac{k^\prime}{(2\pi)^2} \bar{\theta}^{*i}_{\phantom{*i}j} (t,\boldsymbol{k}^\prime) \bar{\theta}^j_{\phantom{j}i} (t,\boldsymbol{k}^\prime) \right]_{k^\prime = k} \nonumber\\
&=& k \int_0^{2\pi} \frac{d\phi}{(2\pi)^2} \bar{\theta}^{*i}_{\phantom{*i}j} (t,\boldsymbol{k}) \bar{\theta}^j_{\phantom{j}i} (t,\boldsymbol{k}).
\end{eqnarray}
Therefore the wavenumber integral in Eq.~\eqref{eq:integratearea} is just the integral of $\bar{\theta}^{*i}_{\phantom{*i}j} \bar{\theta}^{j}_{\phantom{j}i}/(2\pi)^2$ over all of Fourier space. Furthermore, note that integrating over $k\in (0,k_{\text{max}})$ is the same as multiplying by a step function kernel $\Theta(k_{\text{max}}-k)$ and then integrating over $k \in (0,\infty)$, and writing it in this way allows us to see the meaning of the cutoff in position space as follows:
\begin{eqnarray}
&\phantom{=}& \int \frac{d^2k}{(2\pi)^2} \bar{\theta}^{*i}_{\phantom{*i}j} (t,\boldsymbol{k}) \bar{\theta}^j_{\phantom{j}i} (t,\boldsymbol{k}) \Theta(k_{\text{max}}-k) \nonumber\\
&=& \int \frac{d^2k}{(2\pi)^2} \left( \int d^2x\; e^{i\boldsymbol{k} \cdot \boldsymbol{x}} \theta^{i}_{\phantom{i}j} (t,\boldsymbol{x}) \right) \left( \int d^2x^\prime e^{-i\boldsymbol{k} \cdot \boldsymbol{x}^\prime} \theta^j_{\phantom{j}i} (t,\boldsymbol{x}^\prime) \right) \Theta(k_{\text{max}}-k) \nonumber\\
&=& \int d^2x\; d^2x^\prime \left[ \int \frac{d^2k}{(2\pi)^2} \Theta(k_{\text{max}}-k) e^{-i\boldsymbol{k} \cdot (\boldsymbol{x} - \boldsymbol{x}^\prime)} \right] \theta^{i}_{\phantom{i}j}(t,\boldsymbol{x}) \theta^j_{\phantom{j}i}(t,\boldsymbol{x}^\prime) \nonumber\\
&=& \int d^2x\; d^2x^\prime k_{\text{max}} \frac{J_1(\pi k_{\text{max}} |\boldsymbol{x}-\boldsymbol{x}^\prime | )}{|\boldsymbol{x}-\boldsymbol{x}^\prime |} \theta^{i}_{\phantom{i}j}(t,\boldsymbol{x}) \theta^j_{\phantom{j}i}(t,\boldsymbol{x}^\prime) \nonumber\\
&\equiv & \int d^2x\; \theta^i_{\phantom{i}j} (t,\boldsymbol{x}) \left\langle \theta^j_{\phantom{j}i} (t,\boldsymbol{x}) \right\rangle_{\delta x}, \label{eq:meaningpositionspace}
\end{eqnarray}
where $J_1$ is the Bessel function of the first kind, and we have defined $\left\langle \cdot \right\rangle_{\delta x}$ as a spatial coarse-graining operation at scale $\delta x \sim 1/k_{\text{max}}$ (in this case with an isotropic kernel $k_{\text{max}} J_1(\pi k_{\text{max}} |\boldsymbol{x}-\boldsymbol{x}^\prime | ) / |\boldsymbol{x}-\boldsymbol{x}^\prime |$).

We thus arrive at our first concern: the relationship between Eq.~\eqref{eq:meaningpositionspace} and the proposed coarse-graining $A \approx \Sigma_i \sqrt{\gamma(x_i)} (\delta x)^2$ is unclear. The latter is a Riemann sum, which if applied to the Raychaudhuri Eq.~\eqref{eq:integratearea} would yield $\Sigma_i \theta^j_{\phantom{j}l}(x_i) \theta^l_{\phantom{l}j}(x_i) (\delta x)^2 $. The summand could be viewed as a coarse-graining of both factors of the rescaled shear $\theta$, whereas in Eq.~\eqref{eq:meaningpositionspace} only one factor of $\theta$ is coarse-grained. Thus, even if the scaling $A \approx \Sigma_i \sqrt{\gamma(x_i)} (\delta x)^2 \sim (\delta x)^{2-D}$ correctly captures the fractal dimension $D$, it is unclear whether the calculation performed in~\cite{Adams:2013vsa} uses it.

%
%
\subsection{Comparing methods on $1$-dimensional test cases}\label{sec:testcases}

Next, we argue that the Riemann sum approach $A \approx \Sigma_i \sqrt{\gamma(x_i)} (\delta x)^2 \sim (\delta x)^{2-D}$ is a poor fractal dimension estimator. We consider the 1-dimensional version of this, $L \approx \Sigma_i \sqrt{\gamma(x_i)} \delta x \sim (\delta x)^{1-D}$, applied to three different noise curves in the Euclidean plane whose fractal dimensions are known. We refer to this proposed method of determining the fractal dimension of a curve as the `intrinsic metric method'. Despite a strong resemblance, the intrinsic metric method of approximating the length of the curve is distinct from the `compass' or `ruler' method appearing in the pioneering study of coastline lengths~\cite{mandelbrot1967long}, since the former involves approximating the curve by its tangents at the points $x_i$, which are line segments of unequal length and whose end points do not necessarily lie on the curve. Each curve is defined by a function $f(x)$, and therefore has an intrinsic metric induced by the Euclidean metric of its embedding space whose determinant is $1+(\partial_x f)^2$. Thus we can compute coarse-grained versions of the length of the curve as $L \approx L_{\delta x} \equiv \Sigma_i \sqrt{1+(\partial_x f)^2}|_{x=x_i} \delta x$ and then compare with the expected scaling $(\delta x)^{1-D}$. For comparison, we estimate the fractal dimension of the same curves using the madogram method described in~\cite{gneiting2012estimators}. The madogram is defined as $\gamma_1(r) = (1/2) \left\langle \vert f(x) - f(x+r)\vert \right\rangle$, where $\left\langle \cdot \right\rangle$ denotes a spatial average. The madogram is expected to scale as $r^{2-D}$.

\begin{figure}[h!]
\centering
\hbox{\hspace{0cm}\includegraphics[width=\textwidth]{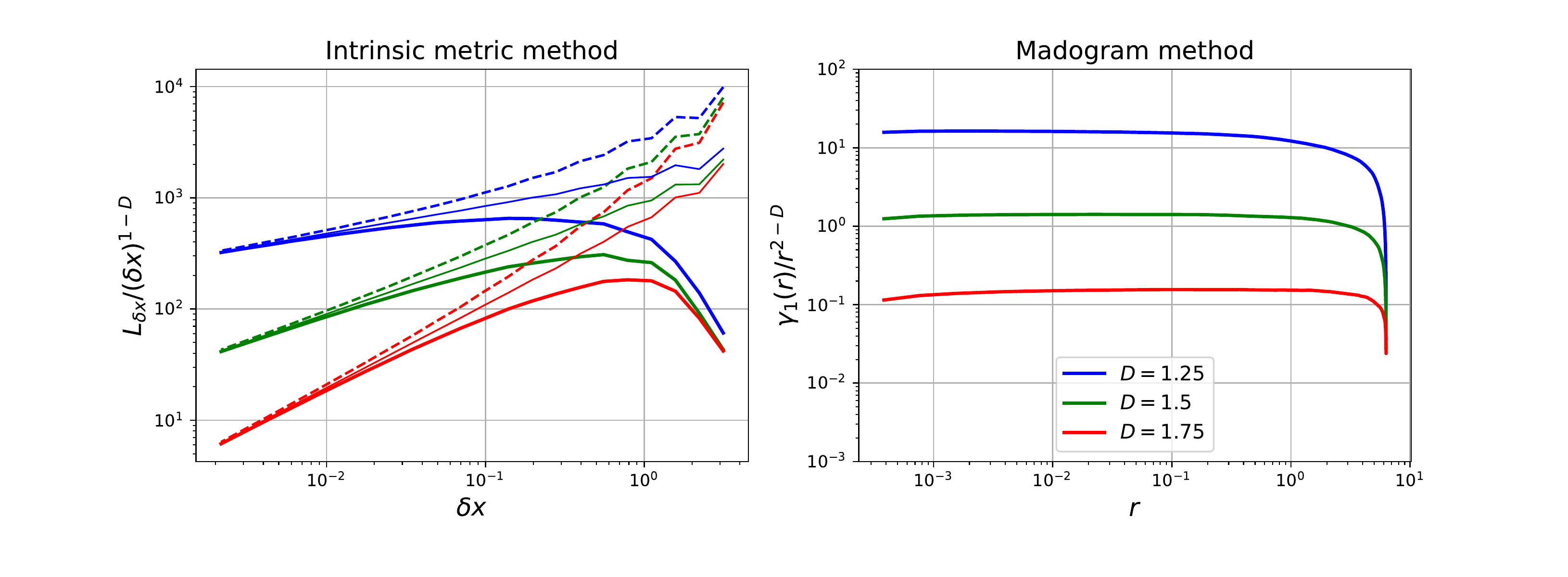}}
\caption{A comparison between the intrinsic metric and madogram methods for estimating the fractal dimension of $1$-dimensional noise with different fractal dimensions $D=1.25$ (blue), $D=1.5$ (green), $D=1.75$ (red). (Left): The coarse-grained length of each noise curve as a function of the coarse-graining scale $\delta x$, using the intrinsic metric method. Each plot is compensated by the expected scaling $(\delta x)^{1-D}$. The thick solid line corresponds to taking the minimum length over $10^3$ shifts of the sampling positions, while the thin solid line corresponds to taking the maximum, and the dashed line corresponds to taking the median. There is no discernible range of $\delta x$ over which the expected scaling is observed, so we conclude that this is method is not an accurate fractal dimension estimator. (Right): By contrast, the madogram $\gamma_1 (r)$ plotted as a function of $r$, compensated by the expected scaling $r^{2-D}$, for the same three noise curves. The expected scaling is clearly evident over a wide range of $r$.} \label{fig:noise_test}
\end{figure}

\begin{figure}[h!]
\centering
\hbox{\hspace{0.9cm}\includegraphics[width=0.9\textwidth]{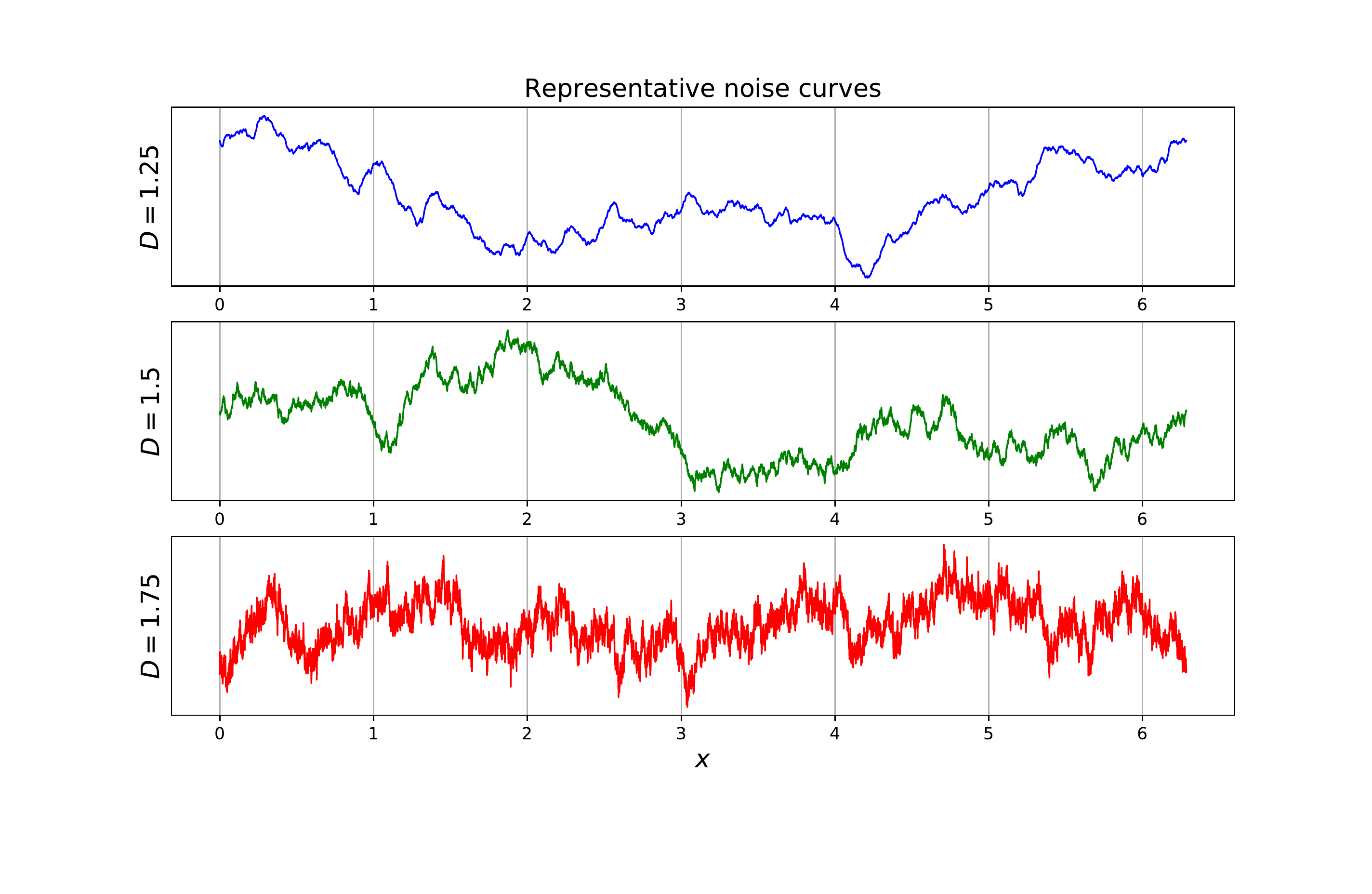}}
\caption{Representative noise curves with fractal dimensions $D=1.25$ (Top), $D=1.5$ (Middle), $D=1.75$ (Bottom). The $D=1.5$ case corresponds to Brownian noise. These curves have power spectra scaling as $k^{-\beta}$ for $\beta = 2.5$, $\beta = 2$, and $\beta = 1.5$, respectively.} \label{fig:noise_curves}
\end{figure}

Fig.~\eqref{fig:noise_test} shows a comparison between the intrinsic metric and madogram methods for estimating the fractal dimension of three noise curves with $D=1.25$ (blue), $D=1.5$ (green), $D=1.75$ (red). Such noise curves have power spectra scaling as $k^{-\beta}$ for $\beta = 2.5$, $2$, $1.5$, respectively. For the range $\beta \in [1,3]$ a topologically $d$-dimensional surface has a fractal dimension $D$ related to the spectral exponent by the approximate relation $D=(2d+3-\beta)/2$~\cite{voss1986}. For $\beta \geq 3$ the surface is sufficiently smooth that the fractal dimension equals its topological dimension, $D=d$, whereas for $\beta \leq 1$ it saturates to $D=d+1$.  In Fig.~\eqref{fig:noise_curves} we display representative curves with fractal dimensions $D=1.25$ (blue, Top), $D=1.5$ (green, Middle), and $D=1.75$ (red, Bottom). 

An ensemble of size $N=100$ is generated for each noise curve, and the estimators $L_{\delta x}$ and $\gamma_1 (r)$ are computed for each member and then averaged over the ensemble. In the intrinsic metric method, it is insufficient to attempt a single set of sampling locations for a given $\delta x$. Instead, we start with the zeroth set of sampling locations $\{ x_i \} = \{0, \delta x, 2\delta x, ... \}$, but also try $999$ additional sets related to the first by a translation $(m/1000)\delta x$ for the $m$th set, for a total of $1000$ sets. This results in $1000$ length estimates $L_{\delta x}$ for each $\delta x$, and we consider taking the minimum, maximum, or median length estimates, shown in Fig.~\eqref{fig:noise_test} (Left) in thick solid, thin solid, and dashed curves, respectively. Such ``shifts'' are an essential part of many fractal dimension estimating algorithms. Box-counting, for example, requires finding the \emph{minimum} number of boxes that cover the object, so many shifts of the box grid must be tried in order to obtain an accurate estimate. \emph{A priori} we do not know whether to take the minimum, maximum, or median estimate of the length $L_{\delta x}$. Different methods for estimating the fractal dimension have different conventions, for example the `compass' or `ruler' dimension~\cite{mandelbrot1967long} takes the maximum length, `box-counting' takes the minimum number of boxes~\cite{barnsley2014fractals}, and 'line transect variogram' methods applied to a surface take the median result from the transects~\cite{gneiting2012estimators}. However, as Fig.~\eqref{fig:noise_test} (Left) shows, none of the three possibilities yield the expected scaling $(\delta x)^{1-D}$ over any discernable range of $\delta x$. We note that taking the average or the median yields nearly identical curves (thin solid).

%
%
\chapter{Results}\label{sec:results}
Using the numerical code described in~\cite{JRWS:2017} and Part~\eqref{part:I}, we evolve a $(2+1)$-dimensional conformal perfect fluid with equation of state $P=\rho/2$ on a $2\pi$-periodic domain with $2048^2$ points. The energy momentum tensor of the fluid is $T_{ab} = (3/2)\rho u_a u_b + (1/2)\rho \eta_{ab}$, with $u^a = \gamma (1, \boldsymbol{v})$ and $\gamma$ the Lorentz factor. The fluid is evolved from rest $\rho=1$, $\boldsymbol{v}=0$, and turbulence is induced and sustained by a random external force with homogeneous, isotropic, Gaussian white-noise-in-time statistics. The external force has support in a narrow band of wavenumbers around $k_f$. Further details can be found in~\cite{JRWS:2017}.

\begin{figure}[h!]
\centering
\hbox{\hspace{0.7cm}\includegraphics[width=0.9\textwidth]{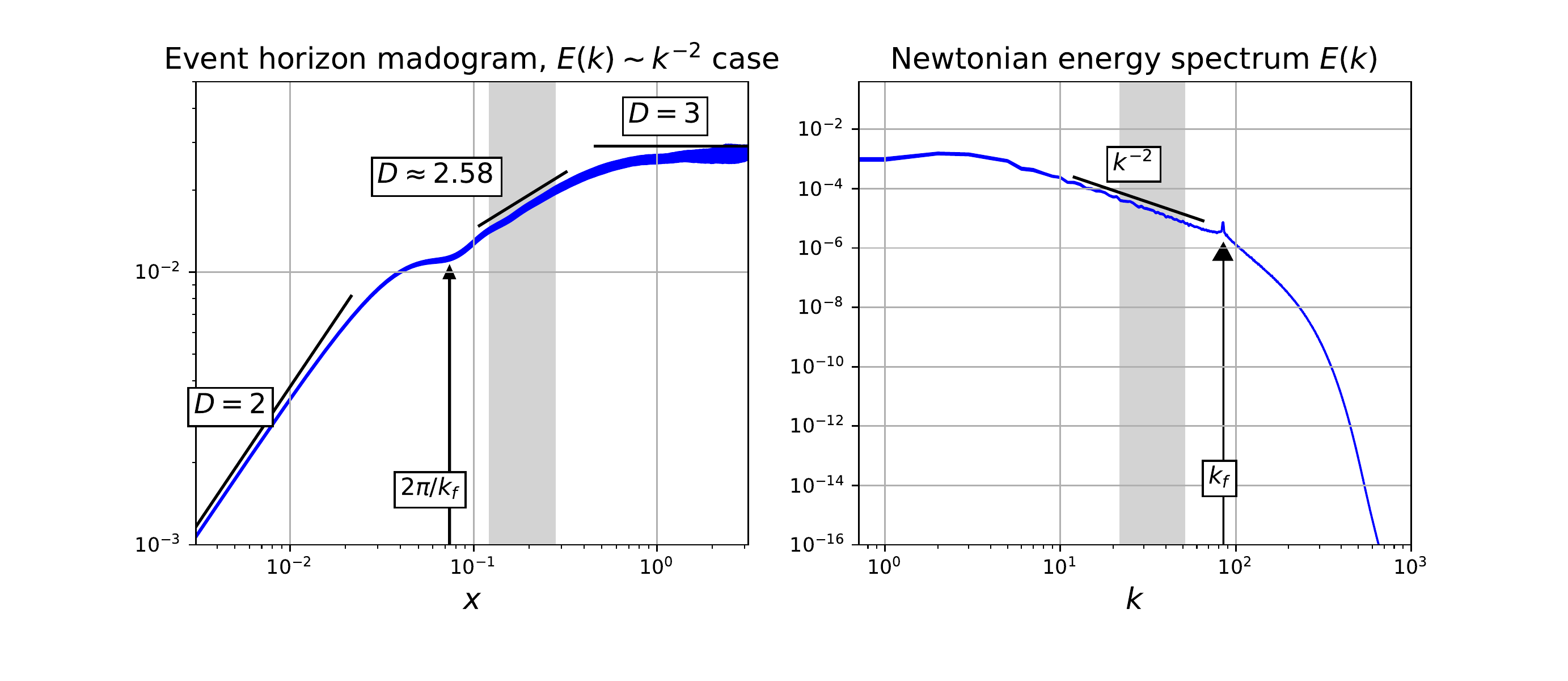}}
\caption{Event horizon madogram (Left) and corresponding boundary fluid isotropic Newtonian kinetic energy spectrum (Right) for the case with $k_f = 85$. The thickness of each plot corresponds to the $\sqrt{N}$ statistical uncertainty. (Left): The madogram yields a fractal dimension of $2$ at small scales, thus agreeing with the topological dimension. This is expected since the horizon is not a true fractal, i.e. it does not exhibit rough structure down to arbitrarily small scales. Above the forcing scale $2\pi/k_f$, a scaling range is observed with $2.584(1)$, where we have indicated the statistical uncertainty in brackets $()$. The range of $x$ over which we fit a power-law is indicated as the shaded grey region, and the corresponding range of wavenumbers is also indicated (Right). At the largest scales, the madogram saturates to $D=3$, which corresponds to the flow resembling white noise there (i.e. $E(k) \sim$ constant). (Right): The isotropic Newtonian specific kinetic energy spectrum $E(k) = \pi \left\langle |\hat{v}|^2 \right\rangle (k)$. A power-law of $k^{-2}$ is shown for reference, and the forcing scale $k_f$ is indicated with an arrow.} \label{fig:scott_madogram}
\end{figure}

\begin{figure}[h!]
\centering
\hbox{\hspace{0.7cm}\includegraphics[width=0.9\textwidth]{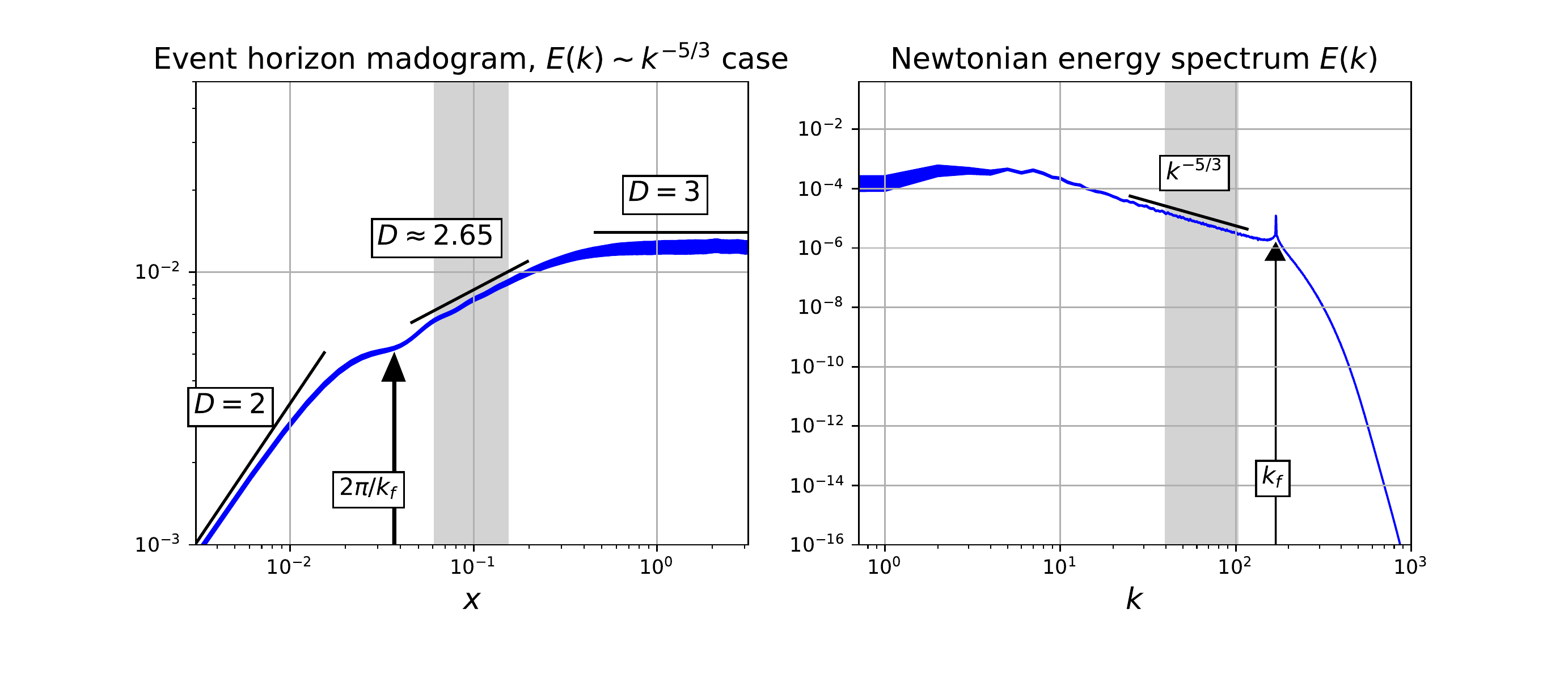}}
\caption{The corresponding plots as in Fig.~\eqref{fig:scott_madogram}, but for the boundary fluid exhibiting Kraichnan-Kolmogorov scaling of the energy spectrum, $E(k) \sim k^{-5/3}$. In this case, the measured fractal dimension is $D=2.645(4)$ over the inverse-cascade range. This value is slightly higher than the case with $E(k) \sim k^{-2}$, which is expected since the flatter spectrum indicates rougher structure. }\label{fig:kraichnan_madogram}
\end{figure}

An inverse-cascade range develops, and since we do not implement any large-scale energy sinks, the resulting flow is referred to as being in a quasi-steady state~\cite{Scott:2007}. For two separate cases with $k_f = 85$ and $k_f = 170$, we generate an ensemble of $20$ flows and perform analysis on snapshots prior to the energy piling up at the scale of the box. As displayed in Fig.~\eqref{fig:scott_madogram} (Right), the $k_f=85$ case yields an isotropic Newtonian specific kinetic energy spectrum $E(k) \sim k^{-2}$, as found in~\cite{Scott:2007} in the incompressible case and confirmed in~\cite{JRWS:2017} for a conformal fluid in the weakly-compressible regime.\footnote{It was found in~\cite{Scott:2007} that the spectrum steepens to $\sim k^{-2}$ when $k_{\text{max}}/k_f \gtrsim 16$, where $k_{\text{max}} \equiv N/3$ and $N$ is the number of points on the grid. In our simulations this would correspond to $k_f \approx 41$, but in their case regular 2nd-order viscosity was used, whereas we use 4th-order dissipation. Thus, we are able to achieve the $k^{-2}$ spectrum with a much larger $k_f$ because our dissipation operates at larger wavenumbers.} The $k^{-2}$ scaling is associated with both a well-resolved direct cascade and an absence of large-scale friction. Since the regime of validity of the fluid-gravity duality is that of an arbitrarily high Reynolds number and no large-scale friction, we argue that this spectrum corresponds to the natural setting for the dual spacetime. However, for comparison we also consider the $k_f = 170$ case, where the force is active deeper into the dissipation range, and which yields the traditional Kraichnan-Kolmogorov scaling $E(k) \sim k^{-5/3}$, as displayed in Fig.~\eqref{fig:kraichnan_madogram}.

We applied the madogram method to $x$- and $y$-transects of the event horizon. The fractal dimension $D$ of the horizon is then obtained by extracting $D_{\text{transect}} \in [1,2]$ from the median madogram of each topologically $1$-dimensional transect, and then writing $D = D_{\text{transect}} +1$. Such a prescription is valid for surfaces exhibiting statistical self-similarity, and its performance was evaluated extensively in~\cite{gneiting2012estimators}. In Figs.~\eqref{fig:scott_madogram} and~\eqref{fig:kraichnan_madogram} (Left) we display the median madogram over all transects of the horizon (herein referred to as the `event horizon madogram'), as applied to the radial coordinate position of the event horizon, $r_{+}(x^c) = 4\pi T(x^c)/3$, for the perturbed boosted AAdS-black brane metric at perfect fluid order,
\begin{eqnarray}
 ds^2 &=& -2 u_a(x^c) dx^a dr  -  \frac{r^2}{R^2}(1  -  \frac{r_{+}^3(x^c)}{r^3})u_a(x^c) u_b(x^c) dx^a dx^b \nonumber\\
 &+& \frac{r^2}{R^2}(\eta_{ab} + u_a(x^c) u_b(x^c)) dx^a dx^b,
\end{eqnarray} 
where the indices $(a,b)$ run over the `boundary' directions $(t,x,y)$ only, $R$ is the AdS length scale (which we set to 1), $u_a$ is the boost $4$-velocity, and $\eta_{ab}$ is the $(2+1)$-dimensional Minkowski metric. For the $(2+1)$-dimensional boundary conformal fluid, $T = \rho^{1/3}$. The perturbations are imagined to be slowly-varying with respect to the boundary directions, which will solve Einstein's equations with arbitrary accuracy in the perfect fluid limit if $u^a$ and $T$ evolve according to conformal hydrodynamics on the boundary. Error estimates have been obtained for solutions constructed from particular boundary fluid data in~\cite{Adams:2013vsa} via direct comparison with full GR simulations, showing agreement at the $1\%$ level (see also~\cite{adams2014dynamical} for error estimates which do not use full GR simulations). 

Fig.~\eqref{fig:scott_madogram} shows the case with $E(k) \sim k^{-2}$ and Fig.~\eqref{fig:kraichnan_madogram} shows the case with $E(k) \sim k^{-5/3}$. The thickness of each plot indicates the $\sqrt{N}$ statistical uncertainty associated with the ensembles. At small scales $x \ll 2\pi/k_f$, the horizons have fractal dimension $2$, which agrees with their topological dimension. This is expected since rough structure does not persist down to arbitrarily small scales. For a range of scales greater than the forcing scale $2\pi/k_f$, power-law behavior is observed in both cases. A least-squares power-law fit over the grey shaded intervals yield a fractal dimension of $D=2.584(1)$ and $D=2.645(4)$ for the cases $E(k) \sim k^{-2}$ and $E(k) \sim k^{-5/3}$, respectively, with $\sqrt{N}$ uncertainties indicated. The corresponding fitting interval in Fourier space is indicated on the plots of the energy spectra (Right). The madograms saturate at $D=3$ at large scales, beyond the inertial range scale, which is due to the flow resembling white noise at those scales (i.e. $E(k) \sim$ constant).

%
%
\section{Covariant construction of fractal dimension}\label{sec:discussion}

Many methods exist for calculating the fractal dimension of a set $F$ embedded in an ambient metric space $(\mathcal{M},d)$, where $d$ is a distance function $d: \mathcal{M}\times\mathcal{M} \rightarrow \mathbb{R}$ which is symmetric $d(x,y)=d(y,x)$ and satisfies $d(x,y) = 0 \Leftrightarrow x=y$ and the triangle inequality $d(x,y)+d(y,z) \geq d(x,z)$.  See eg.~\cite{gneiting2012estimators} for a comparison of many fractal dimension estimator algorithms when the metric space is Euclidean, or~\cite{falconer2004fractal,barnsley2014fractals} for strictly mathematically equivalent definitions. For illustrative purposes, we will focus on the box-counting method in this section, where one defines $N(\epsilon)$ to be the minimum number of boxes of size $\epsilon$ in the embedding space required to completely cover the set $F$, and then computes the fractal dimension as $\lim_{\epsilon \rightarrow 0} \log{(N(\epsilon))}/\log{(1/\epsilon)}$. The box-counting method is known to be diffeomorphism-invariant but not homeomorphism-invariant \cite{ott1984dimension} in the strict $\epsilon \rightarrow 0$ limit. The covering need not use boxes; indeed, any sets $U_i$ of diameter $\epsilon \equiv |U_i| = \sup \{ d(x,y): x,\: y\in U_i \} $ yield the same result~\cite{falconer1986geometry}.

In practical applications one does not take the $\epsilon \rightarrow 0$ limit, but instead fits a power-law to $N(\epsilon)\sim \epsilon^{-D}$ over some finite range $\Delta\epsilon = (\epsilon_{\text{UV}}, \epsilon_{\text{IR}})$. If the covering sets are not constructed covariantly, then any such fitting over $\Delta \epsilon$ would be subject to coordinate ambiguity, since the set in question could be made to appear smooth over the scale $\Delta \epsilon$ via a judicious choice of coordinates. In the example of a Brownian noise curve with $D=1.5$, described by $f(x)$ in Cartesian coordinates in Euclidean space, one could make the coordinate transformation ($\tilde{y} = \left\langle f(x) \right\rangle_{\Delta \epsilon}$, $\tilde{x} = x$), where $\left\langle f(x) \right\rangle_{\Delta \epsilon}$ is $f(x)$ with all modes outside the range of scales $\Delta \epsilon$ filtered out. Counting coordinate boxes over the scales $\Delta \epsilon$ would then give the incorrect result $D\approx 1$. Thus, it is important to construct the covering sets in a diffeomorphism-invariant way when performing a fit over the range of scales $\Delta \epsilon$.

When defining fractal dimension over a Riemannian manifold, one natural covariant choice of covering sets are geodesic balls $B(x,\epsilon/2)$, constructed by taking the union of all geodesics of length $\epsilon/2$ emanating from the point $x$. The event horizon $H\cap \Sigma$ on the slice $\Sigma$ is embedded isometrically both in $(\Sigma,h)$ and $(\mathcal{M},g)$, where $g$ is the full spacetime metric and $h$ is the induced metric on $\Sigma$. Since geodesic paths in $\Sigma$ need not be geodesic in $\mathcal{M}$, one is faced with the choice of whether to cover $H\cap\Sigma$ with geodesic balls in $\Sigma$ or in $\mathcal{M}$\footnote{Since $\mathcal{M}$ has a Lorentzian metric signature, a geodesic ball as defined above would contain the entire light cone of the central point $x$ since the length along null paths is zero. Thus, in this case we can instead define the geodesic ball $B_{\text{SL}}(x,\epsilon/2)$ as the union of only those spacelike geodesic paths of length $\epsilon/2$ emanating from $x$ which intersect $H\cap\Sigma$ at a point $y\neq x$.}. But note that the slice $\Sigma$ itself could be deformed at points off of $H\cap\Sigma$ to yield different geodesic paths while sharing the point set $H\cap \Sigma$. Thus, constructing the geodesic balls in $\Sigma$ would yield a fractal dimension which is not solely a property of $H\cap\Sigma$, but rather dependent on the arbitrary choice of slicing away from $H\cap\Sigma$. For this reason, we advocate using geodesic balls constructed in the full spacetime $\mathcal{M}$ (with a suitable redefinition - see footnote$^2$).

Computed in this way, a geodesic ball-counting procedure over a range of scales $\Delta \epsilon$ would yield a fully covariant estimate of the fractal dimension of any given spatial section of the event horizon. Furthermore, recall that our ensembles of event horizons considered in Chapter~\eqref{sec:results} yield roughly the same fractal dimension. It is often observed that cross-sections of $D$-dimensional fractals or statistically self-similar objects themselves have a fractal dimension of $D-1$ (see eg.~\cite{geologybook} for geological examples). Given our measurement of $D_{H\cap\Sigma}\approx 2.58$ in Chapter~\eqref{sec:results} for the $E(k)\sim k^{-2}$ case, this suggests that in a quasi-steady turbulent state the entire horizon $H$ can be assigned a fractal dimension of $D_H \approx 3.58$ (or $D_H \approx 3.65$ for the Kraichnan-Kolmogorov case $E(k)\sim k^{-5/3}$). However, the mathemetical meaning of this is not clear since $H$ is a null hypersurface embedded in a Lorentzian manifold $\mathcal{M}$, so there is difficulty in defining the diameter of covering sets in a covariant way.

Numerically implementing the procedure described in this section would be expensive, since one would have to integrate a large number of geodesics from a given point to construct a geodesic ball, do so for many geodesic balls to find a covering, and do this for many possible coverings to find the minimal one. In the current work we have not followed a covariant procedure like this. Many others have not either (eg.~\cite{cornish1996time,cornish1996chaos,cornish1997mixmaster,frolov1999chaotic,lehner2011final}), some opting instead to point out the diffeomorphism-invariance of box-counting in the $\epsilon \rightarrow 0$ limit while only fitting over a finite range $\Delta \epsilon$. It would be interesting to see how much these results change when done covariantly. 

Alternatively, it is plausible that the fractal dimension will not depend sensitively on the embedding space if, in a region around the surface, one has well-separated scales over which the surface and the embedding space vary. If this is true, once could obtain an approximate covariant result by embedding the surface isometrically in Euclidean space, and then applying a standard fractal dimension estimator. We attempted to embed the turbulent horizon isometrically in $\mathbb{E}^3$, without success. Indeed, the existence of such a (global) embedding is only guaranteed if the Gaussian curvature is positive over the entire surface, and may or may not exist otherwise. It has been observed~\cite{nollert1996visualization} that even a sufficiently rapidly-rotating Kerr black hole horizon does not have a global embedding in $\mathbb{E}^3$, since the Gaussian curvature becomes negative at the poles. We have computed the Gaussian curvature using our fluid data from Chapter~\eqref{sec:results}, and observed that it changes sign over the domain as rapidly as the external force. Thus, we believe it is highly unlikely that there exists a global embedding into $\mathbb{E}^3$ for arbitrary turbulent horizons in the regime of the fluid-gravity duality, although Euclidean embeddings are guaranteed to exist in sufficiently high dimensions.

%
%

\chapter{Conclusions}
In this work we provided a critical examination of the calculation in~\cite{Adams:2013vsa} which led to the claim that topologically $d$-dimensional turbulent AAdS-black brane horizons $H\cap\Sigma$ embedded in a $(d+1)$-dimensional Riemannian space $\Sigma$ have a fractal dimension $D=d+4/3$, exceeding the upper bound of $d+1$. We offered an alternative numerical computation of $D$ when $d=2$, and discussed issues surrounding the covariance of that quantity. 

In particular, we argued using well-understood test cases of $1$-dimensional noise curves that the proposed definition of fractal dimension in~\cite{Adams:2013vsa}, $A_{\delta x} = \sum_i \sqrt{\gamma(\boldsymbol{x}_i)} (\delta x)^2 \sim (\delta x)^{2-D}$, when specialized to topologically $1$-dimensional objects, performs poorly as a fractal dimension estimator. We emphasize that this is not a proof that the definition fails in the strict $\delta x \rightarrow 0$ limit for genuine fractals, but since the proposed application is on statistically self-similar surfaces which do not exhibit rough structure down to arbitrarily small scales, the performance of this proposal as a fractal dimension estimator is relevant. Furthermore, we argued that the calculation in~\cite{Adams:2013vsa} may not be using their proposed definition at all (so their result of $D=d+4/3$ alone does not necessarily invalidate their proposed definition, hence our separate evaluation of the definition on noise curves of known fractal dimension).

Using simulated turbulent conformal fluid flows in the quasi-steady state regime, we constructed snapshots of the turbulent event horizon using the fluid-gravity duality at perfect fluid order. By applying a line transect madogram method~\cite{gneiting2012estimators} to the event horizon surface $r_{+}(x,y) = 4\pi T(x,y)/3$ in boosted ingoing Finkelstein coordinates, we obtained a fractal dimension for spatial sections of the horizon $H\cap\Sigma$ of $D=2.584(1)$ and $D=2.645(4)$ for the cases with the boundary spectrum $E(k)\sim k^{-2}$ and $E(k)\sim k^{-5/3}$, respectively. We argued that the former scaling, $E(k) \sim k^{-2}$, is a more natural setting for the fluid-gravity duality since it corresponds to the regime of infinite Reynolds number without large-scale dissipation of energy~\cite{Scott:2007,JRWS:2017}.

We also speculated that in the quasi-steady state regime, since the fractal dimension will statistically not depend on the particular time at which a spatial section of the horizon is considered, that the entire horizon $H$ could be assigned a `bootstrapped' fractal dimension of $D_H = D_{H\cap\Sigma} + 1$, although the strict mathematical meaning of this is not clear. Furthermore, we have not shown that $D_{H\cap\Sigma}$ is invariant with respect to deformations of the spatial section of the horizon (when computed over an ensemble of horizons), since we have only considered constant time slices in the ingoing Finkelstein coordinate.


\partkey{II}
\part{Post-Newtonian approach to black hole-fluid systems} \label{part:PN}

\section*{Executive summary}

Many interesting astrophysical systems have a combination of very long characteristic times and important general relativistic effects. For a subset of these systems, \emph{propagating} gravitational degrees of freedom are of subleading importance, and thus allow for approximate descriptions. In this part, we present our progress in implementing and testing a post-Newtonian gravity-hydrodynamics (PNGhydro) code in spherical symmetry. The post-Newtonian (PN) formalism was presented in~\cite{barausse2013post} and is capable of describing the interaction between a black hole and a fluid up to 2.5 PN order. The equations governing the gravitational potentials are all elliptic, owing to the use of the Poisson gauge. This eliminates propagating degrees of freedom from the problem, and thus would be advantageous to implement in a numerical code for the study of systems requiring very long integration times, eg. tidal disruption events. We will also present a general relativistic hydrodynamics (GR fluid) code in spherical symmetry which we developed to provide a benchmark against which to measure the accuracy of the PNGhydro code. Since this is a proof-of-principle, we make no attempt to lower the computational cost of the PNGhydro code with respect to the GR fluid one.

\begin{figure}[h!]
\centering
\hbox{
\hspace{1cm}\includegraphics[width=0.8\textwidth]{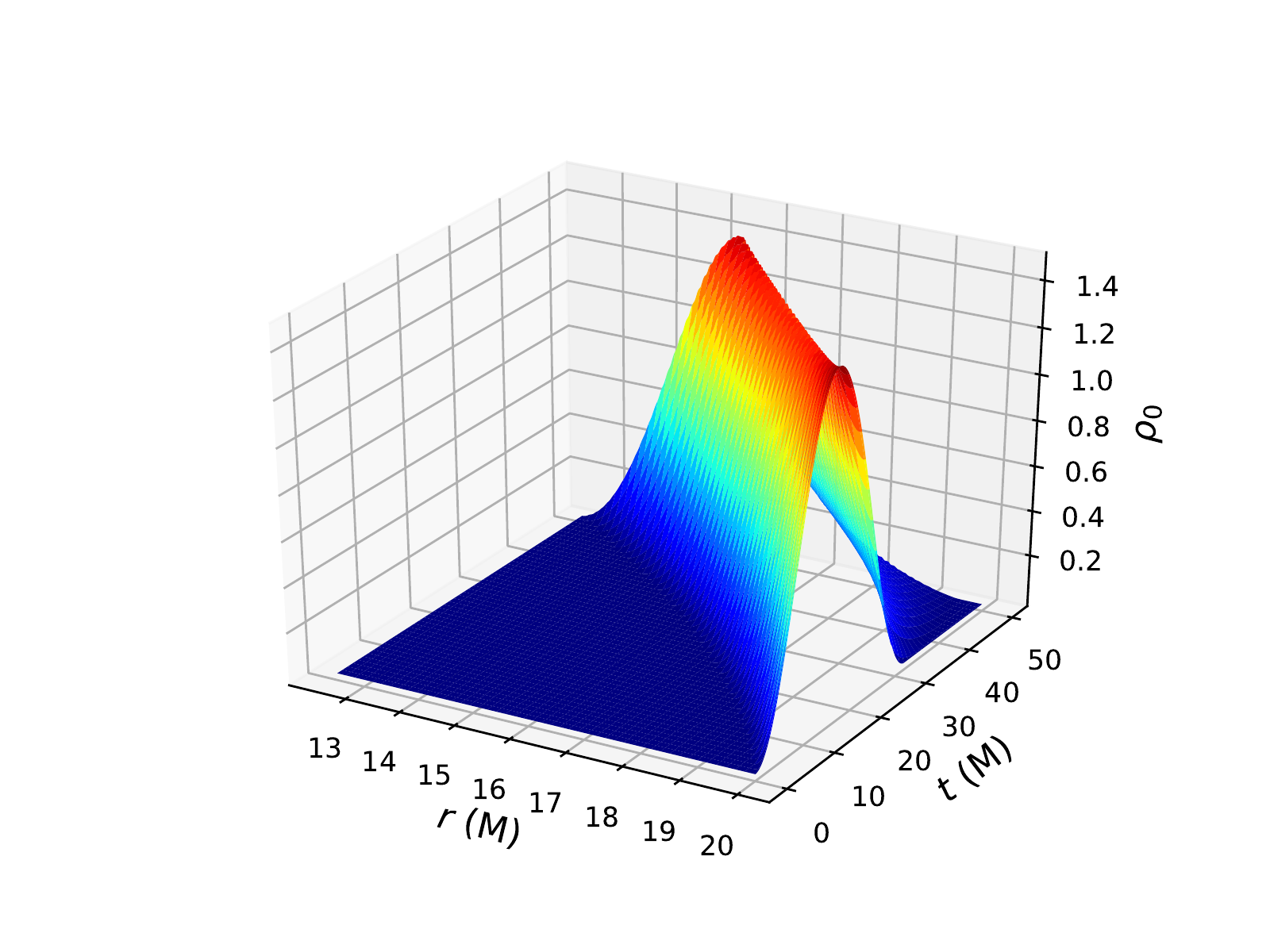}}
\raggedright

Figure II: Preview of the numerical evolution of a fluid pulse, injected into the spacetime of a vacuum black hole of mass $M$ via the outer boundary.

\label{fig:PNmoney}
\end{figure}

\chapter{Introduction}
Many astrophysical systems of interest involve relativistic gravitating fluids in a regime where the propagating degrees of freedom of the gravitational field are of secondary importance compared to the fluid physics and conservative gravitational effects. Among these are core-collapse supernovae (CCSNe) in the early post-bounce regime, where gravitational wave damping of modes excited by core bounce is outdone by other effects~\cite{ott2009gravitational,kotake2013multiple}, and tidal disruption of stars with low compactness by black holes (eg.~\cite{hills1988hyper,rees1988tidal,evans1989tidal}). Even in modern investigations, for the sake of finding cost reductions the former is often treated in Newtonian gravity~\cite{dolence2015two,pan2016two,suwa2015criterion} or with phenomenological general relativistic corrections~\cite{marek2006exploring,o2018two}. Tidal disruption events (TDEs) have the added complication of a black hole, which brings relativistic effects such as an innermost stable circular orbit (ISCO) and spin coupling. ``Pseudo-Newtonian" treatments of black holes have been employed via phenomenological potentials~\cite{paczynsky1980thick,semerak1999pseudo, wegg2012pseudo,tejeda2013accurate}, not restricted to the tidal disruption context, or test-fluid evolutions on non-reactive black hole backgrounds with Newtonian self-gravity of the star~\cite{tejeda2017tidal}.

In the TDE case, there is a large separation of scale between the characteristic speeds of gravity and hydrodynamics: the speed of light and the speed of sound, respectively. Degrees of freedom propagating at the speed of light require smaller time steps to resolve, and if such degrees of freedom are of subleading importance (as they typically are when the star has low compactness), then it is wasteful to resolve them. This problem is exacerbated by the fact that the orbital dynamics in a TDE occur over a large spatial scale, and especially by the fact that long integration times are required. The former can be addressed with mesh refinement techniques, but not the latter. For these reasons, a formalism which captures essential features of gravity, short of propagating effects, is desirable. This can be achieved through a post-Newtonian (PN) approximation.

Note that other approaches to simulating TDEs exist which do not use pseudo- or post-Newtonian treatments of gravity. For example, in~\cite{cheng2013relativistic,cheng2014tidal} the star is followed in Fermi normal coordinates on a fixed black hole background, allowing the tidal effects to be computed at various orders. Another strategy used in~\cite{east2013simulating,east2014gravitational} is to subtract the truncation error due to the dominant black hole background (whose analytic solution is known). While these approaches capture hydrodynamics and gravity, they lack the sophisticated microphysics of the pseudo-Newtonian codes which allow the computation of multi-wavelength electromagnetic emission and nuclear reactions (eg.~\cite{macleod2016optical,kawana2018tidal,tanikawa2018high}).

A PN formalism was presented in~\cite{blanchet1990post} which gives the dynamics in terms of Poisson equations. The purpose of this was for Newtonian gravity codes to easily include relativistic corrections, since they already solve Poisson equations. This formalism did not include black holes, and subsequent numerical implementations used massive point particles as stand-ins for black holes~\cite{ayal2001post,ayal2000tidal} (see~\cite{hayasaki2016circularization} for the inclusion of some spin effects). Producing sufficiently massive stars to represent neutron stars has been reported to be difficult in this formalism~\cite{barausse2013post}.

A similar PN formalism was written down to higher order in~\cite{barausse2013post}, and it can describe a fluid-black hole system. The mass-radius curve of cold equilibrium neutron stars represented by a polystopic perfect fluid with $K=100$, $\Gamma=2$ was reproduced fairly well down to $\sim 10$ km radius in isotropic coordinates. Furthermore, the ISCO radius, when measured in a gauge-invariant manner as $(M\Omega_{\mathrm{ISCO}})^{-2/3}$, is tracked well for spin parameters $\chi \sim [-1,0.25]$, whereas the pseudo-Newtonian potential of~\cite{paczynsky1980thick} does not depend on $\chi$ at all. In this formalism, the post-Newtonian potentials due to the Kerr black hole are given analytically. Owing to the use of the Poisson gauge, the hydrodynamic contribution to the potentials is determined by a set of elliptic partial differential equations, and thus has no propagation speeds. We reiterate that if implemented in a numerical code, this formalism would therefore allow the propagating degrees of freedom of the gravitational field to be ignored. Larger time steps would therefore be possible, reducing the cost of the long integrations required in tidal disruption systems. Furthermore the relativistic effects would be faithfully captured to the prescribed order, and the regime of validity of any integration would be known \emph{a priori}. This is contrasted with pseudo-Newtonian approaches where some of the physics is captured but not others, and errors are not under control due to the lack of a systematic expansion.

In this part, we present our progress towards implementing a proof-of-principle of the formalism in~\cite{barausse2013post} in a spherically-symmetric gravity-hydrodynamics code. We also develop a general relativistic hydrodynamics code (GR fluid) as a benchmark against which to test the post-Newtonian gravity-hydrodynamics (PNGhydro) code. Spherical symmetry eliminates the radiative dynamics and renders a subset of the PN potentials unnecessary, so this is not a full demonstration of the formalism, but it is a logical first step.

\vspace{0.5cm}
This part is organized as follows.

In Chapter~\eqref{ch:PNbg} we provide extensive background material and discussion of our methods. This includes a discussion of the fluid equations in curved spacetime in Secs.~\eqref{ch:3+1fluid},~\eqref{sec:Valencia}, and~\eqref{sec:charstruc}. The formulation of Einstein's equation we use and the derivation of constraint-preserving boundary conditions are discussed in Secs.~\eqref{sec:ECsystem} and~\eqref{sec:CPBC}, respectively. We provide an overview of the PN formalism of~\cite{blanchet1990post,barausse2013post} in Sec.~\eqref{sec:PNformalism}.

In Chapter~\eqref{ch:PNimp} we present details of our numerical implementation. This includes a review of aspects of computational fluid dynamics in Secs.~\eqref{sec:HRSC},~\eqref{ch:ppm}, and~\eqref{sec:contoprim}. In Sec.~\eqref{sec:bdyinjection} we describe our method of injecting fluid matter into the domain through the outer boundary, which eliminates the need for a numerical initial data solver. We describe the current state of our PNGhydro implementation in Sec.~\eqref{sec:PNsolver}.

Chapter~\eqref{ch:codeval} contains a variety of code validations, including ``artificial analytic convergence tests"~\eqref{sec:arttest}, evolutions of a stable TOV star~\eqref{sec:TOVoscillations}, injection of gauge and fluid pulses into a black hole spacetime~\eqref{sec:gaugepulse},~\eqref{sec:hydroandgaugepulse}, and an independent residual test of the PN solver~\eqref{sec:PNtest}.

Finally, we present preliminary results and conclusions in Chapter~\eqref{ch:PNresults}.

%
%
\chapter{Background and methods} \label{ch:PNbg}

\section{3+1 Form of the Fluid Equations} \label{ch:3+1fluid}
In this section, we present the hydrodynamic equations on an arbitrary background geometry in coordinate-free 3+1 form. This is for mainly for an illustration of how to go from covariant expressions to coordinate-free 3+1 form, and we will be very verbose. Although one can obtain a coordinate representation from this form, like the Valencia formulation~\cite{marti1991numerical} we describe in Sec.~\eqref{sec:Valencia} and use in our code implementation, in practice we bypass this step by introducing coordinates right away. One may therefore skip this section without loss of continuity.

\begin{figure}[!h]
\centering
\includegraphics[width=0.8\textwidth]{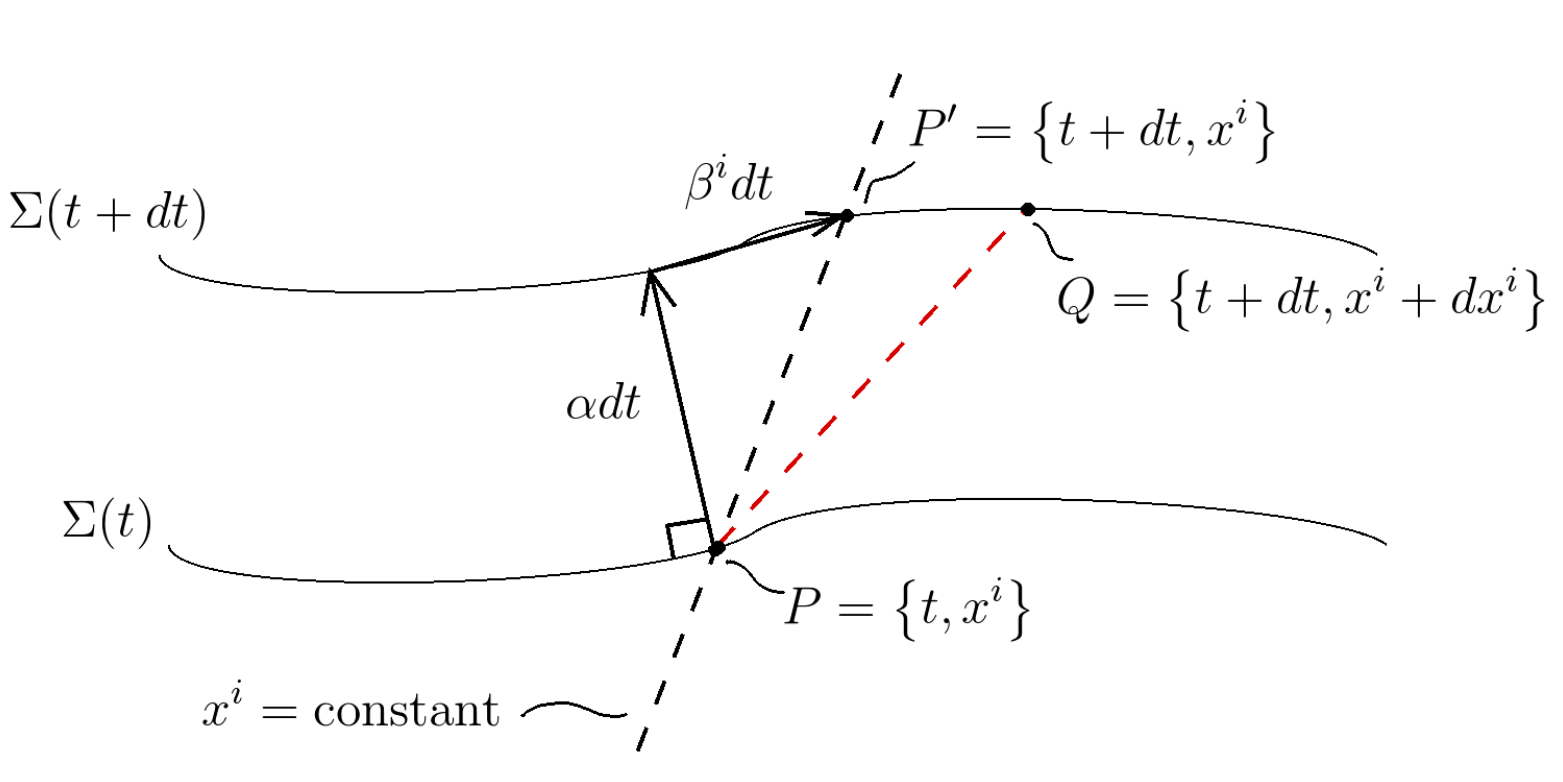}
\caption{Depiction of a spacelike foliation $\Sigma$ of spacetime. Coordinateful quantities like the lapse $\alpha$ and shift vector $\beta^i$ will be used explicitly in later sections.} \label{fig:slices}
\end{figure}

In terms of the stress energy tensor $T^{ab}$, rest mass density $\rho_0$, and fluid 4-velocity $u^a$, we begin with the general covariant form of the equations,
\begin{eqnarray}
\nabla_a T^{ab} &=& 0 \label{eq:stresscons} \\
\nabla_a (\rho_0 u^a) &=& 0 \label{eq:baryoncons},
\end{eqnarray}
where Eqs.~\eqref{eq:stresscons} and~\eqref{eq:baryoncons} express the covariant conservation of stress-energy and baryon number, respectively. We introduce to our spacetime $(\mathcal{M},g_{ab})$ a spacelike foliation described by the timelike unit normal vector field to the hypersurfaces $n_a$, intrinsic metric of the hypersurfaces $\gamma_{ab} = g_{ab} + n_a n_b$, and extrinsic curvature of the hypersurfaces $K_{ab} = -\nabla_a n_b - n_a a_b$. We have chosen the mostly positive signature $(-,+,+,+)$ and defined the acceleration of the fundamental observers as $a_b \equiv n_a \nabla^a n_b$. The foliation and some related quantities (to be used explicitly in later sections) are depicted in Fig.~\eqref{fig:slices}

\FloatBarrier

To obtain the 3+1 form of energy conservation, we must project Eq.~\eqref{eq:stresscons} onto the $n_b$ direction as follows:
\begin{eqnarray}
0 &=&n_b \nabla_a T^{ab} \nonumber\\
&=& \nabla_a ( n_b T^{ab} ) - T^{ab} \nabla_a n_b \nonumber\\
&=& \nabla_a (n_b (\gamma^a_c - n^a n_c)T^{cb}) + T^{ab} K_{ab} + T^{ab} n_a a_b \nonumber\\
&=& \nabla_a (n_b \gamma^a_c T^{cb}) - \nabla_a (n^a n_b n_c T^{cb}) + T^{ab}\gamma_a^c \gamma_b^d K_{cd} + T^{ab} n_a \gamma_b^c a_c \nonumber .
\end{eqnarray}
At this point let us define the energy density $\rho \equiv n_a n_b T^{ab}$, the energy current $j^a = -n_b \gamma^a_c T^{bc}$, and the purely spatial projection of the stress-energy $S^{ab} = \gamma^a_c \gamma ^b_d T^{cd}$. We thus obtain
\begin{eqnarray}
0 &=& - \nabla_a (j^a) - \nabla_a (n^a \rho) + S^{ab} K_{ab} - j^a a_a. \label{eq:rhocons-inter1}
\end{eqnarray}
The first two terms include a full spacetime covariant derivative, which must be further decomposed as follows:
\begin{eqnarray}
 - \nabla_a (j^a) - \nabla_a (n^a \rho) &=&  - \nabla_a (\gamma^{ab} j_b + n^a \rho)) \nonumber\\
 &=& - \gamma^{ab} \nabla_a j_b - j_b \nabla_a \gamma^{ab} -n^a \nabla_a \rho - \rho \nabla_a n^a \nonumber\\
 &=& - D_a j^a - j_b \nabla_a (n^a n^b) - \mathcal{L}_n \rho +\rho K \nonumber\\
 &=& - D_a j^a - j_b (n^a\nabla_a n^b + n^b \nabla_a n^a) - \mathcal{L}_n \rho + \rho K \nonumber\\
 &=& - D_a j^a - j_b a^b - \mathcal{L}_n \rho + \rho K \label{eq:rhocons-inter2}.
\end{eqnarray}
Putting Eqs.~\eqref{eq:rhocons-inter1} and~\eqref{eq:rhocons-inter2} together, we obtain the final form of energy conservation,
\begin{eqnarray}
0 = - D_a j^a - \mathcal{L}_n \rho + \rho K + S^{ab} K_{ab} - 2j^a a_a \label{eq:3+1rhocons}.
\end{eqnarray}
Next, to obtain the 3+1 form of momentum conservation, we apply instead the projector $\gamma_{bc}$ to Eq.~\eqref{eq:stresscons}:
\begin{eqnarray}
0 &=& \gamma_{bc} \nabla_a T^{ab} \nonumber\\
&=& \gamma_{bc} \nabla_a ((\gamma^{ad} - n^a n^d)(\gamma^{be} - n^b n^e)T_{de}) \nonumber\\
&=& \gamma_{bc} \nabla_a ( S^{ab} + n^b j^a + n^a j^b + \rho n^a n^b ) \nonumber\\
&=& \gamma_{bc} (\gamma_{ad} \nabla^d S^{ab} - n_a n_d \nabla^d S^{ab} - j_a K^{ab} - j^b K + n^a\nabla_a j^b + \rho a^b) \nonumber\\
&=& \gamma_{bc} (D_a S^{ab} +S^{ab} n_d \nabla^d n_a - j_a K^{ab} - j^b K + n^a\nabla_a j^b + \rho a^b) \nonumber\\
&=& \gamma_{bc} (D_a S^{ab} +S^{ab} a_a - 2 j_a K^{ab} - j^b K + \mathcal{L}_n j^b + \rho a^b) \label{eq:3+1momcons}.
\end{eqnarray}
Finally, we can also manipulate the scalar expression of baryon number conservation, Eq.~\eqref{eq:baryoncons}, as follows:
\begin{eqnarray}
0 &=& \nabla_a (\rho_0 u^a) \nonumber\\
&=& (\gamma_{ab} - n_a n_b) \nabla^b (\rho_0 (\gamma^{ac} - n^a n^c) u_c) \nonumber
\end{eqnarray}
Defining the Lorentz factor between the fundamental observers and the fluid velocity as $W\equiv -n^cu_c$ and the spatial fluid velocity as $v^a \equiv  \gamma^{ac}u_c/W$, we go on to write
\begin{eqnarray}
0 &=& (\gamma_{ab} - n_a n_b) \nabla^b (\rho_0 W v^a + \rho_0 W n^a) \nonumber\\
&=&  \gamma_{ab} \nabla^b (\rho_0 W v^a ) + \gamma_{ab} \rho_0 W \nabla^b n^a + \rho_0 W v^a n_b \nabla^b n_a + n_b \nabla^b (\rho_0 W) - \rho_0 W n_a n_b \nabla^b n^a \nonumber\\
&=& D_a (\rho_0 W v^a) - \rho_0 W K + \rho_0 W v^a a_a + \mathcal{L}_n (\rho_0 W) \label{eq:3+1baryoncons}.
\end{eqnarray}
This concludes the coordinate-free 3+1 decomposition of the hydrodynamic equations.
%
%
\section{The Valencia formulation of hydrodynamics} \label{sec:Valencia}
We employ the Valencia formulation~\cite{marti1991numerical} of hydrodynamics in our numerical setup. This formulation is flux-conservative, and convenient for shock-capturing methods. We briefly describe the formulation here, following~\cite{Alcubierre:2008}.

We will make use of the identity
\begin{eqnarray}
\nabla_a A^a = \frac{1}{\sqrt{\vert g\vert}} \partial_a \left( \sqrt{\vert g \vert} A^a \right), \label{eq:covdiv1}
\end{eqnarray}
as well as its generalizations to mixed rank-2 tensors,
\begin{eqnarray}
\nabla_a B^a_{\phantom{a}b} = \frac{1}{\sqrt{\vert g\vert}} \partial_a \left( \sqrt{\vert g \vert} B^a_{\phantom{a}b} \right) - \Gamma^c_{ab} B^a_{\phantom{a}c},  \label{eq:covdiv2}
\end{eqnarray}
and contravariant rank-2 tensors,
\begin{eqnarray}
\nabla_a B^{ab} = \frac{1}{\sqrt{\vert g\vert}} \partial_a \left( \sqrt{\vert g \vert} B^{ab} \right) + \Gamma^b_{ac} B^{ac}.  \label{eq:covdiv3}
\end{eqnarray}

We assume a $3+1$-dimensional split of spacetime, where $\alpha$ is the lapse function, $\beta^i$ is the shift vector, and $\gamma_{ij}$ is the spatial metric whose determinant $\gamma$ is related to the full spacetime metric via $\sqrt{\vert g \vert} = \alpha \sqrt{\gamma}$. Recall the Lorentz factor $W=-u^a n_b$. With $n_a = (-\alpha,0,0,0)$ we have $W = \alpha u^t$. The coordinate velocity of the fluid $u^i/u^t$ is related to its velocity in the frame of the fundamental (Eulerian) observers $v^i$ through the lapse and shift as $u^i = u^t (\alpha v^i - \beta^i) = W (v^i - \beta^i/\alpha)$. By applying Eq.~\eqref{eq:covdiv1} to local mass conservation $\nabla_a (\rho_0 u^a) = 0$ and using the aforementioned relations, we obtain
\begin{eqnarray}
0 &=& \partial_t \left( \sqrt{\vert g \vert} \rho_0 u^t \right) + \partial_k \left( \sqrt{\vert g \vert} \rho_0 u^k \right) \nonumber\\
&=& \partial_t \left( \sqrt{\gamma} \rho_0 W \right) + \partial_k \left( \alpha \sqrt{\gamma} \rho_0 W \left( v^k - \beta^k/\alpha \right) \right) \nonumber\\
&=& \partial_t \left( \sqrt{\gamma} D \right) + \partial_k \left( \alpha \sqrt{\gamma} D \left(v^k - \beta^k/\alpha\right)  \right), \label{eq:ValD}
\end{eqnarray}
where we have defined the conservative variable $D\equiv \rho_0 W$.

Similarly for the spatial component of energy-momentum conservation, $\nabla _a T^a_i = 0$. First note that $u_i = g_{ai} u^a = \beta_i u^t + \gamma_{ik} u^k = \beta_i u^t + \gamma_{ik} u^t (\alpha v^k - \beta^k) = W v_i$. Then using Eq.~\eqref{eq:covdiv2} and $T^a_i = \rho_0 h u^a W v_i + P \delta^a_i$ we can write
\begin{eqnarray}
0 &=& \partial_t \left( \alpha \sqrt{\gamma} T^t_i \right) + \partial_k \left( \alpha \sqrt{\gamma} T^k_i \right) - \alpha \sqrt{\gamma} \Gamma^c_{a i} T^a_c \nonumber \\
&=& \partial_t \left( \sqrt{\gamma} \rho_0 h W^2 v_i \right) + \partial_k \left( \alpha \sqrt{\gamma} \rho_0 h W^2 v_i \left( v^k - \beta^k/\alpha \right) + \alpha \sqrt{\gamma} P \delta_i^k \right) - \alpha \sqrt{\gamma} \Gamma^c_{a i} T^a_c \nonumber\\
&=& \partial_t \left( \sqrt{\gamma} S_i \right) + \partial_k \left( \alpha \sqrt{\gamma} S_i \left( v^k - \beta^k/\alpha \right) + \alpha \sqrt{\gamma} P \delta^k_i \right) - \alpha \sqrt{\gamma} \Gamma^c_{a i} T^a_c \label{eq:ValSi},
\end{eqnarray}
where we have defined the conservative variable $S_i \equiv \rho_0 h W^2 v_i$. Written in this way, this equation of motion has a source term coming from the spacetime curvature.

Lastly, the time component of the energy-momentum conservation we treat with indices up, $\nabla_a T^{at} = 0$, and some manipulation is required. First note that $T^{tt} = (1/\alpha^2) (\rho_0 h W^2 - P) \equiv (1/\alpha^2) \mathcal{E}$ and $T^{tk} = (1/\alpha^2) (\rho_0 h W^2 (\alpha v^k -\beta^k) + P \beta^k)$. We prepare for the overall factors of $1/\alpha^2$ by rewriting
\begin{eqnarray}
\partial_a \left( \alpha \sqrt{\gamma} T^{ta} \right) = \frac{1}{\alpha} \partial_a \left( \alpha^2 \sqrt{\gamma} T^{ta}\right) - \sqrt{\gamma} T^{ta} \partial_a \alpha . \nonumber
\end{eqnarray}
Then, applying Eq.~\eqref{eq:covdiv3} to $\nabla_a T^{at} = 0$, we obtain
\begin{eqnarray}
0 &=& \partial_a \left( \alpha^2 \sqrt{\gamma} T^{ta} \right) - \alpha \sqrt{\gamma} T^{ta} \partial_a \alpha + \alpha^2 \sqrt{\gamma} \Gamma^t_{ac} T^{ac} \nonumber\\
&=& \partial_t \left( \sqrt{\gamma} \mathcal{E} \right) + \partial_k \left( \sqrt{\gamma} \left[ \alpha \rho_0 h W^2 \left( v^k - \beta^k/\alpha \right) + P \beta^k \right) \right] - \alpha \sqrt{\gamma} T^{ta} \partial_a \alpha + \alpha^2 \sqrt{\gamma} \Gamma^t_{ac} T^{ac} \nonumber\\
&=& \partial_t \left( \sqrt{\gamma} \mathcal{E} \right) + \partial_k \left( \alpha \sqrt{\gamma} \left[ \mathcal{E} \left( v^k - \beta^k/\alpha \right) + P v^k\right] \right) - \alpha \sqrt{\gamma} T^{ta} \partial_a \alpha + \alpha^2 \sqrt{\gamma} \Gamma^t_{ac} T^{ac} \label{eq:ValEpre},
\end{eqnarray}
where we have defined $\mathcal{E} = \rho_0 h W^2 - P$. It is advantageous to subtract Eq.~\eqref{eq:ValD} from Eq.~\eqref{eq:ValEpre} to obtain an equation of motion for $\tau \equiv \mathcal{E} - D$ instead, since the Newtonian limit of $\tau$ is $\rho_0 \epsilon + \rho_0 v^2/2$~\cite{marti1991numerical} and thus will be numerically more accurate for non-relativistic flows. Doing so results in
\begin{eqnarray}
\!\!\! 0\!\!\! &=& \!\!\! \partial_t \left( \sqrt{\gamma} \tau \right) + \partial_k \left( \alpha \sqrt{\gamma} \left[ \tau \left( v^k - \beta^k/\alpha \right) + P v^k \right] \right) - \alpha \sqrt{\gamma} T^{ta} \partial_a \alpha + \alpha^2 \sqrt{\gamma} \Gamma^t_{ac} T^{ac} \label{eq:Valtau}.
\end{eqnarray}

We furthermore choose to absorb a portion of the spatial metric determinant into the definition of certain variables, resulting in \emph{densitized} variables. With a spatial metric in spherical symmetry given by $\gamma_{ij} dx^i dx^j = \gamma_{rr}\: dr^2 + \gamma_T \: r^2 d\Omega^2$, the factor we absorb is $\sqrt{\gamma_{rr}} \gamma_T$. We densitize the primitive mass density $\rho_0$, resulting in $\tilde{\rho}_0 \equiv \sqrt{\gamma_{rr}} \gamma_T \rho_0$. This change propagates to $\tilde{D} \equiv \sqrt{\gamma_{rr}} \gamma_T D$ and $\tilde{S_i} \equiv \sqrt{\gamma_{rr}} \gamma_T S_i$, via their definitions. Given an ideal fluid equation of state $P = \rho_0 \epsilon (\Gamma -1)$, this also results in $\tilde{P} \equiv \sqrt{\gamma_{rr}} \gamma_T P$ and thus $\tilde{\tau} \equiv \sqrt{\gamma_{rr}} \gamma_T \tau$ as well. For other equations of state not necessarily linear in $\rho_0$, eg. polytropes $P = K \rho_0^{\Gamma}$, care must be taken to undensitize $\tilde{\rho}_0$ in a numerical code before using such a formula.

\subsection{Specialization to spherical symmetry}

In this section we specialize the system of hydrodynamic equations~\eqref{eq:ValD},~\eqref{eq:ValSi},~\eqref{eq:Valtau} to spherical symmetry. We will encounter some regularization issues at the origin, and will fix them with a clever rewriting of the equations.

In spherical symmetry the vector $S_i$ has only a radial component $S_r$, so our densitized hydrodynamic variables are
\begin{eqnarray}
\tilde{\vec{U}} & \equiv & \left( \tilde{D}, \tilde{S_r}, \tilde{\tau}\right) \nonumber\\
&=& \sqrt{\gamma_{rr}}\gamma_T \left( D, S_r, \tau \right).
\end{eqnarray}
The equation of motion~\eqref{eq:ValD},~\eqref{eq:ValSi},~\eqref{eq:Valtau} become, respectively,
\begin{eqnarray}
\partial_t \tilde{D} + \frac{1}{r^2} \partial_r \left( \alpha r^2 \tilde{D} \left( v^r - \beta^r/\alpha\right) \right) &=& 0, \label{eq:ValDsph}\\
\partial_t \tilde{S_r} + \frac{1}{r^2} \partial_r \left( \alpha r^2 \left[ \tilde{S_r} \left( v^r - \beta^r/\alpha \right) + \tilde{P} \right] \right) &=& \mathcal{S}_S, \label{eq:ValSisph} \\
\partial_t \tilde{\tau} + \frac{1}{r^2} \partial_r \left( \alpha r^2 \left[ \tilde{\tau} \left( v^r - \beta^r/\alpha \right) + \tilde{P} v^r \right]\right) &=& \mathcal{S}_\tau, \label{eq:Valtausph}
\end{eqnarray}
where the source terms $\mathcal{S}_S$ and $\mathcal{S}_\tau$ on the right-hand sides are given by
\begin{eqnarray}
\mathcal{S}_S &=& \alpha \left( 2\frac{\tilde{P}}{r} + \frac{\partial_r \gamma_T}{\gamma_T} \tilde{P} - \frac{\partial_r \alpha}{\alpha} \left(\tilde{\tau} + \tilde{D}\right) + \frac{1}{2} \frac{\partial_r \gamma_{rr}}{\gamma_{rr}} \left( \tilde{S_r} v^r + \tilde{P} \right) + \frac{\partial_r \beta^r}{\alpha} \tilde{S_r} \right), \label{eq:ValSisph_src}\\
\mathcal{S}_\tau &=& -\frac{1}{2} \frac{\partial_t \gamma_{rr}}{\gamma_{rr}} \left(\tilde{S_r} v^r + \tilde{P}\right) - \partial_r \alpha \frac{\tilde{S_r}}{\gamma_{rr}} + \frac{\beta^r}{2} \frac{\partial_r \gamma_{rr}}{\gamma_{rr}} \left( \tilde{S_r} v^r + \tilde{P} \right) + \beta^r \frac{\partial_r \gamma_T}{\gamma_T} \tilde{P} \nonumber\\
&-& \frac{\partial_t \gamma_T}{\gamma_T} \tilde{P} + \partial_r \beta^r \left( \tilde{S_r} v^r + \tilde{P} \right). \label{eq:Valtausph_src}
\end{eqnarray}

The first regularization issue concerns the $(1/r^2) \partial_r$ derivative operators, which if evaluated naively at $r=0$ numerically would give an infinite result. We rewrite this operator as $3 \partial_{r^3}$, which by the chain rule is equivalent.

The second regularization issue concerns the term $2\alpha \tilde{P}/r$ in Eq.~\eqref{eq:ValSisph_src}, which also presents a problem at the origin. The key fact to understand with a formally infinite term such as this is that it is merely a coordinate singularity. Thus, the mathematics should work out to eliminate the divergence, we just have to find the appropriate rewriting of the equation. We notice that a piece of the flux term on the left-hand side of Eq.~\eqref{eq:ValSisph} actually cancels the offending source term. Namely, we can rewrite Eq.~\eqref{eq:ValSisph} as
\begin{eqnarray}
\partial_t \tilde{S_r} + \frac{1}{r^2} \partial_r \left( \alpha r^2 \tilde{S_r} \left(v^r - \beta^r/\alpha\right)\right) + \partial_r \left( \alpha \tilde{P}\right) = \mathcal{S}_S - 2\alpha \frac{\tilde{P}}{r}, \label{eq:ValSisph_fluxsplit}
\end{eqnarray}
which eliminates the divergent source term on the right-hand side. This is called the \emph{flux-splitting method}, since we have split the flux into two separate pieces in order to pull out the corresponding divergent term which appears in the source on the right-hand side.
%
%
\section{Characteristic structure of hydrodynamics in curved spacetime} \label{sec:charstruc}
In this section we review the characteristic structure of hydrodynamics in general relativity, specialized to spherical symmetry and an ideal fluid equation of state $P = \rho_0 \epsilon (\Gamma-1)$, which yield significant simplifications. The characteristic structure of the system of Eqs.$\:$(\ref{eq:ValD},~\ref{eq:ValSi},~\ref{eq:Valtau}) in Sec.~\eqref{sec:Valencia} were derived in the special relativistic case in~\cite{donat1998flux}, and then in the general relativistic case in~\cite{ibanez2001riemann}. We follow the latter reference here.

In spherical symmetry, the system of Eqs.$\:$(\ref{eq:ValD},~\ref{eq:ValSi},~\ref{eq:Valtau}) takes the form
\begin{eqnarray}
\partial_t \boldsymbol{U} + \partial_r \boldsymbol{F} = \boldsymbol{S},
\end{eqnarray}
whose principal part gives
\begin{eqnarray}
\partial_t \boldsymbol{U} + \frac{\partial \boldsymbol{F}}{\partial \boldsymbol{U}} \partial_r \boldsymbol{U} = 0. \label{eq:princpart1}
\end{eqnarray}
The factor $\partial \boldsymbol{F}/\partial \boldsymbol{U}$ is a matrix containing the characteristic structure of the system. Its eigenvalues correspond to the characteristic speeds, and for an ideal fluid equation of state are given by
\begin{eqnarray}
\lambda_0 &=& \alpha v^r - \beta^r \nonumber\\
\lambda_{\pm} &=& \alpha \frac{v^r \pm c_s \sqrt{\gamma^{rr}}}{1\pm v^r c_s \sqrt{\gamma_{rr}}} - \beta^r , \label{eq:evals}
\end{eqnarray}
where $c_s$ is the sound speed, given in terms of the specific internal energy as $\sqrt{\Gamma (\Gamma-1) \epsilon/(1+\Gamma\epsilon)}$. The eigenspeed $\lambda_0$ corresponds to the \emph{material wave}, whereas $\lambda_{\pm}$ correspond to \emph{acoustic waves}. Setting $\alpha = \gamma_{rr} = \gamma^{rr} = 1$, $\beta=0$, taking $c_s = \mathcal{O}(v)$, and keeping lowest orders in $v$ yields the Newtonian expressions
\begin{eqnarray}
\lambda_0 &=& v^r\nonumber\\
\lambda_{\pm} &=& v^r \pm c_s + \mathcal{O}(v^3).
\end{eqnarray}
A complete set of right-eigenvectors of $\partial\boldsymbol{F}/\partial \boldsymbol{U}$ corresponding to the eigenvalues in Eqs.~\eqref{eq:evals} are
\begin{eqnarray}
\boldsymbol{r}_0 &=& \left( W^{-1}, v^r \gamma_{rr}, 1-W^{-1} \right)^T \nonumber\\
\boldsymbol{r}_{\pm} &=& \left( 1, hW \sqrt{\gamma_{rr}} \left( v^r \sqrt{\gamma_{rr}} \pm c_s \right), hW \left( 1 \pm c_s v^r \sqrt{\gamma_{rr}}\right) - 1 \right)^T,
\end{eqnarray}
where the enthalpy is $h=1+\Gamma\epsilon$. The corresponding left-eigenvectors (i.e. the rows of the matrix $[\boldsymbol{r}_0,\boldsymbol{r}_{+},\boldsymbol{r}_{-}]^{-1}$) are
\begin{eqnarray}
\boldsymbol{l}_0 &=& \frac{W^2}{h-1} \left( \frac{h}{W}-1, v^r, -1 \right) \nonumber\\
\boldsymbol{l}_{\pm} &=& \frac{W}{2(h-1)} \left(1\mp \frac{v^r c_s \sqrt{\gamma_{rr}}}{\Gamma -1} -\frac{1}{W}, - \left( v^r\mp \frac{ c_s \sqrt{\gamma^{rr}}}{\Gamma -1} \right) ,  1\mp \frac{v^r c_s \sqrt{\gamma_{rr}}}{\Gamma -1}  \right).
\end{eqnarray}
The characteristic variables are obtained as follows. For clarity, first rewrite Eq.~\eqref{eq:princpart1} as
\begin{eqnarray}
\boldsymbol{U}_{t} + A \boldsymbol{U}_{r} = 0,
\end{eqnarray}
where the subscripts denote partial differentiation, and the matrix $A \equiv \partial \boldsymbol{F}/\partial \boldsymbol{U}$. The matrix $A$ can be diagonalized so that $A=T D T^{-1}$, where $D = \text{diag}(\lambda_0,\lambda_+,\lambda_-)$ and $T$ is the matrix whose columns are the right-eigenvectors of $A$, $T=[\boldsymbol{r}_0,\boldsymbol{r}_{+},\boldsymbol{r}_{-}]$. This gives
\begin{eqnarray}
0 &=& \boldsymbol{U}_t + T D T^{-1} \boldsymbol{U}_r \nonumber\\
&=& T^{-1} \boldsymbol{U}_t + D T^{-1} \boldsymbol{U}_r \nonumber\\
&=& (T^{-1} \boldsymbol{U})_t + D (T^{-1} \boldsymbol{U})_r.
\end{eqnarray}
In the last line we brought $T^{-1}$ into the derivatives while ignoring the terms which would ordinarily compensate for this. Since this characteristic analysis assumes linearization on a background solution, those compensating terms $\sim \boldsymbol{U} \partial T^{-1}$ are lower order (in the sense of characteristic analysis) because $T$ is built from the background solution and so no derivatives of the perturbation are present. This justifies ignoring the compensating terms when bringing $T^{-1}$ into the derivatives. 

We have decoupled the equations by transforming to the variables $T^{-1}\boldsymbol{U}$, known as the \emph{characteristic variables}, and within this linear regime they propagate at the characteristic speeds given by Eq.~\eqref{eq:evals}.
In our case of an ideal fluid, the characteristic variables are
\begin{eqnarray}
u_0 &=& \frac{\Gamma -1}{\Gamma} \rho_0 W^2 \nonumber\\
u_{\pm} &=& \frac{\rho_0 W}{2\Gamma} \left( 1\pm v^r c_s \sqrt{\gamma_{rr}} \right). \label{eq:hydrocharvars}
\end{eqnarray}
%
%
%
%
\section{Einstein-Christoffel formulation of general relativity} \label{sec:ECsystem}

In this section we describe the Einstein-Christoffel formulation of general relativity~\cite{anderson1999fixing}, which we use in our simulations. This is a first-order formulation of general relativity, where the shift and densitized lapse are specified \emph{a priori} and thus are treated as non-dynamical. After writing the general equations, we will specialize to spherical symmetry, largely following~\cite{Calabrese:2001kj}, and perfect fluid hydrodynamics. We use a $3+1$-decomposition of spacetime, as described during the course of Secs.~\eqref{ch:3+1fluid} and~\eqref{sec:Valencia}.

We begin with the Hamiltonian and momentum constraint equations, which follow from a full and partial projection of the Einstein equations
\begin{eqnarray}
G_{ab} = 8 \pi T_{ab} \label{eq:EFE}
\end{eqnarray}
onto the normals to the foliation via contraction with $n^a n^b$ and $n^a \gamma^b_c$, respectively~\cite{Alcubierre:2008}:
\begin{eqnarray}
{}^{(3)}R + K^2 - K_{ab} K^{ab} &=& 16 \pi \rho_{\text{ADM}} \label{eq:hcon}\\
D^a \left( K_{ab} - K \gamma_{ab} \right) &=& 8 \pi J_b , \label{eq:mcon}
\end{eqnarray}
where $\rho_{\text{ADM}}\equiv T_{ab}n^a n^b$ and $J_b\equiv -\gamma_b^c T_{ca} n^a$ are the ADM energy density and current coming from matter sources, respectively. Observers moving orthogonally to the spacelike foliation will measure this energy density and current. The evolution equation for the extrinsic curvature of the foliation hypersurfaces is obtained via a full projection of Einstein's equations~\eqref{eq:EFE} onto the hypersurfaces via contraction with $\gamma^a_i \gamma^b_j$:
\begin{eqnarray}
\partial_t K_{ij} &=& \beta^k \partial_k K_{ij} + K_{ki} \partial_j \beta^k + K_{kj} \partial_i \beta^k - D_i D_j \alpha \nonumber\\ 
&+& \alpha \left( {}^{(3)}R_{ij} + KK_{ij} - 2K_{ik}K^k_j \right) + 4 \pi \alpha \left( \gamma_{ij} (S-\rho_{\text{ADM}}) - 2 S_{ij}\right),
\end{eqnarray}
where $S_{ab} \equiv \gamma^c_a \gamma^d_b T_{cd}$ is the spatial energy-momentum tensor and $S = g^{ab} S_{ab}$. The evolution of the intrinsic metric $\gamma_{ij}$ follows from the definition of the extrinsic curvature:
\begin{eqnarray}
\partial_t \gamma_{ij} = -2\alpha K_{ij} + D_i \beta_j + D_j \beta_i .
\end{eqnarray}
Specializing to spherical symmetry and perfect fluid hydrodynamics, we write
\begin{eqnarray}
ds^2 &=& \left(-\alpha^2 + \gamma_{rr}\beta^r \beta^r\right) dt^2 + 2 \gamma_{rr} \beta^r dt dr + \gamma_{rr} dr^2 + \gamma_{T}\: r^2 d\Omega^2 ,\label{eq:ADMmetric}\\
K_{ij} dx^i dx^j &=& K_{rr}\: dr^2 + K_{T}\: r^2 d\Omega ^2,
\end{eqnarray}
where all fields depend on the coordinates $(t,r)$ only. Two auxiliary variables are used to reduce the system to first-order, which are
\begin{eqnarray}
f_{rT} &=& \frac{\gamma_T^\prime }{2} + \frac{\gamma_T}{r}, \label{eq:auxfrT}\\
f_{rrr} &=& \frac{\gamma_{rr}^\prime}{2} + \frac{4 \gamma_{rr} f_{rT}}{\gamma_T}. \label{eq:auxfrrr} 
\end{eqnarray}
The energy-momentum tensor of a perfect fluid reads $T_{ab} = \rho_0 h u_a u_b + P g_{ab}$, which gives the ADM matter sources as
\begin{eqnarray}
\rho_{\text{ADM}} &=& \tau + D, \\
J_r &=& S_r, \\
S_{ij} dx^i dx^j &=& \gamma_{rr} \left( S_r v^r + P \right) dr^2 +  \gamma_T P \: r^2 d\Omega^2 .
\end{eqnarray}
The constraints Eqs.~\eqref{eq:hcon} and~\eqref{eq:mcon} become
\begin{eqnarray}
\!\!\!\!\!\!\!\!\!\!\!\!\!\!\!\!\!\!\!\! 0\! =\! C\! \equiv && \!\!\!\!\!\!\!\!\! \frac{f_{rT}^\prime}{\gamma_{rr} \gamma_{T}} - \frac{1}{2 r^2 \gamma_T} + \frac{f_{rT}}{\gamma_{rr} \gamma_{T}} \left( \frac{2}{r} + \frac{7}{2} \frac{f_{rT}}{\gamma_T} - \frac{f_{rrr}}{\gamma_{rr}} \right)\! -\! \frac{K_T K_{rr}}{\gamma_T \gamma_{rr}}  + \frac{K_T^2}{2\gamma_T^2} + 4\pi \left( \tau + D \right)\! , \\
\!\!\!\!\!\!\!\!\!\!\!\!\! 0\! =\! C_r\! \equiv &&  \!\!\!\!\!\!\!\!\!\! \frac{K_T^\prime}{\gamma_T} + \frac{2K_T}{r\gamma_T} - \frac{f_{rT}}{\gamma_T} \left( \frac{K_{rr}}{\gamma_{rr}} + \frac{K_T}{\gamma_T}\right) + 4\pi S_r. \label{eq:constraints_main}
\end{eqnarray}
The definition of our auxiliary variables Eqs.~\eqref{eq:auxfrT} and~\eqref{eq:auxfrrr} also constitute constraints, i.e.
\begin{eqnarray}
0=C_{rrr} \equiv && \!\!\!\!\!\!\!\! \gamma_{rr}^\prime + \frac{8 \gamma_{rr} f_{rT}}{\gamma_T} - 2 f_{rrr}, \\
0=C_{rT} \equiv && \!\!\!\!\!\!\!\! \gamma_T^\prime + \frac{2\gamma_T}{r} - 2f_{rT}. \label{eq:constraints_auxvars}
\end{eqnarray}
As mentioned earlier, the shift $\beta^r$ and the densitized lapse $\tilde{\alpha} \equiv \alpha/\gamma_{rr}^{1/2} \gamma_T$ are taken to be specified \emph{a priori}. The evolution equations for the intrinsic metric become
\begin{eqnarray}
\partial_t \gamma_{rr} &=& \beta^r \gamma_{rr}^\prime + 2\gamma_{rr} \beta^{r\prime} - 2 \tilde{\alpha} \sqrt{\gamma_{rr}} \gamma_T K_{rr}, \\
\partial_t \gamma_T &=& \beta^r \gamma_T^\prime - 2\tilde{\alpha} \sqrt{\gamma_{rr}} \gamma_T K_T + \frac{2 \beta^r \gamma_T}{r}. \label{eq:eom_metric}
\end{eqnarray}
The evolution equations for the extrinsic curvature become
\begin{eqnarray}
\partial_t K_{rr} =&& \!\!\!\!\!\!\!\! \beta^r K_{rr}^\prime - \frac{\tilde{\alpha} \gamma_T f_{rrr}^\prime}{\sqrt{\gamma_{rr}}} - \tilde{\alpha}^{\prime \prime} \sqrt{\gamma_{rr}} \gamma_T - 6 \frac{ \sqrt{\gamma_{rr}} \tilde{\alpha} f_{rT}^2}{\gamma_T} + 4 \frac{\gamma_T \sqrt{\gamma_{rr}} \tilde{\alpha}^\prime}{r} - 6 \frac{\gamma_T \sqrt{\gamma_{rr}} \tilde{\alpha}}{r^2} \nonumber\\
+&& \!\!\!\!\!\!\!\! 2 K_{rr} \beta^{r\prime} - \frac{\gamma_T \tilde{\alpha} K_{rr}^2}{\sqrt{\gamma_{rr}}} + 2 \sqrt{\gamma_{rr}} \tilde{\alpha} K_{rr} K_T - 8 \frac{ \tilde{\alpha} f_{rT} f_{rrr}}{\sqrt{\gamma_{rr}}} + 2 \frac{\gamma_T \tilde{\alpha} f_{rrr}^2}{\gamma_{rr}^{3/2}} \nonumber\\
+&& \!\!\!\!\!\!\!\! 2 \frac{\gamma_T \tilde{\alpha} f_{rrr}}{r \sqrt{\gamma_{rr}}} - \frac{\gamma_T \tilde{\alpha}^\prime f_{rrr}}{\sqrt{\gamma_{rr}}} + 4 \pi \tilde{\alpha} \gamma_{rr}^{3/2} \gamma_T \left( P - \tau - D - S_r v^r \right), \\
\partial_t K_T =&& \!\!\!\!\!\!\!\! \beta^r K_T^\prime - \frac{\tilde{\alpha} \gamma_T f_{rT}^\prime}{\sqrt{\gamma_{rr}}} + 2 \frac{\beta^r K_T}{r} + \frac{\gamma_T \sqrt{\gamma_{rr}} \tilde{\alpha}}{r^2} + \frac{\tilde{\alpha} \gamma_T K_T K_{rr} }{\sqrt{\gamma_{rr}}} \nonumber\\
-&& \!\!\!\!\!\!\!\! \frac{\gamma_T f_{rT} \tilde{\alpha}^\prime}{ \sqrt{\gamma_{rr}}} - 2 \frac{\tilde{\alpha} f_{rT}^2}{\sqrt{\gamma_{rr}}} + 4 \pi \tilde{\alpha} \sqrt{\gamma_{rr}} \gamma_T^2 \left( P - \tau - D + S_r v^r \right). \label{eq:eom_excurv}
\end{eqnarray}
Our auxiliary variables $f_{rT}$ and $f_{rrr}$ also have evolution equations, given as
\begin{eqnarray}
\partial_t f_{rT} =&& \!\!\!\!\!\!\!\! \beta^r f_{rT}^\prime - \tilde{\alpha} \sqrt{\gamma_{rr}} \gamma_T K_T^\prime + \beta^{r\prime} f_{rT} - \tilde{\alpha}^\prime \sqrt{\gamma_{rr}} \gamma_T K_T + 2 \sqrt{\gamma_{rr}} \tilde{\alpha} K_T f_{rT} \phantom{\frac{1}{1}} \nonumber\\
-&& \!\!\!\!\!\!\!\! \frac{\tilde{\alpha} K_T f_{rrr} \gamma_T}{\sqrt{\gamma_{rr}}} + 2 \frac{ \beta^r f_{rT}}{r}, \\\nonumber\\
\partial_t f_{rrr} =&& \!\!\!\!\!\!\!\! \beta^r f_{rrr}^\prime - \tilde{\alpha} \sqrt{\gamma_{rr}} \gamma_T K_{rr}^\prime - 4 \gamma_{rr}^{3/2} \tilde{\alpha}^\prime K_T + 12 \frac{\gamma_{rr}^{3/2} \tilde{\alpha} K_T f_{rT}}{\gamma_{T}} - 4 \sqrt{\gamma_{rr}} \tilde{\alpha} K_T f_{rrr} \nonumber\\
-&& \!\!\!\!\!\!\!\! \frac{\gamma_T \tilde{\alpha} K_{rr} f_{rrr}}{\sqrt{\gamma_{rr}}} - 10 \sqrt{\gamma_{rr}} \tilde{\alpha} K_{rr} f_{rT} + 3 f_{rrr} \beta^{r\prime} + \gamma_{rr} \beta^{r\prime\prime} -\tilde{\alpha}^\prime \sqrt{\gamma_{rr}} \gamma_T K_{rr} \phantom{\frac{1}{1}} \nonumber\\ 
+&& \!\!\!\!\!\!\!\! 2 \frac{\gamma_T \sqrt{\gamma_{rr}} \tilde{\alpha} K_{rr}}{r} + 8 \frac{ \gamma_{rr}^{3/2} \tilde{\alpha} K_T}{r} + 16 \pi \tilde{\alpha} \gamma_{rr}^{3/2} \gamma_T S_r . \label{eq:eom_auxvars}
\end{eqnarray}
%
%
\section{Constraint-preserving boundary conditions} \label{sec:CPBC}

The freedom in choosing boundary conditions is encapsulated in the choice of ingoing characteristic modes. However, only a subset of the possible choices will also satisfy the constraints of the theory. Without care, the boundary treatment can therefore introduce constraint violating dynamics onto the computational grid, which degrade the quality of the numerical solution. In the general simulation conditions of full $3+1$ dimensionality and arbitrary gauge choices, the negative impact of constraint-violating boundary conditions can be lessened through a variety of means. This includes pushing the boundary farther out into the asymptotically flat regime, which is feasible on a cost basis with mesh refinement techniques, but is still wasteful. Implementing constraint-preserving boundary conditions in such general situations can be quite complex, but in spherical symmetry it is more straightforward. We implement this in our GR fluid code, and describe how to do so in this section. 

We can restrict our choice of boundary conditions to the narrower set of the constraint-preserving ones via an analysis of the characteristics of the system of constraints themselves. The constraints may possess ingoing characteristic modes, and by setting them to zero we can obtain the restrictions that we seek. In this section, we describe this procedure for the Einstein-Christoffel formulation of general relativity in spherical symmetry, following~\cite{Calabrese:2001kj}.

We begin by taking time derivatives of the constraints Eqs.~\eqref{eq:constraints_main} and~\eqref{eq:constraints_auxvars}. On the right-hand sides we can trade time derivatives for space derivatives using the equations of motion of the main variables, Eqs.~\eqref{eq:eom_metric},~\eqref{eq:eom_excurv}, and~\eqref{eq:eom_auxvars}. Then one rewrites the space derivatives of the main variables in terms of the constraints themselves, and their space derivatives. Since we are only interested in the characteristics of the system, we can keep the principal parts and discard lower-order terms throughout this procedure, which greatly simplifies the calculation. The result, up to the principal part, is
\begin{eqnarray}
\partial_t C &=& \beta^r C^\prime - \tilde{\alpha} \frac{\gamma_T}{\sqrt{\gamma_{rr}}} C_r^\prime + l.o. \\
\partial_t C_r &=& -\tilde{\alpha} \sqrt{\gamma_{rr}} \gamma_T C^\prime + \beta^r C_r^\prime + l.o. \phantom{\frac{1}{1}}\\
\partial_t C_{rrr} &=& \beta^r C_{rrr}^\prime + l.o. \phantom{\frac{1}{1}}\\
\partial_t C_{rT} &=& \beta^r C_{rT}^\prime + l.o. \phantom{\frac{1}{1}}.
\end{eqnarray}
This system is of the form $\partial_t \vec{C} = A \vec{C}^\prime + l.o.$ for a matrix $A$, thus the characteristic variables can be found by multiplying $\vec{C}$ on the left by the matrix whose rows are left-eigenvectors of $A$. The characteristic variables and their speeds are
\begin{eqnarray}
C_1 &=& C + \frac{C_r}{\sqrt{\gamma_{rr}}} \;\;\;\;\;\;\;\; (v_1^c = -\beta^r + \tilde{\alpha} \gamma_T) \\
C_2 &=& C - \frac{C_r}{\sqrt{\gamma_{rr}}} \;\;\;\;\;\;\;\; (v_2^c = -\beta^r - \tilde{\alpha}\gamma_T) \\
C_3 &=& C_{rrr} \;\;\;\;\;\;\;\; \;\;\;\;\;\;\;\;\; (v_3^c = -\beta^r )\phantom{\frac{1}{1}} \\
C_4 &=& C_{rT} \;\;\;\;\;\;\;\; \:\;\;\;\;\;\;\;\;\; (v_4^c = -\beta^r ). \phantom{\frac{1}{1}}
\end{eqnarray}
The modes $C_2$, $C_3$, $C_4$ are all guaranteed ingoing for non-negative shift $\beta^r$ (which we will have in our chosen coordinates), so by setting them to zero we obtain the restrictions on our choice of boundary conditions such that no constraint-violating modes are being injected. $C_1$ may be ingoing or outgoing, but in the coordinates we use it will always be outgoing and thus must be left alone.

Setting $C_2$, $C_3$, $C_4$ to zero, writing the resulting expressions in terms of the main variables and their space derivatives, then trading space derivatives for time derivatives using the equations of motion of the main variables, Eqs.~\eqref{eq:eom_metric},~\eqref{eq:eom_excurv}, and~\eqref{eq:eom_auxvars}, yields modified evolution equations which we implement on the boundary. This fixes a subset of the ingoing modes of the main variables. Namely, $C_2=0$ fixes $u_4$, $C_3=0$ fixes $u_1$, and $C_4=0$ fixes $u_2$. The remaining ingoing mode $u_3$ is freely specifiable, and corresponds to a gauge freedom. This is all expressed via modified evolution equations for the main variables, as
\begin{eqnarray}
\partial_t \gamma_{rr} &=& 2 \gamma_{rr} \beta^{r\prime} - 2 \tilde{\alpha} \sqrt{\gamma_{rr}} \gamma_T K_{rr} - 8\frac{\gamma_{rr} f_{rT}}{g_T} \beta^r + 2 f_{rrr} \beta^r , \\
\partial_t \gamma_T &=& -2\tilde{\alpha} \sqrt{\gamma_{rr}} \gamma_T K_T + 2 \beta^r f_{rT}, \phantom{\frac{1}{1}} \\
\partial_t K_T &=& \frac{1}{2} \partial_t u_6 + \frac{3}{4} \frac{\tilde{\alpha} f_{rT}^2}{\sqrt{\gamma_{rr}}} - \frac{1}{4}\frac{\beta^r \sqrt{\gamma_{rr}}}{r^2} - \frac{1}{4} \frac{\beta^r f_{rT}^2}{\sqrt{\gamma_{rr}} \gamma_T} - \frac{1}{2} \frac{\tilde{\alpha} f_{rrr} f_{rT} \gamma_T}{\gamma_{rr}^{3/2}} - \frac{1}{2} \frac{\tilde{\alpha}^\prime f_{rT} \gamma_T}{\sqrt{\gamma_{rr}}} \:\:\:\: \nonumber\\
&+& \frac{\tilde{\alpha} f_{rT} \gamma_T}{r \sqrt{\gamma_{rr}}} + \frac{1}{4} \frac{\tilde{\alpha} \sqrt{\gamma_{rr}} \gamma_T}{r^2} + \frac{1}{2} \frac{\beta^r f_{rT} K_{rr}}{\gamma_{rr}} - \frac{1}{2} \tilde{\alpha} f_{rT} K_T + \frac{1}{2} \frac{\beta^r f_{rT}K_T}{\gamma_T} + \frac{1}{2} \tilde{\alpha}^\prime \gamma_T K_T \nonumber\\
&-& \frac{\tilde{\alpha} \gamma_T K_T}{r} + \frac{1}{2} \frac{\tilde{\alpha} f_{rrr} \gamma_T K_T}{\gamma_{rr}} - \frac{1}{2} \frac{\beta^r K_{rr} K_T}{\sqrt{\gamma_{rr}}} - \frac{1}{4} \tilde{\alpha} \sqrt{\gamma_{rr}} K_T^2 - \frac{1}{4} \frac{\beta^r \sqrt{\gamma_{rr}} K_T^2}{\gamma_T} \nonumber\\
&+& 2\pi \gamma_T \left( \beta^r + \tilde{\alpha} \gamma_T \right) S_r + 2\pi \gamma_T \sqrt{\gamma_{rr}} \left[ \beta^r \left( \tau + D \right) + \tilde{\alpha} \gamma_T \left( P + S_r v^r \right) \right], \phantom{\frac{1}{1}} \\
\partial_t f_{rT} &=& \frac{1}{2} \sqrt{\gamma_{rr}} \partial_t u_6 + \beta^{r\prime} f_{rT} - \frac{3}{4} \tilde{\alpha} f_{rT}^2 + \frac{\beta^r f_{rrr} f_{rT}}{\gamma_{rr}} + \frac{1}{4} \frac{\beta^r \gamma_{rr}}{r^2} - \frac{15}{4} \frac{\beta^r f_{rT}^2}{\gamma_T} + \frac{1}{2} \tilde{\alpha}^\prime f_{rT} \gamma_T \nonumber\\
&-& \frac{\tilde{\alpha} f_{rT} \gamma_T}{r} + \frac{1}{2} \frac{\tilde{\alpha} f_{rrr} f_{rT} \gamma_T}{\gamma_{rr}} - \frac{1}{4} \frac{\tilde{\alpha} \gamma_{rr} \gamma_T}{r^2} - \frac{1}{2} \frac{\beta^r f_{rT}K_{rr}}{\sqrt{\gamma_{rr}}} - \frac{\tilde{\alpha} f_{rT} \gamma_T K_{rr}}{\sqrt{\gamma_{rr}}} + \frac{1}{2} \tilde{\alpha} \sqrt{\gamma_{rr}} f_{rT} K_T \nonumber\\
&-& \frac{1}{2} \frac{\beta^r \sqrt{\gamma_{rr}} f_{rT} K_T}{\gamma_T} - \frac{1}{2} \frac{\tilde{\alpha} f_{rrr} \gamma_T K_T}{\sqrt{\gamma_{rr}}} - \frac{1}{2} \tilde{\alpha}^\prime \sqrt{\gamma_{rr}} \gamma_T K_T + \frac{\tilde{\alpha} \sqrt{\gamma_{rr}} \gamma_T K_T}{r} + \frac{1}{2} \beta^r K_{rr} K_T \nonumber\\
&+& \frac{1}{4} \tilde{\alpha} \gamma_{rr} K_T^2 + \frac{1}{4} \frac{\beta^r \gamma_{rr} K_T^2}{\gamma_T} - 2\pi \gamma_{rr} \gamma_T \left[ \beta^r \left(\tau + D\right) + \tilde{\alpha} \gamma_T \left( P + S_r v^r \right) \right] \nonumber\\
&+& 2\pi \gamma_T \sqrt{\gamma_{rr}} \left( \beta^r + \tilde{\alpha} \gamma_T \right) S_r . \phantom{\frac{1}{1}}
\end{eqnarray}
%
%
\section{Post-Newtonian formalism} \label{sec:PNformalism}
The post-Newtonian (PN) approximation is a simultaneous expansion of the Einstein equations for weak gravity and slow speeds. The weak gravity condition can be understood as a linearization in perturbations on a flat background spacetime, i.e.~when coordinates exist such that
\begin{eqnarray}
g_{ab} \approx \eta_{ab} + h_{ab}, \:\:\: \vert h_{ab}\vert \ll 1, \label{eq:weakgrav}
\end{eqnarray}
where the second condition is understood to hold component-wise. Although in principle one has complete coordinate freedom, it is wise to restrict that freedom to those transformations which preserve Eq.~\eqref{eq:weakgrav} (called \emph{gauge transformations}). PN theory considers in addition to this a slow-motion limit, where a characteristic velocity $v$ of the system in question is small compared to the speed of light.

In this section we give an overview of the PN formalism in~\cite{barausse2013post}. We will focus on the broad strategy followed to 1 PN, with further details to be found in that work for obtaining 2 PN conservative effects + leading order dissipation (i.e. 2.5 PN).

We perturb the flat metric in Cartesian coordinates, and write its components as
\begin{eqnarray}
g_{00} &=& -\left(1 + 2 \frac{\phi}{c^2}\right), \:\:\: g_{0i} = \frac{\hat{\omega_i}}{c^3}, \nonumber\\
g_{ij} &=& \left( 1 - 2 \frac{\psi}{c^2} \right) \delta_{ij} + \frac{\hat{\chi}_{ij}}{c^2}. \label{eq:PNmetric}
\end{eqnarray}
The quantities $\phi, \hat{\omega}_i, \psi, \hat{\chi}$ are the PN potentials. The vector potential $\hat{\omega}_i$ can be decomposed into gradient and divergence-free parts, as
\begin{eqnarray}
\hat{\omega}_i = \partial_i \omega + \omega_i .
\end{eqnarray}
The second term $\omega_i$ obeys $\partial^i \omega_i = 0$. The tensor potential $\hat{\chi}_{ij}$ has its higher-order generalization of this,
\begin{eqnarray}
\hat{\chi}_{ij} = \left( \partial_{i} \partial_{j} - \frac{1}{3} \delta_{ij} \nabla^2 \right) \chi + \partial_{(i} \chi_{j)} + \chi_{ij},
\end{eqnarray}
with $\chi_i$ divergenceless, and $\chi_{ij}$ traceless and obeying $\partial^i \chi_{ij}=0$.

The Poisson gauge is given by
\begin{eqnarray}
\partial^i \hat{\omega_i} = \partial^i \hat{\chi}_{ij} = 0.
\end{eqnarray}
Notice these are 4 conditions, as one expects for a gauge choice. Based on the decompositions of $\hat{\omega}_i$ and $\hat{\chi}_{ij}$, this gauge enforces $\omega = \chi = \chi_i = 0$. Thus we replace $\hat{\omega}_i$ with $\omega_i$ and $\hat{\chi}_{ij}$ with $\chi_{ij}$ in Eqs.~\eqref{eq:PNmetric}.

At linear order the perturbed Einstein equations in this gauge will result in elliptic equations for the potentials $\phi, \psi, \omega_i$ of the form $\nabla^2 \phi = \ldots$ , etc.~with complicated right-hand sides. In spherical symmetry, since we can always choose coordinates which diagonalize the metric and render the spatial part conformally flat, we need only consider $\phi$ and $\psi$. As we mentioned in the introduction, this means the dissipative sector is not active in our implementation, but implementing the formalism in spherical symmetry is a logical first step to take.

The PN equations are obtained by first projecting Einstein's equation onto directions parallel and orthogonal to the fluid 4-velocity $u^a$. We did a similar decomposition in Sec.~\eqref{sec:ECsystem} to obtain Hamiltonian and momentum constraints and evolution equations, using instead the normals to the timelike foliation of the spacetime $n^a$. In the current PN context we impose the Poisson gauge and expand the projected Einstein equations in powers of $1/c$. Further details of this procedure we leave to~\cite{blanchet1990post,barausse2013post}.

At 1 PN order we have a further simplification coming from Einstein's equations, 
\begin{eqnarray}
\psi = \phi + \mathcal{O}\left( \frac{1}{c^2} \right).
\end{eqnarray}
Thus we need only worry about a single potential $\phi$ at this order. It is governed by the elliptic equation
\begin{eqnarray}
\nabla^2 \phi = 4\pi \left( 3 \frac{P}{c^2} + \rho_0(1+\epsilon) \right) + \frac{2}{c^2} \partial_i \phi \; \partial_i \phi + 8\pi \rho_0 (1+\epsilon) \left( \frac{v}{c}\right)^2 - \frac{3}{c^2} \partial_t^2 \phi + \mathcal{O}\left( \frac{1}{c^4} \right)\!\!. \label{eq:PNphieqn}
\end{eqnarray}
Here, $(P,\rho_0,\epsilon,v)$ are respectively the pressure, rest mass density, specific internal energy, and radial velocity of the perfect fluid in our system. Indices are raised and lowered with the flat metric, therefore we ignore the placement of spatial indices (but repeated indices still denote summation).

A black hole is included into this system in the following manner. We consider the limit of vanishing fluid variables, and call the resulting potential $\phi_{\mathrm{BH}}$, which satisfies
\begin{eqnarray}
\nabla^2 \phi_{\mathrm{BH}} = \frac{2}{c^2} \partial_i \phi_{\mathrm{BH}} \; \partial_i \phi_{\mathrm{BH}} - \frac{3}{c^2} \partial_t^2 \phi_{\mathrm{BH}} + \mathcal{O}\left( \frac{1}{c^4} \right). \label{eq:PNphieqn_BHonly}
\end{eqnarray}
We take $\phi_{\mathrm{BH}}$ to be a black hole in vacuum, to the given PN order. The expression for this can be obtained from the Kerr one appearing in~\cite{barausse2013post} by setting the spin parameter to zero, and it reads
\begin{eqnarray}
\phi_{\mathrm{BH}} = -\frac{M}{r} + \frac{1}{c^2}\frac{M^2}{r^2} - \frac{3}{4}\frac{M^3}{r^3 c^4} + \mathcal{O}\left(\frac{1}{c^6}\right) . \label{eq:phiBH}
\end{eqnarray}
One then writes the fluid contribution to the potential as $\phi_{\mathrm{fluid}}$, defined by $\phi \equiv \phi_{\mathrm{fluid}} + \phi_{\mathrm{BH}}$, and plugs this definition of $\phi$ into Eq.~\eqref{eq:PNphieqn} and eliminates terms using Eq.~\eqref{eq:PNphieqn_BHonly}, to obtain
\begin{eqnarray}
\nabla^2 \phi_{\mathrm{fluid}} &=& 4\pi \left( 3 \frac{P}{c^2} + \rho_0(1+\epsilon) \right) +  \frac{2}{c^2} \partial_i \phi_{\mathrm{fluid}} \: \partial_i \phi_{\mathrm{fluid}} + \frac{4}{c^2} \partial_i \phi_{\mathrm{fluid}} \: \partial_i \phi_{\mathrm{BH}} \nonumber\\
 &+& 8\pi \rho_0 (1+\epsilon) \left( \frac{v}{c}\right)^2 - \frac{3}{c^2} \partial_t^2 \phi_{\mathrm{fluid}} + \mathcal{O}\left( \frac{1}{c^4} \right). \label{eq:phieqn}
\end{eqnarray}
The black hole coupling enters through the $(4/c^2) \partial_i \phi_{\mathrm{fluid}} \: \partial_i \phi_{\mathrm{BH}}$ term. 

At the same order in the general case (i.e.~not in spherical symmetry) one also has the split $\omega^i \equiv \omega^i_{\mathrm{fluid}} + \omega^i_{\mathrm{BH}}$ obtained in the same way, and its associated Poisson equation for the fluid part
\begin{eqnarray}
\nabla^2 \omega^i_{\mathrm{fluid}} = 4 \left( 4\pi \rho_0 (1+\epsilon) v^i + \partial_i \partial_t \phi_{\mathrm{fluid}} \right) + \mathcal{O}\left( \frac{1}{c^2} \right). \label{eq:omeeqn}
\end{eqnarray}

Having given a broad overview of the strategy employed in this PN formalism, we leave further details to~\cite{barausse2013post}, including the extension to 2 PN in the conservative sector, as well as an elliptic treatment of the leading-order dissipative effects.

We also warn that the time-derivative terms in Eqs.~\eqref{eq:phieqn} and~\eqref{eq:omeeqn}, particularly the $\partial_t^2 \phi_{\mathrm{fluid}}$ term, have presented numerical difficulties which we have not overcome at this early stage of our implementation. Thus, in all simulations presented in this part we set those terms to zero.

%
%
\chapter{Implementation} \label{ch:PNimp}

\section{High-resolution shock capturing method} \label{sec:HRSC}
The treatment of fluids using finite differences and artificial viscosity that we used in Parts~\eqref{part:I} and~\eqref{sec:fracdim} is not well-suited for handling very large gradients, which occur in shockwaves or near the matter-vacuum interface at the surface of stars. Such a treatment is too dissipative to accurately capture these phenomena. In this part, since we seek to simulate isolated fluid bodies in vacuum, we therefore must use more advanced numerical methods to treat the fluid; namely, high-resolution shock capturing (HRSC) methods.

In HRSC methods one discretizes the fluid into finite volumes or cells. One works with fluid quantities that are averaged over both space and time over each cell, which allows for the representation of large gradients. General resources for these methods include~\cite{leveque1990conservative,leveque2002finite}. The relativistic case brings significant complications, and HRSC applications in special and general relativity are discussed in eg.~\cite{marti2003numerical,font2008numerical} and references therein. Further references on specific aspects will be given in subsequent sections.

\begin{figure}[!h]
\centering
\hspace{0.0cm}\includegraphics[width=0.45\textwidth]{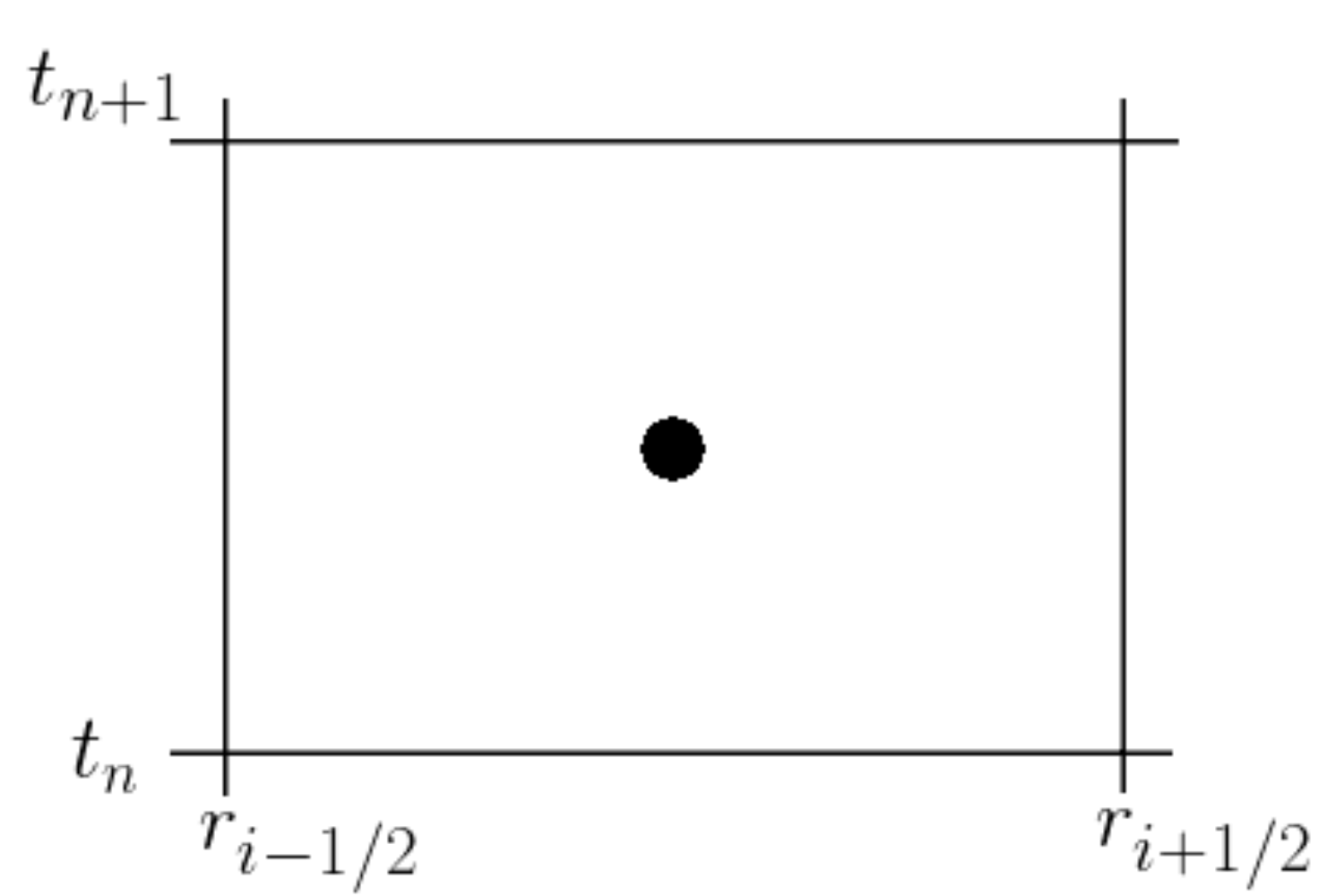}
\caption{A depiction of a finite volume cell, spanning $[r_{i-1/2},r_{i+1/2}]\times [t_n,t_{n+1}]$, where $t_{n+1} = t_n + \Delta t$ and $r_{i+1/2} = r_{i-1/2} + \Delta r$.} \label{fig:cell}
\end{figure}

The HRSC picture begins by imagining the $t-r$ plane covered by rectangular cells whose spatial interfaces are spaced $\Delta r$ apart and reside at half-integer spatial index $i+1/2$, and with temporal interfaces spaced $\Delta t$ apart residing at integer time index $n$. We will work with a schematic equation of the form of Eq.~\eqref{eq:ValSisph_fluxsplit} where there is a split flux,
\begin{eqnarray}
\partial_t \left( r^2 U \right) + \partial_r \left( \alpha r^2 F_1 \right) + r^2 \partial_r \left( \alpha F_2 \right) = r^2 \mathcal{S}, \label{eq:HRSCbegin}
\end{eqnarray}
where we have multiplied through by $r^2$ (which enters the time derivative at no cost), and where $U,F_1,F_2,S$ are understood to be densitized as in Sec.~\eqref{sec:Valencia}. The other hydrodynamic equations~\eqref{eq:ValDsph},~\eqref{eq:Valtausph} are also of this form if one sets the second flux $F_2$ to zero (and sets the source $\mathcal{S}$ to zero in the case of Eq.~\eqref{eq:ValDsph}).

Integrating Eq.~\eqref{eq:HRSCbegin} over an arbitrary cell, we obtain
\begin{eqnarray}
\int_{r_{i-1/2}}^{r_{i+1/2}} \int_{t_n}^{t_{n+1}} \left[ \partial_t \left( r^2 U \right) + \partial_r \left( \alpha r^2 F_1 \right) + r^2 \partial_r \left( \alpha F_2 \right) \right] dt dr = \int_{r_{i-1/2}}^{r_{i+1/2}} \int_{t_n}^{t_{n+1}} r^2 \mathcal{S} dt dr. \label{eq:HRSCfirstint}
\end{eqnarray}
Note that since the fluid is on a curved spacetime, one should use the proper volume element $\alpha \sqrt{\gamma_{rr}} \gamma_T dr dt$. However, we can achieve the same result as Eq.~\eqref{eq:HRSCfirstint} by pre-dividing by the spatial metric determinant $\alpha \sqrt{\gamma_{rr}} \gamma_T$ prior to integration.

We now denote spatial averages with a bar, eg.
\begin{eqnarray}
\bar{U} \equiv \frac{1}{r^2 \Delta r} \int_{r_{i-1/2}}^{r_{i+1/2}} U r^2 dr,
\end{eqnarray}
and temporal averages with a hat, eg.
\begin{eqnarray}
\hat{U} \equiv \frac{1}{\alpha \Delta t} \int_{t_{n}}^{t_{n+1}} U \alpha dt.
\end{eqnarray}
One might object that the prefactor $1/(r^2 \Delta r)$ in the spatial average should instead be $3/(r_{i-1/2}^3 - r_{i+1/2}^3)$, since this is the inverse volume associated with the domain of integration. We will indeed make this replacement later. Although they differ at $\mathcal{O} (\Delta r^3)$, the latter performs better near the origin~\cite{evans1986approach}. 

We can evaluate certain integrals in Eq.~\eqref{eq:HRSCfirstint} using the fundamental theorem of calculus, and the ones we cannot will be spatial or temporal averages so we will denote them using bars and hats. We obtain
\begin{eqnarray}
\left[ \bar{U} r^2 \Delta r \right]_{t_{n}}^{t_{n+1}} + \left[ r^2 \alpha \Delta t \hat{F}_1 \right]_{r_{1-1/2}}^{r_{i+1/2}} + \left[ \overline{\partial_r (\alpha \hat{F}_2)} \Delta t r^2 \Delta r \right] = \Delta t \Delta r \alpha  r^2 \hat{\bar{\mathcal{S}}} \label{eq:HRSCafterint}
\end{eqnarray}
So far, no approximations have been made, but we do so now. We first approximate the spatial average of the split flux term as
\begin{eqnarray}
\overline{\partial_r (\alpha \hat{F}_2)} &=& \frac{1}{r^2 \Delta r} \int_{r_{i-1/2}}^{r_{i+1/2}} \partial_r (\alpha \hat{F}_2) r^2 dr \nonumber\\
&\approx & \frac{1}{r^2 \Delta r} \left[\partial_r (\alpha \hat{F}_2) r^2 \right] \vert_{r_i} \Delta r \nonumber\\
&\approx & \frac{r_i^2}{r^2} \left[\frac{ \alpha_{i+1/2} \hat{F}_{2,i+1/2} - \alpha_{i-1/2} \hat{F}_{2,i-1/2}}{\Delta r}\right],
\end{eqnarray}
where the subscripts $i+1/2$, $i-1/2$ denote evaluation at the corresponding spatial locations. We evaluate spatial averages at cell centres $r_i$, thus eg.
\begin{eqnarray}
\overline{\partial_r (\alpha \hat{F}_2)}_i = \frac{ \alpha_{i+1/2} \hat{F}_{2,i+1/2} - \alpha_{i-1/2} \hat{F}_{2,i-1/2}}{\Delta r},
\end{eqnarray}
as well as evaluate temporal averages at time $t_{n+1/2}$ (also the cell centre). Eq.~\eqref{eq:HRSCafterint} then becomes
\begin{eqnarray}
\bar{U}_i^{n+1} r_i^2 \Delta r - \bar{U}_i^n r_i^2 \Delta r + r_{i+1/2}^2 \alpha^{n+1/2}_{i+1/2} \Delta t \hat{F}^{n+1/2}_{1,i+1/2} - r_{i-1/2}^2 \alpha^{n+1/2}_{i-1/2} \Delta t \hat{F}^{n+1/2}_{1,i-1/2} \nonumber\\
+ \Delta t r_i^2 \left( \alpha^{n+1/2}_{i+1/2} \hat{F}^{n+1/2}_{2,i+1/2} - \alpha^{n+1/2}_{i-1/2} \hat{F}^{n+1/2}_{2,i-1/2} \right) = \Delta t \Delta r \alpha_i^{n+1/2} r_i^2 \hat{\bar{\mathcal{S}}}_i^{n+1/2}.
\end{eqnarray}
Rearranging and replacing $1/(r_i^2 \Delta r) \rightarrow 3/(r_{i+1/2}^3 - r_{i-1/2}^3)$, we finally obtain
\begin{eqnarray}
\bar{U}_i^{n+1} &=& \bar{U}_i^n - \frac{3 \Delta t}{r_{i+1/2}^3 - r_{i-1/2}^3} \biggl( r_{i+1/2}^2 \alpha^{n+1/2}_{i+1/2} \hat{F}^{n+1/2}_{1,i+1/2} - r_{i-1/2}^2 \alpha^{n+1/2}_{i-1/2} \hat{F}^{n+1/2}_{1,i-1/2}  \nonumber\\ 
&+& r_i^2 \left[ \alpha^{n+1/2}_{i+1/2} \hat{F}^{n+1/2}_{2,i+1/2} - \alpha^{n+1/2}_{i-1/2} \hat{F}^{n+1/2}_{2,i-1/2} \right] \biggr) + \Delta t \alpha_i^{n+1/2} r_i^2 \hat{\bar{\mathcal{S}}}_i^{n+1/2}.
\end{eqnarray}

We thus see explicitly here that the evolution of the variable $U$, averaged over the cell, is governed by a flux balance at the cell interfaces and the source terms averaged over the cell. This is why the flux-conservative form of hydrodynamics is convenient for HRSC methods.

Computation of the fluxes is a large subject. Since the variables are defined at cell centres, some interpolation of the variables to the cell interfaces must be done prior to computing the fluxes, and we describe the third-order interpolation we use in the next section (performed on the primitive variables). Let us denote the variables reconstructed from the cell-centred values to the left and right of the cell interface as $U^L,U^R$, respectively. The fluxes themselves are computed using the Harten-Lax-van Leer-Einfeldt~\cite{harten1983upstream,einfeldt1988godunov} (HLLE) formula. If the analytic fluxes themselves are given as functions of $U$ as $F_1(U)$ and $F_2(U)$, then the HLLE formula is given by
\begin{eqnarray}
\hat{F}_{1,i+1/2} = \frac{\lambda_{+} F_1(U^L_{i+1/2}) - \lambda_{-} F_1(U^R_{i+1/2}) + \lambda_{+} \lambda_{-} (U^R_{i+1/2} - U^L_{i-1/2})}{\lambda_{+} - \lambda_{-}},
\end{eqnarray}
where $\lambda_{\pm}$ are the maximum and minimum values of all the characteristic speeds at $r_{i+1/2}$ and $r_{i-1/2}$ and zero,
\begin{eqnarray}
\lambda_{+} = \mathrm{max}(0,\lambda_{0,i+1/2} ,\lambda_{+,i+1/2},\lambda_{-,i+1/2},\lambda_{0,i-1/2} ,\lambda_{+,i-1/2},\lambda_{-,i-1/2}), \nonumber\\
\lambda_{-} = \mathrm{min}(0,\lambda_{0,i+1/2} ,\lambda_{+,i+1/2},\lambda_{-,i+1/2},\lambda_{0,i-1/2} ,\lambda_{+,i-1/2},\lambda_{-,i-1/2}).
\end{eqnarray}
Since we have a split flux and we do not want to have the $\lambda_{+} \lambda_{-} (U^R_{i+1/2} - U^L_{i-1/2})$ term appearing twice, we use for $F_2$ 
\begin{eqnarray}
\hat{F}_{2,i+1/2} = \frac{\lambda_{+} F_2(U^L_{i+1/2}) - \lambda_{-} F_2(U^R_{i+1/2}) }{\lambda_{+} - \lambda_{-}}.
\end{eqnarray}

\section{The piecewise parabolic method} \label{ch:ppm}

The piecewise parabolic method (PPM) is third-order method of determining the hydrodynamic variables at the cell interfaces, which performs well in reproducing sharp gradients such as the surface of a star. PPM was originally presented in~\cite{CW1984}. Other accessible references are~\cite{Mignone2005,Laney1998}, which we follow here. In this section, we will describe the method as applied to a uniform 1-dimensional grid, and the reconstruction will be on the primitive hydrodynamic variables.

The method begins with an estimation of the primitive variables $q^n_{i+1/2}$ at time index $n$ and cell interface position index $i+\frac{1}{2}$ via a quartic polynomial interpolation over the 4 nearest cell centres at position indices $\{i-1$, $i$, $i+1$, $i+2\}$:
\begin{eqnarray}
q^n_{i+1/2} \approx \frac{1}{12} \left( -q^n_{i+2} + 7 q^n_{i+1} +7 q^n_{i} - q^n_{i-1} \right). \label{eq:quarticinterp}
\end{eqnarray}
This approximation is fourth-order accurate. However, slope limiters are applied to ensure that $q^n_{i+1/2}$ is bounded by the two values to the left and right of the interface, $(q^n_{i},q^n_{i+1})$. To achieve this, let us first rewrite Eq.~\eqref{eq:quarticinterp} as
\begin{eqnarray}
q^n_{i+1/2} = \frac{1}{2} \left( q^n_{i+1} + q^n_{i} \right) + \frac{1}{6} \left( \delta q^n_{i} - \delta q^n_{i+1} \right),
\end{eqnarray}
where $\delta q^n_{i} \equiv \frac{1}{2} (q^n_{i+1} - q^n_{i})$. We then bound $q^n_{i+1/2}$ by applying the slope limiter~\cite{VanLeer1997} to $\delta q^n_{i}$ and $\delta q^n_{i+1}$,
\begin{eqnarray}
\overbar{\delta q^n_i} \equiv {\begin{cases} \text{min}\left( \vert \delta q^n_i \vert, 2\vert q^n_i - q^n_{i-1} \vert, 2\vert q^n_i - q^n_{i+1}\vert \right) \text{sign}\left(\delta q^n_i \right) \mbox{      if } ( q^n_{i+1} - q^n_i ) ( q^n_i - q^n_{i-1} ) > 0 \\ 0 \;\;\;\;\;\mbox{ otherwise} \end{cases} },
\end{eqnarray}
which then completes the estimate as
\begin{eqnarray}
q^n_{i+1/2} = \frac{1}{2} \left( q^n_{i+1} + q^n_{i} \right) + \frac{1}{6} \left( \overbar{\delta q^n_{i}} - \overbar{\delta q^n_{i+1}} \right)
\end{eqnarray}

In the next step, we initialize the left and right limiting values at the right interface of cell $i$, written as $q^n_{i+1/2,L}$ and $q^n_{i+1/2,R}$, respectively, as
\begin{eqnarray}
q^n_{i+1/2,L} = q^n_{i+1/2,R} = q^n_{i+1/2}.
\end{eqnarray}
We then apply the following replacements, which ensure a monotonic profile for $q$. The first is a reversion to first-order interpolation if the value in the cell centre, $q^n_{i}$, is a maximum or minimum:
\begin{eqnarray}
q^n_{i+1/2,L} = q^n_{i+1/2,R} = q_i &\mbox{if } (q_{i-1/2,R} - q_i) (q_i - q_{i+1/2,L}) \leq 0.
\end{eqnarray}
The second replacement adjusts one of the interface values in the event that an extremum exists within the cell. The result is that the extremum is moved to the opposite interface, and this is done in such a way that the cell-average value is preserved:
\begin{eqnarray}
\begin{split}
 q^n_{i-1/2,R} = 3 q^n_i - 2 q^n_{i+1/2,L}& \\
\mbox{if } (q^n_{i+1/2,L}-q^n_{i-1/2,R}&) \left(q^n_i - \frac{1}{2}(q^n_{i+1/2,L}+q^n_{i-1/2,R})\right) > \frac{1}{6}(q^n_{i+1/2,L} - q^n_{i-1/2,R})^2 \end{split} \\
\begin{split}
 q^n_{i+1/2,L} = 3 q^n_i - 2 q^n_{i-1/2,R} \\
\mbox{if } (q^n_{i+1/2,L}-q^n_{i-1/2,R})& \left(q^n_i - \frac{q^n_{i+1/2,L}+q^n_{i-1/2,R}}{2}\right) < -\frac{(q^n_{i+1/2,L} - q^n_{i-1/2,R})^2}{6}.
\end{split}
\end{eqnarray}
Due to these modifications, one order of accuracy is lost, and the method is nominally third-order in smooth regions of the flow away from extrema~\cite{Laney1998}.

There exist additional dissipation and jump-discontinuity steepening algorithms, which we omit. The interested reader can consult the original work for those details~\cite{CW1984}.
%
%
\section{Primitive variable recovery} \label{sec:contoprim}

The computation of the hydrodynamic fluxes appearing in Eqs.~\eqref{eq:ValD},~\eqref{eq:ValSi}, and~\eqref{eq:Valtau} requires having the values of the primitive variables. In our numerical scheme the conservative variables are evolved, and after each time step the primitive variables must be recovered from the updated conservative variables. In Newtonian hydrodynamics, the conservative-to-primitive variable transformation is algebraic. In the relativistic case, this recovery is delicate due to the presence of the Lorentz factor, and in general a numerical root-finding algorithm must be employed. Using the equation of state, one can write down a function $f(z)$ of some variable $z$ which depends on the primitive variables, such that the correct values of the primitive variables yield $f(z)=0$. For example, one could use the pressure $z=P$ and then the root-finding algorithm would find the value of $P$ which solves the appropriate equation $f(P) =0$, and following this the rest of the primitive variables can be computed algebraically. 

Most recovery schemes require an initial guess for the root of $f(z)$. In this section we briefly review a \emph{bracketing} scheme presented in the appendix of~\cite{galeazzi2013implementation}, which uses causality and energy conditions to calculated upper and lower bounds on the root of a certain function $f(z)$. Following this, the brackets can be narrowed until the root is known to within a desired tolerance. This method is deterministic, in the sense that the transformation is determined entirely by the hydrodynamic data at the current time step, and thus it is reproducible\footnote{Although one would also have to use the same error handling policy.}. In particular, it does not require an initial guess being provided by eg. the primitive variable values at the previous time step. Here we follow~\cite{galeazzi2013implementation} but specialize to the ideal fluid equation of state $P=(\Gamma-1)\rho_0 \epsilon$ and spherical symmetry.

We choose our root-finding variable to be $z=W\sqrt{\gamma_{rr} v^r v^r} \equiv Wv$. Weighting by the Lorentz factor $W$ provides better handling of highly relativistic velocities $v \sim 1$ since $Wv$ is not bounded from above. Using instead $z=v$ makes the recovery procedure vulnerable to numerical errors when $v \sim 1$.

Begin by defining some auxiliary quantities 
\begin{eqnarray}
a=\frac{P}{\left(\rho_0(1+\epsilon)\right)}, \;\;\; q=\frac{\tau}{D}, \;\;\; r=\sqrt{\gamma_{rr}S^r S^r}, \;\;\; k=\frac{r}{(1+q)}.
\end{eqnarray}
One can show that $z=r/h$, $\rho_0 = D/W$, $W=\sqrt{1+z^2}$, $\epsilon = Wq - zr + W - 1$, and $h=(1+\epsilon) (1+a) = (W-zk)(1+q)(1+a)$. We impose that the pressure cannot exceed the total energy density, that it cannot be negative, and that the speed of sound is real and causal:
\begin{eqnarray}
0 \leq \!\!\! &a& \!\!\! \leq 1 \\
0 \leq \!\!\! &c_s^2& \!\!\! \leq 1.
\end{eqnarray}
However, for the ideal fluid equation of state with $c_s^2 = P\Gamma(\Gamma-1)/\left( \rho_0 (\Gamma-1) + P\Gamma\right)$, these conditions simply mean
\begin{eqnarray}
0 \leq \!\!\! &\epsilon& \!\!\! < \infty \\
0 \leq \!\!\! &\rho_0 & \!\!\! < \infty .
\end{eqnarray}

In the event that the conservative variables have taken on unphysical values, we would like to map them into physical values via an error handling policy. This requires knowing the physical bounds on the conservative variables. For example,
\begin{eqnarray}
q &=& \frac{\tau}{D} \nonumber\\
&=& \frac{\rho_0 h W^2 - P - \rho_0 W}{\rho_0 W} \nonumber\\
&=& W + W\epsilon -1 + W v^2 \frac{P}{\rho_0} \nonumber\\
&\geq & W + W\epsilon -1.
\end{eqnarray}
Since $\epsilon \geq 0$ and $W\geq 1$, we see that $q \geq 0$. 

Next, one can show that $k = v(1+a)/(1+v^2 a)$ and that $\partial_a k \geq 0$. Since $k$ is non-decreasing with $a$, we can explore its minimum values by setting $a=0$, which yields $k=v$. Thus, $0 \leq v \leq k$. By setting $a=1$ we can explore the maximum possible values of $k$. This gives $k=2v/(1+v^2)$, which is non-decreasing in $v$ over the interval $v\in [0,1)$. In this case, to lowest non-trivial order in $v$ we have $k\approx 2v$. Thus we have a final ordering inequality given by
\begin{eqnarray}
0\leq k/2 \leq v \leq k \leq \frac{2v}{1+v^2} < 1. \label{eq:vbracket}
\end{eqnarray}
Therefore $k \leq k_{\text{max}} = 2v_{\text{max}}/(1+v_{\text{max}}^2) <1$. The other inequalities will be used to bracket the root of $f(z)$, to be defined later.

Prior to recovery of the primitive variables, we implement the adjustment policy described in~\cite{galeazzi2013implementation}, which makes the primitive recovery procedure safe but does not necessarily correspond to physical values. The policy is as follows. If $D<\rho_{0,\text{atmos}}$, where $\rho_{0,\text{atmos}}$ is an artificially maintained lower bound on the rest mass density, then all variables are set to the artificial atmosphere values of $v=0$, $\rho_0 = \rho_{0,\text{atmos}}$. We use a polytropic equation of state to determine $\epsilon_{\text{atmos}}$, and set $\epsilon$ to that value as well. Next, if $q<0$ then we reset $q=0$ keeping $k$ and $D$ constant, which amounts to setting $\tau=0$ and adjusting $S_r$ such that $S=\sqrt{\gamma^{rr}S_r S_r} = kD$. Furthermore, if $k>k_{\text{max}}$ then $S$ is rescaled such that $r=k_{\text{max}} (1+q)$, which amounts to setting $S=k_{\text{max}}(D+\tau)$.

For given values of $z$ and conservative variables, we can write $W=\sqrt{1+z^2}$ and then the rest mass density and internal energy as
\begin{eqnarray}
\rho_0 (z) = \frac{D}{W(z)}, \;\;\; \epsilon(z) = W(z)q - zr + W(z) -1. \label{eq:recovery}
\end{eqnarray}
It is numerically more accurate in the Newtonian limit to replace $W-1$ with $z^2/(1+W)$, and we do so. Once $\rho_0$ and $\epsilon$ are obtained, one can obtain $P$ via the equation of state and then the enthalpy $h$. Then the velocity is recovered via
\begin{eqnarray}
v^r = \frac{\gamma^{rr}S_r}{DWh}.
\end{eqnarray}
We find the root of the function
\begin{eqnarray}
f(z) = z - \frac{r}{h},
\end{eqnarray}
which comes from the definitions of our auxiliary quantities above, where we use $h=(1+\epsilon)(1+a)$. During recovery, Eqs.~\eqref{eq:recovery} may yield values outside of the physical range. Thus for a given value of $z$ (and given conservative variables), after computing the candidate values of $\rho_0$ and $\epsilon$ we map them into the physical range via
\begin{eqnarray}
\rho_0(z) & \rightarrow & \text{max}(\rho_0(z),\rho_{0,\text{atmos}}), \nonumber\\
\epsilon(z) & \rightarrow & \text{max}(\epsilon(z),\epsilon_{\text{atmos}}),
\end{eqnarray}
prior to computing $f(z)$. The bracketing Eq.~\eqref{eq:vbracket} implies that the root of $f(z)$ is bounded above and below by
\begin{eqnarray}
z_{-} = \frac{k/2}{\sqrt{1-k^2/4}}, \;\;\; z_{+} = \frac{k}{\sqrt{1-k^2}}.
\end{eqnarray}
With these bounds given, a root-finding algorithm such as bisection or regula falsi will converge to the solution. We use the Illinois algorithm, which is regula falsi with an additional conditional step designed to circumvent slowly-converging situations.

Once the recovery is finished, we impose the atmosphere values on any points with $\rho_0 < \rho_{0,\text{atmos}}$ or $\epsilon < \epsilon_{\text{atmos}}$, and adjust any velocities with $v>v_{\text{max}}$ to $v_{\text{max}}$, where $v_{\text{max}}$ is a specified parameter (0.99 in our case).

\section{Boundary injection of hydrodynamic pulse} \label{sec:bdyinjection}

Since we have done the work necessary to control the boundary conditions, we can obtain fully dynamical evolutions by using stationary black hole solutions as initial data and then inject matter into the domain through the outer boundary. In this section we will see that we can inject a hydrodynamic pulse by setting the velocity $v^r$ to zero at the outer boundary, and control the injection speed with the specific internal energy $\epsilon$.

Looking at the hydrodynamic characteristic speeds Eq.~\eqref{eq:evals} and setting the velocity to zero yields
\begin{eqnarray}
\lambda_0 &=& -\beta^r \nonumber\\
\lambda_{+} &=& \alpha c_s \sqrt{\gamma^{rr}} - \beta^r \nonumber\\
\lambda_{-} &=& -\alpha c_s \sqrt{\gamma^{rr}} - \beta^r . \label{eq:charspeeds_hydro_veq0}
\end{eqnarray}
With this choice, both $\lambda_0$ and $\lambda_{-}$ are ingoing modes. Whether $\lambda_{+}$ is ingoing or outgoing depends on the sound speed $c_s$ (and coordinate conditions). With zero velocity the hydrodynamic characteristic variables Eq.~\eqref{eq:hydrocharvars} reduce to
\begin{eqnarray}
u_0 &=& \frac{\Gamma -1}{\Gamma}\rho_0 \nonumber\\
u_{\pm} &=& \frac{\rho_0}{2\Gamma} .
\end{eqnarray}
Thus in this zero velocity limit the characteristic variables are degenerate, with the amplitude of the injected pulse controlled by $\rho_0$ while the injection speed is controlled by $c_s$. We can control $c_s$ via $\epsilon$ through the relation
\begin{eqnarray}
c_s = \sqrt{\frac{\epsilon\Gamma \left( \Gamma -1\right)}{1+\epsilon\Gamma}}.
\end{eqnarray}

In simulations where the shift vector is zero, we will also prescribe a non-zero value of the velocity $v^r$ in order that the material wave velocity $\lambda_0$ be nonzero and ingoing.
%
%
\section{PN solver} \label{sec:PNsolver}
After a hydrodynamic update, the metric variables are solved for via the PN Eqs.~\eqref{eq:phieqn} and~\eqref{eq:omeeqn}. Since the primitive hydrodynamic variables appear on the right-hand sides, a transformation from conservative to primitive variables is necessary before the PN equations can be solved. To do this, we use the procedure described above in Sec.~\eqref{sec:contoprim}.

However, this transformation itself requires the updated metric component $\gamma_{rr}$ for raising and lowering indices of $S_r$ and $v^r$. Therefore we use the last available values of $\gamma_{rr}$ for the primitive variable recovery.

The PN equations~\eqref{eq:phieqn} and~\eqref{eq:omeeqn} are reduced to first-order in space by defining
\begin{eqnarray}
f_\phi &\equiv & \partial_r \phi_{\mathrm{fluid}} , \nonumber\\
f_\omega &\equiv & \partial_r \omega^r_{\mathrm{fluid}}. \label{eq:1stordaux}
\end{eqnarray}
The Laplacian in spherical symmetry is $\nabla^2 = \partial^2_r +(2/r) \partial_r$, and the PN equations~\eqref{eq:phieqn} and~\eqref{eq:omeeqn} become
\begin{eqnarray}
\partial_r f_{\phi} &=& - \frac{2}{r} f_\phi + 4\pi \left( 3 \frac{P}{c^2} + \rho_0(1+\epsilon) \right) +  \frac{2}{c^2} f_\phi^2 + \frac{4}{c^2} f_\phi \: \partial_r \phi_{\mathrm{BH}} \nonumber\\
 &+& 8\pi \rho_0 (1+\epsilon) \left( \frac{v}{c}\right)^2 - \frac{3}{c^2} \partial_t^2 \phi_{\mathrm{fluid}}, \label{eq:1stordPN} \\
 \partial_r f_{\omega} &=& 4 \left( 4\pi \rho_0 (1+\epsilon) v^i + \partial_t f_\phi \right).\nonumber
\end{eqnarray}

Eqs.~\eqref{eq:1stordaux} and~\eqref{eq:1stordPN} are integrated radially out from the origin using \texttt{LSODA}~\cite{petzold1983automatic}, which is implemented in Fortran in \texttt{ODEPACK}~\cite{hindmarsh1983odepack} and wrapped in Python in the \texttt{Scipy} package~\cite{scipy}. This routine takes internally defined steps, and we use linear interpolations of the right-hand sides to accommodate this. We use boundary conditions at the origin $(\phi_{\mathrm{fluid}},\omega^r_{\mathrm{fluid}},f_\phi,f_\omega) = 0$, since in spherical symmetry these variables have no information about material at greater radii. I.e.~If there is no matter at greater radii, the integration should yield $\phi_{\mathrm{fluid}},\omega^r_{\mathrm{fluid}} = 0 $ everywhere.

With this solution scheme, and treating the time derivative terms in Eqs.~\eqref{eq:1stordPN} with backward stencil finite differences, we experience the development of instabilities over time. This is due in part to the fact that we interpolate the right-hand sides in order to accommodate the internal steps of the solver. Resolving this issue will require more careful discretization of the equations, the use of a lower-order integration method less sensitive to interpolation errors, and possibly evolving the fluid on a finer grid than the PN potentials. Without having this issue resolved at our current stage, we instead turn off the time derivative terms entirely. We monitor those terms, computed during post-processing, to gauge how important they are, although the solution itself in this case is not correct so this monitoring is not necessarily reliable either.

The factor $\partial_r \phi_{\mathrm{BH}}$ in Eq.~\eqref{eq:1stordPN} is given analytically in~\cite{barausse2013post} as
\begin{eqnarray}
\partial_r \phi_{\mathrm{BH}} = \frac{M}{r^2} - \frac{2}{c^2} \frac{M^2}{r^3} + \frac{9}{4} \frac{M^3}{r^4 c^4}.
\end{eqnarray}

Once $\phi_{\mathrm{fluid}}$ and $\omega^r_{\mathrm{fluid}}$ are solved for, they are added to $\phi_{\mathrm{BH}}$ (Eq.~\eqref{eq:phiBH}) and $\omega_{\mathrm{BH}} = 0$. We also have $\psi = \phi_{\mathrm{fluid}} + \phi_{BH} -(7/4)M^2/(r^2 c^2)$. Using these potentials, the ADM variables $\alpha, \beta^r, \gamma_{rr}, \gamma_T$ are computed for use in the next time step in the hydrodynamic equations.
%
%
%
\section{Further code details}

We use a 3rd-order accurate Runge Kutta time stepping scheme that is \emph{total variation diminishing} (TVD)~\cite{shu1988efficient}, which means the point-to-point spatial variability of solutions does not increase in a global sense over time. For some variable $u$ governed by an equation $\partial_t u = L(u)$ for some operator $L$, the scheme is presented as (eg.~\cite{anderson2008simulating})
\begin{eqnarray}
u^{(1)} &=& u^n + \Delta t L(u^n), \nonumber\\
u^{(2)} &=& \frac{3}{4} u^n + \frac{1}{4} u^{(1)} + \frac{1}{4} \Delta t L(u^{(1)}), \\
u^{n+1} &=& \frac{1}{3} u^n + \frac{2}{3} u^{(2)} + \frac{2}{3} \Delta t L(u^{(2)}) \nonumber .
\end{eqnarray}
Here, the superscripts ${}^n$ and ${}^{n+1}$ denote times, whereas ${}^{(1)}$ and ${}^{(2)}$ denote intermediate values at substeps.

Spatial derivatives of graviational variables are performed with 2nd-order centred finite differences which reduce to 1st-order at boundaries such that they satisfy \emph{summation by parts}~\cite{Calabrese:2003vx}. This is required to avoid artificially injecting energy into the grid. In the interior of the grid, the difference operator $D_0$ is
\begin{eqnarray}
\left(D_0 u \right)_j = \frac{u_{j+1} - u_{j-1}}{2 \Delta x},
\end{eqnarray}
while at left and right boundary points the difference operators are, respectively,
\begin{eqnarray}
\left( D_{+} u \right)_j \approx \frac{u_{j+1} - u_j}{\Delta x}, \:\:\: \left( D_{-} u \right)_j \approx \frac{u_{j} - u_{j-1}}{\Delta x}.
\end{eqnarray}
Here the subscripts ${}_j$ indexes spatial points.

For suppressing unresolved high-frequency, short-wavelength noise in the gravitational variables in the GR fluid code, we employ Kreiss-Oliger dissipation~\cite{gustafsson1995time} that is 4th-order in the interior of the grid and 2nd-order at the boundaries~\cite{Calabrese:2003vx}. This is essential for stability of the code. The choice of 4th-order dissipation does not affect the accuracy order of the derivative operator above. The dissipation term is
\begin{eqnarray}
Q u \equiv -\sigma \left( \Delta x \right)^{3} D_{+} D_{+} D_{-} D_{-} u,
\end{eqnarray}
where we fix $\sigma = 0.15$ for all GR fluid simulations. This term is added to the right-hand side of the evolution equations for each variable.
%
%
\chapter{Code validation} \label{ch:codeval}

\section{Artificial analytic convergence test of the hydrodynamic solver} \label{sec:arttest}

In this test, we perform an artificial analytic convergence test of the hydrodynamics solver. This means choosing arbitrary space- and time-dependent functions for the hydrodynamic variables, plugging them into the equations of motion to obtain new source terms, then including those new source terms on the right-hand side of the hydrodynamic evolution equations in the solver. This forces our arbitrary choice of functions to be an analytic solution to the new system, and we can perform convergence tests against this artificial analytic solution. The new source terms are computed independently from the code, so as not to contaminate the test with possible errors in the code itself, and the arbitrary functions we choose are demanding for the solver.

This test can be made as demanding as one wishes, unlike the options available via other analytic solutions which are often time-independent or special in some way. We choose to make the test very demanding by including time- and space-dependence in the fluid variables and the metric.

The artificial solution we choose is written as
\begin{eqnarray}
\rho_0 &=& \phantom{0.2} \exp \left\lbrace \frac{-\left( r - (L/2) \left[ 1 + 0.1 \sin{\omega_r t} \right]\right)^2}{2 (L/40)^2} \right\rbrace \left( 2 + \sin{\omega t} \right) + \rho_{0,\mathrm{atmos}}, \\
v^r &=& 0.2 \exp \left\lbrace \frac{-\left( r - (L/2) \left[ 1 + 0.1 \sin{\omega_r t} \right]\right)^2}{2 (L/40)^2} \right\rbrace \sin{\omega t}, \\ 
\epsilon &=& \rho_0,
\end{eqnarray}
where $L=20M$ is the physical size of the domain, $\omega = 2 (2\pi/T)$ produces an oscillation in amplitude ($T$ is the total integration time $= 5M$), and $\omega_r = (3 e/8) (2\pi/T)$ produces an oscillation in position. These frequencies are chosen to have irrational ratios so that they are not commensurate~\cite{hilborn2000chaos}. As far as the physical appearance of this solution, it is a Gaussian pulse of width $L/40$ whose position is oscillating between $0.9-1.1\times L/2$, and with a rest mass density peak amplitude oscillating between $1-3$. It is important to note that the way the pulse travels is not at all determined by our choice of $v^r$, because a counteracting source term is added to the hydrodynamic equations which cancels its influence (to within the accuracy of the solver).

In order to test the dependence of the densitized variables coming directly from a varying metric, we also need to prescribe a time- and space-dependent metric (see Eqs.~\eqref{eq:ValSisph_src} and~\eqref{eq:Valtausph_src}). We therefore impose that the spacetime metric is that of a Schwarzschild black hole in isotropic coordinates, with a mass varying sinusoidally in time between $0.5-1.5M$ with a frequency $\pi (2\pi/T)$. This turns on terms in the hydrodynamic equations involving $\lbrace \alpha$, $g_{rr}$, $g_T\rbrace$ and their space and time derivatives. 

We also note that since the tails of the pulse approach the atmosphere values, the atmosphere algorithm is employed heavily there, resulting in a particularly demanding test. 

\begin{figure}[!h]
\centering
\hbox{\hspace{0.0cm}\includegraphics[width=1\textwidth]{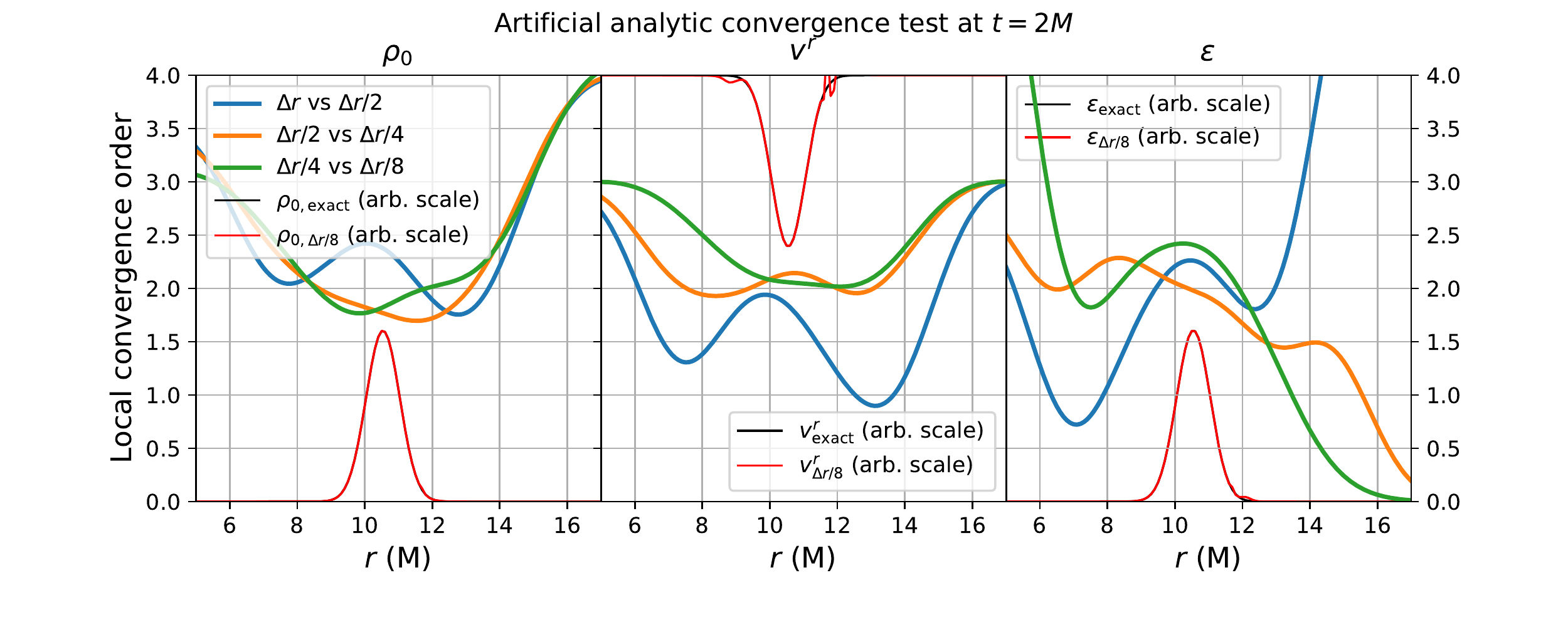}}
\caption{Artificial analytic convergence test of the hydrodynamic solver. The radial profile of the convergence order is plotted at $t=2M$ for three different resolution comparisons (thick lines), and snapshots of the primitive hydrodynamic variables are also shown for reference on an arbitrary linear scale (thin lines). The exact solution is overlaid with the numerical solution at resolution $\Delta r/8$. The convergence order in the bulk of the pulse is $\sim 2$ as one would expect.} \label{fig:art_conv_test}
\end{figure}


We display the results of this artificial analytic convergence test in Fig.~\eqref{fig:art_conv_test}. Local convergence factors are plotted versus radius at a representative time $t=2M$ for the primitive hydrodynamic variables $\rho_0$ (Left), $v^r$ (Centre), and $\epsilon$ (Right). For a given resolution $\Delta r$, the factors are given by
\begin{eqnarray}
\log \left[\frac{ \vert \rho_{0,\Delta r} - \rho_{0,\mathrm{analytic}}\vert }{\vert \rho_{0,\Delta r/2} - \rho_{0,\mathrm{analytic}}\vert }\right] / \log{2}
\end{eqnarray}
for $\rho_0$, as an example, where the subscripts $\Delta r$ and $\Delta r/2$ denote the solution obtained with those resolutions. The absolute value is taken at a specified time. In Fig.~\eqref{fig:art_conv_test} the fiducial resolution is $\Delta r = L/200$. The convergence factors are also smoothed with a Gaussian of width $1.2 M$, which removes uninformative very short-wavelength noise from the plots. We see convergence averaging around an order of $\sim 2$ in the bulk of the fluid pulse, consistent with the numerical methods. In the outer tails of the pulse, the convergence order observed is not representative of the method. We attribute this to the densities being extremely low there, comparable with the atmosphere value, and thus invocations of the atmosphere algorithm are more common.

%
%
It is important to keep in mind that this metric is not physical, but it does not have to be for this test. We are simply manufacturing enough dynamics in order to test the convergence of the hydrodynamic solver in a thorough way. The spacetime evolution equations are not employed in this test, rather the metric is prescribed at each time. The spacetime evolution will be tested in later sections.

\section{TOV oscillations in the Cowling approximation} \label{sec:TOVoscillations}

In this test, we evolve a TOV star with central density $\rho_0\vert_{r=0} = 1.28\times 10^{-3}$, obeying an initial equation of state $P=K \rho_0^\Gamma$ with $K=100$ and adiabatic index $\Gamma = 2$, which yields a $1.4 \: M_\odot$ star with a radius $\sim 14.15 $ km if we take $M_\odot = 1$. This choice of parameters has been used in the literature to test codes and the mode frequencies have been repeatedly obtained, eg.~\cite{font2001axisymmetric,font2002three,baiotti2010new,radice2011discontinuous}. They can also be obtained with perturbative calculations~\cite{yoshida2001quasi}. We perform this test in the Cowling approximation (fixed spacetime) in Schwarzschild-like coordinates. Performing this test with a dynamic spacetime requires regularization of the gravitational equations at the origin, which we have not implemented. In our main application we have the origin excised out of the grid due to a black hole there, so regularization is not an issue. Thus, this test should be regarded as a physical test of the hydrodynamic solver, without the time-dependent metric source terms being activated.

\begin{figure}[!h]
\centering
\hbox{\hspace{0.0cm}\includegraphics[width=1\textwidth]{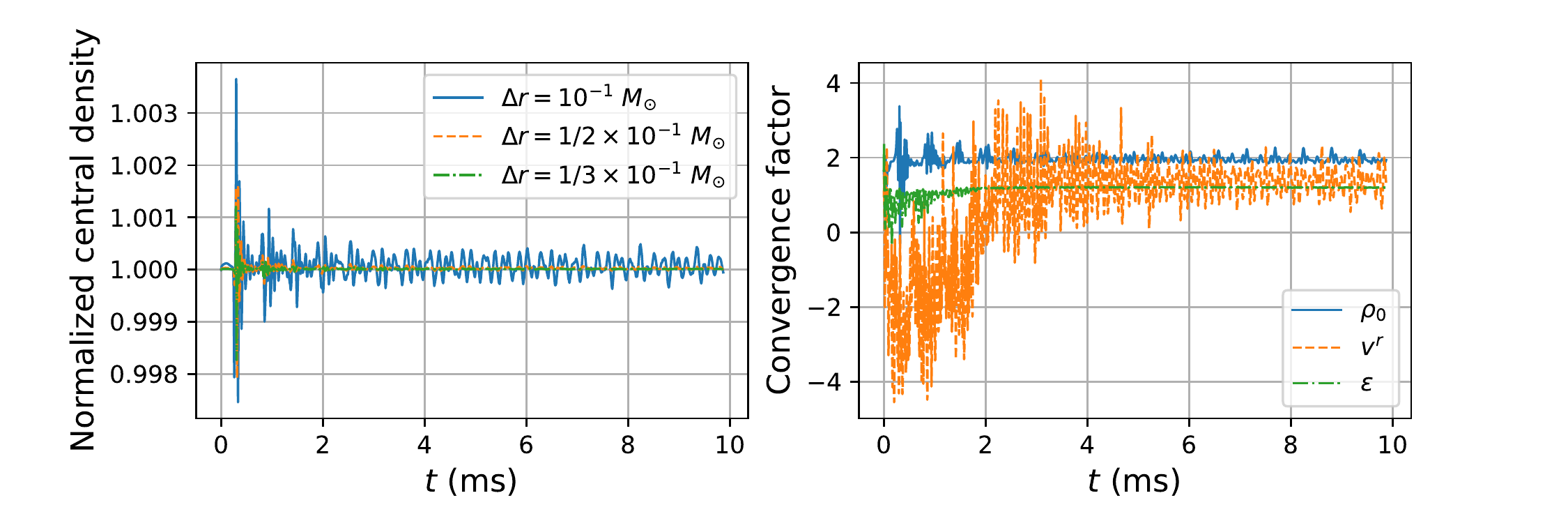}}
\caption{(Left): Time progression of the central density $\rho_0 \vert_{r=0}$ of a stable TOV star with central density $1.28\times 10^{-3}$, $K=100$, and $\Gamma=2$ for three different resolutions $\Delta r = \left\lbrace 1,1/2,1/3 \right\rbrace  \times 10^{-1} M_{\odot}$. The oscillations observed are excited by truncation errors and the interaction of the surface of the star with the artificial atmosphere. (Right): Global convergence factor in an $\mathcal{L}_2$ norm sense for the primitive hydrodynamic variables $\rho_0$, $v^r$, $\epsilon$. The norm is performed over the entire interior of the star. A convergence order of $\sim 2$ is observed for $\rho_0$, but only $\sim 1.2$ for $v^r$ and $\epsilon$ (after a transient phase of apparent non-convergence for $v^r$). This poor convergence is a well-known phenomenon associated with star-vacuum interfaces in finite volume codes (eg.~\cite{kastaun2006high,schoepe2017revisiting}).} \label{fig:TOV_rho0_conv}
\end{figure}
\FloatBarrier

Fig.~\eqref{fig:TOV_rho0_conv} displays the time progression of the normalized central density of the star for different resolutions (Left). Although the TOV solution is static, truncation errors during its numerical evolution excite the normal modes of the star. The global convergence factor in an $\mathcal{L}_2$ norm sense is also displayed (Right) for the primitive hydrodynamic variables $\rho_0$, $v^r$, $\epsilon$. The norm is taken over the entire interior of the star. We observe second-order convergence for the density, and degraded convergence of order $\sim 1.2$ for the remaining variables after $\sim 2$ ms. There is a transient phase during first $2$ ms when the velocity exhibits apparent non-convergence. We attribute this to the fact that the surface of the star is truncated at different radii for different resolutions, and each star must settle down from these slightly different initial configurations. Difficulties with the star-vacuum interface and the 
associated convergence issues are well-known for finite volume methods, see for example~\cite{schoepe2017revisiting}.

\begin{figure}[!h]
\centering
\hbox{\hspace{3.75cm}\includegraphics[width=0.5\textwidth]{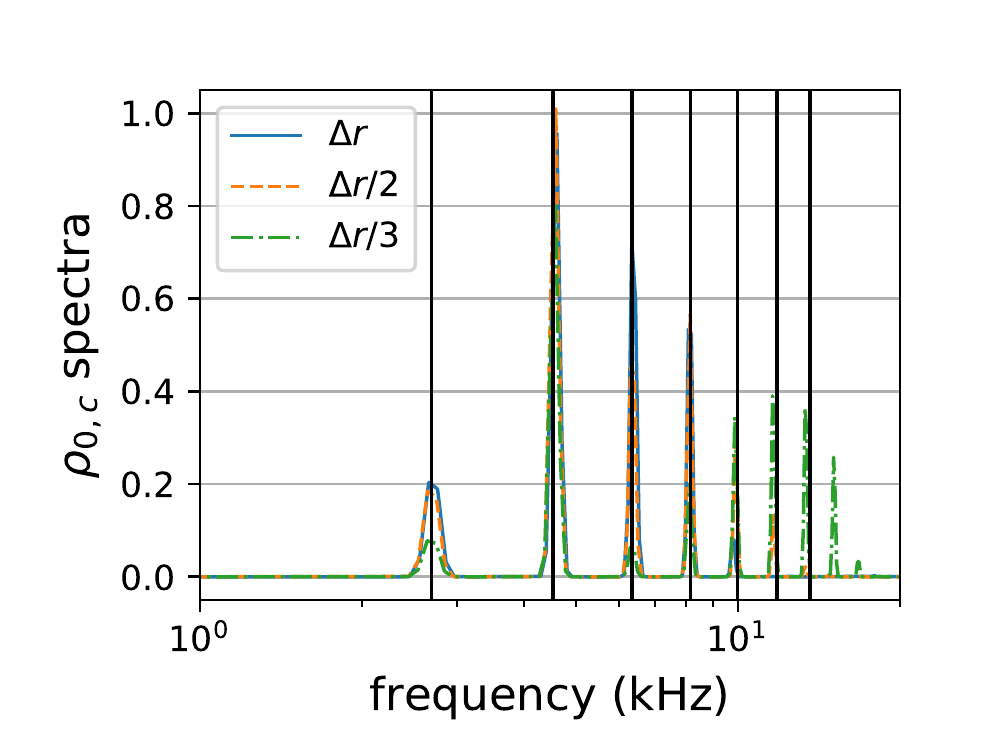}}
\caption{Frequency spectrum of the central density for different resolutions, with the known frequencies of the fundamental mode and first six overtones indicated with vertical lines for comparison. The frequencies were obtained in~\cite{font2002three} as $\sim \lbrace 2.696$, 4.534, 6.346, 8.161, 9.971, 11.806, $13.605 \rbrace $ kHz, respectively. We pre-process $\rho_{0,c}$ by subtracting its average and then applying a Gaussian window $\sim \exp\left\lbrace -(t-6.2 \text{ms})^2/2 (1.5 \text{ms})^2 \right\rbrace$ prior to computing the Fourier transform, which results in cleaner spectra which are less contaminated by the initial burst at $t=0$. The spectra are then normalized.} \label{fig:TOV_spectra}
\end{figure}

In Fig.~\eqref{fig:TOV_spectra} we display the frequency spectra of the central density of the star for different resolutions. To produce these plots, we first subtract from $\rho_{0,c}$ its average value over the whole duration $t=0-10$ ms. Secondly, since there exists an initial burst at $t=0$ visible in Fig.~\eqref{fig:TOV_rho0_conv} (Left), we decrease its contamination of the spectra by multiplying the time series by a Gaussian window function $\sim \exp\left\lbrace -(t-6.2 \text{ms})^2/2 (1.5 \text{ms})^2 \right\rbrace$. The offset of $6.2$ ms and width of $1.5$ ms are chosen simply to retain the central part of the time series while excluding the outer edges in a smooth manner. The windowed time series is more closely represented by a periodic function, and thus produces cleaner Fourier spectra. Similar methods are employed elsewhere, eg.~\cite{agrevz2007dynamics,radice2011discontinuous}. Then we compute the spectra and normalize them to 1. Also overlaid in Fig.~\eqref{fig:TOV_spectra} are vertical lines indicating the frequencies of the fundamental mode $F$ and its first six overtones $H_{1-6}$, which were found in~\cite{font2002three} to be $\sim \left\lbrace 2.696, 4.534, 6.346, 8.161, 9.971, 11.806, 13.605 \right\rbrace$ kHz, respectively.

\FloatBarrier

\section{Ingoing gauge pulse} \label{sec:gaugepulse}

In this test, we inject a gauge pulse into the domain of a static black hole and observe whether the gauge-invariant Misner-Sharp mass is unaffected at the order expected from the numerical method used. This pulse is a purely coordinate degree of freedom, thus no physical quantities should vary.

The gauge pulse consists of a Gaussian superposed on top of the Painlev\'{e}-Gullstrand value of the characteristic variable $u_3$ at the outer boundary. We implement this at the level of the right-hand side of $\partial_t u_3$, which is zero in Painlev\'{e}-Gullstrand coordinates. The modified right-hand side is
\begin{eqnarray}
\partial_t u_3 \vert_{\text{bdy}} = -\frac{2A(t-t_0)}{\tilde{\sigma}^2} e^{-(t-t_0)^2/\tilde{\sigma}^2}
\end{eqnarray}
We choose the pulse parameters to match those chosen in~\cite{Calabrese:2001kj}. Namely, $\tilde{\sigma} = 2M$, $t_0 = 5M$, $A=1$. Furthermore, the boundary is located at $r=30M$ and we evolve until $t_{\text{final}}=150M$, using three different resolutions $\Delta r = M/8$, $M/16$, $M/32$.

\begin{figure}[!h]
\centering
\hbox{\hspace{0.0cm}\includegraphics[width=1\textwidth]{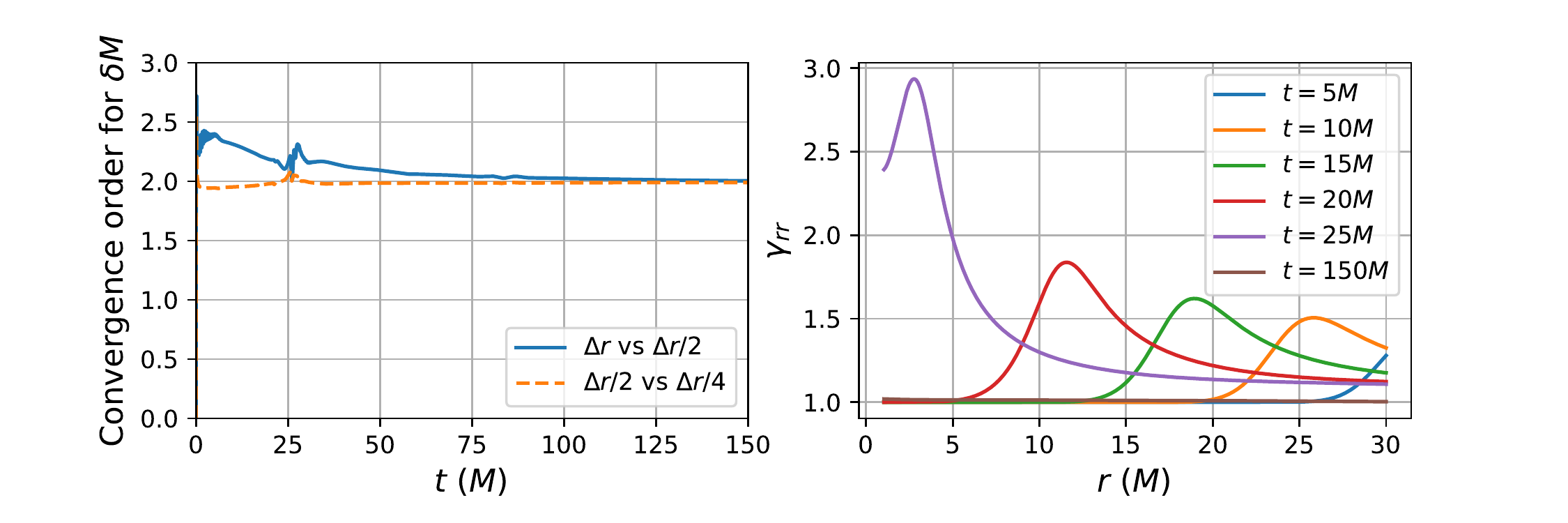}}
\caption{(Left): Convergence order for the mass error $\delta M = \mathcal{L}_2 \left\lbrace M_{\text{MS}} - M\right\rbrace /M$ as a gauge pulse propagates into a static black hole. The fiducial resolution is $\Delta r = M/8$. As resolution is increased the convergence order approaches $2$, consistent with the finite difference method used. (Right): Time progression of the radial component of the spatial metric $\gamma_{rr}$ from the same simulation ($\Delta r = M/32$ case shown). } \label{fig:gauge_mass_grr}
\end{figure}
\FloatBarrier

In Fig.~\eqref{fig:gauge_mass_grr} we plot the global convergence order for the Misner-Sharp mass in an $\mathcal{L}_2$ norm sense (Left), and a time progression of $\gamma_{rr}$ from the most resolved run. The fiducial resolution is $\Delta r = M/8$ in the (Left) plot. The convergence order is observed to approach $2$.

\begin{figure}[!h]
\centering
\hbox{\hspace{0.0cm}\includegraphics[width=1\textwidth]{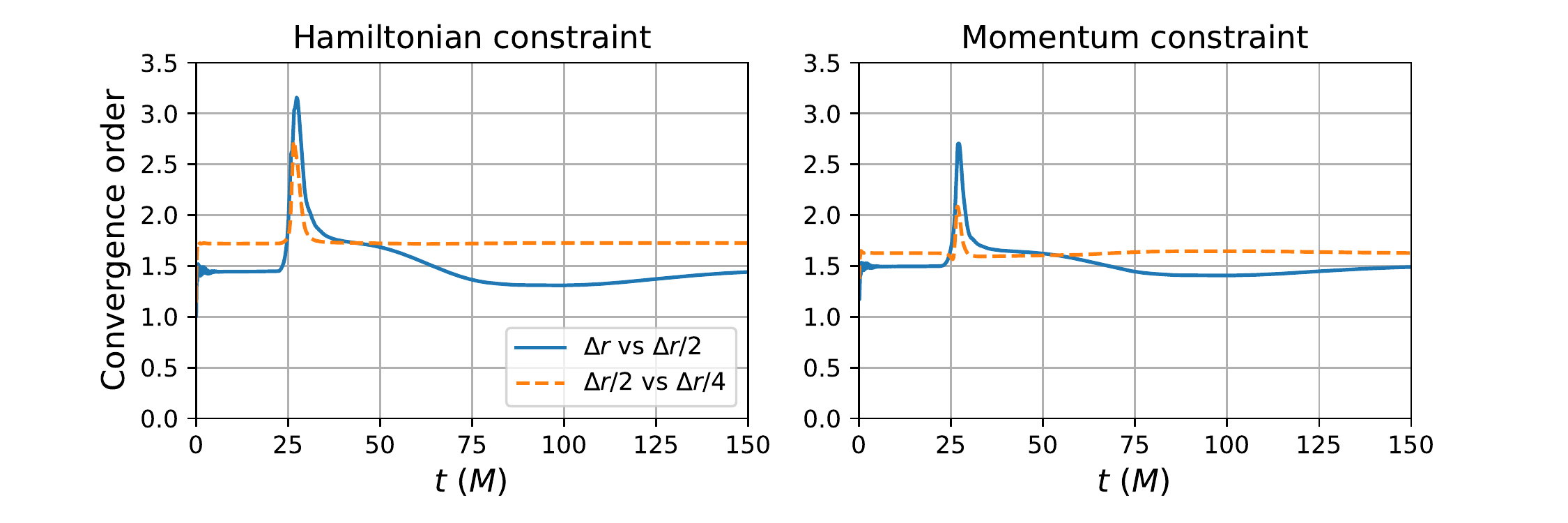}}
\caption{Convergence order for the Hamiltonian constraint (Left) and momentum constraint (Right) for a gauge pulse propagating into a static black hole. The pulse exits the domain through the excision surface around $T=25M$, where gradients are large and the error spikes. The fiducial resolution is $\Delta r = M/8$. A convergence order of $\sim 1.7$ is obtained in the most-resolved case. } \label{fig:gauge_constraints}
\end{figure}
\FloatBarrier

In Fig.~\eqref{fig:gauge_constraints}, we display the convergence order of the constraints in an $\mathcal{L}_2$ norm sense, yielding $\sim 1.7$ in the most-resolved case. A spike occurs when the gauge pulse exits the domain through the excision surface, where gradients are high and the errors grow.

\section{Ingoing hydrodynamic and gauge pulses} \label{sec:hydroandgaugepulse}
In this test, we place the outer boundary at $r=10M$ and evolve a black hole of initial mass $M$ until $t_{\text{final}}=30M$. A hydrodynamic pulse is injected through the outer boundary by imposing
\begin{eqnarray}
\rho_0 \vert_{\text{bdy}} &=& 
\begin{cases}
	A_{\rho_0} \sin^2( \frac{2\pi}{T} t )  & 0 < t \leq T/4 \\
	A_{\rho_0} & T/4 < t \leq T/4 + t_{\text{gap}} \\
	A_{\rho_0} \sin^2( \frac{2\pi}{T} (t-t_{\text{gap}}) ) & T/4 + t_{\text{gap}} < t \leq T/2 + t_{\text{gap}} \\
	0 & \text{otherwise} \label{eq:rho0bdy}
\end{cases},
\end{eqnarray}
which is a pulse profile with $\sim \sin^2(t)$ rise and fall, with a constant duration at max amplitude for a time $t_{\text{gap}}$. The specific internal energy is chosen to be
\begin{eqnarray}
\epsilon \vert_{\text{bdy}} = 10^2 \rho_0 \vert_{\text{bdy}},
\end{eqnarray}
and the pressure is determined from the ideal gas equation of state $P=\rho_0 \epsilon (\Gamma-1)$ with $\Gamma=2$, and $v^r=0$ for simplicity. For these tests, we choose $T=8M$ and $t_{\text{gap}}=M$, which gives a rise duration of $2M$, a constant duration of $M$, followed by a fall duration of $2M$. The amplitude is chosen as $A_{\rho_0} = 1.28 \times 10^{-4} M^{-2}$. This pulse results in a final black hole mass of $\sim 1.21 M$, a 21\% increase.

In one case, we inject only the hydrodynamic pulse. In another more demanding case, we also inject the same gauge pulse defined in Sec.~\eqref{sec:gaugepulse}, but delayed so that the peak occurs at $t=10M$. This delay is done to reduce the overlap of the hydrodynamic and gauge pulses at the boundary, although the overlap is still non-zero there since the gauge pulse is Gaussian. The pulse is also different than the previous test in that it is injected at $r=10M$ rather than $r=30M$, as in Sec.~\eqref{sec:gaugepulse}.

\begin{figure}[htbp]
\centering
\hbox{\hspace{0cm}\includegraphics[width=1\textwidth]{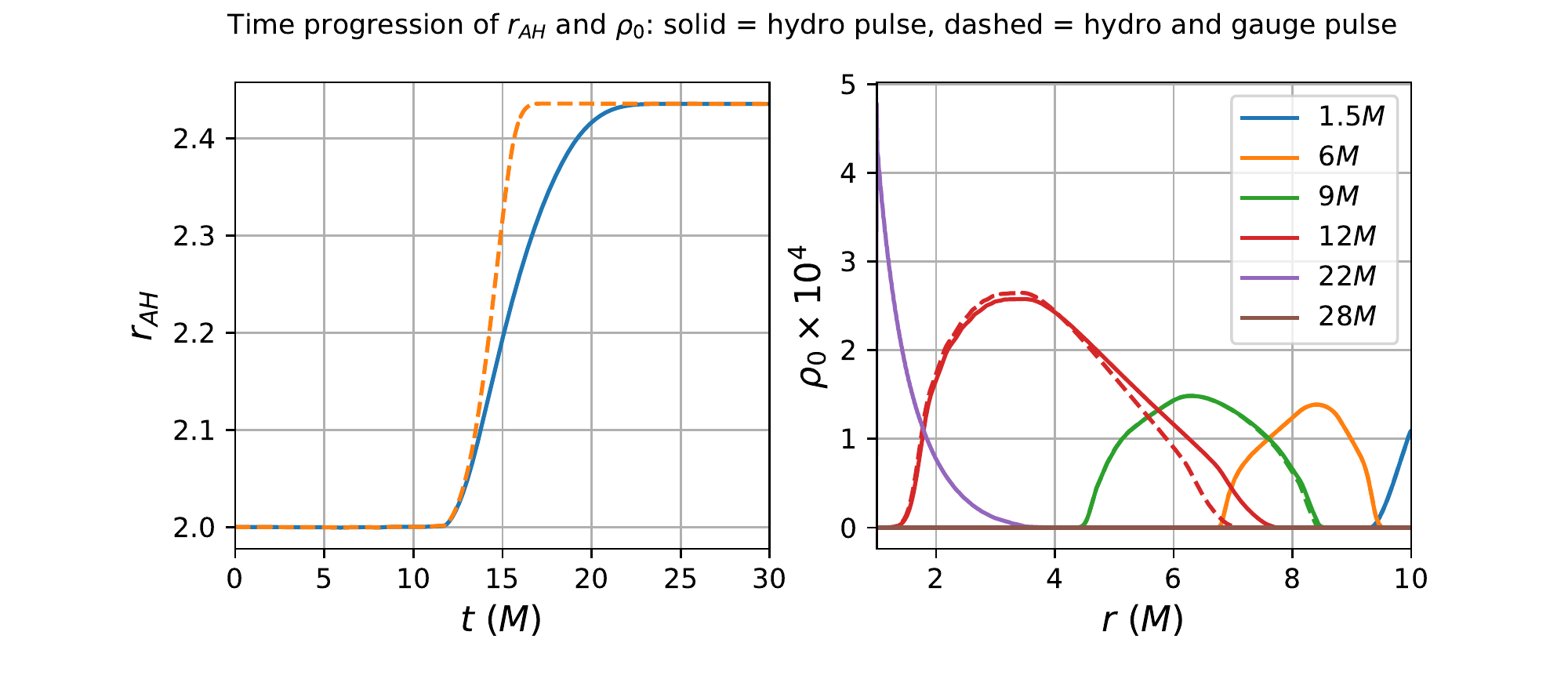}}
\caption{Time progression of the apparent horizon areal radius $r_{\text{AH}}$ (Left) and the fluid rest mass density $\rho_0$ (Right). Solid lines correspond to the case with a hydrodynamic pulse, dashed lines correspond to the case with both hydrodynamic and gauge pulses. The gauge pulse causes the radial coordinate to evolve differently, as well as the hypersurface, contributing to a slightly different appearance of the hydrodynamic pulse profile while the gauge pulse propagates through the grid, as well as a different evolution of the coordinate-dependent apparent horizon. The difference in the pulse profile is also affected by the fact that the gauge pulse partly overlaps with the hydrodynamic pulse at the outer boundary, which affects the hydrodynamic characteristic speeds there and thus the hydrodynamic pulse profile itself. Both cases settle down to the same apparent horizon area at late times.} \label{fig:hydro_gauge_AH_rho0}
\end{figure}

Fig.~\eqref{fig:hydro_gauge_AH_rho0} shows a time progression of the areal radius of the apparent horizon (Left) and the rest mass density of the fluid (Right). Solid lines correspond to the case with an ingoing hydrodynamic pulse only, while dashed lines correspond to the case with consecutive hydrodynamic and gauge pulses. The apparent horizon radius is defined implicitly by $f_{rT} - \sqrt{\gamma_{rr}} K_T = 0$, and is coordinate-dependent. The gauge pulse affects the evolution of the coordinates, so it is expected that it would produce a modified evolution of the apparent horizon. However, after the black hole has absorbed everything, both cases settle down to the same black hole mass (Left). There is a slight discrepancy apparent in the hydrodynamic pulse profile at $t=12M$ between the two cases (Right). This is partly due to a disagreement between the radial coordinates in the two cases; plotting $\rho_0$ vs the areal radius for both cases increases the coincidence of the two pulse profiles. However this does not completely account for it, and the remaining disagreement we attribute to a combination of different hypersurface evolution and a physical difference between the injected hydrodynamic pulses arising from the non-zero overlap between the hydrodynamic and gauge pulses at the boundary (since the latter is a Gaussian and so never vanishes). Such an overlap modifies the lapse and $\gamma^{rr}$ at the outer boundary, which from Eq.~\eqref{eq:charspeeds_hydro_veq0} affects the fluid characteristic speeds there, which can result in a slightly different pulse profile. This small difference would magnify as the fluid enters the black hole around $t=12M$, where volume decreases and the fluid velocity reaches $\sim 0.1 c$.

\FloatBarrier

\begin{figure}[htbp]
\centering
\hbox{\hspace{0cm}\includegraphics[width=1\textwidth]{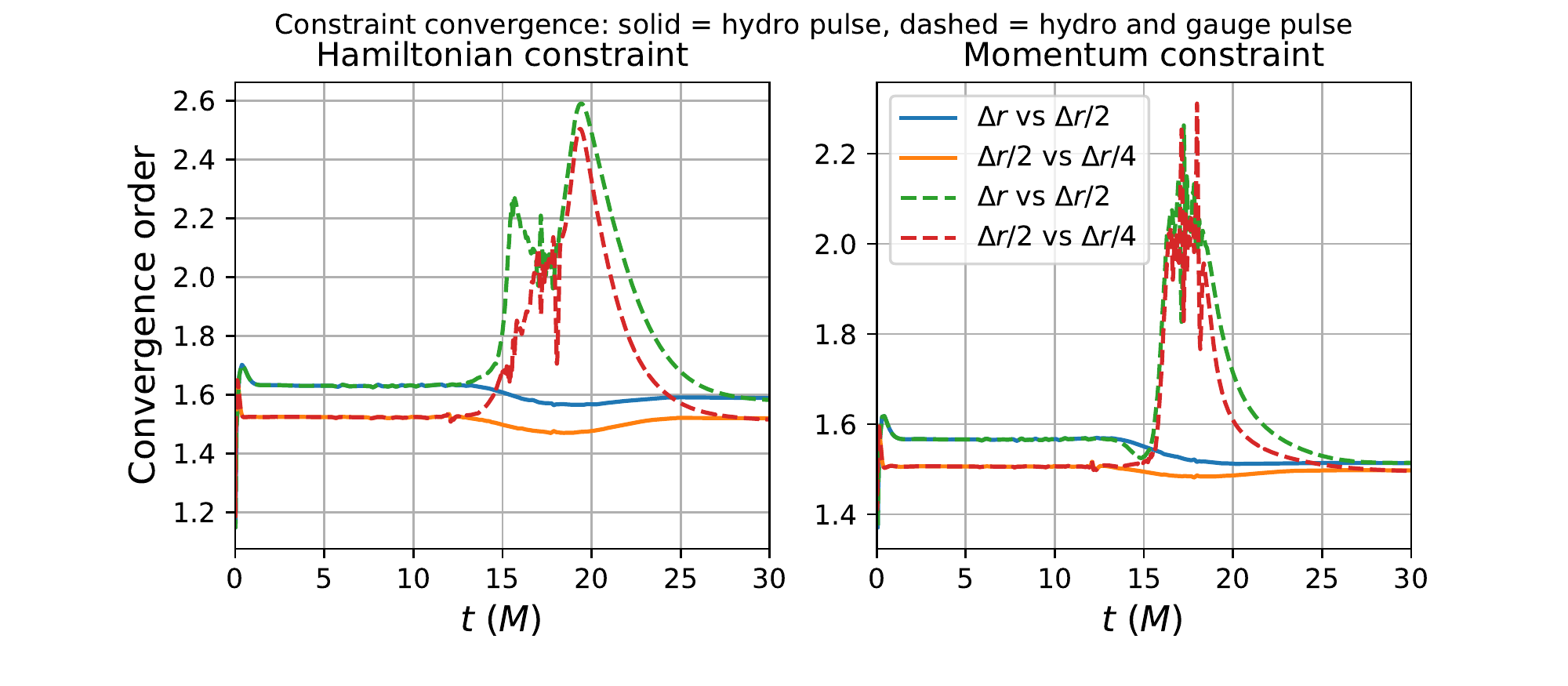}}
\caption{Global  $\mathcal{L}_2$ norm convergence order for the Hamiltonian constraint (Left) and momentum constraint (Right) for the case with an ingoing hydrodynamic pulse (solid lines) and both ingoing hydrodynamic and gauge pulses (dashed lines). The fiducial resolution is $\Delta r = M/20$. A convergence order of $\sim 1.5$ is obtained while the error is dominated by the hydrodynamic solver, whereas a convergence order of $\sim 2$ is obtained while the the gauge pulse propagates through the fluid and into the black hole.} \label{fig:hydro_gauge_constraints}
\end{figure}

In Fig.~\eqref{fig:hydro_gauge_constraints} we plot the global convergence order in an $\mathcal{L}_2$-norm sense for the constraints. The analytic solution would have vanishing constraints, so this is an analytic convergence test rather than a self-convergence test. Again, solid lines correspond to the case with a hydrodynamic pulse only, whereas dashed lines also include the gauge pulse. When the error is dominated by the hydrodynamic solver, we obtain the expected convergence order of $\sim 1.5$. We observe that when the gauge pulse propagates through the grid, the error changes significantly such that between $t\sim 15-20M$ the convergence order increases to $\sim 2$. Note that this does not mean that the \emph{errors} have decreased, only that the convergence \emph{order} has increased; one expects that the errors are larger during this time, and some other truncation error term beyond the lowest one is momentarily dominant.

\FloatBarrier

\begin{figure}[htbp]
\centering
\hbox{\hspace{3.0cm}\includegraphics[width=0.6\textwidth]{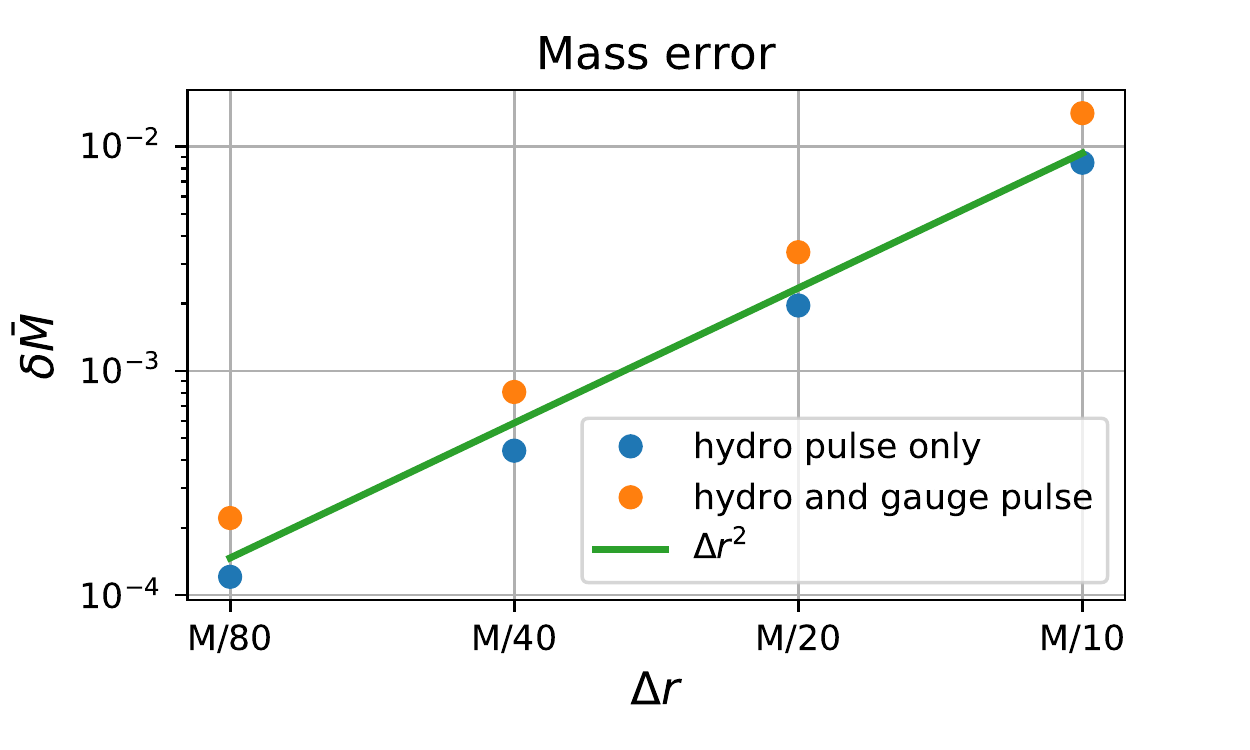}}
\caption{A convergence test for the change in black hole mass. Either a hydrodynamic pulse is injected through the outer boundary, or both hydrodynamic and gauge pulses are injected through the outer boundary consecutively, and absorbed by a black hole of initial mass $M$. The gauge pulse is injected after the hydrodynamic pulse, and passes through it on its way into the black hole. Plotted are the differences $\delta \bar{M}$ between the change in mass of the black hole, obtained via the Misner-Sharp mass evaluated at the outer boundary $r=10 M$ at the final time $t=30 M$, and the total proper ADM mass of the hydrodynamic pulse evaluated at the moment it clears the outer boundary, $t=5M$. The initial black hole mass is $M$. Second-order convergence is observed.} \label{fig:hydro_gauge_masserr}
\end{figure}

Finally, in Fig.~\eqref{fig:hydro_gauge_masserr} we display second-order convergence of the change in black hole mass between the beginning and end of the simulations. This is a test to see whether the black hole grows by the expected amount, based on how much fluid it absorbs. The difference we are plotting is 
\begin{eqnarray}
\delta \bar{M} = \biggr{\vert} M_{\text{MS}}\vert_{r=10M,t=t_{\text{final}}} - M - \int \rho_{\text{ADM}}\vert_{t=5M} \sqrt{\gamma}d^3r \biggr{\vert} ,
\end{eqnarray}
i.e. a comparison between the change in black hole mass according to the Misner-Sharp mass $M_{\text{MS}}$ evaluated at the outer boundary $r=10M$ at the initial and final times, and the total proper ADM mass of the hydrodynamic pulse,
\begin{eqnarray}
M_{\mathrm{ADM}} = \int \rho_{\mathrm{ADM}} \sqrt{\gamma} d^3 r,
\end{eqnarray}
at $t=5M$ when it has first completely cleared the outer boundary. As expected, the errors are larger for the more demanding case with a large gauge pulse (orange circles).

\FloatBarrier

%
%
\section{Hydrodynamic pulse with PN gravity} \label{sec:PNtest}
In this section we test the PNGhydro code with a fluid pulse initialized on the computational grid. We have a ``PN black hole" of mass $M$ present (i.e.~a black hole approximated to some PN order), and the computational domain extends out to $20 M$. The black hole PN potentials are obtained from~\cite{hergt2008higher,barausse2013post} by setting the spin parameter to zero, since we are in spherical symmetry. This yields
\begin{eqnarray}
\phi_{\mathrm{BH}} &=& -\frac{M}{r} + \frac{1}{c^2} \frac{M^2}{r^2} - \frac{3}{4} \frac{M^3}{r^3 c^4} + \mathcal{O}\left( \frac{1}{c^6} \right), \label{eq:BHPNpotentials}\\
\psi_{\mathrm{BH}} &=& \phi_{\mathrm{BH}} - \frac{7}{4} \frac{M^2}{r^2 c^2} + \mathcal{O}\left(\frac{1}{c^4} \right) \nonumber\\
\omega^r_{\mathrm{BH}} &=& \mathcal{O}\left(\frac{1}{c^5} \right), \nonumber\\
\chi_{rr, \mathrm{BH}} &=& -2\chi_{T,\mathrm{BH}} = \mathcal{O}\left( \frac{1}{c^6} \right). \nonumber
\end{eqnarray}
The initial profile of the fluid corresponds roughly to an ingoing pulse~\cite{neilsen1999extremely}, which is conveniently specified and easily implemented in terms of the conservative variables as
\begin{eqnarray}
D &=& \exp \left( -\frac{(r-7.5)^2}{2 (1/3)^2} \right) + \rho_{0,\mathrm{atmos}} \\
\tau &=& D \\
S_r &=& - \tau .
\end{eqnarray}
The choice of $S_r = -\tau$ is the crucial aspect of this choice which reduces the outgoing part of the pulse. This was observed in~\cite{neilsen1999extremely} in simulations, where an initially stationary pulse splits into ingoing and outgoing pieces with $S_r \sim -\tau$ and $S_r \sim \tau$, respectively. One expects this would be reflected at the level of the characteristic variables as well, as in Sec.~\eqref{sec:charstruc}.

We evolve this initial setup for a total time $T=3 M$ and perform an \emph{independent residual test} on the simulation output. This means that we write an independent script, without any reference to the simulation code, which evaluates the left- and right-hand sides of the PN equations based on the code's output. The difference between the left- and right-hand sides is analytically known to be zero by virtue of the equations. Thus, when computing convergence factors in an independent residual test, one only needs two resolutions. We denote the residual in eg.~the $\phi_{\mathrm{fluid}}$ case as
\begin{eqnarray}
R_\phi = \mathrm{LHS}(\phi_{\mathrm{fluid}}) - \mathrm{RHS}(\phi_{\mathrm{fluid}}),
\end{eqnarray}
and the convergence factor will be
\begin{eqnarray}
\log{ \left[ \frac{\mathcal{L}_2 (R_\phi \vert_{\Delta r} - R_{\phi,\mathrm{analytic}})}{\mathcal{L}_2 ( R_\phi \vert_{\Delta r/2} - R_{\phi,\mathrm{analytic}})} \right] } / \log{2},
\end{eqnarray}
where $R_{\phi,\mathrm{analytic}} = 0$ by virtue of Eq.~\eqref{eq:phieqn}. This is therefore a form of an analytic convergence test. Our fiducial resolution is $\Delta r = 0.1 M$. We also only solve for the PN potential $\phi_{\mathrm{fluid}}$, which means we are restricting to coordinates where $\omega^r_{\mathrm{fluid}} = 0$ (which can always be chosen in spherical symmetry).

\begin{figure}[htbp]
\centering
\hbox{\hspace{-0.5cm}\includegraphics[width=1.05\textwidth]{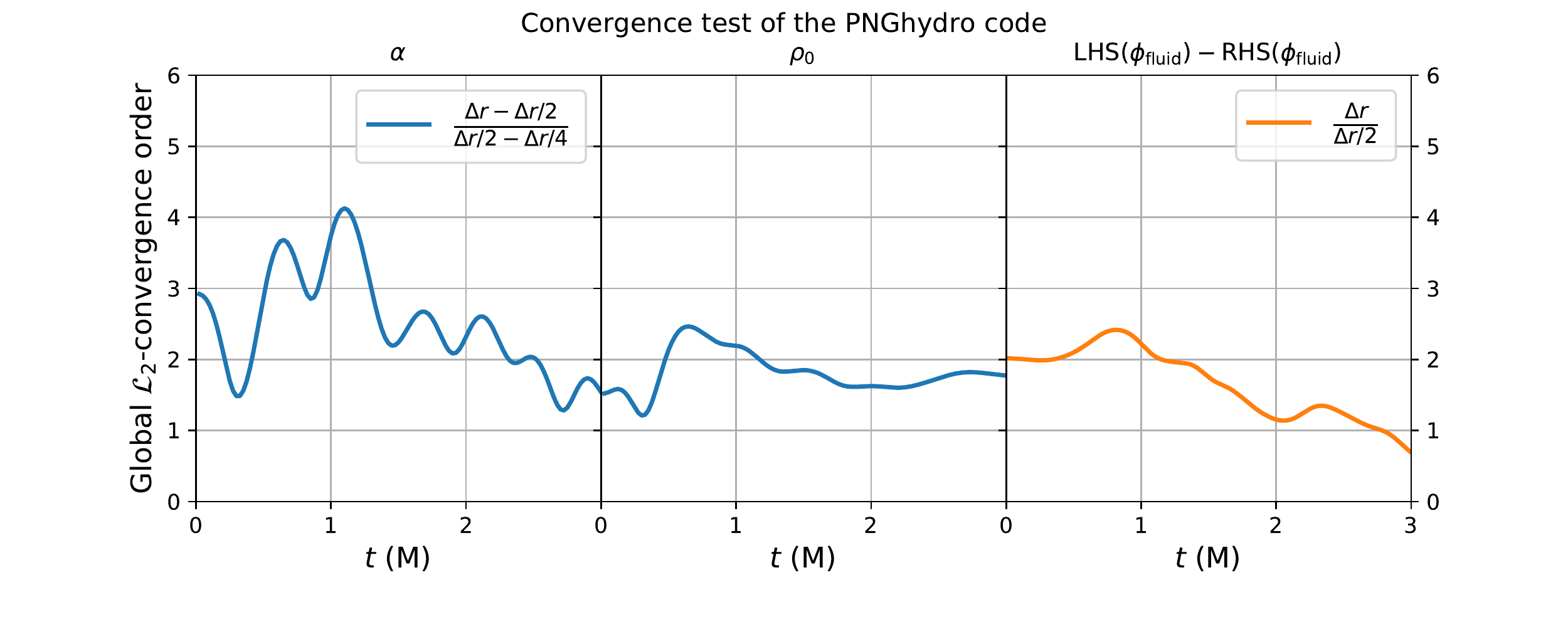}}
\caption{Convergence test of the PNGhydro code. (Right): Global $\mathcal{L}_2$-convergence factor for an independent residual of the equation governing $\phi_{\mathrm{fluid}}$, Eq.~\eqref{eq:phieqn} (without the time-derivative term $\propto \partial_t^2 \phi_{\mathrm{fluid}}$). (Left \& Centre): Global self-convergence factors of the lapse $\alpha$ and rest mass density $\rho_0$, for comparison.} \label{fig:PN_test}
\end{figure}

Fig.~\eqref{fig:PN_test} shows the results of this test: (Left \& Centre) show global self-convergence of the lapse $\alpha$ and rest mass density $\rho_0$ to indicate a reference level of convergence. Self-convergence requires three resolutions, and we use $\lbrace \Delta r, \Delta r/2, \Delta r/4 \rbrace$. (Right) shows the independent residual test, yielding a convergence order $\sim 1-2$.

Thus, although our solution method interpolates the right-hand side of Eq.~\eqref{eq:phieqn} for the internal steps taken by ODEPACK (which introduces errors), we obtain reasonable results. We will improve the solution method in future efforts, but we present preliminary results with the current code below.

\FloatBarrier

%
%
\chapter{Preliminary results \& future work} \label{ch:PNresults}
Here we present preliminary results, wherein we simply compare the evolutions of the same system in both the GR fluid and PNGhydro codes. Since the black hole PN potentials in Eqs.~\eqref{eq:BHPNpotentials} are just truncated $M/r$ expansions of a Schwarzschild black hole in isotropic coordinates, we use the exact black hole in isotropic coordinates as the initial data for the GR fluid run. We place the outer boundary at $r= 20M$, use a resolution $\Delta r = 0.0125$, and evolve for a total time of $50M$. 

The rest mass boundary data as a function of time is given by Eq.~\eqref{eq:rho0bdy}, except with no constant duration at max amplitude (i.e.~$t_{\mathrm{gap}}=0$), and with amplitude $A_{\rho_0}= 1.28\times 10^{-5} M^{-2}$. This again is a $\sim \sin^2(t)$ rise and fall, and the parameter $T=64 M$ yields rise and fall times of $16 M$, so that the pulse clears the boundary at $t=32 M$. The specific internal energy at the boundary is chosen to be
\begin{eqnarray}
\epsilon \vert_{\mathrm{bdy}} = 10^{-1} \rho_0 \vert_{\mathrm{bdy}}(t).
\end{eqnarray}

As mentioned at the end of Sec.~\eqref{sec:bdyinjection}, since the shift vector is zero in isotropic coordinates, the material wave velocity ($\lambda_0 = \alpha v^r - \beta^r$) will be zero unless we choose a non-zero $v^r$. We therefore prescribe the boundary value of $v^r$ in units of $c$ as
\begin{eqnarray}
v^r \vert_{\mathrm{bdy}} =
\begin{cases}
	-0.1  & 0 < t \leq 16 M \\
	-0.1 \frac{\rho_0 \vert_{\mathrm{bdy}}(t)}{A_{\rho_0}} & t > 16 M  \label{eq:vbdy}
\end{cases},
\end{eqnarray}
which thus turns off in a $\sim \sin^2(t)$ fashion commensurate with $\rho_0 \vert_{\mathrm{bdy}}$ beginning at $t=16 M$ when the boundary rest mass reaches maximum amplitude. These boundary parameters result in the injection of a fluid pulse of total mass $\sim 0.1 M$.

As in Sec.~\eqref{sec:PNtest}, we set $\omega^r_{\mathrm{fluid}} =0$, which corresponds to a coordinate choice. Since our goal is to test the PN formalism in a numerical implementation, we would ideally not make this choice, so that we can test all of the PN equations. However, given that our GR fluid implementation uses an \emph{a priori}-specified shift vector, we cannot possibly specify the shift in accordance with what the PNGhydro evolution is representing since we do not know what it is. This difficulty only exists for this simple side-by-side comparison of the evolutions; in future explorations we will compare coordinate-invariant quantities so that these issues do not arise.

\begin{figure}[htbp]
\centering
\hbox{\hspace{-1.5cm}\includegraphics[width=1.2\textwidth]{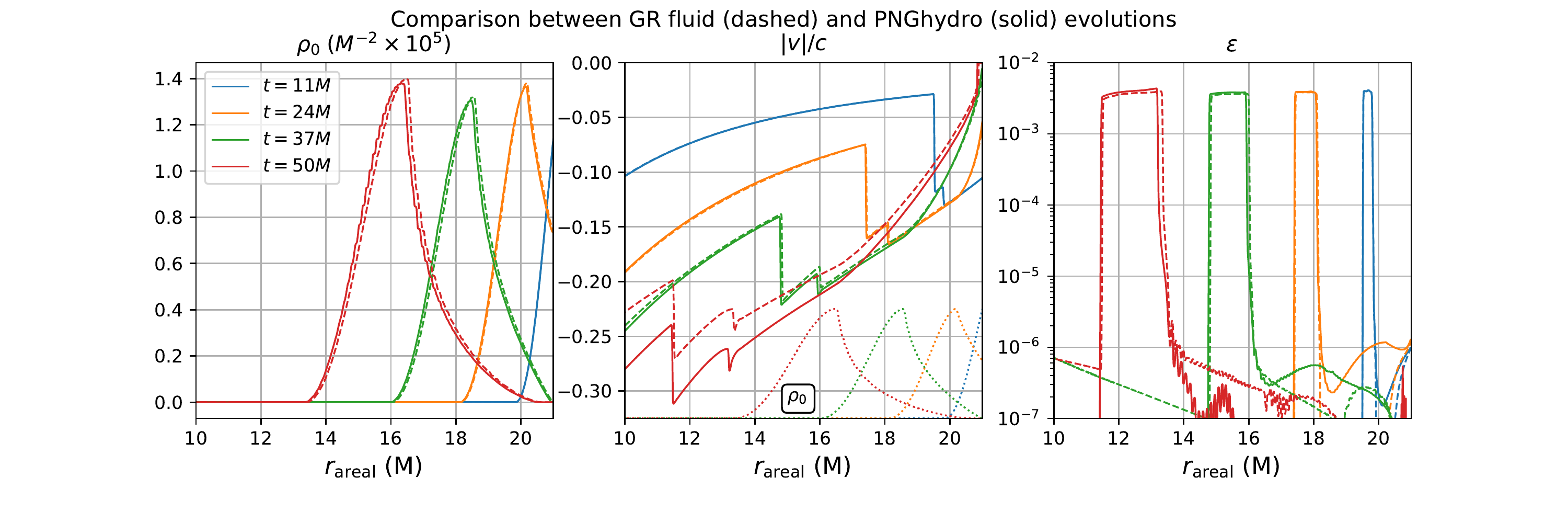}}
\caption{Comparison of the evolution of the primitive variables in the GR fluid code (dashed lines) and the PNGhydro code (solid lines). Everything is plotted vs the areal radial coordinate. Differences are seen to grow with time in all cases. (Centre): We display the $\rho_0$ profiles at the bottom of plot in order to indicate the important portions of the velocity. The kinked features in $|v|$ correspond to a small shockwave at the bow of the pulse which is only visible in $\rho_0$ on a logarithmic scale. Between the pulse and the black hole, the atmosphere accretes at an every-increasing velocity, which accounts for the larger velocities towards smaller radii. (Right): The specific internal energy $\epsilon$ in the body of the pulse experiences significant invocations of our error handling policy described in Sec.~\eqref{sec:contoprim} and in~\cite{galeazzi2013implementation}, but the shock heating at the bow of the pulse agrees well between the GR fluid and PNGhydro runs.} \label{fig:PN_GR_evo_compare}
\end{figure}

In Fig.~\eqref{fig:PN_GR_evo_compare} we display the temporal progression of the hydrodynamic quantities $\rho_0$ (Left), $|v|=\sqrt{\gamma_{rr} v^r v^r}$ (Centre), and $\epsilon$ (Right). All quantities are plotted versus the areal radial coordinate, $r_{\mathrm{areal}} = \sqrt{\gamma_T} r$. In (Centre), we also display the rest mass density profiles at the bottom of the plot in dotted lines, so as to indicate where the bulk of the fluid is at each time. The kinked portions of the velocity occur at the bow of the pulse, where there exists a small shockwave with rest mass amplitude near the atmosphere level $\rho_{0,\mathrm{atmos}} = 10^{-13}$. At lesser radii the velocity increases, which is simply due to the atmosphere accreting towards the black hole. In (Right), the specific internal energy $\epsilon$ is observed to have poor agreement between the GR fluid and PNGhydro runs in the bulk of the pulse. However, the larger values in the shock-heating region of the aforementioned small shockwave agree well (large rectangular features).


\begin{figure}[htbp]
\centering
\includegraphics[width=0.6\textwidth]{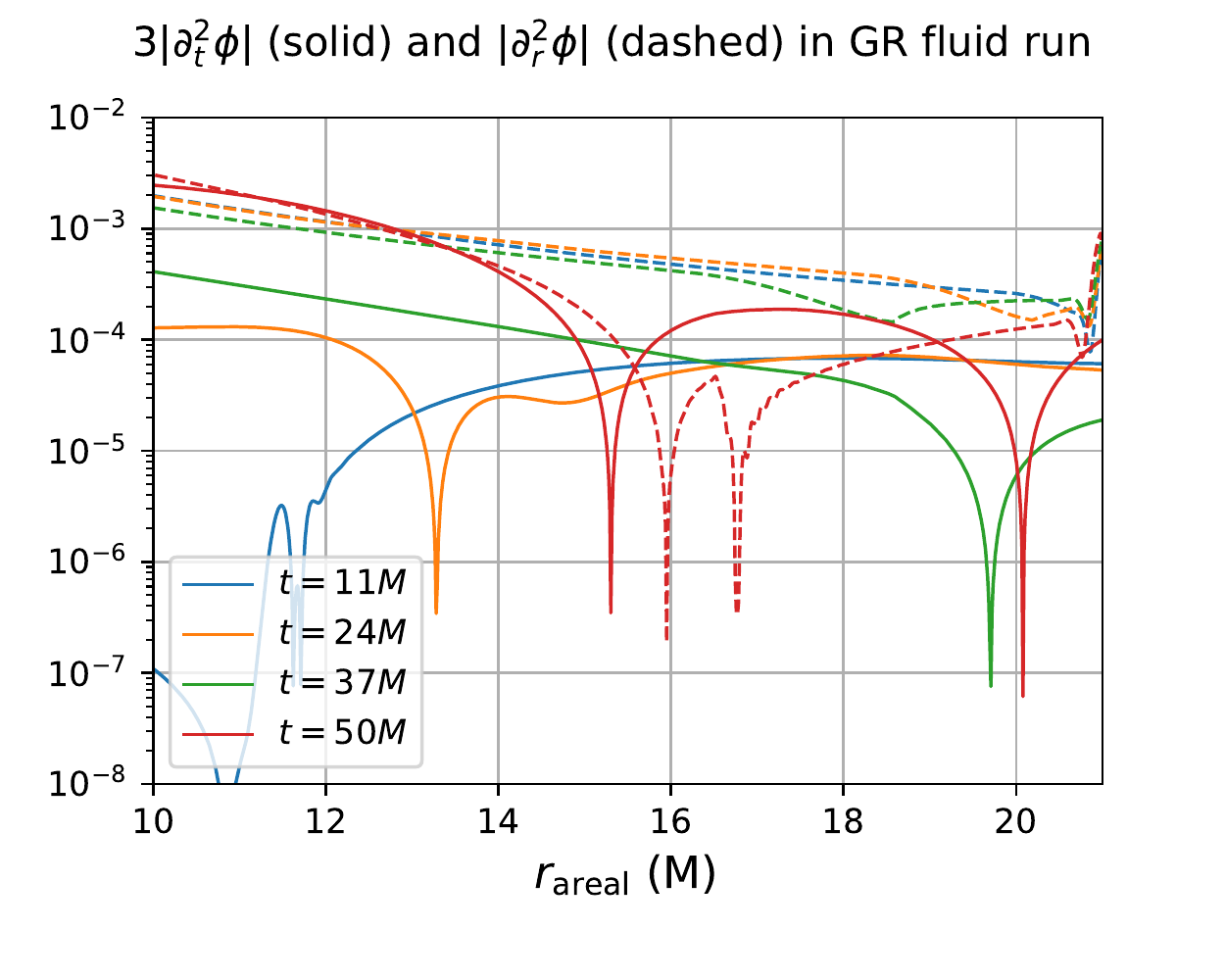} 
\caption{Comparison of the magnitudes of the 2nd derivatives of the PN potential $\phi$ for the GR fluid run. Recall that in the PNGhydro evolution the term $\sim 3 \partial^2_t \phi$ has been set to zero on the right-hand side of Eq.~\eqref{eq:phieqn}. We plot $3| \partial^2_t \phi |$ (solid) and $| \partial^2_r \phi |$ (dashed) at several times. The GR fluid run serves as an estimate of whether this term is accurately captured by our numerical PN solution. We see that the spatial derivatives are $\sim$ a few times to $\sim 10$ times larger than the time derivatives, except at the final time $t=50M$ where the situation is reversed. Thus neglecting the $\sim 3 \partial^2_t \phi$ term in the PNGhydro evolution is reasonable for the earlier times.} \label{fig:PN_GR_deriv_compare}
\end{figure}

\begin{figure}[htbp]
\centering
\includegraphics[width=0.6\textwidth]{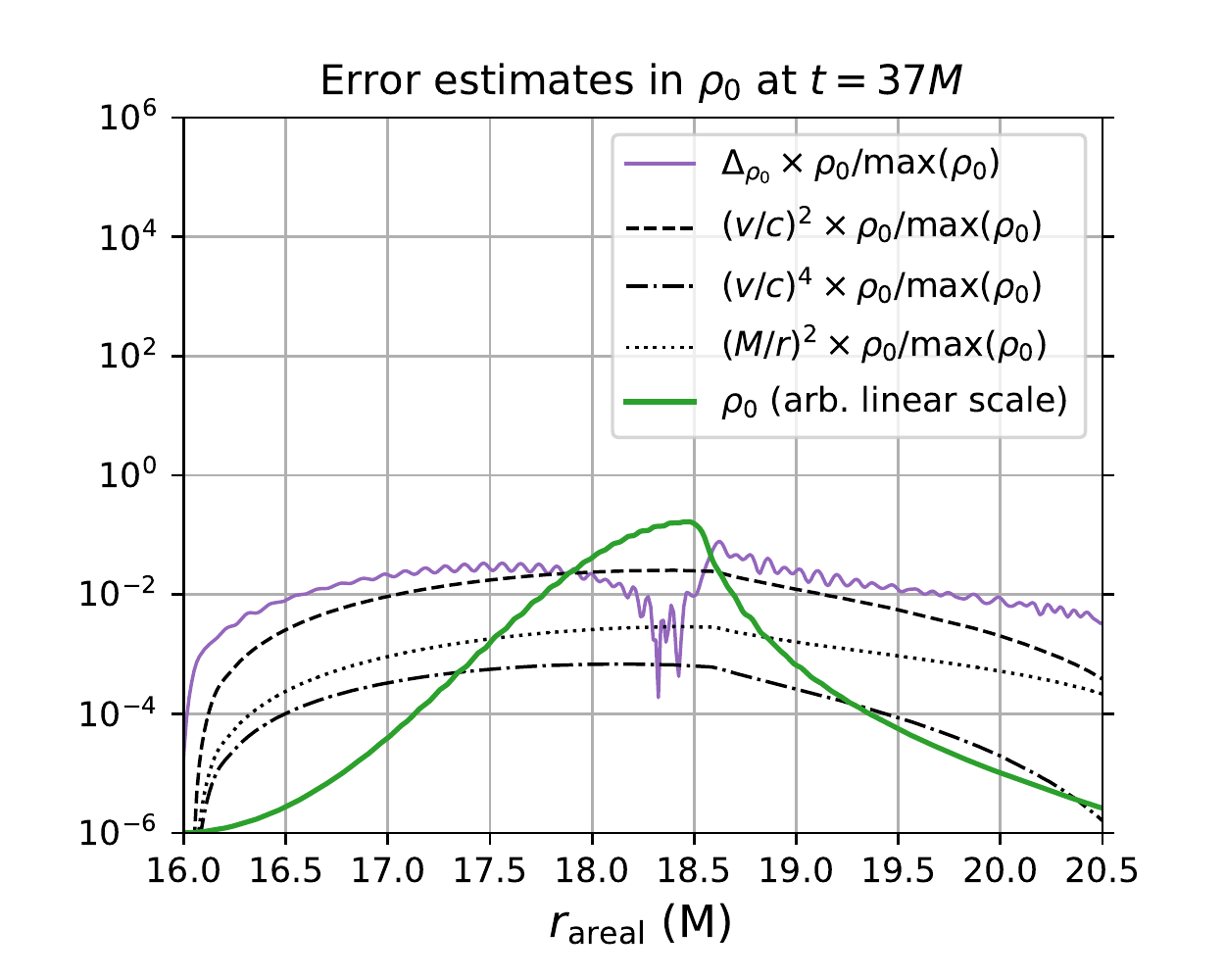} 
\caption{Comparison of the errors in $\rho_0$ at $t=37M$, all normalized by $\rho_0(r)/\mathrm{max}(\rho_0)$ in order accentuate the most important areas of the pulse. The relative error $\Delta_{\rho_0}$ between the GR fluid and PNGhydro runs is shown (purple, thin, solid). Error estimates based on velocity $\sqrt{\gamma_{rr} v^r v^r}/c$ and compactness $M/r$ to different powers are displayed in black: $(v/c)^2$ (0PN error, dashed), $(v/c)^4$ (1PN error, dot-dashed), $(M/r)^2$ (1PN error, dotted). We also include the pulse profile $\rho_0$ (red, thick solid) on an arbitrary linear scale, for orientation. The error is seen to approach the expected level for a 1PN-accurate calculation (dominated by the $(M/r)^2$ error level) in the vicinity of the pulse peak, around $r\sim 18.3M$.} \label{fig:PN_GR_error_compare}
\end{figure}

As we explained in Sec.~\eqref{sec:PNsolver}, we have set the $-3\partial_t^2 \phi_{\mathrm{fluid}}$ term to zero on the right-hand side of Eq.~\eqref{eq:phieqn} due to numerical difficulties. In Fig.~\eqref{fig:PN_GR_error_compare} we assess the degree to which this has affected the evolutions we obtained in this section. Using the GR fluid run, we transform from ADM metric variables Eq.~\eqref{eq:ADMmetric} to the PN potentials Eq.~\eqref{eq:PNmetric}. This yields PN potentials which are more accurate than the PNGhydro output would yield, since they come from solving the full Einstein equations without any PN approximations. We then numerically compute $3\vert \partial_t^2 \phi_{\mathrm{fluid}}\vert$ and $\vert \partial_r^2 \phi_{\mathrm{fluid}}\vert$ using 2nd-order centred finite differences, in order to compare their magnitudes. We plot the absolute value of these derivatives in Fig.~\eqref{fig:PN_GR_error_compare}, and observe that at all times except the final time $t=50M$, the spatial derivatives have a larger magnitude than the time derivatives by a factor of $\sim 10$. This gives a sense of how large the omitted $-3 \partial_t^2 \phi_{\mathrm{fluid}}$ term is on the right-hand side of Eq.~\eqref{eq:phieqn} for this particular evolution.

Lastly, in Fig.~\eqref{fig:PN_GR_error_compare} we compare the relative error in $\rho_0$ at $t=37M$, defined via
\begin{eqnarray}
\Delta_{\rho_0} = \frac{ \vert \rho_{0,\mathrm{GR}} - \rho_{0,\mathrm{PN}} \vert }{\rho_{0,\mathrm{GR}}},
\end{eqnarray}
and the expected error coming from the order at which the PN expansion has been truncated. Since the PN expansion is a simultaneous expansion in $M/r$ and $v/c$, we plot various contributions in black: the 0PN error $(v/c)^2$ (i.e.~the error expected if computations are done at 0PN), the 1PN error $(v/c)^4$, and the 1PN error\footnote{$M/r$ is $\mathcal{O}(v/c)^2$ in the context of circular orbital motion. However, our fluid is undergoing spherical collapse, so $M/r$ can differ significantly from $(v/c)^2$.} $(M/r)^2$. All errors are normalized by $\rho_0 (r)/\mathrm{max}(\rho_0(r))$ in order to suppress the less important errors at the tails of the pulse, which can be quite large otherwise. The pulse profile itself is also shown on an arbitrary linear scale, to indicate its position and where the peak is. We see that the error in the pulse due to the PN treatment is well-matched by the 1PN compaction error $(M/r)^2$ near the peak of the pulse around $r\sim 18.3 M$. Elsewhere the error is larger, which we will investigate during future efforts to improve our PNGhydro code.

Once we have overcome the numerical difficulties in treating the time-derivative terms in the PN equations~\eqref{eq:phieqn},~\eqref{eq:omeeqn}, we will proceed to implement the higher order 2.5PN equations derived in~\cite{barausse2013post}. It is these higher order approximations which have yielded a marked improvement in eg.~the mass-radius curve for cold neutron stars~\cite{barausse2013post}. Thus the prospect for improving pseudo-Newtonian simulation codes via this PN formalism remains an exciting possibility.
\FloatBarrier

%
%

%
%
\partkey{III}
\part{Mode structure of rotating core-collapse supernovae} \label{part:IV}

\section*{Executive summary}
In this part, we show that asteroseismology of rotating core-collapse supernovae is possible with a multimessenger strategy. Using a standard stellar model with $20\: M_\odot$ zero-age main sequence mass with $1.0$ rad/s pre-collapse central rotation, we show that an $l=2$, $n\gtrsim 2$ oscillation mode of the newly-born rotating proto-neutron star with frequency $\lesssim 280$ Hz is responsible for a correlated peak emission frequency in both gravitational waves and neutrino luminosities. We achieve this by first identifying the mode in the non-rotating model via an eigenfunction matching procedure between the full nonlinear simulation and a perturbative calculation. Then we follow the mode along a sequence of rotating models by establishing the continuity of its eigenfunction. For an initial pre-collapse rotation rate of $1$ rad/s, the correlated frequency of gravitational wave and neutrino luminosity emerges. We show that the dominant angular harmonics of the emission pattern of neutrinos on the sky are consistent with the mode energy in a shell around the neutrinospheres, where $r\sim 60-80$ km. Thus, the neutrinos carry information about the mode in the outer region of the proto-neutron star where $r\sim 60-80$ km, whereas the gravitational waves probe the deep inner core $r \lesssim 30$ km.

\begin{figure}[!h]
\centering
\hbox{\hspace{0.45cm}\includegraphics[width=0.95\textwidth]{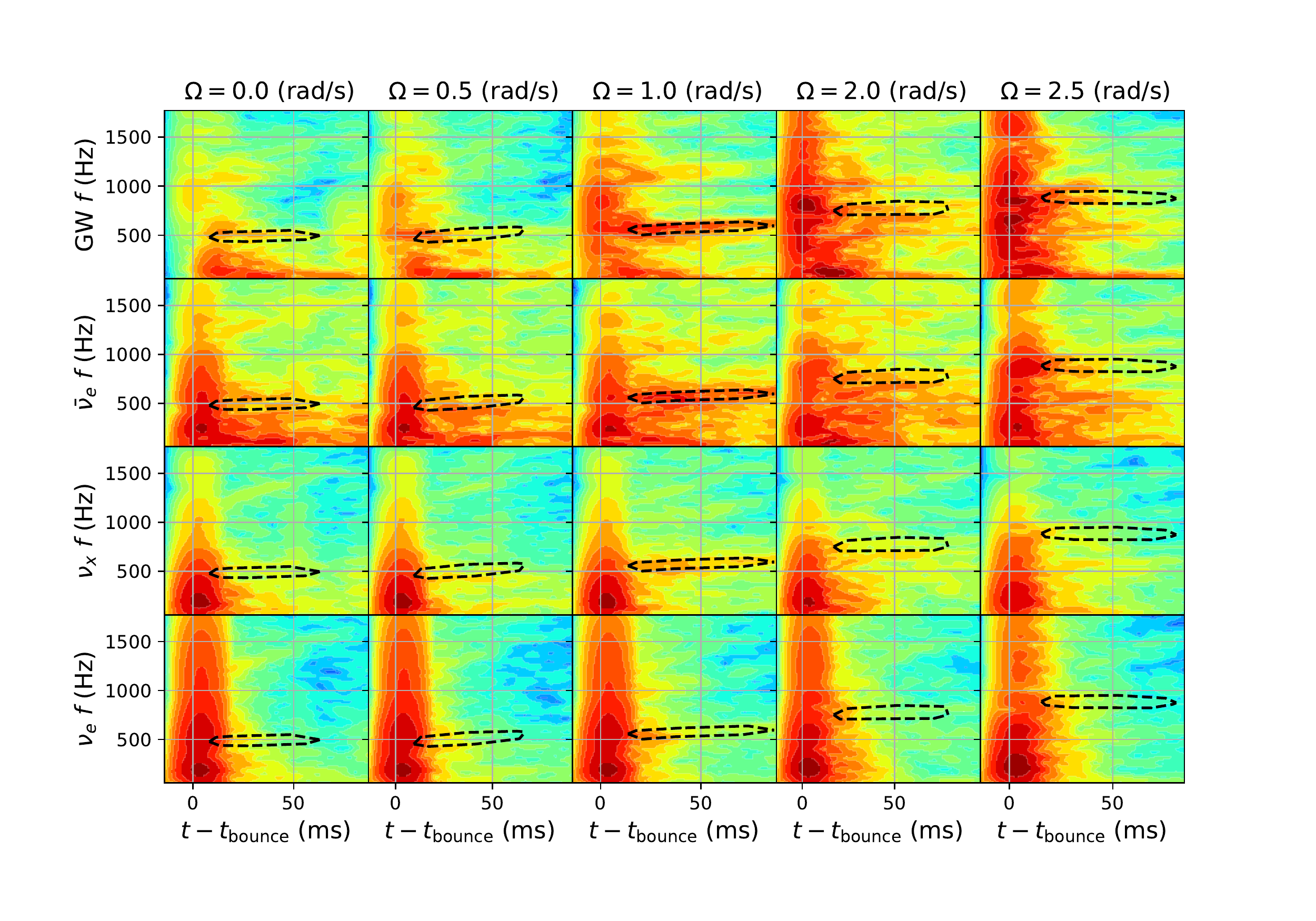}}
\raggedright
Figure III: Preview of the $l=2$, $m=0$, $n=2$ mode, whose frequency mask is overlaid on the gravitational wave and neutrino spectrograms from our sequence of rotating models. The models are parameterized by pre-collapse rotation $\Omega = \lbrace 0.0,0.5,1.0,2.0,2.5\rbrace$ rad/s. Here we see that the mode is coincident with dominant bands in the $\Omega = 1.0$ rad/s case across all messengers.
\label{fig:CCSNmoney}
\end{figure}

\chapter{Introduction}

The collapse and bounce of the iron cores of massive ($M\gtrsim 10 M_{\odot}$) stars and the possible ensuing supernovae are expected to produce detectable gravitational waves and neutrinos if they occur within or very near our galaxy. Indeed, neutrinos have already been detected from such an event, named SN1987A~\cite{hirata1987observation,bionta1991observation}. Events with a successful explosion are called core-collapse supernovae (CCSNe). The electron-degenerate iron core collapses once it exceeds the Chandrasekhar limit, and unless it is very massive the collapse will be halted once the core reaches nuclear densities $\rho \sim 10^{14}$ g and its equation of state stiffens. This sudden halting imparts momentum to the supersonically infalling stellar material, causing a powerful outward shockwave. Whether and how this shockwave and the subsequent dynamics results in a successful explosion is a central theme of research in this area.

In the event of a successful explosion, as in SN1987A, photons are also detectable.  In contrast with photons, which are heavily reprocessed before freely streaming to an observer, the intervening stellar material between the core and an observer are transparent to gravitational waves. The star is also largely transparent to neutrinos, except for example within several dozens of kilometres of the centre of the core and within tens of milliseconds after core bounce, where scattering is still strong. Neutrinos and gravitational waves therefore offer deep probes of the central engine of a CCSN.

Interestingly, correlated frequencies between gravitational waves and neutrino luminosities have been observed in simulations within tens of milliseconds after core bounce in~\cite{ott2012correlated}. Similar correlations at later times have been reported in~\cite{kuroda2017correlated}, likely due to the growth of the standing accretion-shock instability (SASI). These observations point towards a wealth of opportunity to probe specific aspects of the central dynamics occurring at different times, from tens of ms to several seconds and beyond.

Asteroseismology is the study of the interior structure of stars inferred from observations of its seismic oscillations. There have been some very recent efforts to use gravitational waves to do the same with CCSNe, so-called \emph{gravitational wave asteroseismology}~\cite{fuller2015supernova,torres2017towards, morozova2018gravitational,torres2018towards}. The main strategy is to use data from numerical simulations as input for a perturbative mode calculation, where the simulated data serves as a background solution. The key point to note is that there is a separation of scale between the period of the modes of interest and the time scale over which the post-bounce CCSN background changes significantly. For example, towards the pessimistic end, a $200$ Hz mode has a period of $5$ ms, whereas the CCSN background changes over a timescale of several tens of ms. Therefore one expects to be able to treat the CCSN background as stationary for the purposes of a perturbative calculation at any instant of time. In~\cite{fuller2015supernova} this was done using a perturbative Newtonian hydrodynamic scheme in an effort to generate a qualitative understanding of the gravitational wave emission due to the oscillations of a rotating proto-neutron star that were excited at bounce. Subsequently,~\cite{torres2017towards} presented a similar effort using perturbative hydrodynamic calculations in the relativistic Cowling approximation. Shortly thereafter, and during the course of this work,~\cite{morozova2018gravitational} partially relaxed the Cowling approximation by allowing the lapse to vary, governed by the Poisson equation. They claimed an improved coincidence between their perturbative mode frequencies and certain emission features in the gravitational wave spectrograms from simulations. The Cowling approximation was then relaxed even further in~\cite{torres2018towards}, where the conformal factor of the spatial metric was allowed to vary as well, leaving only the shift vector fixed.

All of these studies, however, attempt to identify specific modes of oscillation of the system primarily by coincidence between perturbative mode frequencies and peaks in gravitational wave spectra, across time. This is potentially problematic for a number of reasons. Firstly, the approximations used in the perturbative calculations introduce errors in mode frequencies that can be quite significant, eg.~several tens of percent in the case of lower order modes in the Cowling approximation. Secondly, any partial relaxation of the Cowling approximation presents difficulties with the interpretation of results, since this strategy neglects some terms at a given order but not others, and thus is not \emph{a priori} under control. Thirdly, the approximations used in the simulations themselves introduce their own frequency errors. For example, as had motivated our efforts in Part~\eqref{part:PN} to improve gravity treatments in some codes, gravity is often treated in a pseudo-Newtonian manner in CCSN simulations by modifying the potential to mimic relativistic effects, as in~\cite{morozova2018gravitational,o2018two,pan2018equation}. The mode population in the vicinity of a given frequency bin in a gravitational wave spectrum can be rather dense, therefore all these sources of error serve to lower the significance of any given observed coincidence between perturbative and simulated mode frequencies. Indeed, the purported identification of a $g$-mode in~\cite{pan2018equation} required a \emph{post hoc} modification of its frequency formula when matching to gravitational wave spectrograms from simulations.

We opt to take a different approach. Rather than looking for coincidence in mode frequencies, we look for coincidence in \emph{mode functions}. This means comparing mode functions, obtained from perturbative calculations, with the velocity data from simulations, which are heavily post-processed using spectral filters and vector spherical harmonic decompositions. We will find that if a mode has an adequate excitation, its matching with candidate perturbative mode functions is unambiguous. Since this strategy does not use frequency-matching, we will also discover a shortcoming of our simulation code: the frequencies observed in our simulations are significantly overestimated, with the true values being at least $60$\% lower. This illustrates the power of matching in mode functions rather than mode frequencies, and bears out our concerns with focusing only on mode frequency coincidence as in~\cite{torres2017towards,morozova2018gravitational,torres2018towards}.

We use the perturbative scheme in the relativistic Cowling approximation from~\cite{torres2017towards}, which applies only to spherical systems. Therefore we will only apply it to a non-rotating model in order to identify modes of oscillation that are excited at bounce and ring for $\sim 10-100$ ms. This identification serves to label the corresponding modes in the rotating models~\cite{friedman2013rotating} whose mode functions deformed continuously with increasing rotation, picking up a mixed character in angular harmonics. We also simulate a sequence of rotating models with progressively larger pre-collapse rotations of $\Omega = \lbrace 0.0, 0.5, 1.0, 2.0, 2.5 \rbrace$ rad/s. In order to follow the modes along this sequence, we take inspiration from the works of~\cite{font2001axisymmetric,dimmelmeier2006non,gaertig2008oscillations}. In~\cite{font2001axisymmetric}, knowledge of the modes of the non-rotating star were combined with continuity in frequency to follow modes across such a sequence, whereas in~\cite{dimmelmeier2006non,gaertig2008oscillations} continuity in mode functions was used. Continuity in mode function is more powerful than continuity in frequency, since separate modes can have very similar frequencies and thus would be degenerate in a continuity analysis. Thus we will chiefly use mode function continuity to follow modes along our sequence of rotating models.

A powerful method in~\cite{dimmelmeier2006non} called \emph{mode recycling} was used to converge toward the mode function of rotating stars. One simulates the star with an initial perturbation corresponding to an educated guess for the mode function of interest, which due to its inaccuracy will excite several unwanted modes. By applying spectral filters to the velocity field in the star, a more accurate trial mode function for the target mode can be extracted and used as an initial perturbation in a second simulation. This process is repeated until the initial perturbation results in a clean excitation of the target mode, with unwanted modes highly suppressed. The strategy of mode recycling is not available in our context, since our target modes are excited by core bounce, which we do not attempt to manipulate. Thus we rely strictly on mode function continuity, but we will employ a suite of known facts about the possible deformations of modes in the slow-rotation regime in order to bolster our analysis.

\emph{Our chief result will be the implication of an $l=2$, $m=0$ mode with at least 2 radial nodes in the production of correlated gravitational waves and neutrinos in the $\Omega = 1.0$ rad/s model.} In contrast with~\cite{ott2012correlated} where the neutrino treatment did not supply information about the emission pattern on the sky, our treatment does allow this. We will relate the dominant angular harmonics of the emission to the dominant energy harmonics in the $l=2$ mode function in the vicinity of the neutrinospheres. The causal explanation for the oscillations in the neutrino luminosity will be that the $l=2$ mode of the proto-neutron star, which in the rotating $\Omega = 1.0$ rad/s model has a mixed character in $l$, is producing $l=2$ and $l=0$ variations in the shape of the neutrinospheres. Since the neutrinospheres are roughly the boundary between trapped and free-streaming neutrinos, this means that the region producing free-streaming neutrinos is undergoing variations in shape in accordance with the activity of the mode in the vicinity of the neutrinospheres, at $r\sim60-80$ km. This causes the oscillations in neutrino luminosity registered by an observer far away. We therefore find that detailed asteroseismology of CCSN is possible with joint detection of gravitational waves and neutrinos, where the neutrinos supply information from the neutrinosphere region $r \sim 60-80$ km, and the gravitational waves supply information from deeper in, around $r\lesssim 30$ km where the bulk of the mode energy resides.

\vspace{0.5cm}
This part is organized as follows. In the next section~\eqref{sec:backgroundterminology} we will review the necessary background material and terminology from the theory of oscillations of stars required to understand this work. Chapter~\eqref{ch:CCSNimp} will contain a description of our simulation code in Sec.~\eqref{sec:FLASHdescribe} and the perturbative schemes of~\cite{torres2017towards,morozova2018gravitational} in Sec.~\eqref{sec:pertschemes}. We will describe the process of extracting modes from our simulations in Sec.~\eqref{sec:modeextract}. Mode tests of our perturbative schemes and our simulation code will be performed on a stable TOV star in Secs.~\eqref{sec:CCSN_test_pert_schemes} and~\eqref{sec:FLASHTOV_test}, respectively. We will find that the scheme of~\cite{morozova2018gravitational}, which we dub the \emph{partially relaxed Cowling approximation} or simply the  \emph{partial Cowling approximation}, is significantly less accurate than the Cowling approximation itself for fundamental mode frequencies, and even fails to reproduce the correct radial order of high-order modes. In Chapter~\eqref{ch:CCSNresults} we present our results concerning the $l=2$, $m=0$, $n\gtrsim 2$ mode and its associated signatures in gravitational waves and neutrinos for the model with pre-collapse rotation $\Omega = 1.0$ rad/s. We conclude in Chapter~\eqref{ch:CCSNconc}, and display mode-matching for many other modes in the non-rotating model in Appendix~\eqref{app:CCSNapp}.

%
%
\section{Background and terminology} \label{sec:backgroundterminology}

In this section we review some basic facts and terminology surrounding the mode structure of stars which we will use in our analysis to characterize modes. We will follow~\cite{friedman2013rotating}.

Hydrodynamic perturbations consist of scalar and vector parts, which can be decomposed angularly using spherical harmonics $Y_{lm}$. Scalar perturbations admit expansions in the spherical harmonic functions themselves, whereas vector perturbations admit expansions in \emph{vector spherical harmonics}. The vector spherical harmonic basis is
\begin{eqnarray}
\lbrace Y_{lm} \hat{r},\: \vec{\nabla} Y_{lm},\: \hat{r} \times \vec{\nabla} Y_{lm} \rbrace.
\end{eqnarray}
In general the last two members of this basis will have both $\hat{\theta}$ and $\hat{\phi}$ components, but in axisymmetry ($m=0$) the situation simplifies: the vector spherical harmonic basis splits up into $\lbrace \hat{r}, \hat{\theta}, \hat{\phi} \rbrace$ components,
\begin{eqnarray}
\lbrace Y_{lm} \hat{r},\: \partial_\theta Y_{lm} \hat{\theta},\: \partial_\theta Y_{lm} \hat{\phi} \rbrace.
\end{eqnarray}

In spherical symmetry, the modes of stars split up into separate spherical harmonics $(l,m)$, with an additional radial order specified by $n$, which is a count of the number of nodes in the radial part of the mode function. An $l$-mode with no nodes ($n=0$) is called a \emph{fundamental mode}, and its $n>0$ companions are called \emph{overtones}. Furthermore, in spherical symmetry modes are degenerate in $m$ in the sense that modes with the same $(l,n)$ but different $m$ have the same frequency. Rotation breaks this degeneracy, and mode functions no longer split up into different $(l,m)$ but rather become superpositions of arbitrarily many different spherical harmonics. The possible deformations of axisymmetric modes of rotating stars (relative to the same modes in the non-rotating star) are tamed via their parity-definiteness, which we explain next.

There are different notions of parity which should be understood. A spherical harmonic function $Y_{lm}$ behaves under a parity transformation $\vec{r} \rightarrow -\vec{r}$ (or $\theta \rightarrow \pi - \theta, \phi \rightarrow \phi + \pi$) as
\begin{eqnarray}
Y_{lm} (\theta,\phi) = (-1)^l\: Y_{lm} (\pi - \theta, \phi+ \pi).
\end{eqnarray}
The first notion of parity is whether a sign change occurs under parity inversion. Thus, we say that $Y_{lm}$ has \emph{even parity} for even $l$ and \emph{odd parity} for odd $l$. However, the even or odd parity of $Y_{lm}$ is not preserved when moving to the vector spherical harmonics; $Y_{lm} \hat{r}$ and $\vec{\nabla} Y_{lm}$ will preserve the even or odd parity of $Y_{lm}$, whereas $\hat{r} \times \vec{\nabla} Y_{lm}$ will invert the parity due to the presence of the curl operation. Thus, for example, in axisymmetry if a mode function is a superposition of even-$l$ terms $\propto \hat{r}, \hat{\theta}$ and odd-$l$ terms $\propto \hat{\phi}$, then it has even parity. Conversely, if a mode function is a superposition of odd-$l$ terms $\propto \hat{r}, \hat{\theta}$ and even-$l$ terms $\propto \hat{\phi}$, then it has odd parity.

The second notion of parity is relative to $Y_{lm}$. Namely, if a member of the vector spherical harmonic basis for a given $(l,m)$ \emph{has the same parity as} $Y_{lm}$, then it is said to have \emph{polar parity}. Conversely, if the member has the  opposite parity of $Y_{lm}$, then it is said to have \emph{axial parity}. Thus, for example, in axisymmetry a mode with definite even or odd parity could be a superposition of both polar and axial terms, but all axial terms will be $\propto \hat{\phi}$ and all polar terms will be $\propto \hat{r}, \hat{\theta}$.

As mentioned above, an important fact we will use in our analysis is that axisymmetric modes have definite even or odd parity. Thus, an axisymmetric mode of a rotating star cannot be a mixture of even-$l$ terms $\propto \hat{r}, \hat{\theta}$ and $\propto \hat{\phi}$, for example. In our simulations below, axisymmetry is enforced exactly, so we can use the parity-definiteness of axisymmetric modes to help determine whether two $l$-terms are part of the same mode function.

There is a convention for labelling modes of rotating stars. First note that modes of rotating stars deform continuously as a function of the rotation rate $\Omega$. In most cases, a mode of a rotating star deforms to a single spherical harmonic in the $\Omega \rightarrow 0$ limit\footnote{So-called \emph{hybrid modes} have frequencies which go to zero in the non-rotating limit, while the eigenfunction is a superposition of different $Y_{lm}$s~\cite{friedman2013rotating}}. The convention is to label the mode in the rotating case according to what it deforms to in the non-rotating limit. For example, the ``fundamental $l=0$ mode" of a rotating star (a so-called \emph{quasi-radial} mode) may in fact have $l=2$ terms $\propto \hat{r}, \hat{\theta}$ in its mode function, which would allow it to radiate gravitational waves, and may in fact have radial nodes -- but in the non-rotating limit it is purely $l=0$ and $\propto \hat{r}$ and nodeless. Note that strictly speaking one must specify a sequence of rotating stars in order to talk about a mode's deformation as a function of $\Omega$, for example a sequence of fixed rest mass or of fixed central density, and one can observe different qualitative behavior of the modes along such sequences.

The deformations of mode functions of rotating stars are further restricted under the condition of slow rotation. Polar $l$-modes of non-rotating stars receive deformations by axial $l^\prime$-terms at $\mathcal{O}(\Omega)$ such that $l-l^\prime$ is odd, and receive deformations at the next order $\mathcal{O}(\Omega^2)$ by polar $l^{\prime \prime}$-terms such that $l-l^{\prime \prime}$ is even. Conversely for axial $l$-modes of non-rotating stars: polar deformations at $\mathcal{O}(\Omega)$ with $l-l^\prime$ even, and axial deformations at $\mathcal{O}(\Omega^2)$ with $l-l^{\prime \prime}$ odd. Notice these deformations preserve even or odd parity. 

Lastly, axial modes of non-rotating stars ($\propto \hat{\phi}$ in axisymmetry) have zero frequency (stationary flows), and their frequency increases linearly with the rotation rate of the star in the slow-rotation limit. Such modes are called $r$-modes since their oscillations are rotationally supported by the Coriolis force. In our simulations, such modes cannot be excited in the non-rotating case because the velocity in the $\hat{\phi}$ direction is exactly zero, but they may be excited in rotating cases.

%
%

\chapter{Implementation}\label{ch:CCSNimp}

\section{\texttt{FLASH} implementation~\cite{o2018two}} \label{sec:FLASHdescribe}
We will be performing multi-physics simulations, which is necessary for the study of CCSN. To this end we use a mature, modular code base which has applications in many areas, which we describe now.

We use the \texttt{FLASH}\footnote{http://flash.uchicago.edu} code~\cite{fryxell2000flash,dubey2009extensible} version 4, which is a publicly available massively parallel modular code which simulates compressible Newtonian flows. It supports adaptive mesh refinement, and is applied in many distinct settings with various extensions. The development of \texttt{FLASH} has focused on modularity and extensibility, with application-specific enhancements made possible without an alteration of the core code.

In the hydrodynamics solver, the variable reconstruction at cell interfaces uses the piecewise parabolic method, as in Sec.~\eqref{ch:ppm}. Since different flux formulae have different strengths and weaknesses, the fluxes are treated with a hybrid method which reverts to HLLE as the lower-accuracy option.

The gravity and hydrodynamics treatments are Newtonian, however an ``effectively" general relativistic potential obtained through phenomenological considerations and tested in $1$-dimensional CCSN evolutions has been introduced in~\cite{Keil1997,rampp2002radiation,marek2006exploring} and implemented in \texttt{FLASH} in~\cite{o2018two}. This is a correction to the monopole term in the potential, and helps to recover the structure of relativistic stars in spherical symmetry. 

The neutrino physics is too costly to implement without approximations, since it involves solving the Boltzmann problem for a very large number of particles. Such a calculation is limited by the large dimensionality of phase space, and many different approximate treatments have been employed in the literature. We use one of the better treatments available, the \emph{M1 scheme}~\cite{o2015open}, which evolves the moments of the neutrino distribution function with an analytic closure in which the first two moments (average and variance) carry all of the information. This treatment evolves neutrino fields on the computational grid, and thus allows us to extract the directional emission of neutrinos, which will play an important role in this work. Directional emission of neutrinos was not available in similar earlier studies due to the use of cruder neutrino treatments which yield only isotropic emission (the so-called \emph{leakage scheme})~\cite{ott2012correlated}.

A full description of this code is beyond our scope, so we refer to the references~\cite{fryxell2000flash,dubey2009extensible,o2015open,o2018two} for more details.

Seeing as our simulations use pseudo-Newtonian gravity, gravitational waves (GWs) are not actually present on the grid. We instead extract a GW signal using the quadrupole formula~\cite{thorne1980multipole}. This approach has been tested in the context of rotating stellar core collapse in~\cite{reisswig2011gravitational}.

\subsection{Stellar models} \label{sec:models}

The models we evolve come from the widely used stellar evolution calculations of Woosley \& Heger~\cite{woosley2007nucleosynthesis}; we use a star with $20 M_\odot$ zero-age main sequence mass. The SFHo equation of state~\cite{steiner2013core} is used, which is a modern tabulated nuclear equation of state compatible with the constraints available from eg. astrophysical observations of neutron stars.

The Woosley \& Heger models were obtained from 1-dimensional evolutions. We will be studying a sequence of rotating models, which are initialized with pre-collapse rotation by hand. The rotation law imposed is via the same prescription as in~\cite{zwerger1997dynamics,ott2004gravitational, ott2012correlated}, namely
\begin{eqnarray}
\Omega(\varrho) = \frac{\Omega_{c,\mathrm{initial}}}{ 1 + \left( \frac{\varrho}{A}\right)^2 }, \label{eq:rotlaw}
\end{eqnarray}
where $\varrho$ is the cylindrical radial coordinate. This profile results in constant angular velocity on cylindrical shells. The parameter $A$ is roughly the radius at which differential rotation becomes important, and throughout our rotating models we set it to $800$ km. This profile will result in constant specific angular momentum on cylindrical shells with radius $\varrho \gg A$, which is a condition expected to hold for electron-degenerate cores~\cite{tassoul2015theory} such as the iron core of a CCSN progenitor. Such cores would have a size comparable to the Earth, $\sim 10^4$ km, and thus the value of the parameter $A=800$ km is well within the core.

Our rotating models are then parameterized by the single quantity $\Omega_{c,\mathrm{initial}}$, which is the central value of the angular velocity. We will denote this parameter simply as $\Omega$, and our sequence of rotating models consists of the values $\Omega = \{0.0,0.5,1.0,2.0,2.5\}$ rad/s.

\section{Perturbative schemes} \label{sec:pertschemes}

Here we summarize the perturbative system of equations of~\cite{morozova2018gravitational}, which uses a \emph{partial Cowling approximation}; the derivation can be found in that work. The system in the Cowling approximation (from~\cite{torres2017towards}) can be obtained by setting the metric perturbation to zero. 

The spacetime is assumed to be spherically-symmetric and presented in spatially conformally flat form with zero shift,
\begin{eqnarray}
ds^2 = -\alpha^2 dt^2 + \psi^4 \delta_{ij} dx^i dx^j,
\end{eqnarray}
where $\alpha$ is the lapse, $\psi^4$ is the conformal factor, and $\delta_{ij}$ is the flat Euclidean metric in $3$ dimensions. A relativistic perfect fluid is perturbed on this spacetime, with energy-momentum tensor $T^{ab} = \rho h u^a u^b + P g^{ab}$. The lapse is the only metric function permitted to vary, hence we refer to it as a \emph{partial Cowling approximation}. By relating the gravitational potential $\Phi$ to the lapse via $\alpha = e^\Phi$, the Poisson equation is taken to relate the Eulerian perturbations of $\alpha$ and $\rho$:
\begin{eqnarray}
\nabla^2 \left( \frac{\delta\alpha}{\alpha} \right) = 4 \pi \delta \rho. \label{eq:Poisson}
\end{eqnarray}
Only polar perturbations are considered (no $\hat{\phi}$ part), and the perturbed quantities are taken to be spherical harmonics with harmonic time dependence,
\begin{eqnarray}
\delta P = \delta \hat{P} (r) Y_{lm} e^{-i \sigma t}, & \;\; & \delta \alpha = \delta \hat{\alpha}(r) Y_{lm} e^{-i \sigma t}, \nonumber\\
\xi^r = \eta_r (r) Y_{lm} e^{-i \sigma t}, & \;\; & \xi^\theta = \frac{\eta_\theta (r)}{r^2} \partial_\theta Y_{lm} e^{-i \sigma t}. \label{eq:pertansatz}
\end{eqnarray}

The perturbations are taken to be adiabatic, so that the Lagrangian perturbations $\Delta P$ and $\Delta \rho$ are related by $\Delta P/\Delta \rho = h c_s^2 = P \Gamma_1 /\rho$. Here, $\Gamma_1$ is the adiabatic index and $c_s$ is the speed of sound.

Eq.~\eqref{eq:Poisson} is reduced to two first-order equations by defining $f_\alpha \equiv \partial_r(\delta\hat{\alpha}/\alpha)$, which in spherical symmetry results in
\begin{eqnarray}
\!\!\!\! \!\!\!\! \!\!\!\! \partial_r f_\alpha \!\! &=& \!\! -\frac{2}{r} f_\alpha - 4\pi \left( \partial_r \rho - \frac{\rho}{P \Gamma_1} \partial_r P \right) \eta_r + \frac{4\pi\rho}{P\Gamma_1} \frac{\rho h \psi^4}{\alpha^2} \sigma^2 \eta_\theta - \left( \frac{4\pi\rho^2 h}{P\Gamma_1 \alpha} - \frac{l(l+1)}{\alpha r^2} \right) \delta\hat{\alpha},\label{eq:faux}\\
\!\!\!\! \!\!\!\! \!\!\!\! \partial_r \delta \hat{\alpha} \!\! &=& \!\! f_\alpha \alpha + \frac{\partial_r \alpha}{\alpha} \delta \hat{\alpha}.\label{eq:deltaLaphat}
\end{eqnarray}
The equations for the perturbations $\eta_r$, $\eta_\theta$ are given by
\begin{eqnarray}
\!\!\!\! \!\!\!\! \!\!\!\! \partial_r \eta_r \!\! &=& \!\! - \left(\frac{2}{r} + \frac{\partial_r P}{\Gamma_1 P} + 6\frac{\partial_r \psi}{\psi}\right) \eta_r - \frac{\psi^4}{\alpha^2 c_s^2} \left( \sigma^2 - \mathcal{L}^2 \right) \eta_\theta + \frac{\delta\hat{\alpha}}{\alpha c_s^2}, \label{eq:etar}\\
\!\!\!\! \!\!\!\! \!\!\!\! \partial_r \eta_\theta \!\! &=& \!\! \left(1-\frac{\mathcal{N}^2}{\sigma^2}\right) \eta_r - \left( \partial_r\; \mathrm{ln} \frac{\rho h \psi^4}{\alpha^2} - \frac{\partial_r P}{\rho h} \left(1 + \frac{1}{c_s^2}\right) \right) \eta_\theta + \frac{\rho h}{\alpha \partial_r P} \frac{\mathcal{N}^2}{\sigma^2} \delta \hat{\alpha}, \label{eq:etaperp}
\end{eqnarray}
where $\mathcal{N}$ is the Brunt-V\"{a}is\"{a}l\"{a} frequency 
\begin{eqnarray}
\mathcal{N}^2 = \frac{\alpha^2 \partial_r P}{\psi^4 \rho h} \left( \frac{\partial_r (\rho(1+\epsilon))}{\rho h} - \frac{\partial_r P}{\Gamma_1 P} \right)
\end{eqnarray}
and $\mathcal{L}$ is the Lamb frequency
\begin{eqnarray}
\mathcal{L}^2 = \frac{\alpha^2 c_s^2}{\psi^4} \frac{l(l+1)}{r^2}.
\end{eqnarray}

The Brunt-V\"{a}is\"{a}l\"{a} frequency is interpreted as the frequency with which a fluid element will oscillate in position after a small displacement, and thus is imaginary in regions of the fluid that are unstable to convective motion (i.e.~where a small displacement results in fluid elements running away exponentially from their initial position). The Lamb frequency is related to the sound crossing time of the star, with a dependence on angular harmonic $l$.

For a given frequency $\sigma = 2\pi f$, the system of equations Eqs.~\eqref{eq:faux},~\eqref{eq:deltaLaphat},~\eqref{eq:etar}, and~\eqref{eq:etaperp} are integrated outwards from $r=0$. The background is a spherically-averaged snapshot from full nonlinear simulations. Using the ODEPACK package~\cite{hindmarsh1983odepack}, we start the integration from the first point $r=\Delta r$ where we choose $\eta_r\vert_{\Delta r}$ to be a small number ($10^{-5}$ in our units). However, note that if $\eta_r\vert_{\Delta r}$ is chosen too small, we find that the quality of solution can degrade for low $l$. Quantities on the right-hand side are interpolated linearly when their value is requested between grid points. At $r=\Delta r$, $\eta_\theta$ is determined by the regularity condition
\begin{eqnarray}
\eta_\theta = \frac{r \eta_r}{l}.
\end{eqnarray}
Note this is different from the (incorrect) regularity condition $\eta_r = l \eta_\theta$ reported in~\cite{torres2017towards,morozova2018gravitational}, which was taken from the study~\cite{reisenegger1992new} whose definition of $\eta_\theta$ differed by a factor of $r$. It was subsequently corrected in~\cite{torres2018towards}. We checked that the mode frequencies for a TOV star do not depend sensitively on this error, yielding at most a $1-2$ Hz difference for the frequencies reported in Table~\eqref{tab:TOVCowling} of Sec.~\eqref{sec:CCSN_test_pert_schemes} below. It also does not significantly affect the mode functions we compute on the CCSN background in Chapter~\eqref{ch:CCSNresults} and Appendix~\eqref{app:CCSNapp}, up to a normalization.

The final specification is the outer boundary condition. Only those values of the frequency $\sigma$ which satisfy the outer boundary condition will be regarded as eigenmodes of the system. In~\cite{morozova2018gravitational}, they imposed that the Lagrangian perturbation of the pressure vanishes at the PNS surface, $\Delta P = 0$, corresponding to a free surface. This is equivalent to
\begin{eqnarray}
\frac{\rho h \psi^4}{\alpha^2} \sigma^2 \eta_\theta - \frac{\rho h}{\alpha} \delta\hat{\alpha} + \partial_r P \eta_r = 0.
\end{eqnarray}
This boundary condition is reasonable in the case of the study~\cite{morozova2018gravitational}, where the focus is on timescales of hundreds of milliseconds after bounce when the PNS surface is narrowed to within a $\sim 5$ km range where the density drops rapidly and the anisotropic velocity~\cite{takiwaki2012three},
\begin{eqnarray}
v_{\mathrm{aniso}} = \sqrt{\frac{\left\langle \rho \left[(v_r - \left\langle v_r\right\rangle )^2 + v_\theta^2 \right] \right\rangle}{\left\langle \rho \right\rangle}},
\end{eqnarray} 
rises rapidly (eg.~\cite{morozova2018gravitational,pan2018equation}). Here, $\left\langle . \right\rangle$ denotes spherical averaging. In our case we instead focus on tens of milliseconds after bounce when the accretion rate is larger and thus the PNS surface is not as well-defined. Thus we opt to instead use a different outer boundary condition, namely
\begin{eqnarray}
\eta_r \vert_{\mathrm{outer\;boundary}} = 0.
\end{eqnarray}
This was imposed at the spherically-averaged location of the stalled shockwave in~\cite{torres2017towards}. For ease of analysis, we impose this condition at $r=100$ km. We checked how sensitive our results are to this choice by instead placing the outer boundary at the shockwave location $\sim 150$ km, defined as where the derivative of the radial velocity is maximally negative. This yielded only a $\sim$few Hz change in frequency and an increase in the number of radial nodes by $1-2$. The main effect is to move radial nodes in the outer regions $r\gtrsim 60$ and add $1-2$ in the latter $100-150$ km. We will be working with density-weighted velocity mode functions, therefore these outer regions are suppressed and our analysis is largely insensitive to the location of the outer boundary (modulo the node counts being underestimated by $1-2$).

\section{Mode extraction from simulated data} \label{sec:modeextract}

To compare mode functions, the velocity field of the star is run through spectrogram filters which cut out the time-varying frequency peaks. In more detail: for each point in the star, and for each component of the velocity, one has a 1-dimensional time series from which a spectrogram is calculated. By looking at spectrograms from various points in the star, persistent frequency peaks are identified and masks are drawn to filter them. For each mask, the entire velocity field of the star is filtered. The resulting field is then decomposed angularly using vector spherical harmonics, which allows us to view temporal snapshots of the mode eigenfunctions, possibly contaminated by additional modes and nonlinear motions. These velocity snapshots can be compared to the mode eigenfunctions computed perturbatively. Although the perturbative eigenfunctions are displacement fields, their assumed harmonic time dependence in the ansatz~\eqref{eq:pertansatz} ensures that they will be proportional to their time derivatives (which is what we actually extract from the velocity data).

After identifying the modes in the non-rotating case, we can repeat the mode extraction in the rotating cases. This allows us to observe the progressive change in the mode eigenfunctions across the different rotation cases~\cite{friedman2013rotating,dimmelmeier2006non}.

We identify candidate mode frequency bands visually from the spectrograms of $v_r$, $v_\theta$, $v_\phi$ at various points within the proto-neutron star (PNS). Most bands have the majority of their energy concentrated within $r \lesssim 30$ km. To accentuate where the energy resides in space, we analyse the mode functions with a density weight $\sqrt{\rho}$. This also has the benefit of suppressing the low-energy regions of the mode functions, where a reliable comparison with velocity data from our simulations becomes strained due to the presence of noise and/or nonlinear dynamics.

When identifying frequency bands, it is essential to sample many different points in the core, since different mode functions can have very weak activity at points in the star near their nodes. In addition, often a frequency band for a given rotation case is difficult to identify without also considering neighbouring rotation cases, where the band may be more easily identified. When a band is identified, a mask is drawn around it for the purpose of filtering the rest of the velocity field out, allowing for a targeted analysis of that band. Note that we do not draw band masks based on the gravitational wave spectrograms nor the neutrino luminosity spectrograms. This allows us to see whether a given feature in the signal spectrograms is due to a single band or a mixture of them, which is important for extracting detailed source information from the detected signals. This process of identifying bands and drawing masks around them is done manually, and thus relies on human judgement. Ideally one would design an algorithm to do this, for the sake of reproducibility. However, our conclusions below appear to be sound, and being the first to do this analysis we will leave further automation for future improvements.

In what follows, we will focus on several frequency bands which are coincident with observable features in the gravitational wave or neutrino luminosity spectrograms. We will attempt to implicate specific modes by following them along our sequence of rotating models. Continuity of the modes will be argued in part from a combination of:
\begin{enumerate}
\item Reasonable shifts in frequency,
\item Minimized deformations in mode functions,
\item Consistent $l$-mixing based on the definite even or odd parity of axisymmetric modes,
\item Consistent decay rates among different $l$ which belong to the same mode,
\item Consistent frequency among different $l$ which belong to the same mode, when the average frequency in the band mask is not changing too much in time.
\end{enumerate}

All energies displayed as functions of time will be smoothed using a Gaussian window with $0.5$ ms width, which is an adequate window width for eliminating the oscillations of the energies since the window is wider than the longest mode period. This allows for a cleaner inspection of the relative energy levels among the different $l$ and their decay rates. Occasionally the velocity data will be similarly smoothed in space when appropriate (and this will be explicitly pointed out).

It is important to note that, although we will be referring to bands using their approximate frequencies from the simulation, based on our tests in Sec.~\eqref{sec:FLASHTOV_test} the true frequencies of all modes under consideration are expected to be lower. We will quote the Cowling frequency of the best-fit mode function when available, and that frequency should be regarded as closer to the true one (but still overestimated). This will have important implications for the detectability of the signal features we identify, since in general the lower frequency will be easier to detect, both for gravitational wave detectors with LIGO-like sensitivity as well as neutrino detectors which require higher event rates to resolve higher frequency oscillations in neutrino luminosity. 

For a given $l$ and time in the simulations, all perturbative mode functions are computed in the range of frequencies $\sim 20-2000$ Hz. The best match between all of the candidate perturbative mode functions and the velocity data can be found by normalizing them  and measuring their difference. Suppose ${}^l \vec{v}_{\sigma,\mathrm{pert}}$ is the perturbative mode function for some $l$ and frequency $\sigma$ and ${}^l \vec{v}_{\mathrm{sim}}$ is the $l$-component of the spectrogram-filtered velocity field from the simulation (for the band mask of interest). We compute a measure of difference via
\begin{eqnarray}
\Delta \equiv \Sigma \left( {}^l \vec{v}_{\sigma,\mathrm{pert}} - {}^l \vec{v}_{\mathrm{sim}} \right)^2, \label{eq:Delta}
\end{eqnarray}
where the sum is over radial points on the numerical grid, and then minimize $\Delta$ over the perturbative modes (parameterized by a discrete set of frequencies $\sigma$ which yield mode functions which satisfy the boundary conditions).

The snapshots from our simulations which we input into Eq.~\eqref{eq:Delta} are taken from within a $\sim$few ms neighborhood of the time at which the perturbative calculation is performed. The snapshots are chosen to yield peak amplitude of the oscillation for $\sqrt{\rho}\eta_r$ and positive value near the origin. The moment of peak amplitude of the oscillation gives the sharpest picture of the mode function, which facilitates an accurate comparison with perturbation theory, and the positive value near the origin is in accordance with our choice of initial conditions at the first radial point $r=\Delta r$ in the perturbative calculations (see Sec.~\eqref{sec:pertschemes}). Both the velocity snapshots and the perturbative mode functions are normalized to facilitate a comparison of their shape.

Since our perturbative schemes are valid only for non-rotating models, we abandon it when establishing mode function continuity across different rotation models. For this, we also use a difference of the form of Eq.~\eqref{eq:Delta}, except computed between simulation snapshots from neighbouring rotation cases (i.e.~between ${}^l \vec{v}_{\mathrm{sim, 0.0 rad/s}}$ and ${}^l \vec{v}_{\mathrm{sim, 0.5 rad/s}}$, for example). Further details of this procedure will be described in Chapter~\eqref{ch:CCSNresults}, since the procedure is more easily understood in the immediate context of the results.

\section{Tests of perturbative schemes} \label{sec:CCSN_test_pert_schemes}

\begin{figure}[htbp]
\centering
\hbox{\hspace{3.75cm}\includegraphics[width=0.5\textwidth]{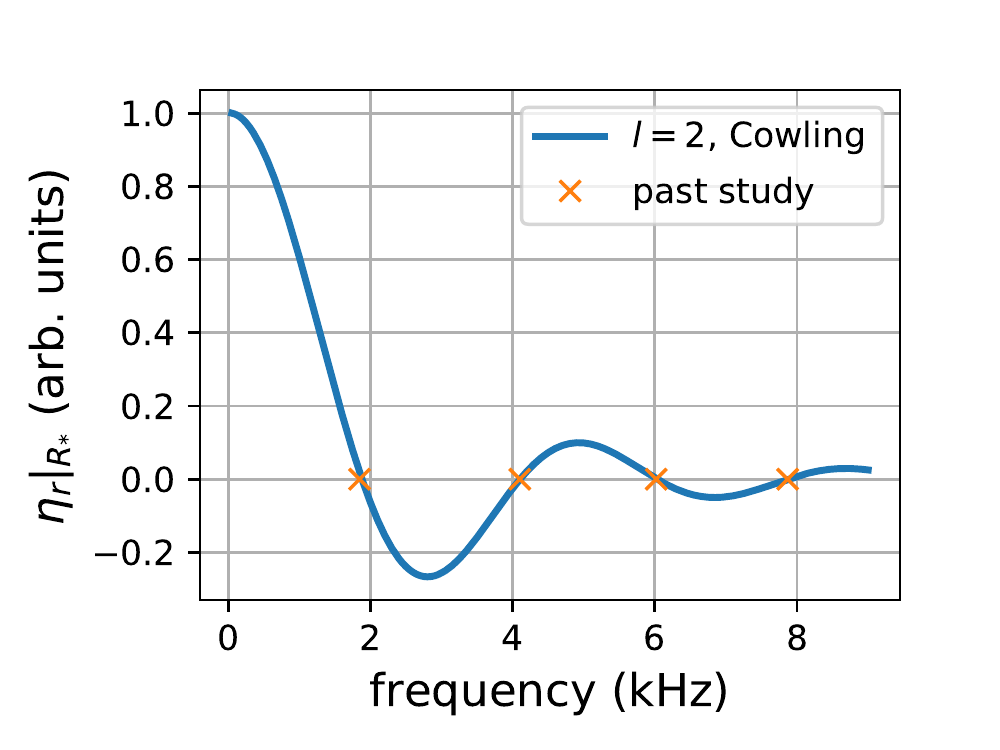}}
\caption{An illustration of how the eigenfrequencies of the star are obtained. For a chosen frequency, the equations are integrated out from the origin to the surface of the star $R_{*}$. If the radial displacement $\eta_r$ vanishes at $r=R_{*}$, then the outer boundary condition for eigenmodes is satisfied. We find it adequate in this test to sample the function $\eta_r \vert_{R_{*}}$ every $50$ Hz, shown here, then use a rootfinding algorithm on a cubic interpolation of the function. This gives rise to the values in Tables~\eqref{tab:TOVCowling} and~\eqref{tab:TOVconformalflat}. The eigenfrequencies from~\cite{font2001axisymmetric} are indicated in the plot for comparison. Using bisection and no interpolation gives similar results.} \label{fig:pertTOV_lastetar}
\end{figure}

In this section we test our implementation of the perturbative schemes of~\cite{torres2017towards} and~\cite{morozova2018gravitational} on a stable TOV star with polytropic equation of state $P = K \rho^\Gamma$ with $\Gamma=2$, $K=100$, and central density $\rho_c = 1.28 \times 10^{-3}$ in geometrized units. This star has been used extensively as a test bed for numerical codes, eg.~\cite{font2001axisymmetric,font2002three,baiotti2010new,radice2011discontinuous}.  The boundary condition we impose is $\eta_r \vert_{R_*} = 0$, where $R_*$ is the stellar surface (using instead $\Delta P \vert_{R_*} = 0$ yields $\lesssim 1\%$ difference in mode frequencies). Finding such solutions amounts to a rootfinding problem in the frequency variable, i.e.~finding the frequency whose corresponding mode satisfies the boundary condition. One can visualize this by plotting $\eta_r \vert_{R_*}$ as a function of frequency, as in Fig.~\eqref{fig:pertTOV_lastetar}. The blue curve we obtained from our perturbative calculations, and the eigenfrequencies from the study~\cite{font2001axisymmetric} are indicated in orange crosses, coincident with the roots of $\eta_r \vert_{R_*}$.

\FloatBarrier

\begin{center}
\begin{table}[]
\centering
\scalebox{0.8}{%
\begin{tabular}{l|l|l|l|l|l|l|l|l|l|l|l|l|}

 & ${}^1 f$ & ${}^1 p_1$ & ${}^1 p_2$ & ${}^1 p_3$ & ${}^2 f$ & ${}^2 p_1$ & ${}^2 p_2$ & ${}^2 p_3$ & ${}^3 f$ & ${}^3 p_1$ & ${}^3 p_2$ & ${}^3 p_3$ \\ \hline
From~\cite{font2001axisymmetric} (kHz)                                                                                   & 1.335    & 3.473      & 5.335      & 7.136      & 1.846    & 4.100      & 6.019      & 7.867      & 2.228    & 4.622      & 6.635      & 8.600      \\ \hline
From~\cite{gaertig2008oscillations} (kHz)                                                                                & -        & -          & -          & -          & 1.890    & 4.130      & -          & -          & -        & -          & -          & -          \\ \hline
Current work (kHz)                                                                                                                    & 1.376 & 3.469 & 5.336 & 7.141 & 1.881 & 4.104 & 6.028 & 7.866           & 2.255 & 4.640 & 6.647 & 8.535           \\ \hline \hline
\% diff. with~\cite{font2001axisymmetric}    & 3.7         & 0.12           & 0.019 & 0.070 & 2.4 & 0.096 & 0.15 & 0.013 & 1.2 & 0.39 & 0.18 & 0.76           \\ \hline
\% diff. with~\cite{gaertig2008oscillations} & -        & -          & -          & -          & 0.48 & 0.63 & -          & -          & -        & -          & -          & -         
\end{tabular}}
\caption{A comparison between the mode frequencies we obtain with boundary condition $\eta_r \vert_{r=R^*} = 0$ and those obtained in~\cite{font2001axisymmetric,gaertig2008oscillations} using different methods, for a $\Gamma =2$, $K=100$, $\rho_{0,c}=1.28\times 10^{-3}$ TOV star in the Cowling approximation. The TOV background is generated using the same code that was used in Sec.~\eqref{sec:TOVoscillations}. The dominant error in the frequency is in the specification of location of the star surface, at which the boundary condition is imposed; changing it by one grid point yields a possible modification of the frequencies by $\sim 1$ Hz.} \label{tab:TOVCowling}
\end{table}
\end{center}

Relative to the full Cowling approximation, the corrections obtained for the fundamental ($n=0$) mode frequencies for $l=2,3,4$ in~\cite{morozova2018gravitational} using the partially relaxed Cowling approximation are all upward, whereas the true frequencies have been found to be at lower values for a wide variety of modes and stars, eg.~\cite{font2002three,cerda2008new,zink2010frequency, chirenti2015fundamental,mendes2018new}. We also checked that this presumably erroneous upward correction for the fundamental modes occurs for $l=1,5,6,7,8$ as well. On the other hand, for all modes with $n>0$ that we checked for $l=1-8$, the correction to the mode frequency supplied by the partial Cowling approximation is downward, as one would expect. This is shown for $l=2,4$ in Table~\eqref{tab:TOVconformalflat}, where the overtone frequencies are more correct than the fundamental ones, in comparison to the full Cowling approximation. The partial Cowling approximation also begins to fail to reproduce the correct number of nodes for modes with increasing $l$ and $n$, and even to fail to reproduce modes at all (i.e.~none satisfying the boundary condition $\eta_r =0$ at the surface of the star).

The erroneous upward correction of the fundamental mode frequencies in the partial Cowling approximation potentially calls into question the conclusion of~\cite{morozova2018gravitational} that the dominant gravitational wave emission is due to the fundamental $l=2$ (and $m=0$) mode, since that conclusion is based solely on the coincidence over time between that mode's frequency (as computed in the partial Cowling approximation) and the peak emission in the gravitational wave spectrograms from simulations. A comparison between the perturbative eigenfunctions and the velocity data from simulations would conclusively identify the dominantly radiating mode. Hence we follow this eigenfunction-matching strategy in this work.

Our simulations extend only to $\sim 100$ ms after bounce, which is much shorter than the range of $\sim 1.5$ s considered in~\cite{morozova2018gravitational}, so we cannot attempt to reproduce their mode identifications. However, rather than identify modes via coincident frequencies while ignoring the mode functions, we take the opposite approach by identifying modes via mode function matching while ignoring the mode frequencies. Given the errors introduced by the Cowling and partial Cowling approximations, as well as possible errors in the simulations arising from approximate treatments of the hydrodynamics and/or gravity, we advocate this approach since we believe it is less likely that these approximations would confuse modes at the level of the mode functions.

\begin{table}[]
\centering
\begin{tabular}{l|l|l|l|l|}

                         & ${}^2 f$ & ${}^2 p_1$ & ${}^4 f$ & ${}^4 p_1$ \\ \hline
From~\cite{dimmelmeier2006non}, GR CFC (kHz) & 1.586    & 3.726      & 2.440    & 4.896      \\ \hline
Current work, partial Cowling (kHz)       & 2.496    & 3.777      & 3.047    & 4.999      \\ \hline
Current work, Cowling (kHz) & 1.881 & 4.104 & 2.565 & 5.112      \\ \hline \hline
\% diff.~\cite{dimmelmeier2006non} vs partial Cowling                 & 57      & 1.4        & 25     & 2.1 \\ \hline
\% diff.~\cite{dimmelmeier2006non} vs Cowling & 19 & 10 & 5.1 & 4.4 \\ \hline
\end{tabular}
\caption{A comparison between the mode frequencies we obtain perturbatively using the partially relaxed Cowling approximation of~\cite{morozova2018gravitational} and those obtained in~\cite{dimmelmeier2006non} using full numerical simulations in the conformal flatness approximation, for the same $\Gamma =2$, $K=100$, $\rho_{0,c}=1.28\times 10^{-3}$ TOV star. The TOV background is generated using the same code that was used in Sec.~\eqref{sec:TOVoscillations}. The conformal flatness approximation is regarded as quite accurate for these modes~\cite{friedman2013rotating}. The agreement with~\cite{dimmelmeier2006non} is worsened considerably for the fundamental modes, but improved for the overtones shown, in comparison to the frequencies obtained in the full Cowling approximation. It has been observed repeatedly that the Cowling approximation tends to overestimate the true frequencies, see eg.~\cite{font2002three,cerda2008new,zink2010frequency, chirenti2015fundamental,mendes2018new}.}
\label{tab:TOVconformalflat}
\end{table}

\section{TOV mode test in the \texttt{FLASH} implementation~\cite{o2018two}} \label{sec:FLASHTOV_test}

In this study we use the \texttt{FLASH}~\cite{fryxell2000flash,dubey2009extensible} implementation of~\cite{o2018two}, which uses Newtonian hydrodynamics and a phenomenological effective gravitational potential developed in~\cite{Keil1997,rampp2002radiation,marek2006exploring}, designed to mimic general relativity in spherical symmetry. The neutrino physics is also implemented using the M1 scheme~\cite{o2015open}. However, in this section we test only the Newtonian hydrodynamics and effective gravitational treatment in the context of the modes of oscillation of a stable TOV star. This test is particularly relevant to our study since we will be extracting the dominant modes of oscillation in core-collapse supernova simulations within this implementation.

A TOV \emph{migration test} was carried out in~\cite{o2018two}, where a TOV star on the unstable branch is observed to migrate to the stable branch. The ensuing oscillations were observed to have a frequency $\sim 50 \%$ higher than in the general relativistic case. In this section we extract the fundamental radial ($l=0$) and axisymmetric quadrupolar ($l=2$, $m=0$) modes $\lbrace F$, ${}^2 f \rbrace$ and their overtones $\lbrace H_1$, $H_2$, ${}^2 p_1$, ${}^2 p_2$, ${}^2 p_3 \rbrace$ from the same stable TOV star studied in Secs.~\eqref{sec:TOVoscillations}~\eqref{sec:CCSN_test_pert_schemes} (central density $\rho_c = 1.28 \times 10^{-3}$, $\Gamma =2$, $K=100$). To extract both radial and quadrupolar modes, we initialize the simulation with a velocity perturbation 
\begin{eqnarray}
v_\theta = 0.001 \sin(\pi r/R)\sin{\theta}\cos{\theta}
\end{eqnarray}
(in units $c=1$), where $R$ is the radius of the star. This perturbation was used in~\cite{dimmelmeier2006non} as a trial eigenfunction for the excitation of quadrupolar modes in rotating and non-rotating stars, but we find it also excites radial modes in our case. It comes from the $l=2$ vector spherical harmonic for $\hat{\theta}$, $\partial_\theta Y_{20} \propto \partial_\theta (3 \cos^2 \theta -1)$.

\begin{figure}[htbp]
\centering
\hbox{\hspace{-1.6cm}\includegraphics[width=1.2\textwidth]{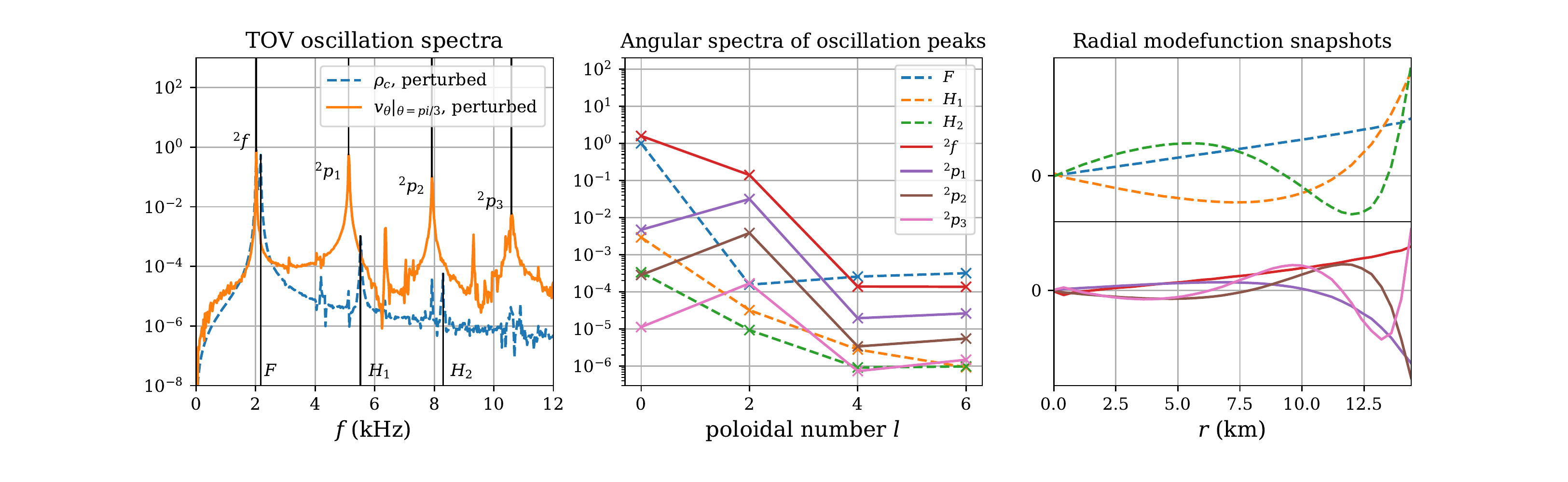}}
\caption{Mode identification from simulations of a $\rho_c = 1.28 \times 10^{-3}$, $\Gamma=2$, $K=100$ stable TOV star using the FLASH implementation of~\cite{o2018two}, which employs Newtonian hydrodynamics and the effective gravitational potential of~\cite{marek2006exploring} (case A). (Left): Frequency spectra are shown from points within the star initialized with the quadrupolar perturbation $v_\theta = 10^{-3}\sin(\pi r/R) \sin{\theta} \cos{\theta}$ (in units $c=1$). The frequencies of the first three radial modes $\lbrace F$, $H_1$, $H_2 \rbrace$ and the first four quadrupolar modes $\lbrace {}^2 f$, ${}^2 p_1$, ${}^2 p_2$, ${}^2 p_3 \rbrace$ are manifested as peaks and are indicated with vertical black lines. The frequencies are shown and compared with full GR and Cowling values in Table~\eqref{tab:FLASHTOV_mode_test}. The FLASH implementation overestimates the frequencies with respect to full GR in all cases, and with respect to the Cowling approximation in the case of non-radial modes. (Center): Angular power spectra of the velocity field band-passed around the frequency peaks, integrated over time. The peak energy is in $l=0$ for the radial modes and in $l=2$ for the for quadrupolar overtone modes. What we claim is the ${}^2 f$ mode has a similar level of $l=2$ excitation as the ${}^2 p_1$ mode, but is also contaminated by a dominant $l=0$ part. We checked that the ${}^2 f$ frequency peak indeed corresponds to a linear mode by increasing the perturbation strength by $10\times$ and observing the excitation change linearly, but the $l=0$ contamination did not diminish in relative strength. (Right): Snapshots of the normalized modefunctions for the $l=0$ (Top) and $l=2$ (Bottom) modes, displaying their respective node counts of $n=0,1,2$ for $\lbrace F,H_1,H_2\rbrace$ and $n=\lbrace 0,1,2,3\rbrace$ for $\lbrace {}^2 f,{}^2 p_1, {}^2 p_2, {}^2 p_3 \rbrace$.} \label{fig:FLASHTOV_test}
\end{figure}

Mode identification proceeds via band-passing the velocity field around dominant frequency peaks in the simulations. The frequency peaks for the radial modes are identified from the central density evolution. Its average is subtracted and then the resulting time series is windowed using a Gaussian function, which enhances the peaks in the Fourier transform. The frequency peaks for the quadrupolar modes are instead extracted from the evolution of $v_\theta$. We first subtract a Gaussian-smoothed version of itself to eliminate any secular variation, and the result is again windowed with a Gaussian and then its spectra are computed and averaged along a line with $\theta = \pi/3$. These spectra are displayed in Fig.~\eqref{fig:FLASHTOV_test} (Left), with peaks labelled and indicated with vertical black solid lines. 

After band-passing the velocity field about these frequency peaks, an angular decomposition into vector spherical harmonics is performed. This allows us to compute the total energy (integrated over time and space) in each poloidal number $l$, i.e. the angular spectra of the frequency peaks. These spectra are displayed in Fig.~\eqref{fig:FLASHTOV_test} (Center), and serve to show the concentration of energy in $l=0$ for the radial modes and $l=2$ for the quadrupolar modes. A lone exception is the mode which we claim is the fundamental quadrupolar one, ${}^2 f$. It is contaminated by a strong $l=0$ part which appears nodeless. We verified that this frequency peak corresponds to a linear mode by increasing the perturbation strength by $10\times$ and observing the peak amplitude change linearly in response. However, the $l=0$ contamination did not disappear, and its energy scaled by the same factor as the $l=2$ part. This may have to do with the close proximity of the $F$-mode frequency. Alternatively, it could mean that our \texttt{FLASH} implementation incorrectly reproduces the pure-$l$ character of mode functions of spherical stars. All the quadrupolar overtones, on the other hand, have dominant energy in $l=2$.

The vector spherical harmonic decomposition also allows for the observation of the radial modefunction associated with a given $l$ and given frequency peak, and snapshots of these are displayed in Fig.~\eqref{fig:FLASHTOV_test}. These radial mode functions reveal the node counts of the modes, which helps justify our mode identifications.

With the modes identified, the frequencies can now be compared with those obtained in the Cowling approximation and those obtained in full GR (in the conformal flatness approximation in the case of the non-radial modes). These comparisons are shown in Table~\eqref{tab:FLASHTOV_mode_test}. The mode frequencies extracted from the \texttt{FLASH} simulations are overestimated with respect to full GR in all cases. In comparison to the Cowling approximation, the \texttt{FLASH} simulations yield an improvement in the fundamental radial mode frequency (i.e.~downward correction) only, but a worsening in the case of all other modes (i.e.~upward correction).

These observations help to interpret the results of the current work, where we will be identifying the modes of oscillation in rotating CCSN simulations in \texttt{FLASH} with the expectation that the true frequencies are significantly lower than those observed in our simulations. We will see that the Cowling frequencies are lower than the simulation by a factor $\sim 0.6$. As we saw in Sec.~\eqref{sec:Valencia}, the relativistic hydrodynamic fluxes contain factors of the lapse $\alpha$, which would serve to slow fluid motions inside a gravitational well. The \texttt{FLASH} implementation uses Newtonian hydrodynamics and therefore the lapse is absent. However, it seems that the overestimates of mode frequencies cannot be totally accounted for by the lapse, which only reaches a minimum of $\sim 0.8$ at the origin. This suggests the effective gravitational potential, also used in the \texttt{FLASH} implementation, is responsible for a large fraction of the error.

\begin{table}[]
\centering
\begin{tabular}{l|l|l|l|l|l|l|l|}

                         & $F$ & $H_1$ & $H_2$ & ${}^2 f$ & ${}^2 p_1$ & ${}^2 p_2$ & ${}^2 p_3$ \\ \hline
From~\cite{dimmelmeier2006non} \&~\cite{font2002three}, GR CFC (kHz) & 1.442    & 3.955  & 5.916      & 1.586    & 3.726 & - & -     \\ \hline
Current work, Cowling (kHz)       & 2.696    & 4.534 & 6.346     & 1.881    & 4.104  &  6.028 &  7.866    \\ \hline
Current work, \texttt{FLASH} (kHz) & 2.174 & 5.522 & 8.295 & 2.024 & 5.122 & 7.920 & 10.593      \\ \hline \hline
\% diff. \texttt{FLASH} vs GR CFC                 & +51      & +40        & +40     & +28 & +37 & - & - \\ \hline
\% diff. \texttt{FLASH} vs Cowling & -19 & +22 & +31 & +8 & +25 & +31 & +35 \\ \hline
\end{tabular}
\caption{A comparison between the mode frequencies we obtain from \texttt{FLASH} simulations and those obtained in the Cowling approximation and in full GR in the conformal flatness approximation (GR CFC), for the same $\Gamma =2$, $K=100$, $\rho_{0,c}=1.28\times 10^{-3}$ TOV star. The \texttt{FLASH} simulations yield frequencies overestimated with respect to full GR in all cases. We observe an improvement in the fundamental radial mode frequency with respect to the Cowling approximation (i.e.~a downward correction), whereas all other mode frequencies obtain an erroneous upward correction.}
\label{tab:FLASHTOV_mode_test}
\end{table}

\chapter{Results \& Discussion} \label{ch:CCSNresults}

%
%
\section{$l=2$, $n\gtrsim2$ mode around $\sim 280$ Hz (Cowling)}
In this section we implicate an $l=2$, $n \gtrsim 2$ mode in producing prominent gravitational wave frequency peaks in all $\Omega = \lbrace0.0,0.5,1.0,2.0,2.5\rbrace$ rad/s pre-collapse rotation cases, with coincident frequency peaks in the neutrino luminosities in a subset of cases. We will see that this mode picks up $l=1,3$ deformations consistent with expected leading order effects in rotation. The mode's frequency in the simulation steadily rises across the sequence of rotating models, from $\sim 500$ Hz to $\sim 900$ Hz in the $\Omega = 2.5$ rad/s case where it is thoroughly mixed with many other $l$ and possibly other modes. \emph{The chief result will be the clean identification of peak frequencies of oscillation in gravitational waves \emph{and} neutrino luminosities, both caused by the same $l=2, n\gtrsim 2$ mode whose frequency in the Cowling approximation is $\sim 280$ Hz, about $60\%$ lower than the simulation (and likely still overestimated).}


\subsection{$\Omega = 0.0$ rad/s model}
\begin{figure}[htbp]
\centering
\hbox{\hspace{-1.6cm}\includegraphics[width=1.2\textwidth]{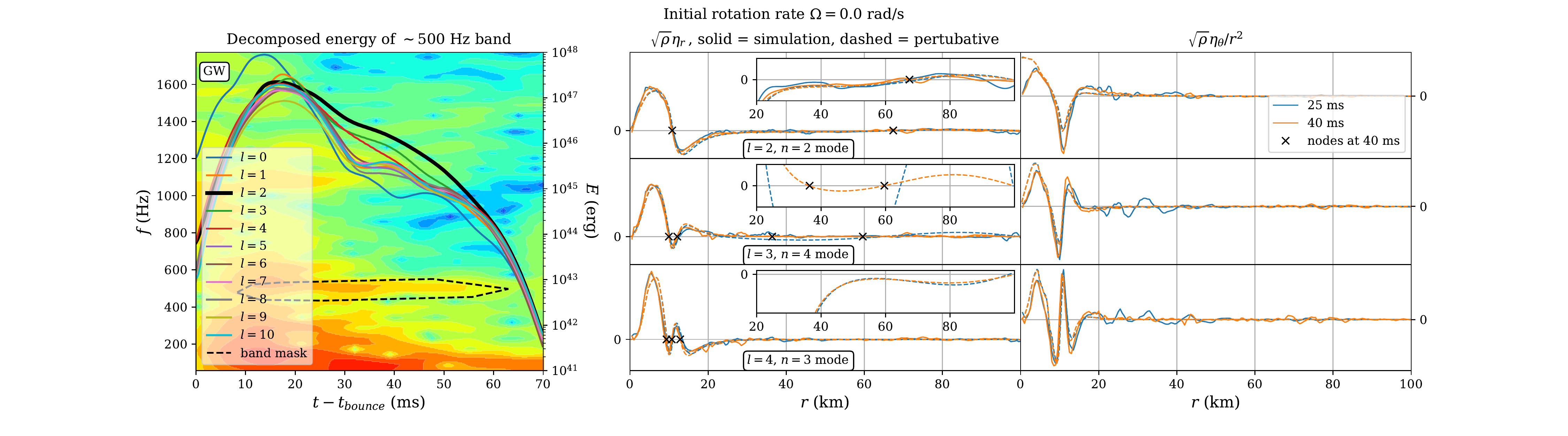}}
\caption{Mode data and best-fits in the Cowling approximation for the $\Omega = 0.0$ rad/s model. (Left): Gravitational wave spectrogram with the $\sim 500$ Hz band mask overlaid, as well as the angularly-decomposed energy in the corresponding filtered velocity field. The band mask is coincident with an emission feature in the spectrogram. We see $l=2,3,4$ energies standing out. (Centre \& Right): Snapshots of the $\sqrt{\rho}$-weighted vector spherical harmonic-decomposed and spectrogram-filtered velocity field from the simulation (solid lines) as well as the best-fit perturbative mode functions in the Cowling approximation (dashed lines). The agreement is striking. Insets zoom in on the low-energy tails for $r \gtrsim 20$ km, showing the nodes in the perturbative mode function (black crosses). In the $l=2$ case, the inset also shows the velocity data, smoothed with a Gaussian of width $5$ km, since in that case the noise level is low enough to allow a meaningful comparison with the perturbative mode function.} \label{fig:500Hz_top3modes}
\end{figure}

This frequency band mask around $\sim 500$ Hz has most of its energy stored in $l=2$ between $25-50$ ms, with $l=3$ and $l=4$ containing second and third most energy (see Fig.~\eqref{fig:500Hz_top3modes} (Left)). We will argue that the $l=2$ component is its own mode, and that it is responsible for signal features in the rotating models as well as the non-rotating model. We will follow all the dominant modes in the band for as long as we can, and the $l=2$ will be followed across the entire sequence of models.

Firstly, we interpret these three dominant components $l=2,3,4$ as different modes based on a few known facts and observations which we already outlined in Secs.~\eqref{sec:backgroundterminology} and~\eqref{sec:modeextract}:
\begin{enumerate}
\item Modes of spherical stars are pure spherical harmonics, so no mixing of $l$'s is expected. Although this simulation is not enforcing spherical symmetry exactly, the condition of zero rotation \emph{is} enforced exactly, which means deformations of the proto-neutron star away from spherical are small. Thus, one does not expect any given mode to have $l$-mixing so strong as to have comparable energy in different $l$, as we observe in Fig.~\eqref{fig:500Hz_top3modes} (Left).
\item The $l=3$ component is forbidden to be part of the $l=2$ or $l=4$ modes since it has odd parity in this non-rotating model (velocity has no $\hat{\phi}$ part) whereas the $l=2,4$ components have even parity.
\item The $l=4$ component, although having parity consistent with the $l=2$ component, does not manifest the same decay rate as the $l=2$ component in Fig.~\eqref{fig:500Hz_top3modes} (Left).
\end{enumerate}

Next, we note that the gravitational wave emission observed within the band mask in Fig.~\eqref{fig:500Hz_top3modes} (Left) is certainly dominated by the $l=2$ component, since the gravitational wave signal itself was extracted using the quadrupole formula. Physically we expect this to be true as well, since the $l=2$ component has the most energy and is a vastly more efficient radiator of gravitational wave energy, being enhanced with respect to $l=4$ by a factor $\sim c^2$~\cite{thorne1969nonradial}.

In Fig.~\eqref{fig:500Hz_top3modes} (Center \& Right) we make direct comparisons between the velocity data and the perturbative mode functions computed at $25$ and $40$ ms on the CCSN background using the Cowling approximation. The first column compares the density-weighted radial mode functions $\sqrt{\rho} \eta_r$, and the second column compares the non-radial functions $\sqrt{\rho} \eta_\theta/r^2$. We observe a small change in both the perturbative and simulated mode functions between $25$ and $40$ ms, which illustrates the separation of scale between the oscillation period of the modes and the characteristic time over which the background changes. The inset zooms in on the low-energy region $r \gtrsim 20$ km to show any nodes which may be present there in the perturbative mode functions. In the case of the $l=2$ mode, we also display the velocity data smoothed using a Gaussian window of width $5$ km, since in this case the noise level is low enough to show a meaningful agreement with the perturbative mode function.

\begin{figure}[htbp]
\centering
\includegraphics[width=0.8\textwidth]{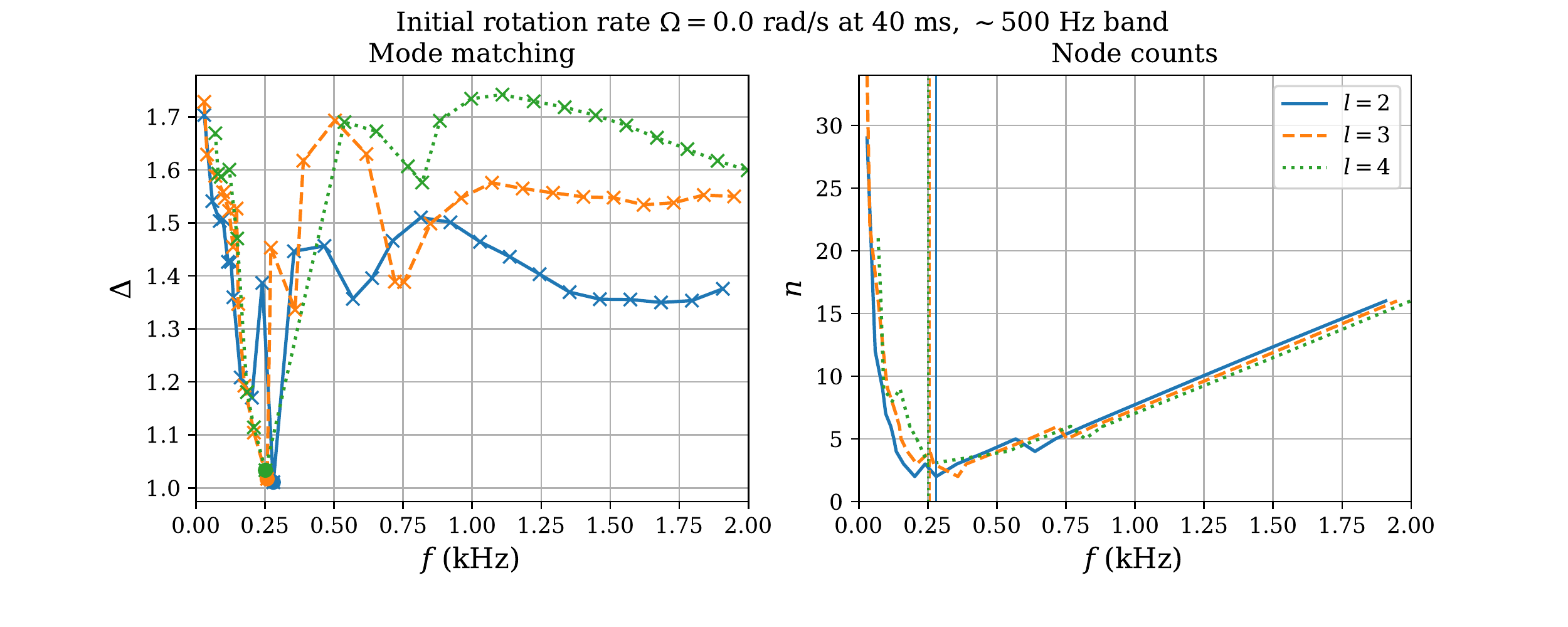}
\caption{(Left): The normalized difference $\Delta$ defined in Eq.~\eqref{eq:Delta} plotted for all candidate perturbative mode frequencies for the $l=2,3,4$ cases under consideration in this section. The best-fit mode functions occur at the minima of $\Delta$, and are indicated as circles. (Right): The node counts $n$ of each candidate perturbative mode frequency for the same cases, with the best-fit mode frequencies indicated with vertical lines. The best-fit modes occur in the vicinity of the minimum node counts, and thus cannot be definitively identified as $g$-modes ($n$ decreasing as frequency increases) or $p$-modes ($n$ increasing as frequency increases) on the basis of this plot.} \label{fig:500Hz_modematching}
\end{figure}

The $l=2$ mode displayed in Fig.~\eqref{fig:500Hz_top3modes} (Right) has $\Delta \sim 1.0$, as compared with most other possibilities with $\Delta \sim 1.4$, where the difference $\Delta$ is defined in Eq.~\eqref{eq:Delta} and its surrounding discussion. We display the value of $\Delta$ for all candidate modes for $l=2,3,4$ as a function of their frequencies in Fig.~\eqref{fig:500Hz_modematching} (Left). The circles indicate the minima of $\Delta$, which occur for the best-fit modes.

In Fig.~\eqref{fig:500Hz_modematching} (Right) we plot the node counts for all candidate modes as a function of their frequencies. The best-fit mode frequencies are indicated in vertical lines. Typically one expects $p$-modes to have increasing frequency as $n$ increases, and the opposite behavior for $g$-modes, with the fundamental mode living at the boundary of these two mode classes (see eg.~\cite{torres2017towards,torres2018towards}). Since the best-fit modes occur near the boundary between these two behaviors in Fig.~\eqref{fig:500Hz_modematching} (Right), we are unable to identify the modes as $p$- or $g$-modes on this basis.

\FloatBarrier

The best-matching $l=2$ perturbative mode at $40$ ms has a Cowling frequency of $\sim 280$ Hz, roughly $58\%$ lower than that observed in the simulation. We already showed in Sec.~\eqref{sec:FLASHTOV_test} that our \texttt{FLASH} implementation overestimates the frequencies of all modes with respect to the Cowling approximation (except the $F$-mode). Furthermore, the Cowling approximation itself tends to overestimate the true frequency of modes. Thus, \emph{the Cowling value of} $280$ Hz \emph{should be taken to be an upper bound on the true frequency of this mode}. As discussed in Sec.~\eqref{sec:FLASHTOV_test}, the lapse reaches a minimum value at the origin of $\sim 0.8$, so its absence in the hydrodynamic equations does not completely account for the discrepancy. This suggests the effective gravitational potential accounts for a large fraction of it.

\subsubsection{Neutrino emission for $\Omega = 0.0$ rad/s model}

\begin{figure}[htbp]
\centering
\includegraphics[width=1\textwidth]{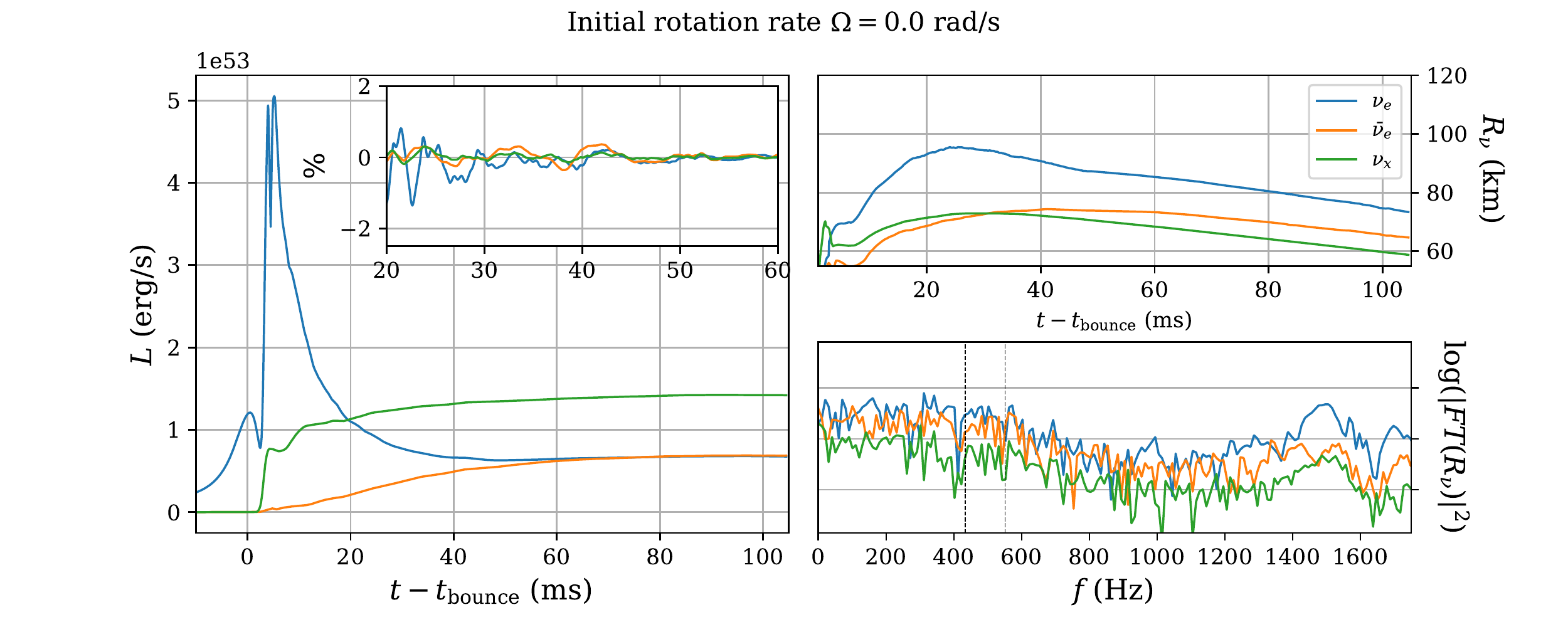}
\caption{(Left): Sky-averaged neutrino lightcurves for the species $\nu_e$ (blue), $\bar{\nu}_e$ (orange), $\nu_x$ (green) extracted at $500$ km. (Left, Inset): The same lightcurves with a Gaussian smoothing subtracted, and normalized by their respective neutrino luminosities at $40$ ms. The amplitude therefore represents a percentage fluctuation on the ambient luminosity. (Right, Top): Spherically-averaged neutrinosphere radii $R_\nu$ for each species. (Right, Bottom): Fourier spectra of $R_\nu$ with a Gaussian smoothing subtracted and windowed with another Gaussian centred at $60$ ms with a width $\sigma = 20$ ms. Vertical dashed black lines are the upper and lower limits of the $\sim 500$ Hz band mask. No prominent peak in this frequency range is apparent in this non-rotating model.} \label{fig:rot00_l2mode_Lnu_Rnu}
\end{figure}

In Fig.~\eqref{fig:rot00_l2mode_Lnu_Rnu} we display the sky-averaged neutrino lightcurves $L_\nu$ for the $\Omega = 0.0$ rad/s model (Left) along with the spherically-averaged neutrinosphere radii $R_\nu$ (Right, Top) and their spectra (Right, Bottom). The spectra are computed after subtracting a smoothing of $R_\nu$, so as to accentuate any fluctuations about the mean that may be present. We observe no prominent frequency peak within the limits of the $\sim 500$ Hz band mask at $40$ ms (shown in black vertical lines).

\FloatBarrier

\subsection{$\Omega = 0.5$ rad/s model}
\begin{figure}[htbp]
\centering
\hbox{\hspace{-1.6cm}\includegraphics[width=1.2\textwidth]{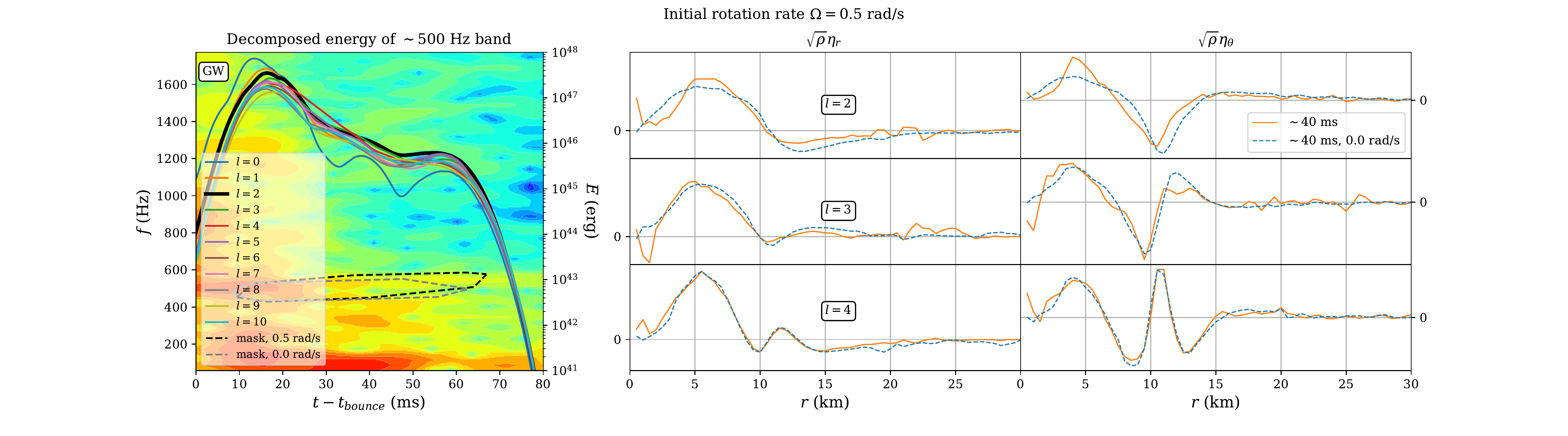}}
\caption{Similar to Fig.~\eqref{fig:500Hz_top3modes}, for the $\Omega = 0.5$ rad/s model. This time, only the velocity data snapshots are compared between the $\Omega=0.0 - 0.5$ rad/s models, since the modes are already identified and we are only trying to establish continuity along the sequence of rotating models (we also do not have a perturbative scheme for rotating models).} \label{fig:500Hz_rot05_top3modes}
\end{figure}

In the $\Omega = 0.5$ rad/s model, we have a slight change in the frequency mask as displayed in Fig.~\eqref{fig:500Hz_rot05_top3modes} (Left), but without well-separated energies between the different $l$'s. Nonetheless, comparison between the velocity data in the $\Omega = 0.5$ rad/s case and the perturbative mode functions from the $\Omega = 0.0$ rad/s case at $40$ ms still display good agreement. 

In more detail: we begin with the mode function snapshots at $\sim 40$ ms from the $\Omega = 0.0$ rad/s model displayed in Fig.~\eqref{fig:500Hz_top3modes} (Centre \& Right). For each band mask defined on the $\Omega = 0.5$ rad/s model, we do a search for the best-fit mode function snapshots for each of the $l=2,3,4$ modes separately, over a time interval of width $12$ ms centred on $40$ ms. This part of the search is required since there is no reason to expect the phase of each mode to remain the same across different rotation cases. Once the best-fit snapshots are found within each band mask, the best-fit \emph{across the band masks} is computed, also using $\Delta$ as described around Eq.~\eqref{eq:Delta}. In this way, we obtain the band mask that is most likely to contain the mode, based on continuity of its mode function. We stress that this is done separately for each mode, since different mode frequencies will depend on $\Omega$ differently and so there is no reason to expect them all to be contained in the same band mask as we move along the sequence of rotation models.

\begin{figure}[htbp]
\centering
\includegraphics[width=0.5\textwidth]{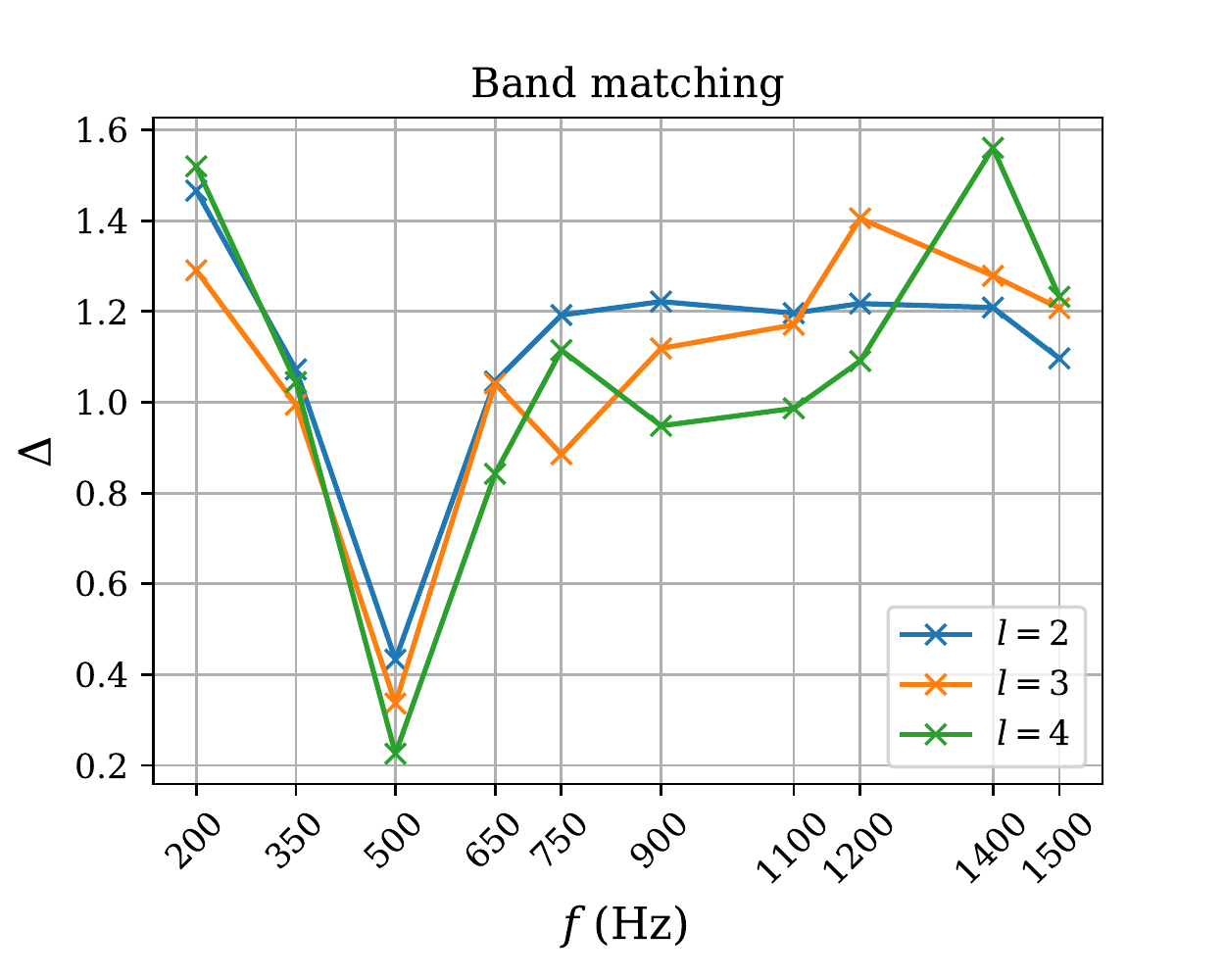}
\caption{The degree of fit between the $\sim 40$ ms snapshots from the $\Omega = 0.0$ rad/s model and the best-fit snapshots from each band mask defined in the $\Omega=0.5$ rad/s model, using $\Delta$ defined in the text. Minima of $\Delta$ over the band masks occur at the $\sim 500$ Hz mask, which means each of the $l=2,3,4$ modes happen to have landed in the same band mask around $\sim 500$ Hz in the $\Omega = 0.5$ rad/s model.} \label{fig:rot00_to_05_l2mode_modematching}
\end{figure}
\FloatBarrier

We display the best-fit values over the aforementioned time window for each band mask in the $\Omega = 0.5$ rad/s model in Fig.~\eqref{fig:rot00_to_05_l2mode_modematching}. Each tick label on the horizontal axis represents a band mask, and we see that $\Delta$ is minimized distinctly for each of the $l=2,3,4$ modes at the $\sim 500$ Hz band mask. This means that each of those modes has happened to land in that mask for the $\Omega = 0.5$ rad/s model, although this was not guaranteed.

\subsubsection{Neutrino emission for $\Omega = 0.5$ rad/s model}

The neutrino luminosity spectrograms appear to have the beginnings of a frequency band coincident with the one currently under analysis. This will be more prominent in the next rotation rate of $\Omega = 1.0$ rad/s.

\begin{figure}[htbp]
\centering
\includegraphics[width=1\textwidth]{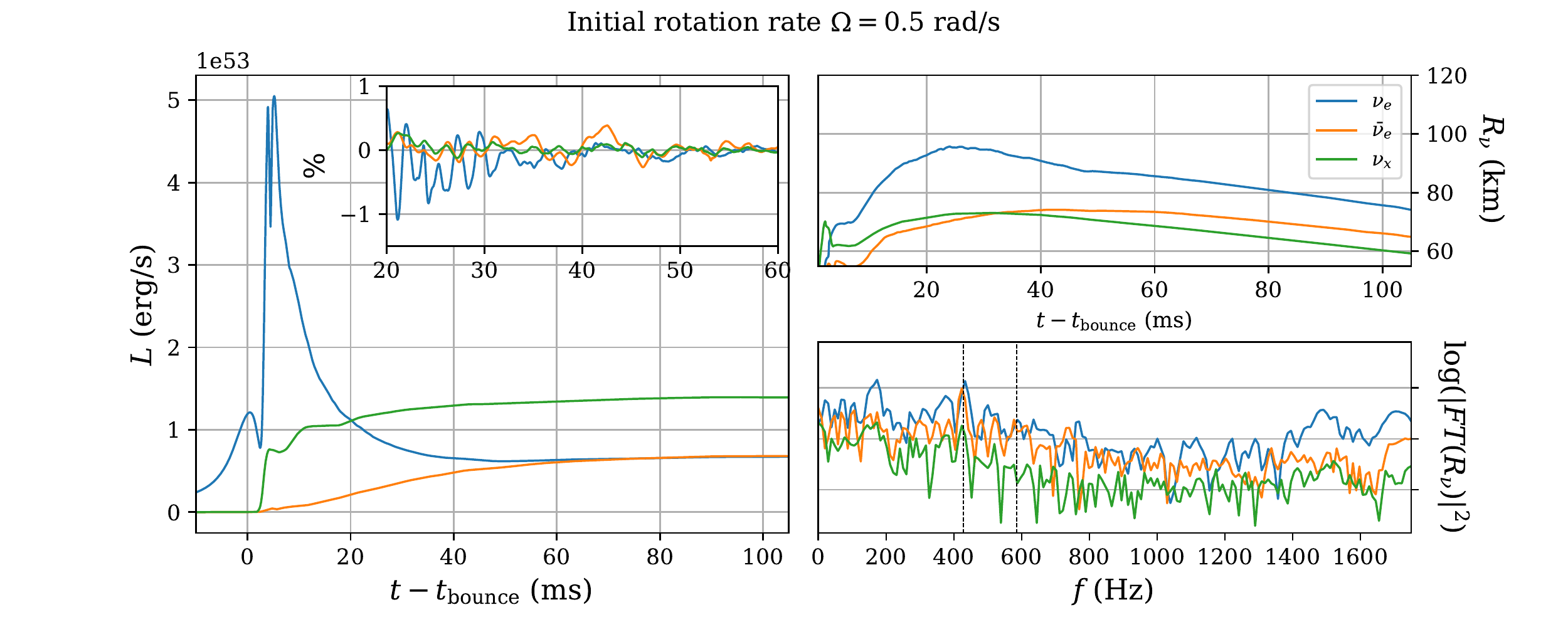}
\caption{Same as in Fig.~\eqref{fig:rot00_l2mode_Lnu_Rnu}, except for the $\Omega = 0.5$ rad/s case. (Right, Top): These neutrinosphere radii also exhibit oscillations, suggesting that the cause for the luminosity oscillations is a temporal variation in the free-streaming volume around the PNS. (Right, Bottom): A local maxima occurs within the mask, although a more prominent peak occurs on the edge of its lower frequency limit.} \label{fig:rot05_l2mode_Lnu_Rnu}
\end{figure}
\FloatBarrier

\subsection{$\Omega = 1.0$ rad/s model}

\begin{figure}[htbp]
\centering
\includegraphics[width=0.5\textwidth]{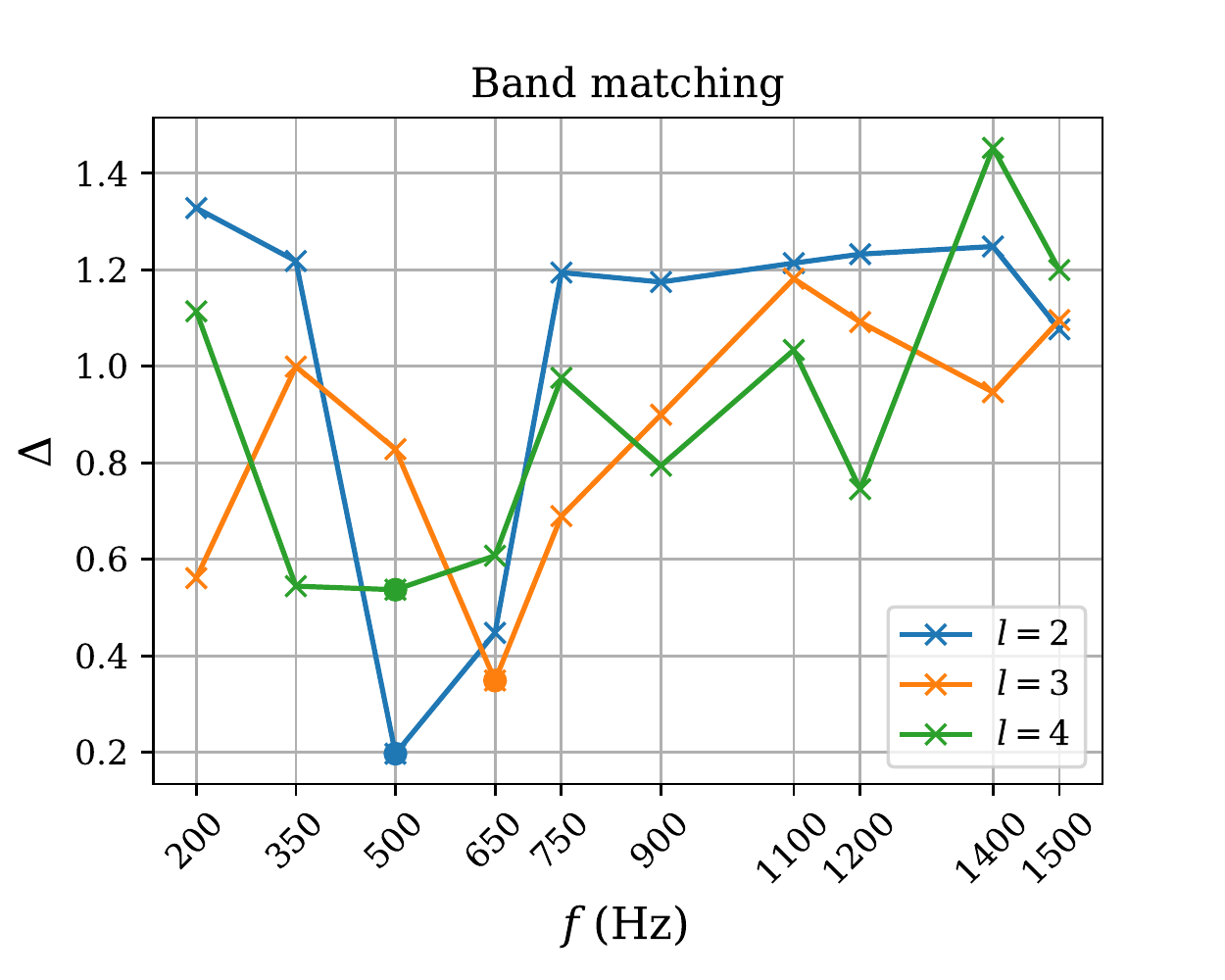}
\caption{Band matching between the $\Omega = 0.5,1.0$ rad/s models, as described in the text. The $l=2$ mode continues to be matched by the  $\sim 500$ Hz mask in the $\Omega = 1.0$ rad/s model, whereas the $l=3$ mode has moved to the $\sim 650$ Hz mask and the $l=4$ mode has similar quality fits in multiple band masks, signifiying that we have lost track of the $l=4$ mode.} \label{fig:rot05_to_10_l2mode_modematching}
\end{figure}

In Fig.~\eqref{fig:rot05_to_10_l2mode_modematching} we display the band matching for the $l=2,3,4$ modes going from the $\Omega = 0.5 \rightarrow 1.0$ rad/s models. The $l=2$ mode continues to be best-fit by the $\sim 500$ Hz band, whereas the $l=3$ mode has moved to the $\sim 650$ Hz band where it is a subdominant component. We thus stop following the $l=3$ mode. The $l=4$ mode has similar levels of fit between multiple bands, which signals that we have lost track of it. This occurs since, not only do modes deform as $\Omega$ varies, but the excitation of them by the core bounce changes as well since the bounce proceeds differently. For example, the bounce occurs at the poles before the equator, and at different radii, since there is rotational support at the equator but not at the poles. This is at odds with the work of eg.~\cite{dimmelmeier2006non} upon which our analysis is based, where the applied excitation of the star is under control and designed to excite specific modes. We thus also stop following the $l=4$ mode, and focus solely on the $l=2$ mode. 

\begin{figure}[htbp]
\centering
\hbox{\hspace{-1.6cm}\includegraphics[width=1.2\textwidth]{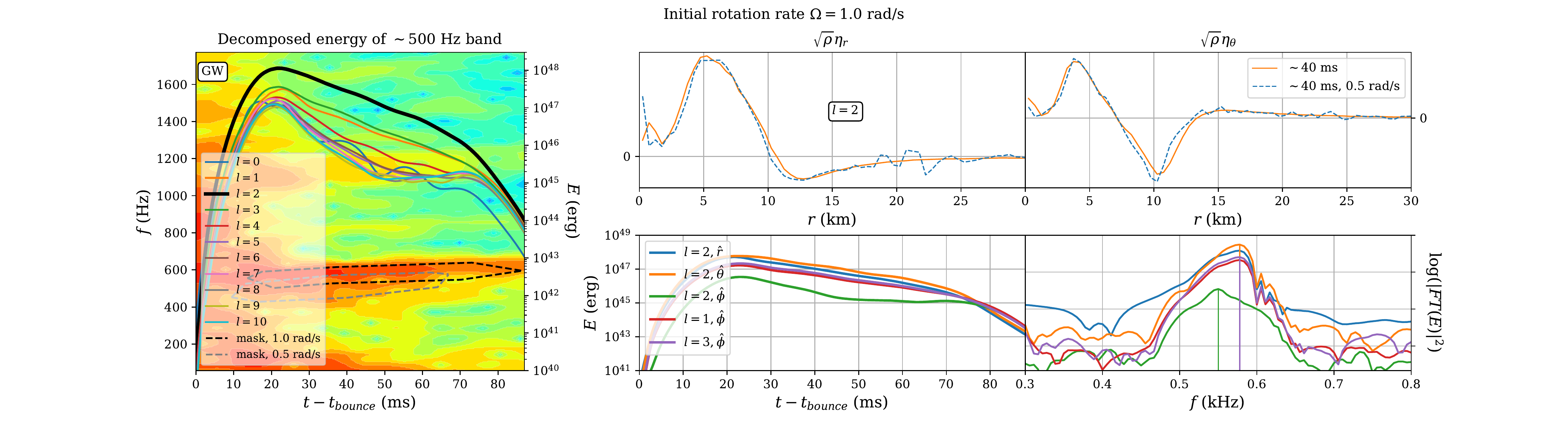}}
\caption{Similar to Fig.~\eqref{fig:500Hz_rot05_top3modes}, for the $\Omega = 1.0$ rad/s model. We focus on the $l=2$ mode here, since we lost track of the $l=4$ mode and the $l=3$ mode landed in a different band mask in which it is subdominant in energy. (Right, Bottom Row): The energy of specific components of the mode (Left), where we see the even parity pieces of $l=1,2,3$ exhibiting commensurate decay rates, to be contrasted with the odd parity $l=2$ piece ($\hat{\phi}$). We also show the frequency spectra of the same components (Right), where we observe coincident peaks for all the even parity components. We argue that the $(l=1:\hat{\phi})$ and $(l=3:\hat{\phi})$ components are deformations of the $l=2$ mode we are following (see text).} \label{fig:500Hz_rot10_top1modes}
\end{figure}
\FloatBarrier

In the $\Omega = 1.0$ rad/s model, we continue to observe convincing continuity of the $l=2$ mode function in Fig.~\eqref{fig:500Hz_rot10_top1modes} (Right, Top Row). The frequency mask has shifted more than when going from $\Omega = 0.0 \rightarrow 0.5$ rad/s, but it is still a modest shift with the masks overlapping in Fig.~\eqref{fig:500Hz_rot10_top1modes} (Left). At this rotation rate we observe a clear separation in the $l=2$ energy, $l=1,3$ energy, and the remaining poloidal numbers. We will now argue that the $l=1,3$ components are deformations of the $l=2$ mode we are following.

If the $l=1,3$ components are deformations of the $l=2$ mode, they must have the same (even) parity. Since $l=1,3$ are odd, only their $\hat{\phi}$ components have even parity. We display the energy vs time in the $\hat{\phi}$ components of $l=1,3$ in Fig.~\eqref{fig:500Hz_rot10_top1modes} (Center, Bottom), along with the $\hat{r},\hat{\theta}$ components of $l=2$ (even parity) and $\hat{\phi}$ component of $l=2$ (odd parity) for comparison. We observe that the even parity pieces have energies evolving with a unified exponential decay rate, whereas the odd parity $l=2$, $\hat{\phi}$ piece (which cannot be part of the mode) has noticeably different behaviour. Furthermore, since the band mask displayed in Fig.~\eqref{fig:500Hz_rot10_top1modes} (Left) has many frequency bins in it, one can check to see whether the even parity $l=2$ pieces ($\hat{r},\hat{\theta}$) have the same frequency of oscillation as the even parity $l=1,3$ pieces ($\hat{\phi}$). This is confirmed, and displayed in Fig.~\eqref{fig:500Hz_rot10_top1modes} (Right, Bottom); we plot the frequency spectrum of the same energies, which without the smoothing performed in Fig.~\eqref{fig:500Hz_rot10_top1modes} (Center, Bottom) exhibit oscillations at twice the frequency of the underlying motion (and we compensate for this by rescaling the horizontal axis). We observe that the even parity parts of the $l=1,2,3$ components have peaks occurring at the same frequency (indicated with color-matching vertical lines). The odd parity $l=2$, $\hat{\phi}$ component, however, peaks $\sim 20$ Hz away.

Lastly, axial $l=1,3$ deformations ($\hat{\phi}$) are the type of deformations of a polar $l=2$ ($\hat{r},\hat{\theta}$) mode we would expect at first order in the rotation rate $\Omega$~\cite{friedman2013rotating}, as we described in Sec.~\eqref{sec:backgroundterminology}. Rotational orders occur in powers of the ratio $\Omega/\Omega_K$, where $\Omega_K$ is the Keplerian orbital frequency $\sim \sqrt{GM/r^3}$. For the current model at $40$ ms, this ratio is $\sim 0.18$ when averaged over the inner $30$ km, where the bulk of the mode energy resides (see Fig.~\eqref{fig:500Hz_rot10_top1modes}). The energy ratio between the $(l=1,3: \hat{\phi})$ and $(l=2: \hat{r},\hat{\theta})$ parts of the mode at $40$ ms averages to a commensurate value, $\sim 0.19$. These observations taken together tell us that we have correctly followed the $l=2$ mode from $\Omega = 0.5-1.0$ rad/s and that it has picked up axial $l=1,3$ deformations, and when following the mode further along our sequence of rotating models we will do the best-fit search according to snapshots of all the $(l=2:\hat{r},\hat{\theta})$ and $(l=1,3:\hat{\phi})$ components of the mode function.

The frequency mask is coincident with the peak gravitational wave emission, as seen in Fig.~\eqref{fig:500Hz_rot10_top1modes} (Left). It is also coincident with prominent frequency bands in the neutrino luminosity spectrograms, as we will describe next. 

\subsubsection{Neutrino emission for $\Omega = 1.0$ rad/s model}

\begin{figure}[htbp]
\centering
\includegraphics[width=1\textwidth]{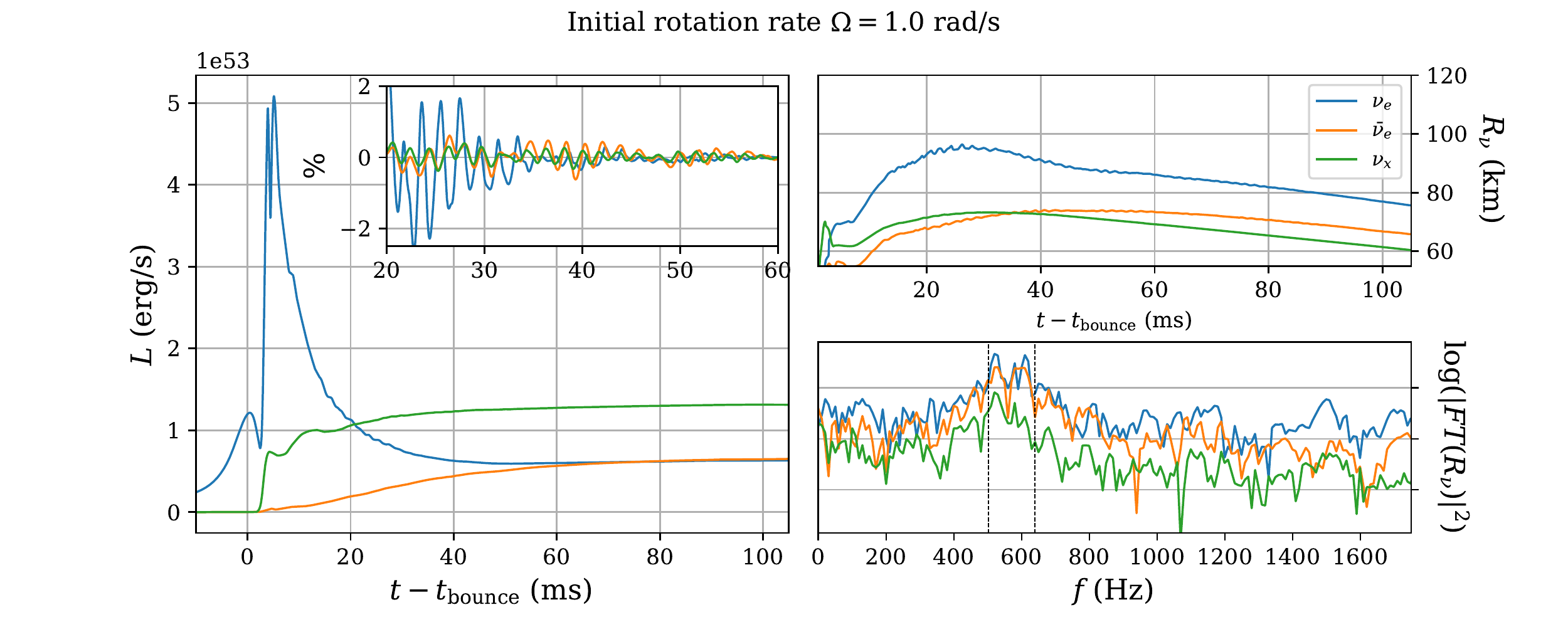}
\caption{Same as in Fig.~\eqref{fig:rot05_l2mode_Lnu_Rnu}, but for the rotation model $\Omega = 1.0$ rad/s. Vertical dashed black lines in (Right, Bottom) are the upper and lower limits of the $\sim 500$ Hz band mask at $40$ ms. The prominent peak amplitude occurs within the mask, confirming spectral coincidence with the mode of oscillation under consideration in this section. A second prominent peak continues to occur just within the lower frequency of the mask.} \label{fig:rot10_l2mode_Lnu_Rnu}
\end{figure}

In Fig.~\eqref{fig:rot10_l2mode_Lnu_Rnu} (Left) we plot the sky-averaged neutrino luminosities for electron neutrinos $\nu_e$ (blue), electron anti-neutrinos $\bar{\nu}_e$ (orange), and the remaining flavours muon and tau are combined and denoted as $\nu_x$ (green). The muon and tau flavours are grouped together due to their similar behaviour at the typical energies of a CCSN system. Namely, their scattering cross-section with material in the system is significantly lower than for electron-flavour neutrinos since the muon and tau particles are too heavy to be produced in large numbers with the energy available.

The neutrino luminosities exhibit oscillatory behaviour, as shown in the inset of Fig.~\eqref{fig:rot10_l2mode_Lnu_Rnu} (Left). There we have subtracted a Gaussian smoothing and normalized by the luminosity values at $40$ ms. The inset therefore displays the oscillations as a percentage of the luminosities. In Fig.~\eqref{fig:rot10_l2mode_Lnu_Rnu} (Right, Top) we plot the spherically-averaged neutrinosphere radii vs time, defined as the radius at which the optical depth to infinity is $2/3$. Roughly speaking, the stellar material is considered transparent to neutrinos outside of the neutrinosphere, although this is a heuristic notion because the transparency varies continuously and the neutrinosphere radius in fact depends on the energy of the neutrinos. Nonetheless, the neutrinosphere radius can be considered as tracking the the region producing free-streaming neutrinos. We see that the neutrinosphere radii also exhibit oscillations, and in Fig.~\eqref{fig:rot10_l2mode_Lnu_Rnu} (Right, Bottom) we display Fourier spectra of them (with a Gaussian smoothing subtracted). Spectral peaks are seen within the range of frequencies covered by our $\sim 500$ Hz band mask, indicated in black dashed vertical lines. The coincidence in frequency between the neutrino luminosity oscillations, the neutrinosphere radii oscillations, and the mode under consideration suggest a cause for the former. Namely, that the $l=2$, $n\gtrsim2$ mode of the system is causing oscillations in the free-streaming neutrino emitting volume, which is impacting the neutrino lightcurves for an observer far away. If this is true, one expects an $l=2$ pattern of oscillatory emission on the sky, which we will show next.

\FloatBarrier

\begin{figure}[htbp]
\centering
\hbox{\hspace{-1.3cm}\includegraphics[width=1.15\textwidth]{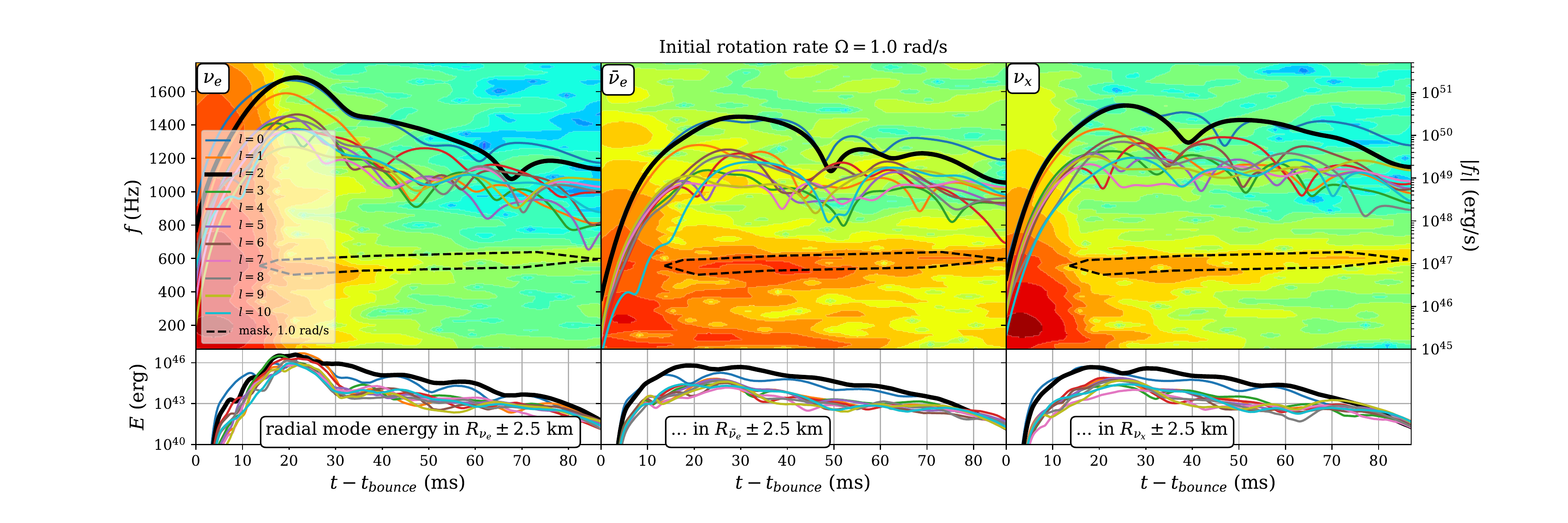}}
\caption{(Top Row): Spherical harmonic decompositions of the neutrino luminosities on a sphere at $500$ km for species $\nu_e$ (Left), $\bar{\nu}_e$ (Centre), $\nu_x$ (Right) plotted on top of spectrograms of the sky-averaged neutrino lightcurves. The current band mask is again displayed, and is coincident with a prominent emission feature in the spectrograms. The neutrino lightcurves along each direction on the sky have had a Gaussian smoothing subtracted, and underwent the same spectrogram filtering as we applied to the velocity field using the band mask shown. The resulting time series were then decomposed angularly to obtain spherical harmonic coefficients $f_l$ at each time. The absolute value $|f_l|$ is then smoothed with a Gaussian of width $10$ ms before plotting. (Bottom Row): The corresponding energy in the $l=2$ mode of the PNS, integrated over a $5$ km width radial shell centred on the respective neutrinospheres.} \label{fig:rot10_l2mode_Lnusky}
\end{figure}

In Fig.~\eqref{fig:rot10_l2mode_Lnusky} we display spectrograms of the neutrino luminosities with a smoothing subtracted (Top Row) for $\nu_e$ (Left), $\bar{\nu}_e$ (Centre), and $\nu_x$ (Right), with the band mask of the $l=2$ mode overlaid. Also overlaid on the spectrograms are the spherical harmonic coefficients in absolute value $|f_l|$ for the neutrino luminosity pattern on the sky, Gaussian-smoothed in time with a $10$ ms window to eliminate oscillations. We observe that the emission pattern indeed has a strong $l=2$ component, along with a competing $l=0$ part. For comparison, we plot the energy in the radial part of the $l=2$ mode integrated over a $5$ km thick shell centred on the neutrinospheres (Bottom Row). We plot only the radial part since that is what would produce changes in the volume of the neutrino-emitting region. We see a similar level of excitation between the $l=2$ and $l=0$ components as in the neutrino emission pattern. This supports the notion that the $l=2$ mode we are following is producing an oscillatory variation in the neutrino-emitting volume, which then imprints on the neutrino lightcurves. We will also see that the $l=2$ mode in the next model in our sequence ($\Omega = 2.0$ rad/s) has an $l=0$ deformation clearly identifiable at $\mathcal{O}(\Omega^2)$, which may explain the prevalence of the $l=0$ part in Fig.~\eqref{fig:rot10_l2mode_Lnusky} in the outer $\gtrsim 60$ km of the PNS where the neutrinosphere surfaces reside.

We have therefore implicated this $l=2,n\gtrsim 2$ mode as imprinting itself upon both the gravitational waves and the neutrino luminosities, producing peak oscillatory emission in both cases. The neutrinos carry information about the outer tails of the mode where $r\gtrsim 60$ km, whereas the gravitational waves probe the inner core.

\FloatBarrier

\subsection{$\Omega = 2.0$ rad/s model}

\begin{figure}[htbp]
\centering
\includegraphics[width=0.5\textwidth]{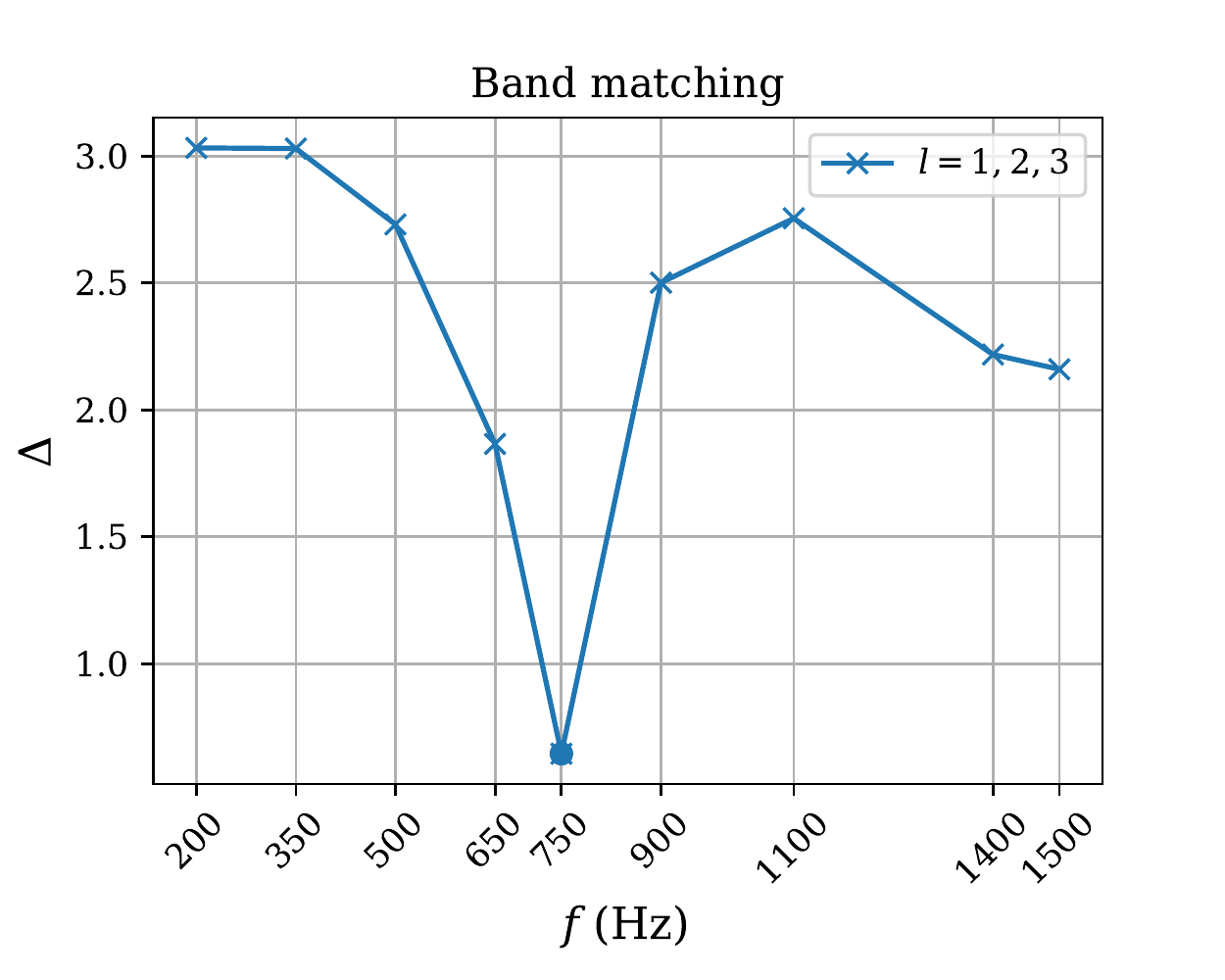}
\caption{Band matching between the $\Omega = 1.0,2.0$ rad/s models, as described in the text. The $l=2$ mode has migrated to the $\sim 750$ Hz mask in the $\Omega = 2.0$ rad/s model.} \label{fig:rot10_to_20_l2mode_modematching}
\end{figure}

In Fig.~\eqref{fig:rot10_to_20_l2mode_modematching} we display the band matching from $\Omega = 1.0 \rightarrow 2.0$ rad/s models for the combination of the $(l=2:\hat{r},\hat{\theta})$ and $(l=1,3:\hat{\phi})$ parts, which to first-order in $\Omega$ constitute the deformed $l=2$ mode that we are following. The mode has migrated from the $\sim 500$ Hz band to the $\sim 750$ Hz band, where $\Delta$ is minimized distinctly.

\begin{figure}[htbp]
\centering
\hbox{\hspace{-1.6cm}\includegraphics[width=1.2\textwidth]{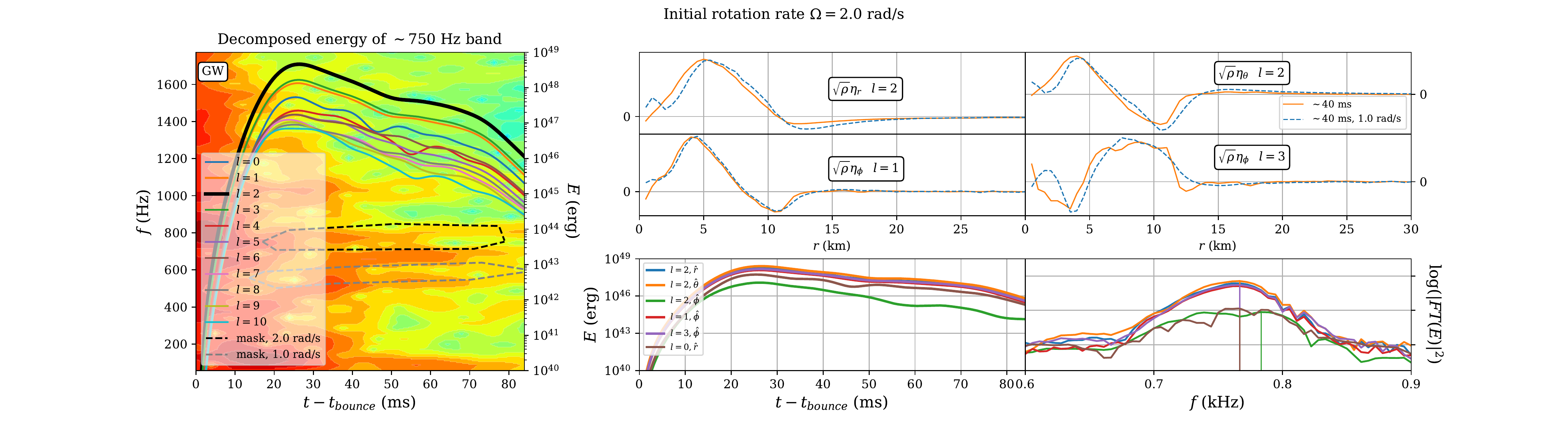}}
\caption{Same as Fig.~\eqref{fig:500Hz_rot10_top1modes}, for $\Omega = 2.0$ rad/s. We continue to see the polar $l=2$ mode with axial $l=1,3$ deformations at $\mathcal{O}(\Omega)$. An $l=0$ correction at $\mathcal{O}(\Omega^2)$ is also emerging, and exhibits a commensurate decay rate (Centre, Bottom) and peak frequency (Right, Bottom).} \label{fig:750Hz_rot20_l2mode}
\end{figure}
\FloatBarrier

In Fig.~\eqref{fig:750Hz_rot20_l2mode} (Left) we continue to see a similar distribution of energy as in the last rotation model, with $l=2$ dominant with $l=1,3$ deformations, and all other poloidal numbers subdominant. The current $\sim 750$ Hz mask continues to be coincident with an emission feature in the gravitational wave spectrogram, but the most dominant emission seems to occur at frequencies just below the mask. In Fig.~\eqref{fig:750Hz_rot20_l2mode} (Centre, Bottom) we again see the energies of the components of this deformed $l=2$ mode evolving in unison, as compared with the $(l=2:\hat{\phi})$ energy which is forbidden to be part of this mode by parity selection rules. As well we see again coincident frequency peaks in Fig.~\eqref{fig:750Hz_rot20_l2mode} (Right, Bottom), quite distinct from the $(l=2:\hat{\phi})$ spectrum. The average ratio of the $(l=1,3: \hat{\phi})$ and $(l=2, \hat{r},\hat{\theta})$ energies at $40$ ms has grown to $\sim 0.33$, still commensurate with $\Omega/\Omega_K \sim 0.37$.

We even see evidence that an $l=0$ deformation has entered at $\mathcal{O} (\Omega^2)$. The $l=0$ peak frequency in Fig.~\eqref{fig:750Hz_rot20_l2mode} (Right, Bottom) is aligned with the rest of the components of the mode, and its energy evolution in Fig.~\eqref{fig:750Hz_rot20_l2mode} (Center, Bottom) is also commensurate. Furthermore, its energy at $40$ ms is $\sim 0.13$ of the energy in $(l=2: \hat{r},\hat{\theta})$, which compares well with the expected order $(\Omega/\Omega_K)^2 \sim 0.14$. Polar deformations such as this are expected at second order in rotation for the polar $l=2$ axisymmetric mode we are following~\cite{friedman2013rotating}. The remaining poloidal numbers have energies about one order less than the $(l=2: \hat{r},\hat{\theta})$ energy at $\sim 0.05$ relative strength.

\subsubsection{Neutrino emission for $\Omega = 2.0$ rad/s model}

\begin{figure}[htbp]
\centering
\includegraphics[width=1\textwidth]{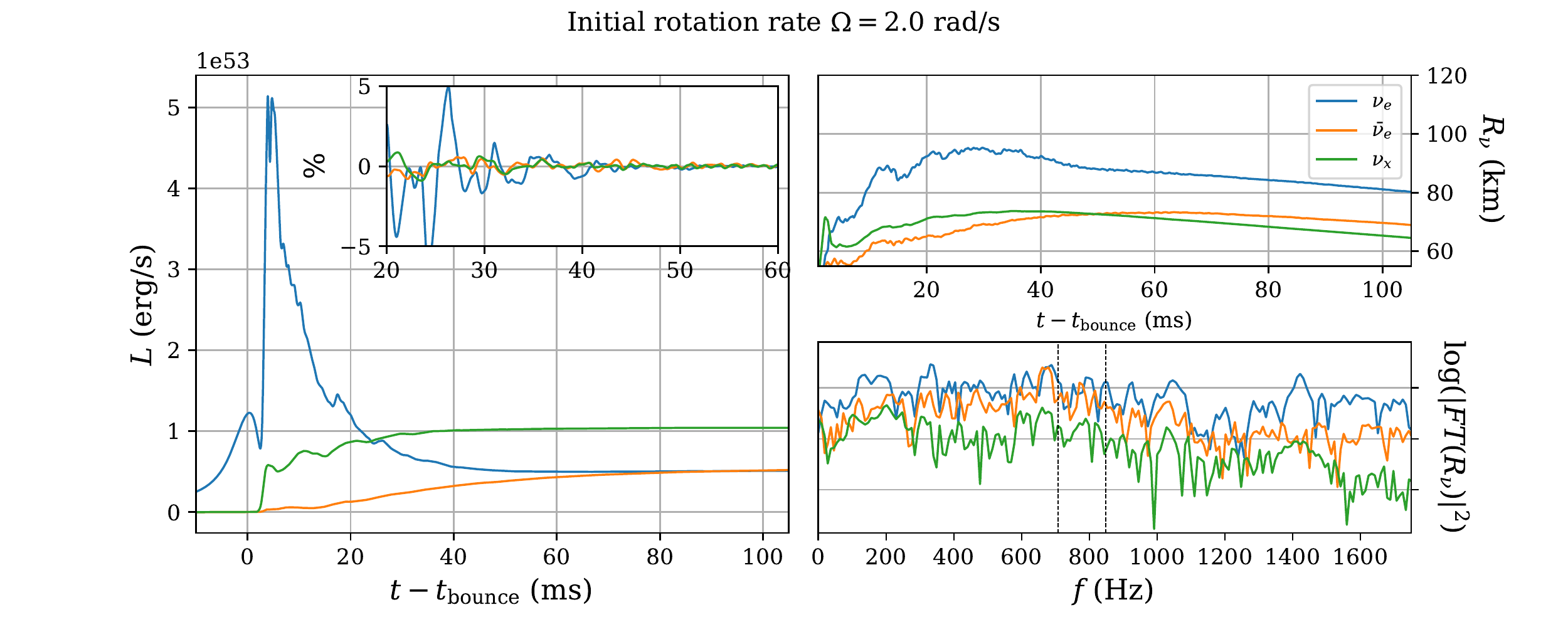}
\caption{Same as in Fig.~\eqref{fig:rot05_l2mode_Lnu_Rnu}, but for the rotation model $\Omega = 2.0$ rad/s. Vertical dashed black lines in (Right, Bottom) are the upper and lower limits of the $\sim 750$ Hz band mask at $40$ ms. Peak amplitudes are less distinguished in general, but one occurs just out of the mask towards lower frequencies.} \label{fig:rot20_l2mode_Lnu_Rnu}
\end{figure}

While the neutrino emission continues to exhibit oscillations in the $\Omega = 2.0$ rad/s model, it appears whiter in frequency than the $\Omega = 1.0$ rad/s model. This can be seen in the time domain in Fig.~\eqref{fig:rot20_l2mode_Lnu_Rnu} (Left, Inset), and is also reflected in the frequency spectra of the neutrinosphere radii in (Right, Bottom). The $\sim 750$ Hz mask frequency bounds are again displayed in vertical black lines, where no remarkable peaks are apparent. Thus this mode does not appear to be coincident with a prominent modulation in the neutrino luminosities for this model.

\FloatBarrier

\begin{figure}[htbp]
\centering
\hbox{\hspace{-1.3cm}\includegraphics[width=1.15\textwidth]{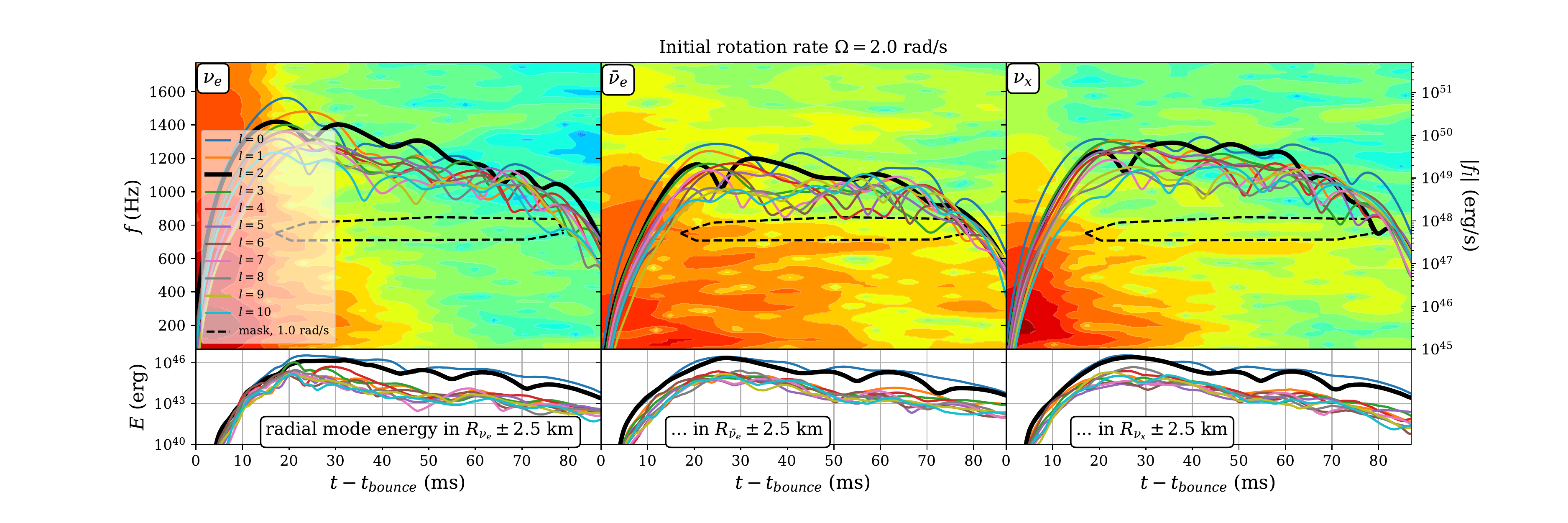}}
\caption{Same as in Fig.~\eqref{fig:rot10_l2mode_Lnusky} but for the rotation model $\Omega = 2.0$ rad/s. (Top Row): The $l=2$ and $l=0$ parts of the neutrino emission pattern on the sky enjoy less dominance over the other $l$ than in the $\Omega = 1.0$ rad/s model, but that systematic trend is still evident. The radial mode energy around the neutrinosphere continues to have dominant $l=2,0$ components. The frequency mask clearly misses an emission feature in the spectogram of $L_{\nu_x}$, just below the mask (Top, Right).} \label{fig:rot20_l2mode_Lnusky}
\end{figure}

This is further demonstrated in Fig.~\eqref{fig:rot20_l2mode_Lnusky}, where the band mask does not surround any dominant emission, with the possible exception of $\bar{\nu}_e$. Nonetheless, the emission that does exist within the band mask still has a slight dominance in $l=2$ and $l=0$ (Top Row), in concert with the radial mode energy around the neutrinospheres (Bottom Row).

\FloatBarrier

\subsection{$\Omega = 2.5$ rad/s model}

\begin{figure}[htbp]
\centering
\includegraphics[width=0.5\textwidth]{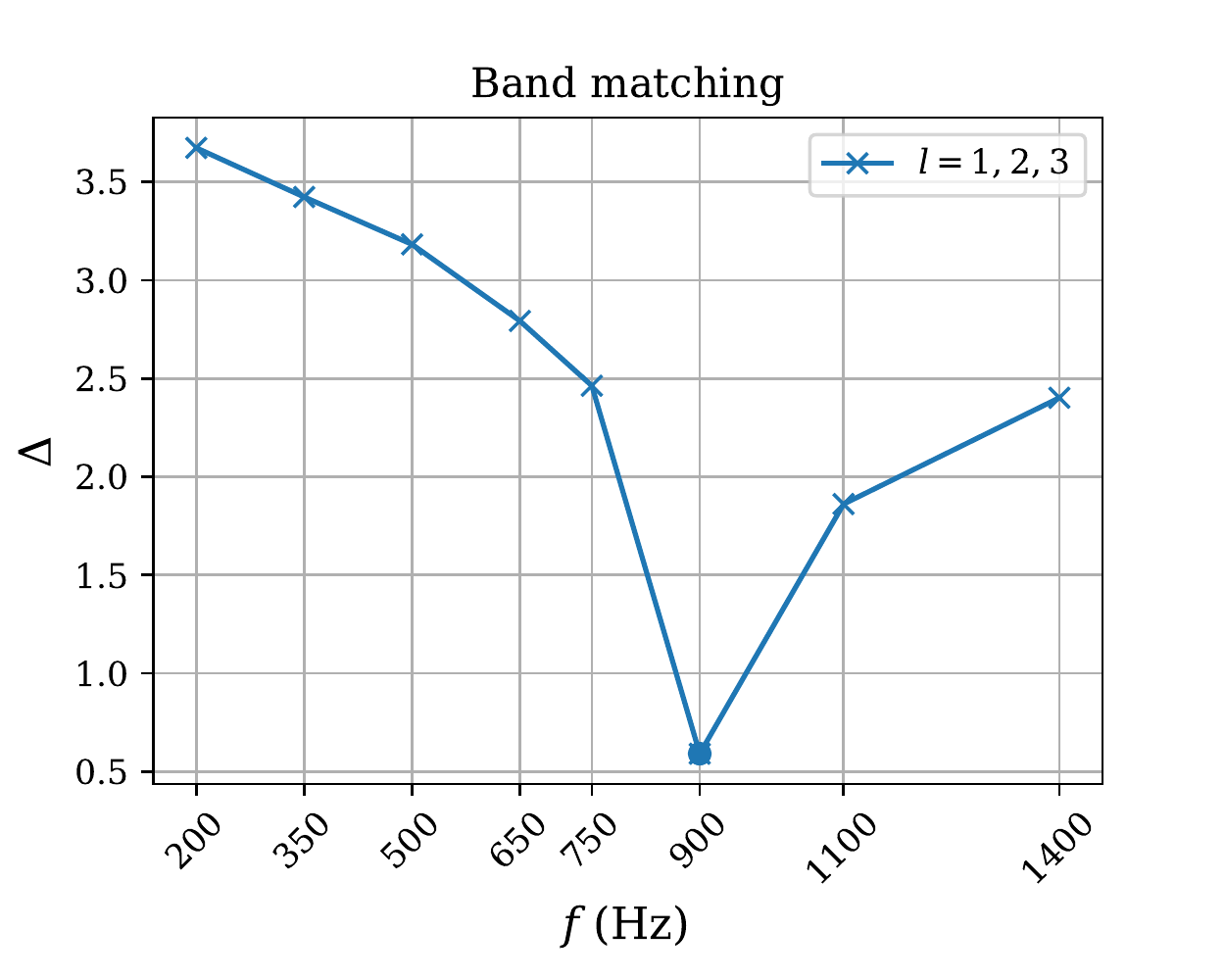}
\caption{Band matching between the $\Omega = 2.0,2.5$ rad/s models, as described in the text. The $l=2$ mode has migrated further still to the $\sim 900$ Hz mask in the $\Omega = 2.5$ rad/s model.} \label{fig:rot20_to_25_l2mode_modematching}
\end{figure}

\begin{figure}[htbp]
\centering
\hbox{\hspace{-1.6cm}\includegraphics[width=1.2\textwidth]{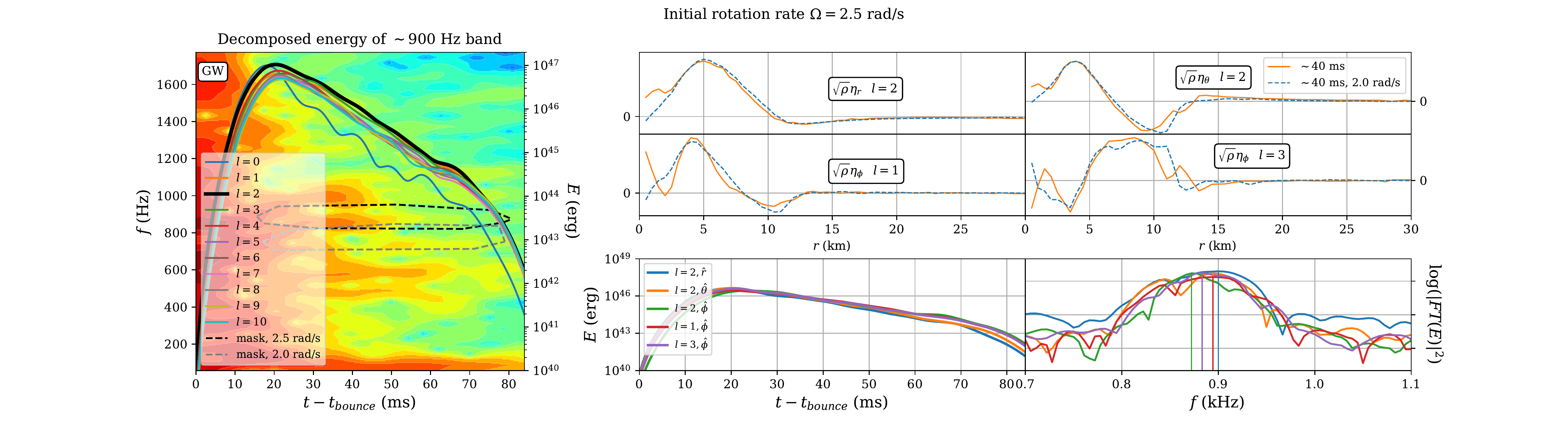}}
\caption{Same as in Fig.~\eqref{fig:750Hz_rot20_l2mode}, for the $\Omega = 2.5$ rad/s model. Here we see strong energy mixing in $l$ (Left). Peak frequencies are beginning to split apart (Right, Bottom), signalling that we are beginning to lose track of this mode.} \label{fig:900Hz_rot25_l2mode}
\end{figure}

In Fig.~\eqref{fig:rot20_to_25_l2mode_modematching} we continue to see this deformed $l=2$ mode migrate to higher frequencies, now falling within the $\sim 900$ Hz band for our most rapidly spinning $\Omega = 2.5$ rad/s model. However, as shown in Fig.~\eqref{fig:900Hz_rot25_l2mode} (Left), the energies in many different $l$ are becoming comparable. The ratio $\Omega/\Omega_K$ averaged over the inner $30$ km for this model is $\sim 0.42$. We see in Fig.~\eqref{fig:900Hz_rot25_l2mode} (Centre, Bottom) that even the $(l=2:\hat{\phi})$ energy is evolving in unison with the mode we are following (although it cannot be part of the mode, by parity), and in Fig.~\eqref{fig:900Hz_rot25_l2mode} (Right, Bottom) we see the frequency peaks beginning to split apart. 

Both of these facts are signalling that we are beginning to lose track of this mode, as other distinct oscillations are apparently excited as well. We interpret the unified evolution of the energy as arising from a common source, the strong rotational kinematics of the star. In order to separate the many modes that are present, one would need to do a perturbative calculation on this rapidly rotating background to obtain the mode functions, and do a large search for a best-fit combination of modes and amplitudes. This would be quite challenging, and is beyond our scope here.

\FloatBarrier

\subsubsection{Neutrino emission for $\Omega = 2.5$ rad/s model}

\begin{figure}[htbp]
\centering
\includegraphics[width=1\textwidth]{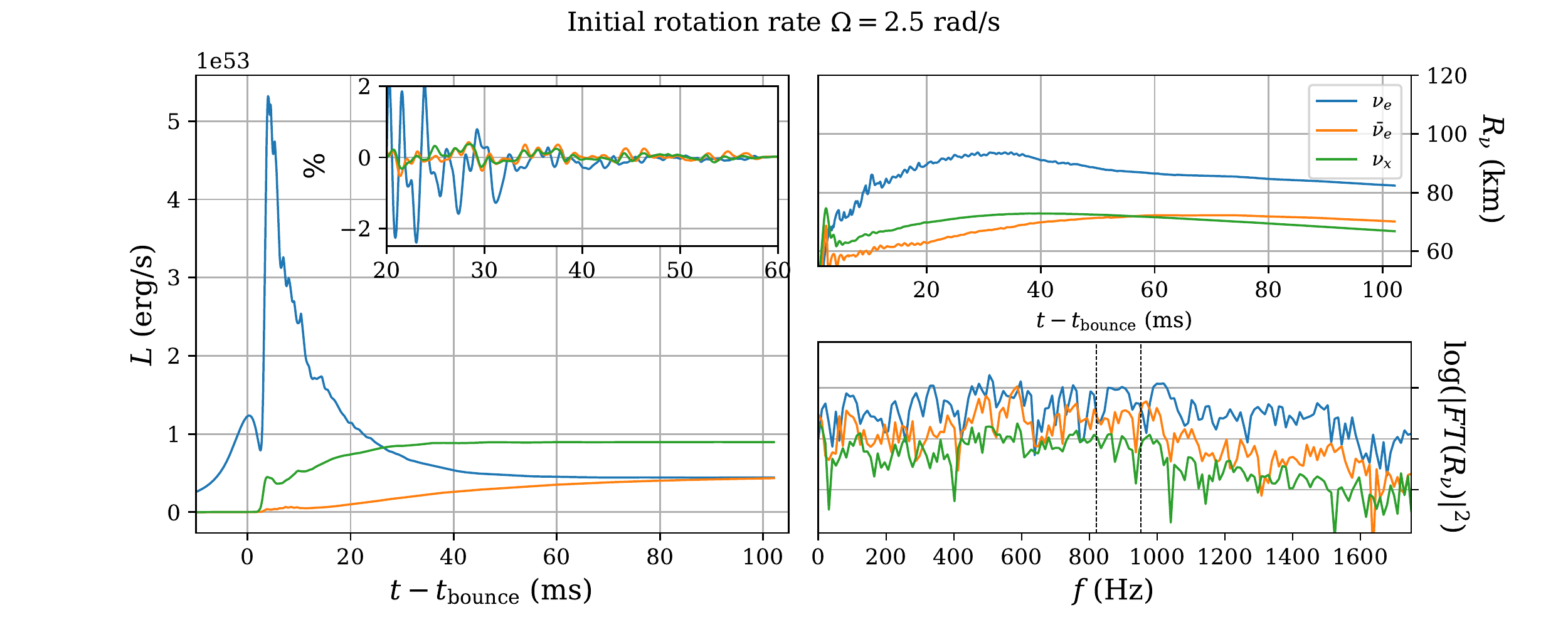}
\caption{Same as in Fig.~\eqref{fig:rot10_l2mode_Lnu_Rnu}, but for the rotation model $\Omega = 2.5$ rad/s. Vertical dashed black lines in (Right, Bottom) are the upper and lower limits of the $\sim 900$ Hz band mask. Peak amplitudes are even less distinguished than the $\Omega = 1.0$ rad/s model.} \label{fig:rot25_l2mode_Lnu_Rnu}
\end{figure}

The neutrino emission continues to have modulations with an incoherent appearance in time, and there are few if any distinguished peaks in the spectra of the neutrinosphere radii (see Fig.~\eqref{fig:rot25_l2mode_Lnu_Rnu}.

\begin{figure}[htbp]
\centering
\hbox{\hspace{-1.3cm}\includegraphics[width=1.15\textwidth]{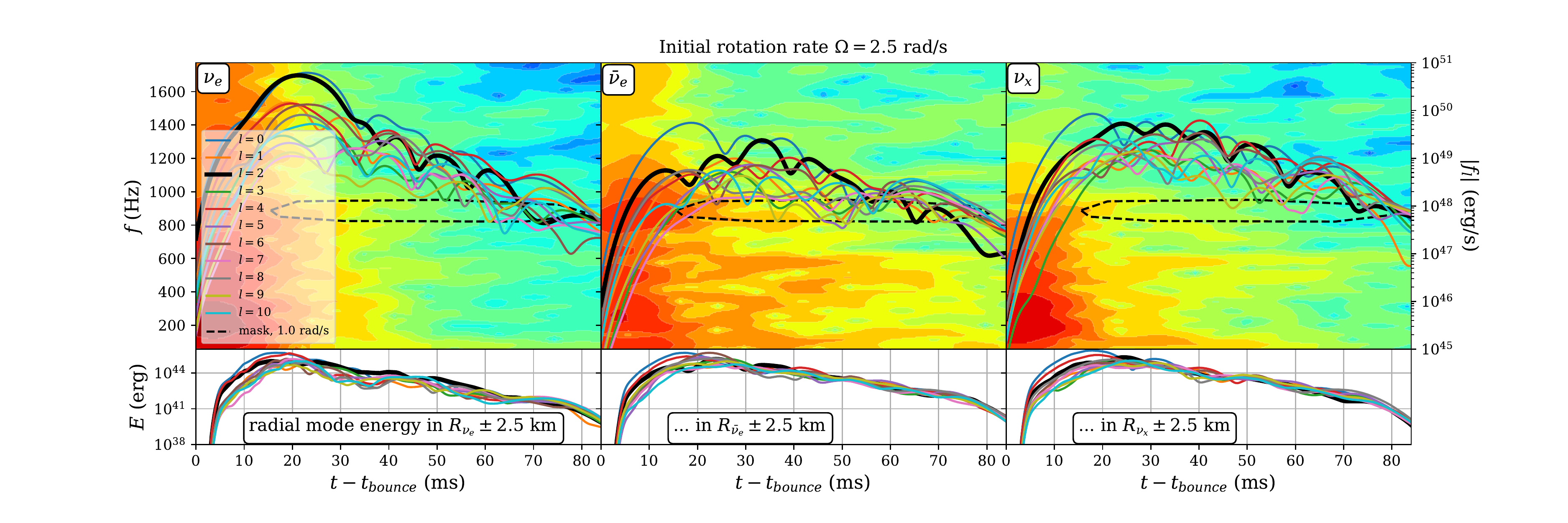}}
\caption{Same as in Fig.~\eqref{fig:rot10_l2mode_Lnusky} but for the rotation model $\Omega = 2.5$ rad/s. (Top Row): The neutrino emission pattern on the sky is more mixed in $l$ than in the $\Omega = 2.0$ rad/s model, but the band mask does not enclose any particularly prominent emission features in the neutrino luminosity spectrograms, except partially for $\bar{\nu}_e$. The radial mode energies around the neutrinosphere are very well-mixed in $l$, much like the total energy in Fig.~\eqref{fig:900Hz_rot25_l2mode}. Thus no connection between the mode and the neutrino emission can credibly be claimed in this model.} \label{fig:rot25_l2mode_Lnusky}
\end{figure}

In Fig.~\eqref{fig:rot25_l2mode_Lnusky} we see the $\sim 900$ Hz band mask for this model not enclosing anything remarkable in the neutrino spectrograms, with a partial exception for $\bar{\nu}_e$ (Top Row). The emission pattern on the sky is well-mixed in $l$ (Top Row), and the radial mode energy around the neutrinosphere even more so (Bottom Row). Thus we cannot credibly claim any connection between the mode and the neutrino emission in this rapidly-rotating model.

\FloatBarrier

\subsection{$l=2, n\gtrsim2$ mode frequency for all models}

\begin{figure}[htbp]
\centering
\hspace{0cm}\includegraphics[width=0.5\textwidth]{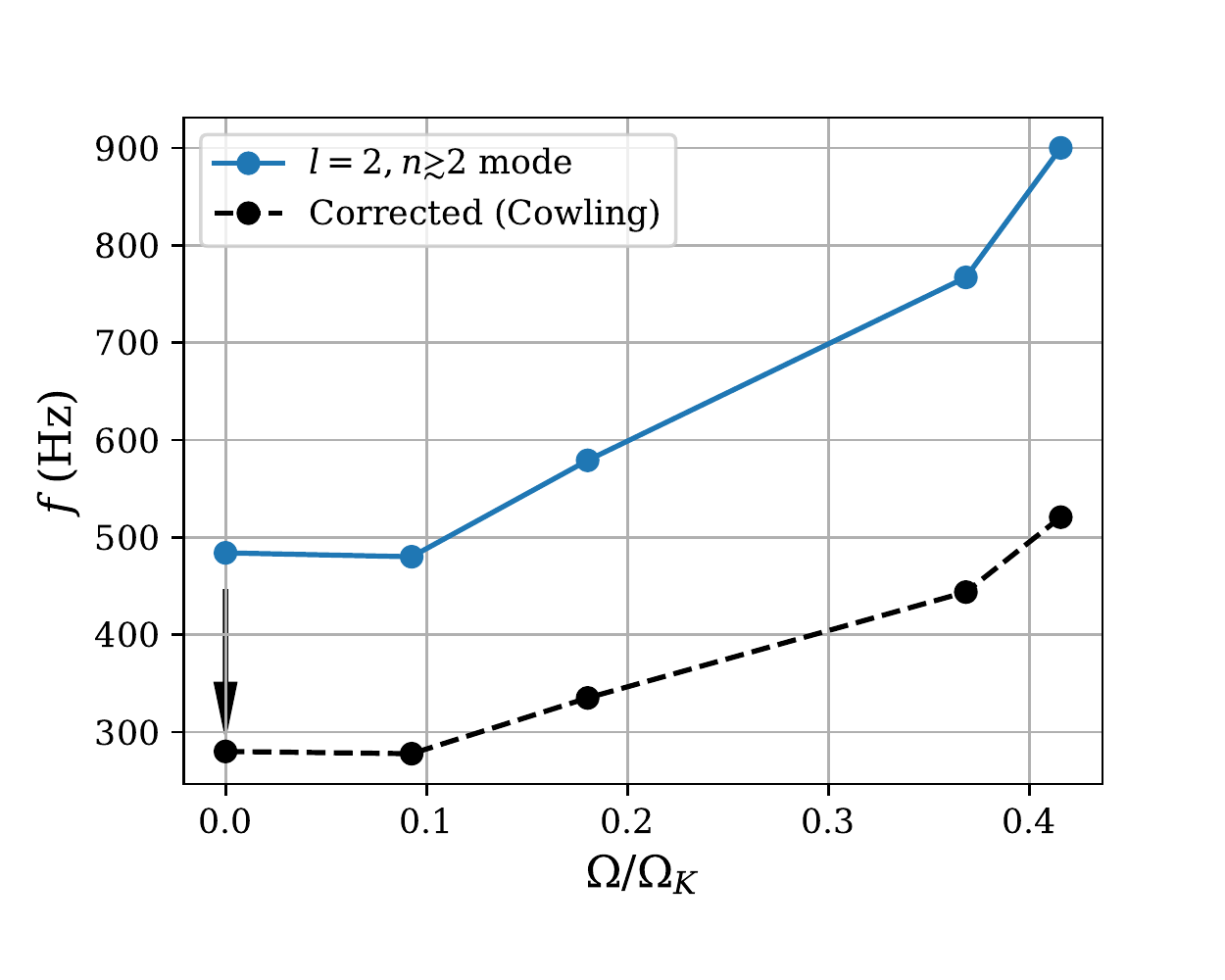}
\caption{The frequency of the $l=2,n\gtrsim 2$ mode across the entire sequence of rotating models. $\Omega/\Omega_K$ is computed at $40$ ms and averaged over the innermost $30$ km. The frequencies (blue) are extracted from the $(l=2:\hat{r})$ component in each model. Corrected frequencies (black) are also shown, where we have scaled them down to $\sim 58\%$ of their values in order to reproduce the best-fit mode frequency obtained in the Cowling approximation for the non-rotating model.} \label{fig:l2mode_f_vs_Omega}
\end{figure}

Finally, we present the frequency of the $l=2,n\gtrsim 2$ mode, as extracted from the $(l=2, \hat{r})$ component, across all models in our sequence of rotating stars in Fig.~\eqref{fig:l2mode_f_vs_Omega}. The ratio $\Omega/\Omega_K$ is computed at $40$ ms and averaged over the innermost $30$ km. The frequency of the mode decreases slightly from the $\Omega = 0.0 \rightarrow 0.5$ rad/s model (first two points), and then rises monotonically along the sequence from the $\Omega = 0.5$ rad/s model onwards. The increase may be surprising, since the central density $\rho_c$ at $40$ ms decreases monotonically from $3.94 \rightarrow 3.29 \times 10^{14}\: \mathrm{g}/\mathrm{cm}^3$ across the $\Omega=0.0\rightarrow 2.0$ rad/s models, and one expects mode frequencies to scale with $\sqrt{G \rho_c}$ (and therefore decrease)~\cite{friedman2013rotating}. The opposite trend we see may be due to an additional limitation of our \texttt{FLASH} implementation to accurately reproduce mode frequencies of stars with rapid rotation.

We also plot a corrected version of the mode frequency trend in Fig.~\eqref{fig:l2mode_f_vs_Omega} (black), where we have multiplied all the frequencies by a factor $\sim 0.58$, which is the correction required to scale the mode frequency in the non-rotating model to its best-fit Cowling value. We reiterate that the Cowling value is most likely an overestimate of the true frequency as well, as we saw in Sec.~\eqref{sec:CCSN_test_pert_schemes}. Together with the \emph{a priori} expectation that the mode frequency should be decreasing along a sequence of models with decreasing central density, one ought to regard the corrected curve in Fig.~\eqref{fig:l2mode_f_vs_Omega} as as upper bound on the true frequency for all models on the sequence.

\FloatBarrier

\chapter{Conclusion} \label{ch:CCSNconc}
In this part, we showed using a $20 M_\odot$ ZAMS model that detailed asteroseismology of the proto-neutron star at the centre of a rotating core-collapse supernovae is possible with both gravitational waves and neutrinos. These two messengers are complementary in that they carry information about certain linear modes of the core at different radii, with neutrinos probing the outer $60-80$ km and gravitational waves probing deeper in, $r \lesssim 30$ km. The mechanism by which the linear modes of the core imprint themselves on the neutrino lightcurves appears to be that the neutrino-emitting volume undergoes deformations in time according to the frequency and angular harmonics of the modes in the vicinity of the neutrinospheres. The dominant angular harmonics are then reflected in the emission pattern of neutrinos on the sky.

As we saw in Chapter~\eqref{ch:CCSNresults}, many other spectral peaks exist both in gravitational wave and neutrino luminosity spectrograms along our entire sequence of rotating models. In Appendix~\eqref{app:CCSNapp} we show numerous additional modes matched between perturbation theory and our non-rotating model. The many modes that are active offer to explain the additional spectral peaks we observe along our rotating sequence. 

We followed a strategy of mode function matching, rather than mode frequency matching as in~\cite{torres2017towards,morozova2018gravitational, torres2018towards}. This we believe is a more robust approach that is less susceptible to mode misidentification, especially given the approximations employed in both simulations and perturbative schemes. It is somewhat laborious work, but clearly a wealth of information can be extracted from any given model. A fully automated suite of algorithms would be desirable to hasten analysis and increase reproducibility.

Further investigation is necessary to fully understand the limitations of the \texttt{FLASH} implementation used in this work. We tested the accuracy of the implementation in reproducing modes of a non-rotating stable TOV star in Sec.~\eqref{sec:FLASHTOV_test} and found a systematic overestimation of the mode frequencies (except for the $F$-mode). A question we were not able to answer is whether the modes (but particularly the ${}^2 f$-mode) are pure spherical harmonics. This can be answered by repeated rounds of mode recycling, which we have not done. Furthermore, the question of whether the \texttt{FLASH} implementation provides even lower quality of representation of mode frequencies for rapidly rotating stars we left completely open. However, we have some indication that the implementation performs poorly in this regime based on the trend of increasing frequency observed in Fig.~\eqref{fig:l2mode_f_vs_Omega}, which is the opposite of what one would expect based on the decreasing central density of the proto-neutron star along the sequence of rotating models. Confirmation of this would require repeating a study such as~\cite{dimmelmeier2006non} in our implementation, where the modes of stable rotating stars are extracted from nonlinear simulations with the aid of mode recycling. Confirming this would also bolster our tracking of the mode across the model sequence, since it would explain why the frequency would rise and by how much.

\begin{appendices}

\chapter{Other modes in $\Omega = 0.0$ rad/s case} \label{app:CCSNapp}
For the interested reader, in this appendix we show comparisons between the velocity data and perturbative mode functions for many other frequency bands we were able to identify. The purpose of this deluge of figures is to convey the following:
\begin{itemize}
\item Identification of modes via a matching procedure between perturbative mode functions and velocity data from simulations is \emph{robust and repeatable}, across a wide range of frequency spanning at least $\sim 350-1500$ Hz.
\item The quality of a given mode match depends on its level of excitation, and the below figures have instances of very exceptional matches and rather poor ones. This is expected, since noise or nonlinear motion in the simulations would have a certain amplitude which the mode would need to rise above in order to be clearly identified. Thus the figures below should convey the range of quality of mode matches that can occur in a CCSN simulation.
\item The spectrograms in Fig.~\eqref{fig:CCSNmoney} have many additional features which we have not pinned on any mode. However, given our results, they very likely \emph{are} due to other modes. The figures below show a wide range of possible candidates. In future work, we will repeat the analysis of Chapter~\eqref{ch:CCSNresults} on the modes shown in the figures below, for the purpose of even more detailed asteroseismology.
\item For the $l=2$ mode we focused on in Chapter~\eqref{ch:CCSNresults}, we were not able to characterize it as a $p$- or $g$-mode (because its node count $n\gtrsim 2$ occurred so close to the boundary case). But in some instances below we can clearly characterize modes as $p$-modes., particular towards the higher end of the frequency range we consider.
\end{itemize}
The figures below are of the same format as those presented (with more explanation) in Chapter~\eqref{ch:CCSNresults}, so we refer to that chapter as a prerequisite for understanding the ones below.

\subsubsection{$\sim 350$ Hz band:}

\begin{figure}[htbp]
\centering
\hbox{\hspace{-1.6cm}\includegraphics[width=1.2\textwidth]{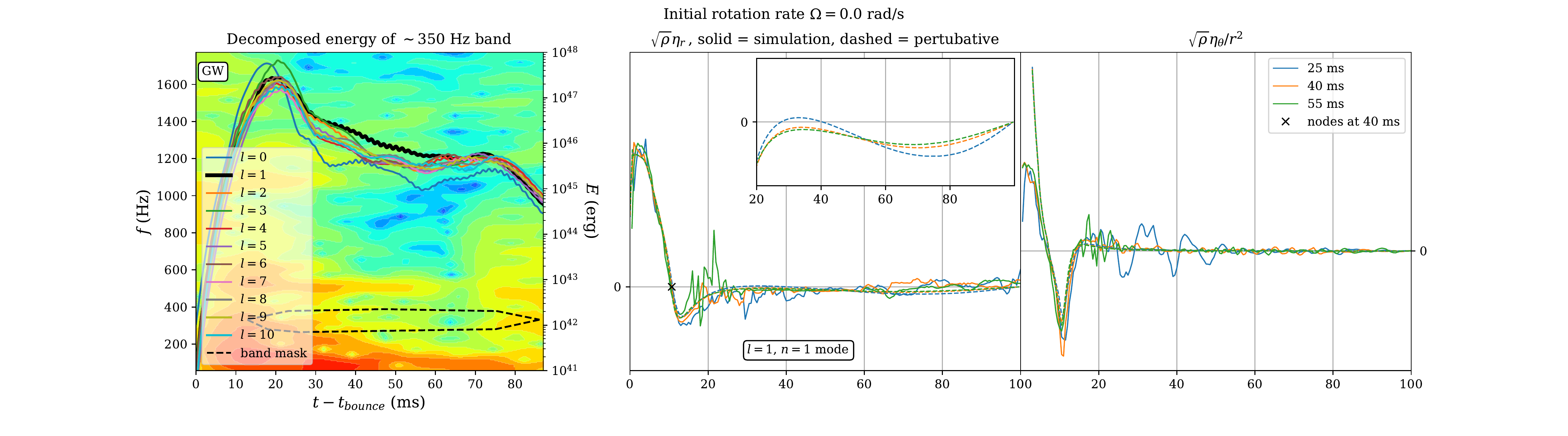}}
\caption{Analysis in the $\sim 350$ Hz band mask. The most dominant energy is $l=1$, albeit slightly, and as a result the simulation snapshots (Centre \& Right) are rather noisy. Nonetheless, meaningful agreement is evident with a perturbative $l=1,n=1$ mode. } \label{fig:350Hz_top1modes}
\end{figure}
\FloatBarrier

\begin{figure}[htbp]
\centering
\includegraphics[width=1\textwidth]{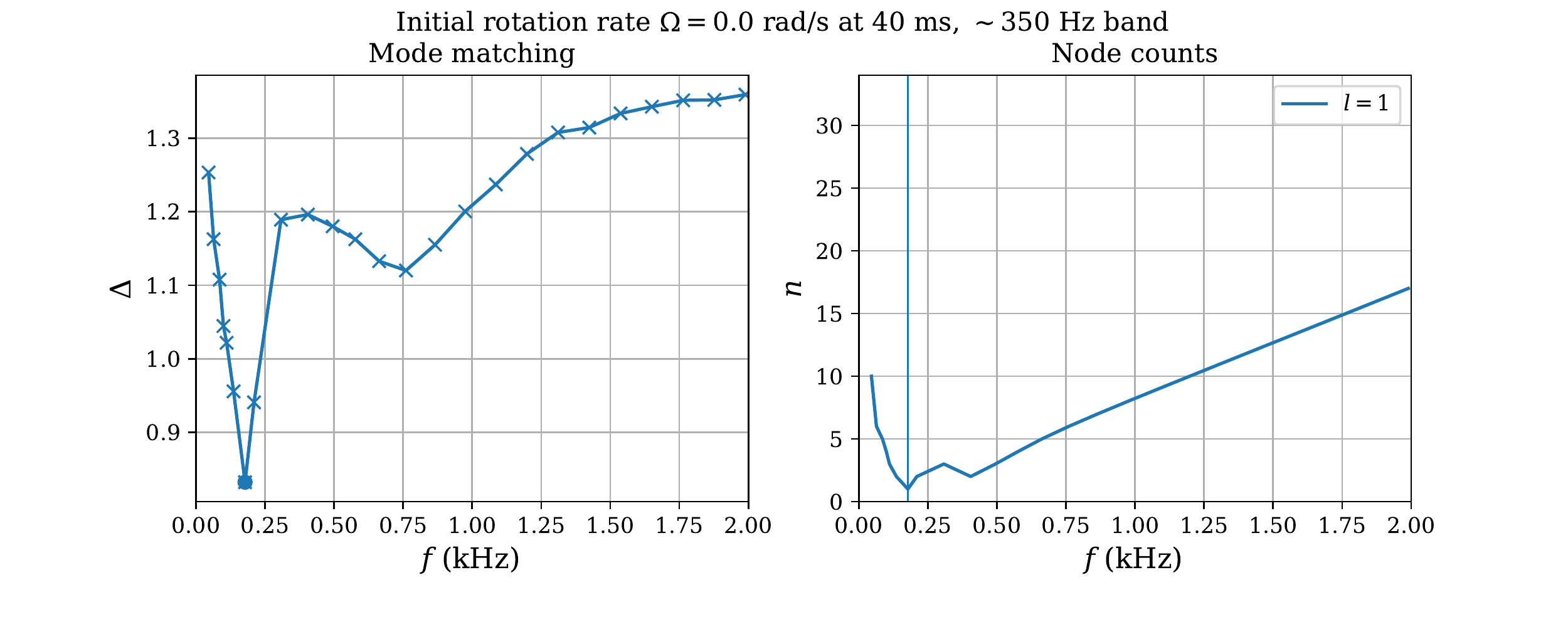}
\caption{Mode-matching in the $\sim 350$ Hz band. The best-fit modes are indicated} \label{fig:350Hz_modematch}
\end{figure}
\FloatBarrier

\subsubsection{$\sim 650$ Hz band:}

\begin{figure}[htbp]
\centering
\hbox{\hspace{-1.6cm}\includegraphics[width=1.2\textwidth]{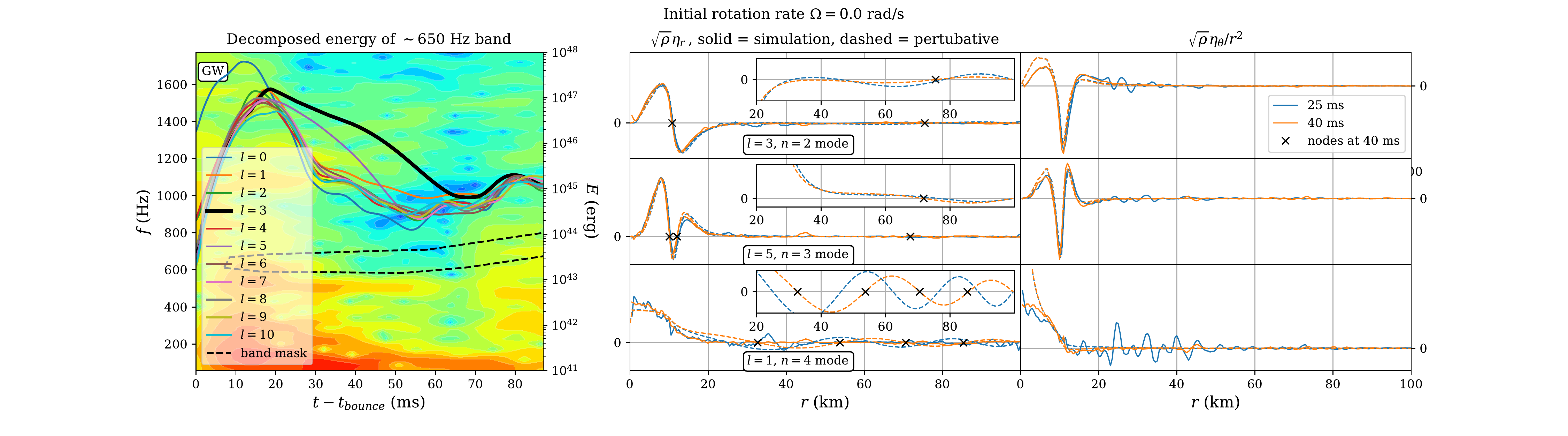}}
\caption{Analysis in the $\sim 650$ Hz band mask. The $l=1$ mode is an example of a poor fit, corresponding to its weak excitation.} \label{fig:650Hz_top3modes}
\end{figure}
\FloatBarrier

\begin{figure}[htbp]
\centering
\includegraphics[width=1\textwidth]{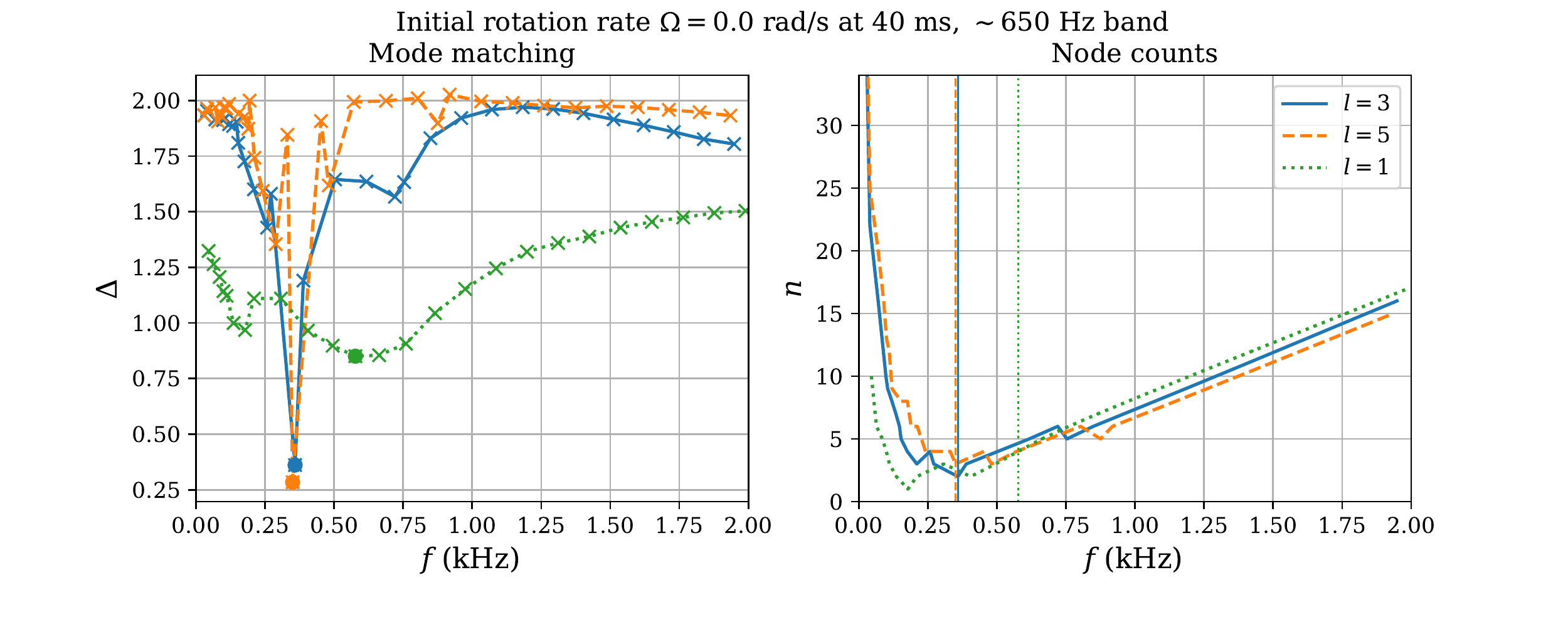}
\caption{Mode-matching in the $\sim 650$ Hz band. The $l=1$ mode is an example of a poor fit, corresponding to its weak excitation.} \label{fig:650Hz_modematch}
\end{figure}
\FloatBarrier

\subsubsection{$\sim 750$ Hz band:}

\begin{figure}[htbp]
\centering
\hbox{\hspace{-1.6cm}\includegraphics[width=1.2\textwidth]{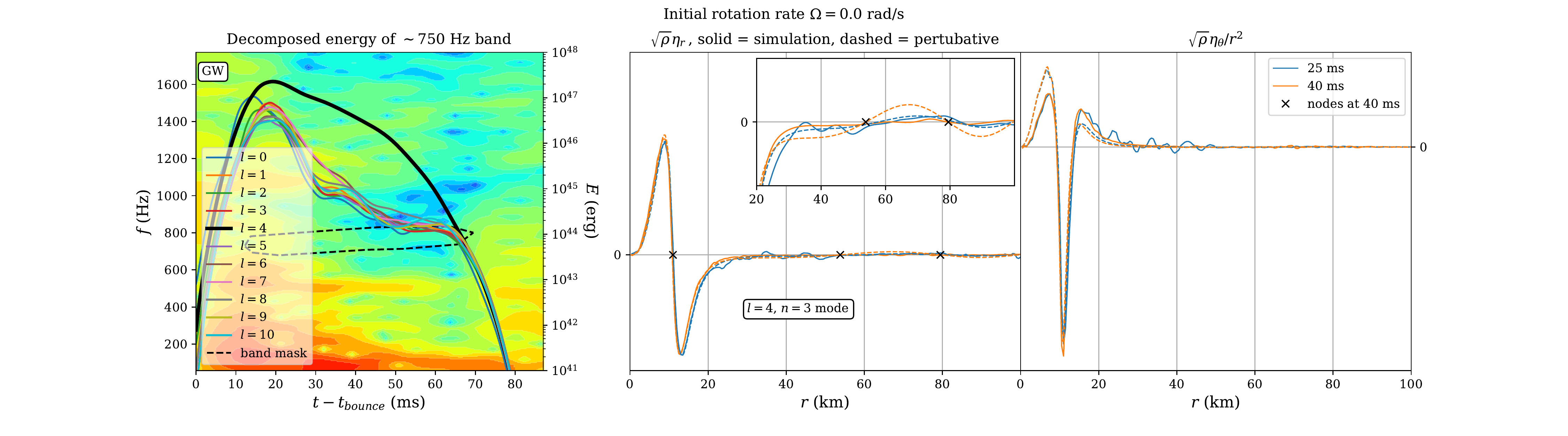}}
\caption{Analysis in the $\sim 750$ Hz band mask. This band has a very clean excitation of an $l=4$ mode, with meaningful matching occuring even on the tail of the mode (Centre, inset).} \label{fig:750Hz_top1modes}
\end{figure}
\FloatBarrier

\begin{figure}[htbp]
\centering
\includegraphics[width=1\textwidth]{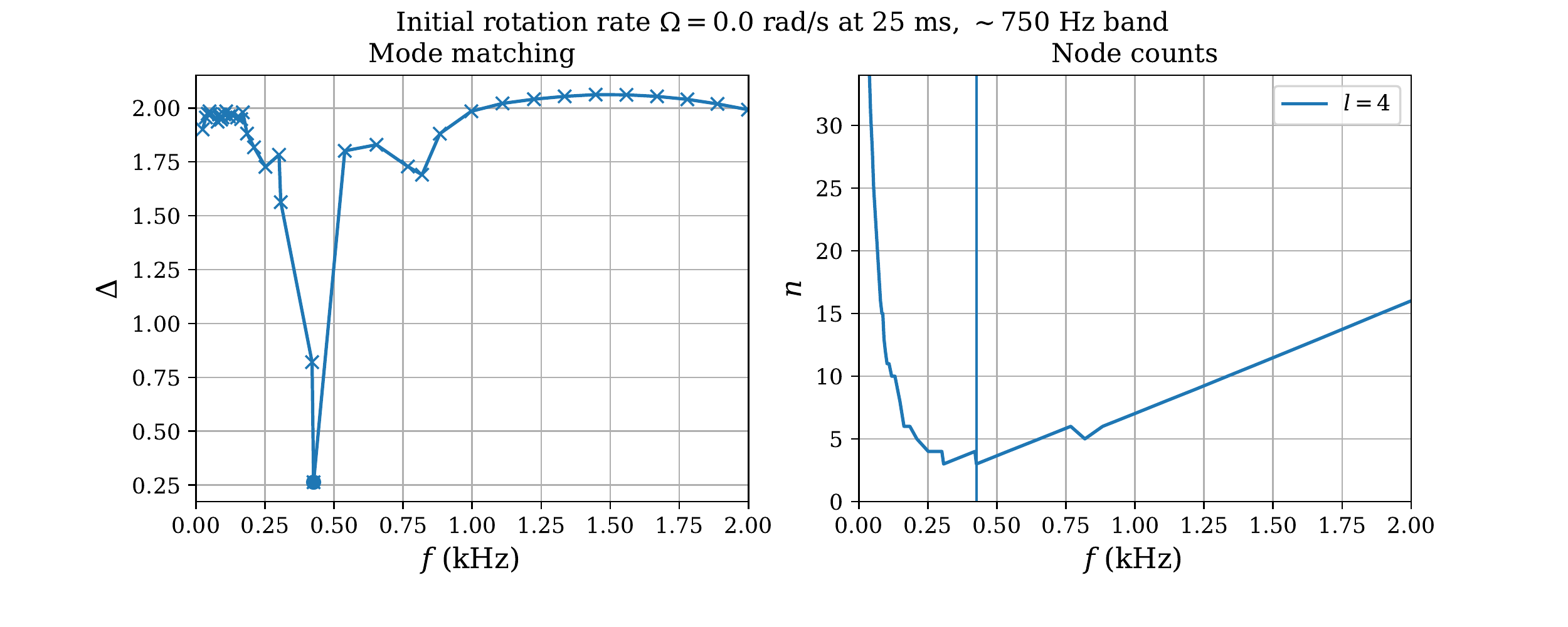}
\caption{Mode-matching in the $\sim 750$ Hz band.} \label{fig:750Hz_modematch}
\end{figure}
\FloatBarrier

\subsubsection{$\sim 850$ Hz band:}

\begin{figure}[htbp]
\centering
\hbox{\hspace{-1.6cm}\includegraphics[width=1.2\textwidth]{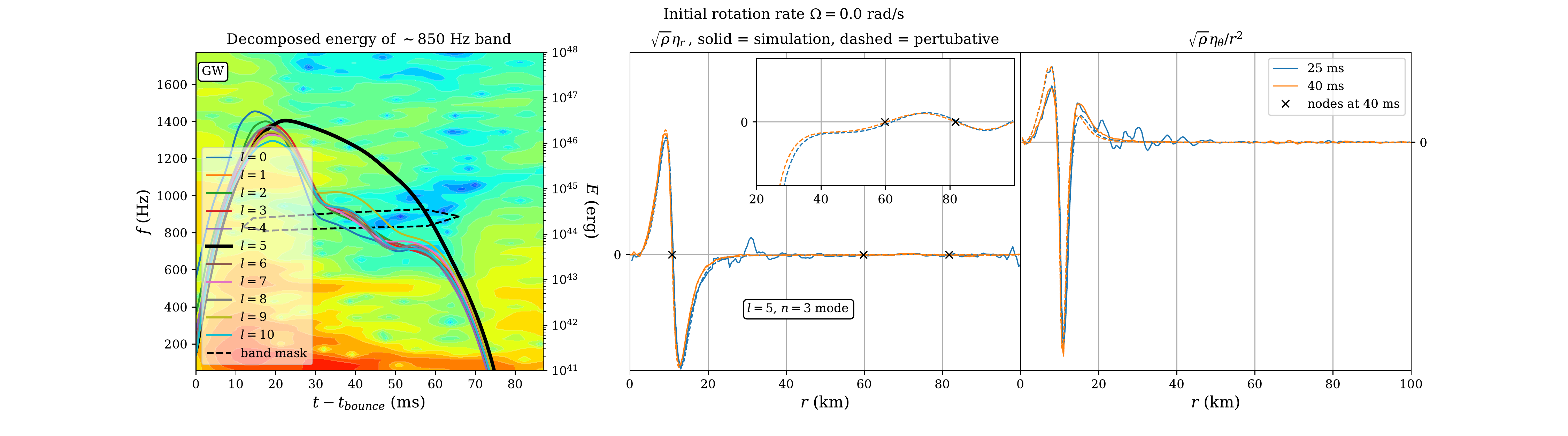}}
\caption{Analysis in the $\sim 850$ Hz band mask. An $l=5$ mode is excited here.} \label{fig:850Hz_top1modes}
\end{figure}
\FloatBarrier

\begin{figure}[htbp]
\centering
\includegraphics[width=1\textwidth]{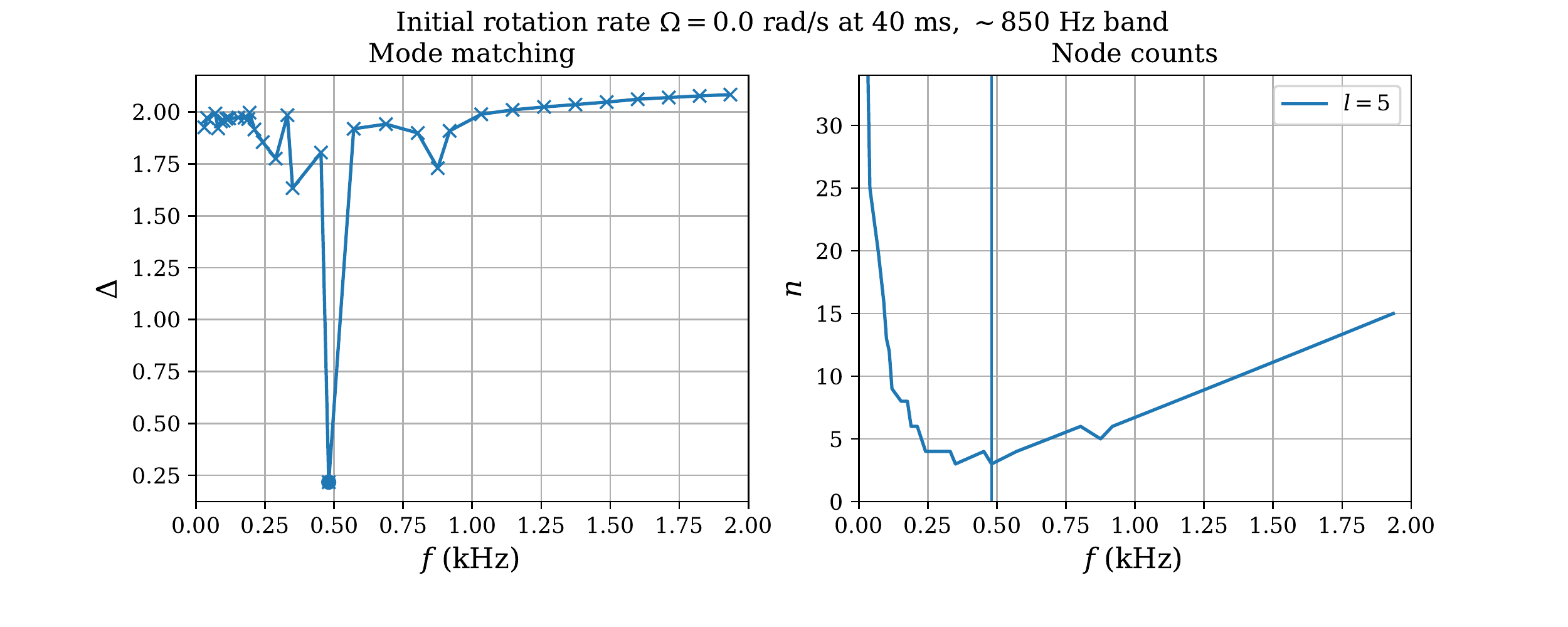}
\caption{Mode-matching in the $\sim 850$ Hz band.} \label{fig:850Hz_modematch}
\end{figure}
\FloatBarrier

\subsubsection{$\sim 900$ Hz band:}

\begin{figure}[htbp]
\centering
\hbox{\hspace{-1.6cm}\includegraphics[width=1.2\textwidth]{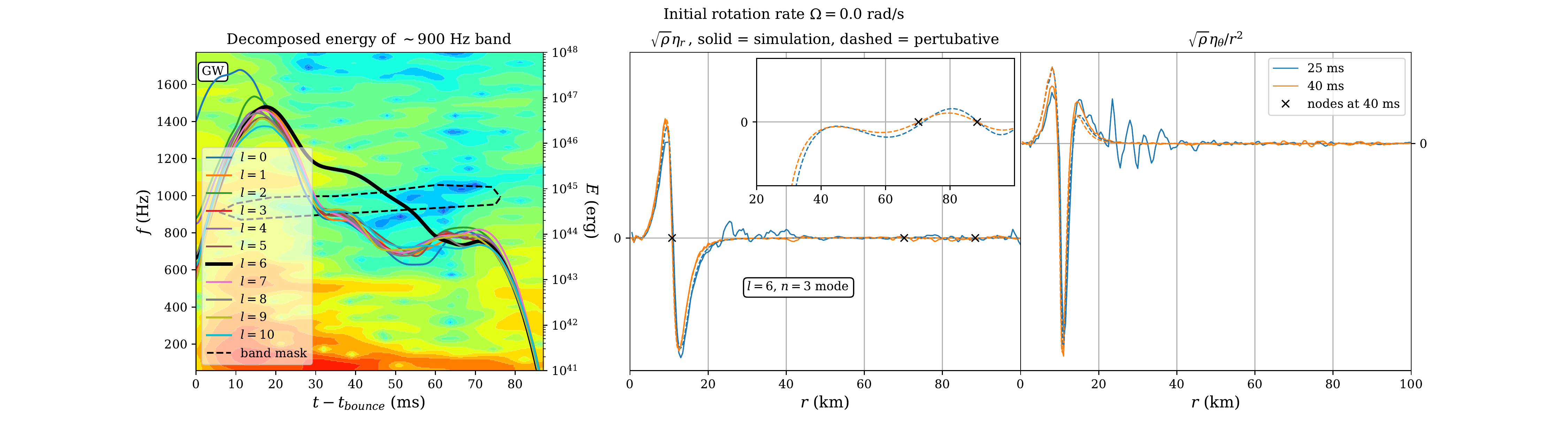}}
\caption{Analysis in the $\sim 900$ Hz band mask. An $l=6$ mode is excited here.} \label{fig:900Hz_top1modes}
\end{figure}
\FloatBarrier

\begin{figure}[htbp]
\centering
\includegraphics[width=1\textwidth]{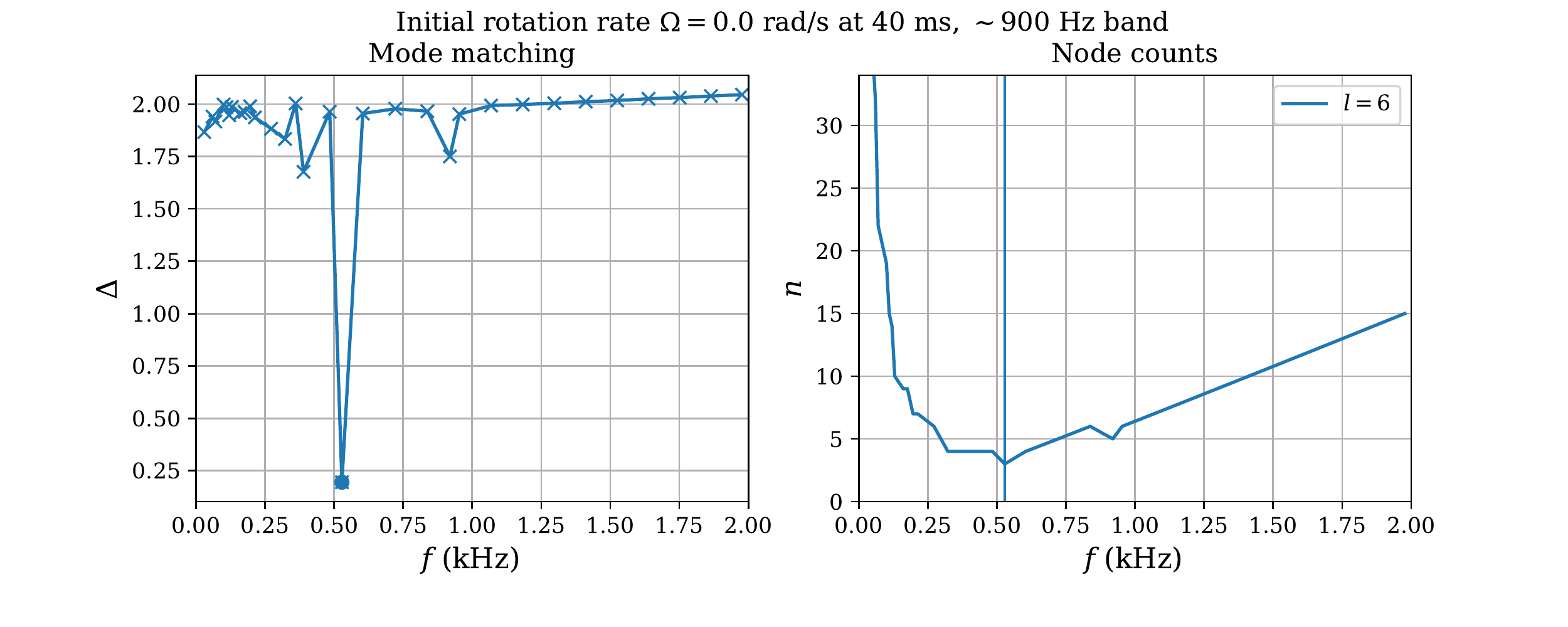}
\caption{Mode-matching in the $\sim 900$ Hz band.} \label{fig:900Hz_modematch}
\end{figure}
\FloatBarrier

\subsubsection{$\sim 1100$ Hz band:}

\begin{figure}[htbp]
\centering
\hbox{\hspace{-1.6cm}\includegraphics[width=1.2\textwidth]{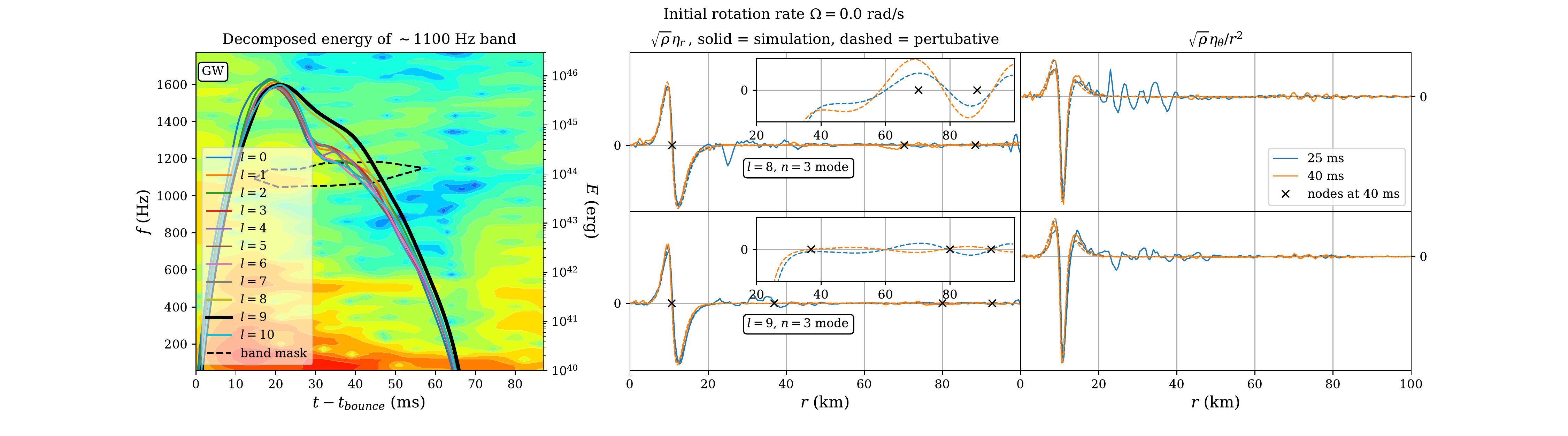}}
\caption{Analysis in the $\sim 1100$ Hz band mask. $l=8$ and $l=9$ modes are excited.} \label{fig:1100Hz_top2modes}
\end{figure}
\FloatBarrier

\begin{figure}[htbp]
\centering
\includegraphics[width=1\textwidth]{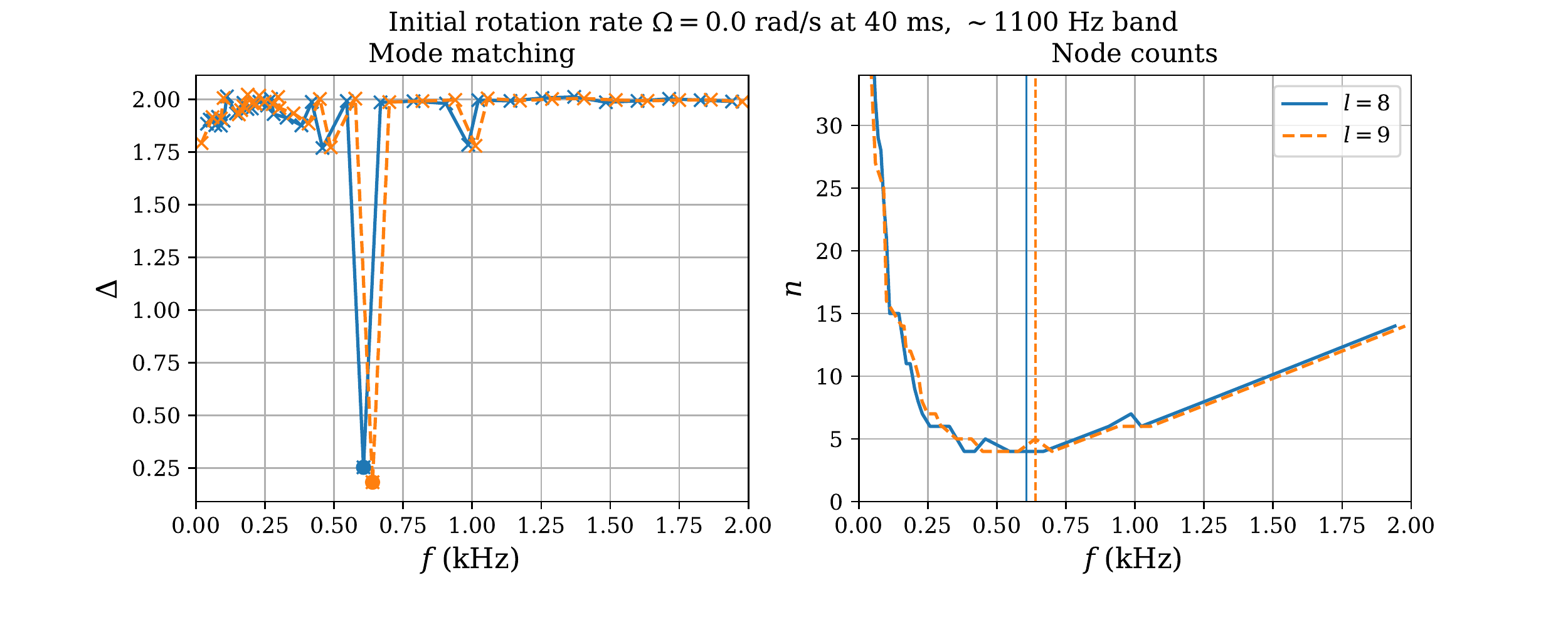}
\caption{Mode-matching in the $\sim 1100$ Hz band.} \label{fig:1100Hz_modematch}
\end{figure}
\FloatBarrier

\subsubsection{$\sim 1200$ Hz band:}

\begin{figure}[htbp]
\centering
\hbox{\hspace{-1.6cm}\includegraphics[width=1.2\textwidth]{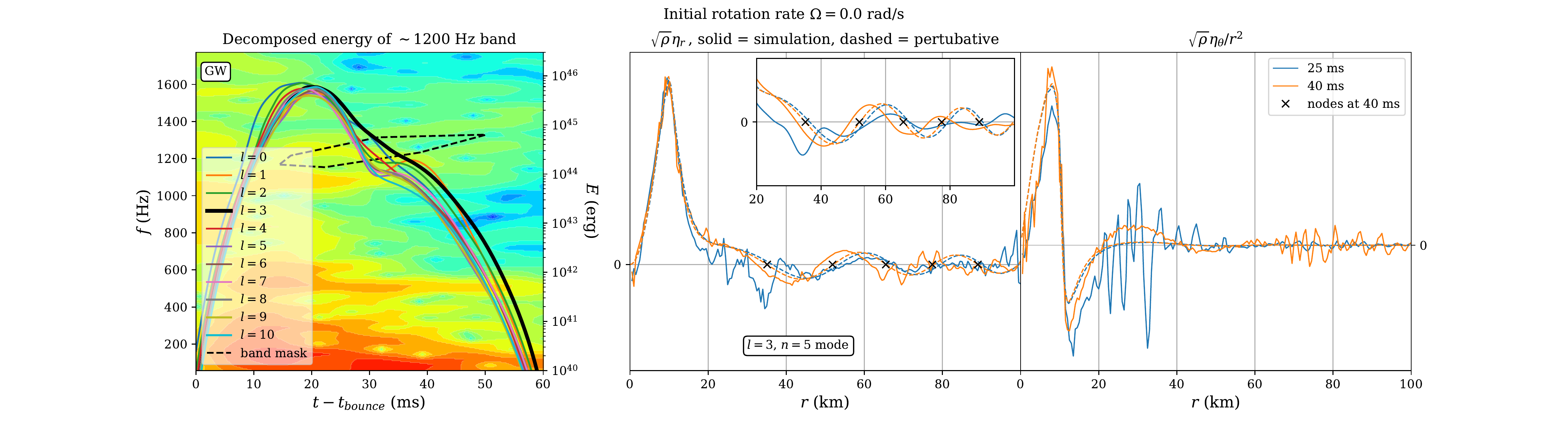}}
\caption{Analysis in the $\sim 1200$ Hz band mask. An $l=3$ mode is excited.} \label{fig:1200Hz_top2modes}
\end{figure}
\FloatBarrier

\begin{figure}[htbp]
\centering
\includegraphics[width=1\textwidth]{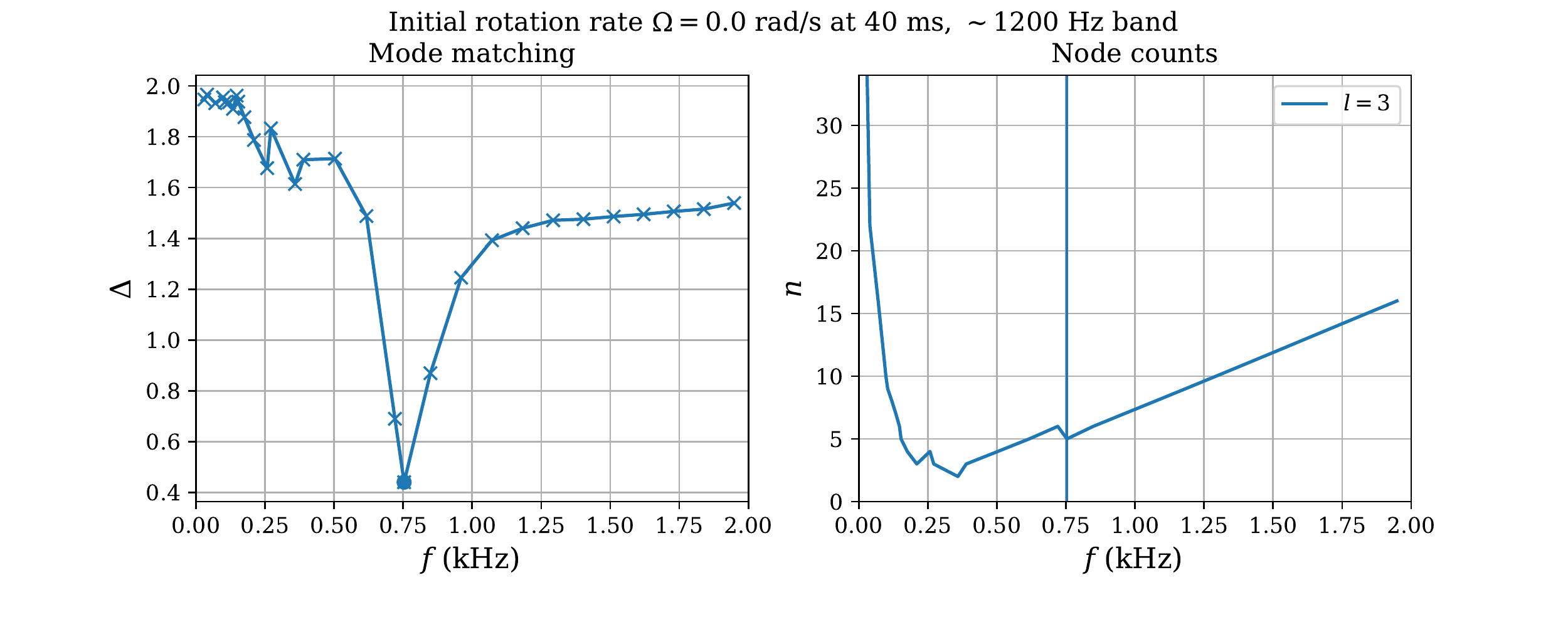}
\caption{Mode-matching in the $\sim 1200$ Hz band. The node count occurs on the $p$-mode side of the curves (Right), i.e.~to the right of the minimum.} \label{fig:1200Hz_modematch}
\end{figure}
\FloatBarrier

\subsubsection{$\sim 1400$ Hz band:}

\begin{figure}[htbp]
\centering
\hbox{\hspace{-1.6cm}\includegraphics[width=1.2\textwidth]{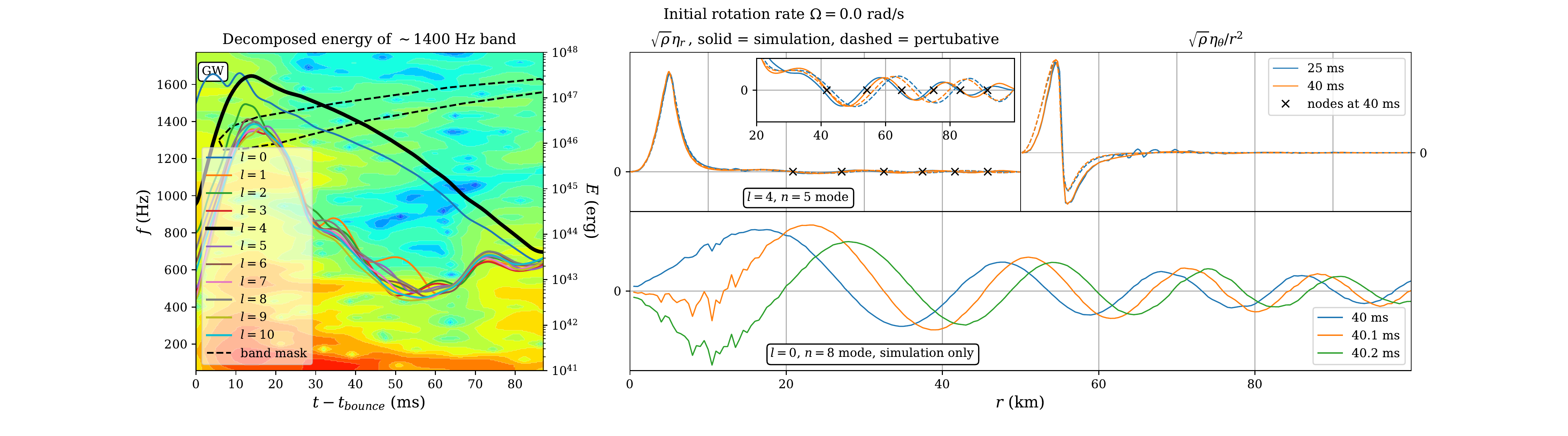}}
\caption{Analysis in the $\sim 1400$ Hz band mask. Both $l=4$ and $l=0$ modes are cleanly excited, with node counting possible in the $l=4$ case without reference to the perturbative mode function (which matches somewhat poorly at large radii). We do not have $l=0$ perturbative mode functions. We display snapshots at $t=\lbrace 40,40.1,40.2\rbrace$ ms in order to convey that the $l=0$ mode has the form of an outward travelling mode.} \label{fig:1400Hz_top2modes}
\end{figure}
\FloatBarrier

\begin{figure}[htbp]
\centering
\includegraphics[width=1\textwidth]{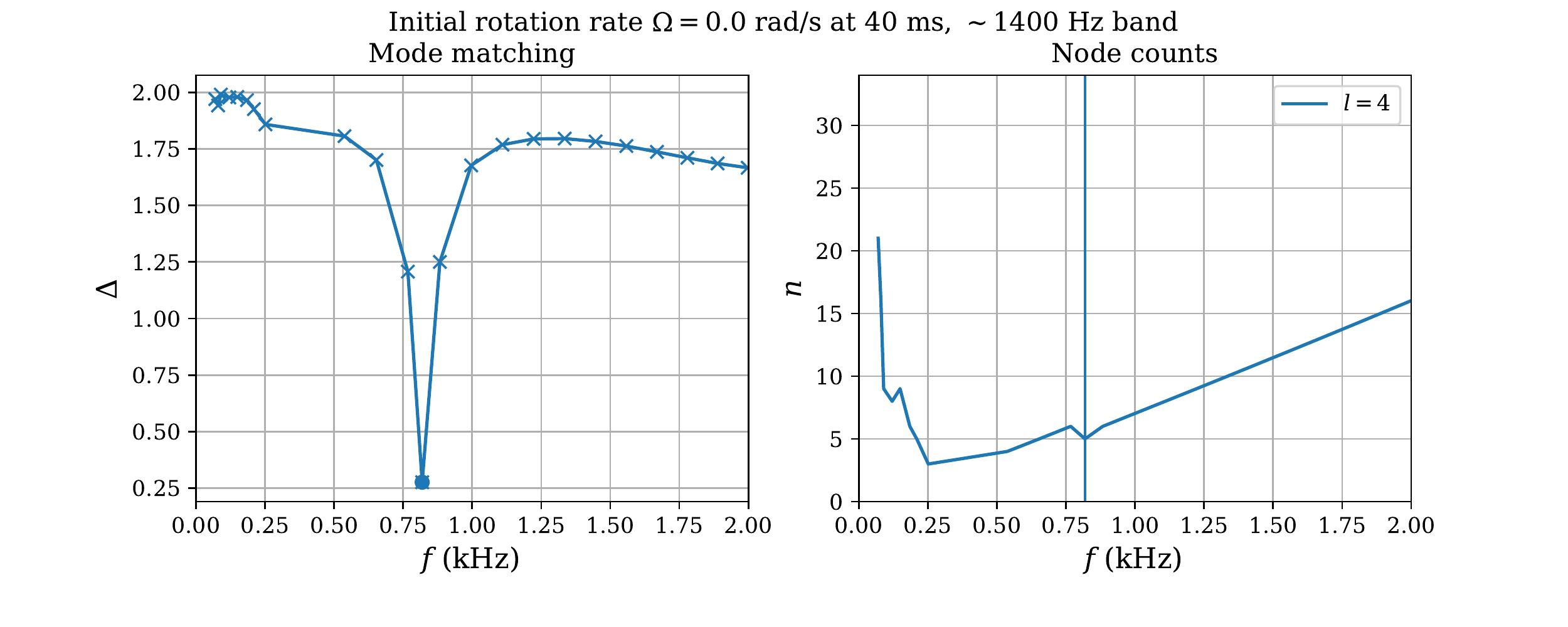}
\caption{Mode-matching in the $\sim 1400$ Hz band. The node count occurs on the $p$-mode side of the curves (Right), i.e.~to the right of the minimum.} \label{fig:1400Hz_modematch}
\end{figure}
\FloatBarrier

\subsubsection{$\sim 1500$ Hz band:}

\begin{figure}[htbp]
\centering
\hbox{\hspace{-1.6cm}\includegraphics[width=1.2\textwidth]{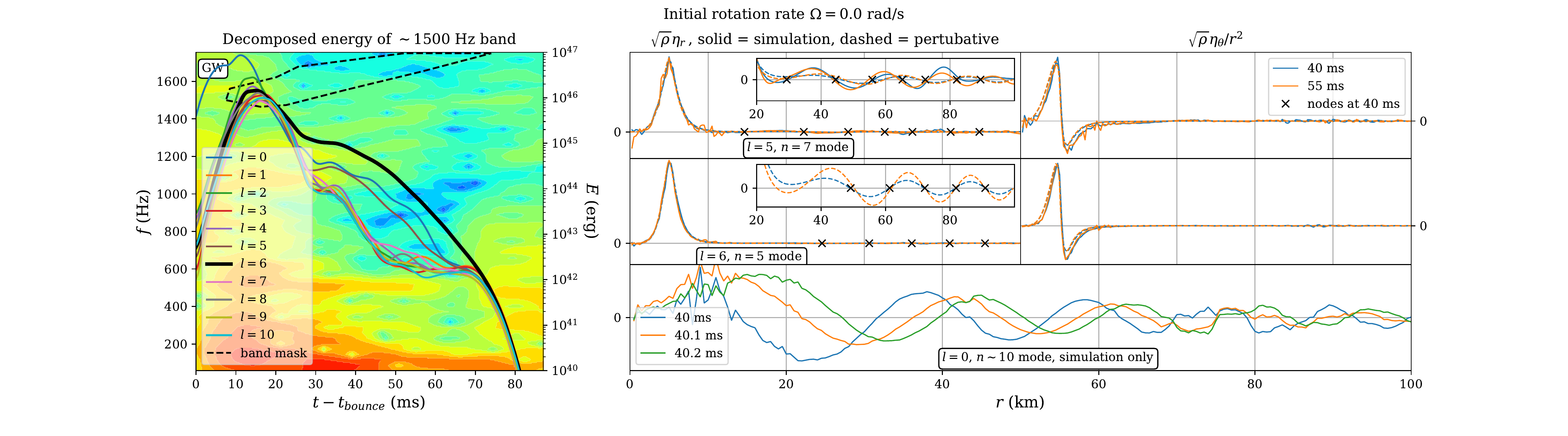}}
\caption{Analysis in the $\sim 1500$ Hz band mask. Similar to the previous mask, we have excitation of high-$n$ modes, including an outgoing $l=0$ travelling mode.} \label{fig:1500Hz_top3modes}
\end{figure}
\FloatBarrier

\begin{figure}[htbp]
\centering
\includegraphics[width=1\textwidth]{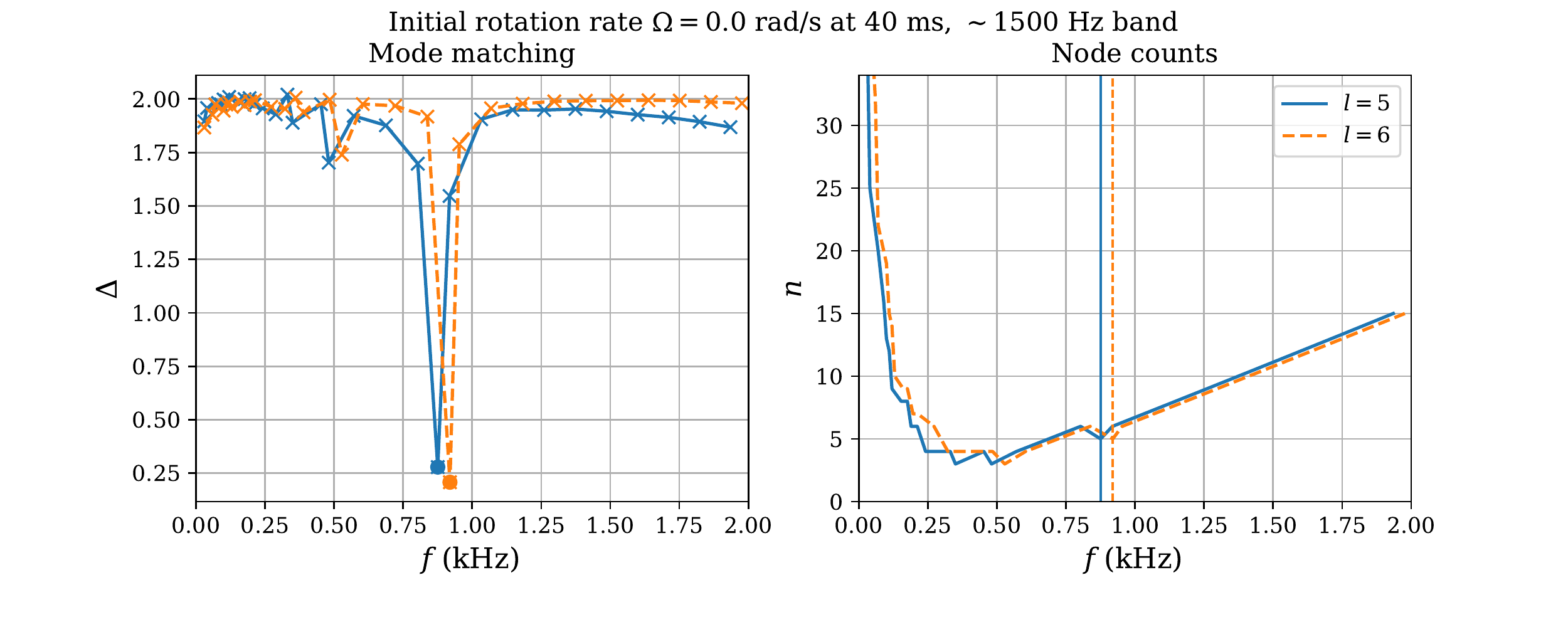}
\caption{Mode-matching in the $\sim 1500$ Hz band. The node counts are clearly on the $p$-mode side of the curves (Right), i.e.~to the right of the minima.} \label{fig:1500Hz_modematch}
\end{figure}
\FloatBarrier
\end{appendices}

\renewcommand\bibname{References}

\bibliographystyle{apsrev4-1}
\bibliography{fluidbib}

\end{document}